\documentclass{amsart}
\usepackage{graphicx}
\usepackage{amssymb}
\usepackage{amsmath}
\usepackage{amscd}
\usepackage{thmdefs}
\usepackage{float}
\usepackage{subfigure}
\usepackage{subfig}

\theoremstyle{definition}
\theoremstyle{remark}
\numberwithin{equation}{section}

\input{tcilatex}

\begin{document}
\title[Exponential Levy Processes]{Asymptotics for Exponential Levy
Processes and their Volatility Smile:\ Survey and New Results}
\author{Leif Andersen}
\address{Bank of America Merrill Lynch}
\email{leif.andersen@baml.com}
\author{Alexander Lipton}
\address{Bank of America Merrill Lynch}
\email{alex.lipton@baml.com}
\date{June 1st, 2012}
\thanks{This paper is in final form and no version of it will be submitted
for publication elsewhere.}
\maketitle

\begin{abstract}
Exponential L\'{e}vy processes can be used to model the evolution of various
financial variables such as FX rates, stock prices, etc. Considerable
efforts have been devoted to pricing derivatives written on underliers
governed by such processes, and the corresponding implied volatility
surfaces have been analyzed in some detail. In the non-asymptotic regimes,
option prices are described by the Lewis-Lipton formula which allows one to
represent them as Fourier integrals; the prices can be trivially expressed
in terms of their implied volatility. Recently, attempts at calculating the
asymptotic limits of the implied volatility have yielded several expressions
for the short-time, long-time, and wing asymptotics. In order to study the
volatility surface in required detail, in this paper we use the FX
conventions and describe the implied volatility as a function of the
Black-Scholes delta. Surprisingly, this convention is closely related to the
resolution of singularities frequently used in algebraic geometry. In this
framework, we survey the literature, reformulate some known facts regarding
the asymptotic behavior of the implied volatility, and present several new
results. We emphasize the role of fractional differentiation in studying the
tempered stable exponential Levy processes and derive novel numerical
methods based on judicial finite-difference approximations for fractional
derivatives. We also briefly demonstrate how to extend our results in order
to study important cases of local and stochastic volatility models, whose
close relation to the L\'{e}vy process based models is particularly clear
when the Lewis-Lipton formula is used. Our main conclusion is that studying
asymptotic properties of the implied volatility, while theoretically
exciting, is not always practically useful because the domain of validity of
many asymptotic expressions is small.
\end{abstract}

\tableofcontents

\section{Introduction\label{Intro}}

In the classical Black-Scholes-Merton (BSM) European option pricing model
(see \cite{bs} and \cite{merton-1}), asset processes are assumed to be
strictly diffusive in nature and characterized by a single (log-normal)
volatility $\sigma $. In practice, no actual option market conforms with
this framework, so to make the BSM formula work, practitioners are forced to
make the volatility argument in this formula depend on option maturity $%
(\tau )$ and strike $(K)$. Indeed, it is common practice in virtually all
option markets to maintain a strike- and maturity-dependent \textit{implied
volatility surface}, $\sigma _{imp}(t;\tau ,K)$, such that a call option on
an asset $F$ paying $(F(T)-K)^{+}$ at expiration time $T\geq t$ has a time $%
t $ undiscounted price $C(t;\tau ,K)$ given by 
\begin{equation}
\frac{C(t;\tau ,K)}{F(t)}=\Phi (d_{+})-e^{k}\Phi (d_{-}),
\label{eq:BS_formula}
\end{equation}%
where $k\triangleq \ln (K/F(t))$, $\tau \triangleq T-t$, $\Phi (\cdot )$ is
the cumulative Gaussian distribution function, and 
\begin{equation}
d_{\pm }=\frac{-k\pm \frac{1}{2}\sigma _{imp}(t;\tau ,K)^{2}\tau }{\sigma
_{imp}(t;\tau ,K)\sqrt{\tau }}.  \label{eq:d_pm}
\end{equation}%
Here and below, as usual,%
\begin{equation*}
\left( x\right) ^{\pm }=\pm \max \left( \pm x,0\right) .
\end{equation*}%
Note that our version of the BSM formula assumes that the asset price $F$ is
a risk-neutral martingale, an assumption that is easily relaxed or, in any
case, justified if we consider $F$ a forward process. At time $t$, the $t$%
-observable function $\sigma _{imp}(t;\tau ,K)$ can be implied (with
assistance of interpolation and extrapolation techniques) from quoted call
option prices at multiple maturities and strikes. For later use, notice that
the term 
\begin{equation}
\Delta =\Phi (d_{+}),  \label{eq:delta}
\end{equation}%
is known as the option's (forward) \textit{delta}\emph{.}

Below we often use the time value of a call option defined as follows%
\begin{equation}
\delta C\left( t;\tau ,K\right) =C(t;\tau ,K)-\left( F\left( t\right)
-K\right) ^{+},  \label{eq:time_value_0}
\end{equation}%
or, equivalently,%
\begin{equation}
\frac{\delta C(t;\tau ,K)}{F(t)}=\Phi (d_{+})-e^{k}\Phi (d_{-})-\left(
1-e^{k}\right) ^{+}.  \label{eq:time_value_1}
\end{equation}

For future reference, it is convenient to introduce the following
non-dimensional function $\mathsf{C}^{BS}\left( v,k\right) $ of the
annualized variance $v$ and log-strike $k$:%
\begin{equation}
\mathsf{C}^{BS}\left( v,k\right) =\Phi \left( \mathsf{d}_{+}\right)
-e^{k}\Phi \left( \mathsf{d}_{-}\right) ,  \label{eq: BS non-dim}
\end{equation}%
where $0<v<\infty $, $-\infty <k<\infty $, and,%
\begin{equation*}
\mathsf{d}_{\pm }=\frac{-k\pm \frac{1}{2}v}{\sqrt{v}}.
\end{equation*}%
In the limiting cases we have%
\begin{equation*}
\mathsf{C}^{BS}\left( 0,k\right) =\left( 1-e^{k}\right) ^{+},\ \ \ \ \ 
\mathsf{C}^{BS}\left( \infty ,k\right) =1.
\end{equation*}%
It is clear that%
\begin{equation*}
\frac{C(t;\tau ,K)}{F(t)}=\mathsf{C}^{BS}\left( \sigma _{imp}(t;\tau
,K)^{2}\tau ,\ln \left( \frac{K}{F(t)}\right) \right) .
\end{equation*}

In most real markets, the market implied volatility function $\sigma
_{imp}(t;\tau ,K)$ differs very significantly from a constant, and may have
considerable slope and convexity as a function of $K$ for both very small
and very large values of $\tau $. To understand and explain this phenomenon,
several alternative models have been proposed in the literature. see, e.g., 
\cite{merton-2}, \cite{dupire}, \cite{heston}, \cite{bates}, \cite{and-and}, 
\cite{lipton-2} among others. Broadly speaking, the available approaches can
be categorized as follows:

\begin{itemize}
\item Local volatility (LV) models, where the instantaneous volatility $%
\sigma $ of $F$ is a deterministic function of time and $F$.

\item Stochastic volatility (SV) models, where $\sigma $ is a random
variable, possibly correlated with $F$.

\item Jump models, where the process for $F$ is assumed to be a purely
discontinuous jump-process.

\item Universal volatility (UV) models, where local volatility, stochastic
volatility, and jump processes are combined.
\end{itemize}

Full-blown UV models are rarely used in practice, and instead markets tend
to converge around a simpler model that ultimately becomes a \textit{de facto%
} market standard. For instance, LV models are popular in the field of
equity derivatives, and jump-free combinations of LV and SV (known as LSV,
or local-stochastic volatility, models) are dominant in the FX options
arena. Such usage of simplified models is, however, rarely motivated by
empirical facts, but instead are done for practical reasons in recognition
of the fact that models that combine LV, SV, and jumps are often highly
complicated to calibrate and implement. One particularly thorny issue is the
question of how precisely to mix LV, SV, and jumps, a problem that is made
especially vexing by the fact that even a simple model like LV can, on its
own, match essentially any arbitrage-free option volatility surface (see,
e.g., \cite{dupire}).

In UV models, there are several potential strategies for attacking the
\textquotedblleft mixing\textquotedblright\ problem. As LV, SV, and jumps
give rise to different dynamic evolution of the volatility surface over
time, one idea is to embed observations of the volatility dynamics into the
model calibration problem (see, e.g., \cite{and-pit} for a discussion). A
closely related approach is to incorporate market-observable exotic option
prices that are sensitive to the evolution of volatility, e.g. variance
options, barrier options, and similar; this approach is commonly used in FX
markets to mix LV and SV into an LSV model. Yet another idea is to examine
various extremes of the volatility surface (short maturity, long maturity,
large strikes, small strikes) and attempt to understand which type of model
feature (SV/LV/jumps) will control the asymptotic model behavior most
strongly and most \textquotedblleft naturally\textquotedblright . This
information, in turn, could then be used to inspire the model building
process. To give an example, consider that the convexity of $\sigma
_{imp}(t;\tau ,K)$ around $K=F(t)$ can often be observed to be substantially
higher for small option maturities than for larger ones$.$ A pure LV model
would model this by letting the convexity (in $F$) of the local volatility
function decay rapidly as a function of time. Such a model would, however,
be highly non-stationary, an undesirable model feature that can be avoided
by introducing jumps or (mean-reverting) stochastic volatility.

The idea of considering volatility function asymptotics is particularly
attractive from an analytical perspective, as one can often work out simple
closed-form expressions for these asymptotics, even for complex models. For
SV, LV, and LSV models the procedure for generating such asymptotics is
well-understood (see e.g. \cite{henry} for a survey of recent results), and
the range of validity for the resulting expansions has been examined
closely, generally with fairly satisfying results. Recently, similar
attempts have been undertaken for jump processes, especially for processes
in the so-called \textit{exponential L\'{e}vy class}. More precisely, these
models write 
\begin{equation}
F(t)=F(0)e^{X(t)},  \label{eq:S}
\end{equation}%
where $X(t)$ is a L\'{e}vy process. In this setup, particular emphasis has
been put on the small-time asymptotic of $C(t;\tau ,K)$ for $\tau
\rightarrow 0$ and $K$ fixed at some level different from $F(t)$, see for
instance \cite{levendorskii-1}, \cite{roper}, \cite{lopez-forde}, \cite%
{lopez et al} among many others. A typical result in this area of research
is that 
\begin{equation}
\delta C\left( t;\tau ,K\right) \sim \tau ,\quad K\neq F(t),
\label{eq:nonatmas}
\end{equation}%
where $\sim $ indicates the leading order term as $\tau \rightarrow 0$. For
this relation to be true, the corresponding implied volatility must explode
in the short-time limit. This line of research is closely related in
studying small time asymptotics for the densities of L\'{e}vy processes,
see, e.g., \cite{leandre}, \cite{picard}, \cite{ruschendorf}, \cite{lopez-1}%
, \cite{lopez-2}, \cite{marchal}, \cite{bentata}. Small-time results for
at-the-money (ATM) options where $K=F(t)$ are scarcer, but in \cite%
{carr-wu-1} Tanaka's formula is used to demonstrate that 
\begin{equation}
C\left( t;\tau ,F(t)\right) \sim \left\{ 
\begin{array}{cc}
\tau , & \sigma =0, \\ 
\sqrt{\tau }, & \sigma \neq 0,%
\end{array}%
\right.  \label{eq:atmas}
\end{equation}%
when $X$ is a mixed jump-diffusion consisting of a Brownian motion with
volatility $\sigma $ and a finite variation\footnote{%
See Section \ref{ELPs} for a definition of various attributes of Levy
processes.} L\'{e}vy jump process. \cite{tankov} further elaborates on this
result in implied volatility space, and demonstrates that for a finite
variation L\'{e}vy process (necessarily without a diffusion component) 
\begin{equation}
\sigma _{imp}\left( t;\tau ,F(t)\right) \sim \sqrt{2\pi \tau }\max \left\{
\int_{\mathbb{R}}\left( e^{x}-1\right) ^{+}\nu \left( dx\right) ,\int_{%
\mathbb{R}}\left( 1-e^{x}\right) ^{+}\nu \left( dx\right) \right\} ,
\label{eq:tankov}
\end{equation}%
where $\nu (\cdot )$ is the L\'{e}vy measure of $X.$ In the presence of a
diffusion component with constant volatility $\sigma $, \cite{tankov} shows
that if 
\begin{equation*}
\int_{\mathbb{R}}x^{2}\nu (dx)<\infty ,
\end{equation*}%
then 
\begin{equation*}
\sigma _{imp}\left( t;\tau ,F(t)\right) \sim \sigma
\end{equation*}%
for $\tau \rightarrow 0$. Other relevant papers about short-time option
prices in models with jumps include \cite{alos et al}, \cite{medvedev}, and 
\cite{muhle-karbe}.

As for long-maturity and extreme strike asymptotics, the literature is
generally quite sparse. Under mild regularity conditions, \cite{tehranchi}, 
\cite{rogers-1} use large deviation principles to demonstrate that the
implied volatility surface will always flatten out for large enough
maturities and finite strikes, irrespective of the underlying process
assumptions, which agrees with earlier observations by \cite{backus}. For
exponentiated L\'{e}vy processes, the authors list explicit formulas tying
the long-time volatility asymptote to moments of $F$. Extreme strike
asymptotics for the specific case of L\'{e}vy processes follow from the
moment formulas of \cite{lee}, \cite{benaim}, \cite{benaim1} and \cite%
{gulisashvili1}; see \cite{tankov} for a brief discussion.

Overall, while some progress has been made in recent years, the available
asymptotic results for jump processes are limited when compared to the
literature for LV and SV diffusion processes. In particular, little is known
about the range of validity of jump process asymptotics, in part due to the
difficulty of establishing accurate numerical option prices for L\'{e}vy
processes, especially in the short-time limit. In this paper, we add to the
literature in several ways. First and foremost, we completely characterize
the $\tau $ and $K$ volatility surface asymptotes for the important class of
(exponentiated) tempered stable L\'{e}vy processes, with and without a
diffusive component. In addition, we also list a number of asymptotic
expansions for the Merton jump-diffusion and the NIG (Normal Inverse
Gaussian) process. Note that our results cover not only the level of the
volatility surface, but also its slope and convexity -- properties that are
highly relevant when calibrating a model to observed option prices\footnote{%
Indeed, in FX markets it is standard to effectively quote directly on smile
slope and convexity, see Section \ref{FXMarket}.}. We mainly work in Fourier
space using the Lewis-Lipton call option representation \cite{lewis-2}, \cite%
{lipton-2}, and our primary tools for asymptotic expansion are the dominated
convergence theorem (for the short-time limits) and classical complex
variable techniques, such as the saddlepoint theorem (for long-dated
options) and high-frequency Fourier asymptotics (for large and small strike
limits). For testing and illustration purposes, we present several
closed-form option prices for special cases of the tempered stable L\'{e}vy
process, and also list some results for the computation of Green's function.
Further, we draw attention to the applicability of \emph{fractional
differentiation} in characterizing the PIDE (partial integro-differential
equation) that governs option prices in the tempered stable model class;
these results allows one to draw on modern numerical algorithms when pricing
both vanilla and exotic options. As an example, we develop an operator
splitting method which is $O\left( N^{2}\right) $ complex and second order
accurate in time and space. The idea is to split jumps into positive and
negative and treat them separately on each step; the problem then boils down
to the inversion of Hessenberg matrices which may be accomplished via the
generalized Thomas algorithm in $O\left( N^{2}\right) $ operations. A
regular diffusion can be added as an additional iteration, as needed.

To test and motivate our asymptotic results, our primary focus in this paper
is on FX options markets. These markets are of interest to us for two
reasons: i) very short-dated options actively trade in this market (it is
not uncommon for dealers to quote on options with a maturity being just a
few hours); and ii) volatility quoting conventions in these markets use the
option delta $\Delta $ (see (\ref{eq:delta})), rather than the strike $K$.
At first glance the delta quotation standard may appear to be a nuisance,
since additional translation from the delta space into the strike space is
needed, but on second thought it becomes apparent that this construct has
deep mathematical roots. Specifically, quoting volatility as a function of
delta is closely related to the so-called \emph{resolution of singularities}
frequently used in algebraic geometry and other mathematical disciplines
(see \cite{arnold}). Our tests disprove a number of \textquotedblleft urban
myths\textquotedblright\ about L\'{e}vy process asymptotics, especially
regarding the range of applicability of short- and long-time asymptotics.

Our paper is organized as follows. Section \ref{FXMarket} provides a very
brief introduction to FX volatility quotation standards, and introduce the
concepts of risk reversals and straddles (also known as butterflies). We
examine some representative market data for implied volatility, and
highlight how the short-dated asymptotics are unnatural in a diffusion
setting. Section \ref{ELPs} gives an overview of L\'{e}vy processes, with an
emphasis on the tempered stable class and the fractional derivative
representation of the corresponding option pricing equation. It also
discusses both conventional and novel numerical methods for the tempered
stable class which capitalize on their interpretation as fractional
derivatives. Section \ref{ExamplesELPs} discusses some representative
examples. Section \ref{LLFormula} introduces the Lewis-Lipton formula which
is the main working tool for establishing all our asymptotic results. It
also discusses calibration of the corresponding stochastic processes to
market data introduced in Section \ref{FXMarket}. Section \ref{BSAsymptotics}
discusses the asymptotic behavior of the BSM formula which is used
subsequently in order to analyze the behavior of implied volatility.
Sections \ref{AsymptoticsLong}, \ref{AsymptoticsShort} study long-time and
short-time asymptotics, respectively, and Section \ref{AsymptoticsWing}
deals with wing asymptotics. Besides deriving theoretical results, Sections %
\ref{AsymptoticsLong}, \ref{AsymptoticsShort}, \ref{AsymptoticsWing} also
contain a series of numerical tests aimed at establishing their practical
relevance. Finally, conclusions are drawn in Section \ref{Conclusions}.
Appendices contain some of the more elaborate proofs and other useful
material.

\section{Background on FX Market\label{FXMarket}}

The FX options market is one of the largest over-the-counter options markets
in the world, yet its conventions are quite idiosyncratic and differ
markedly from those used in other derivatives markets (e.g., interest rates
and equities). Moreover, almost every concept of importance can be
interpreted differently, often depending on the currency pair in question,
which makes systematic analysis and comparison of FX options particularly
difficult. There are both historical and financial reasons for the existence
of FX quotation styles, the most obvious being that FX\ transactions, unlike
many other financial transactions, are inherently symmetric in nature, in
the sense that units of currency are exchanged into units of currency (see,
e.g., \cite{lipton-book}).

In FX markets, it is standard practice to represent the volatility smile in
terms of the option delta, rather than in terms of the option strike. We may
define 
\begin{equation}
\Sigma (t;\tau ,\Delta )=\sigma _{imp}\left( t;\tau ,K\right)
\label{eq:sigma_delta}
\end{equation}%
where $\Delta $ is the delta defined in (\ref{eq:delta}). The map between $%
\Delta $ and $K$ is monotonic, so (\ref{eq:sigma_delta}) is always
well-defined. We should note that several other definitions of delta than (%
\ref{eq:delta}) exist in the FX market, not all of which are monotonic in
strike\footnote{%
For instance, the so-called \emph{premium-adjusted} (forward)\ delta $\Phi
(d_{-})K/F$ is not mononic in $K$.}; for our purposes, we ignore this
complication and just refer to \cite{reiswich} and \cite{clark}, among many
other sources, for additional information on various delta definitions.

In the FX options markets, the function $\Sigma (t;\tau ,\Delta )$ is
normally liquidly quoted at only three different levels of delta: 0.25,
0.50, and 0.75. Somewhat confusingly, only $\Sigma (t;\tau ,0.5)$ is
directly quoted (the at-the-money volatility\footnote{%
In real FX options markets, the ATM strike may actually deviate slightly
from $K=F$, as additional conventions govern the choice of the at-the-money
strike. For instance, a common alternative is to use the \emph{delta-neutral
strike} $K=F\exp (\sigma _{imp}^{2}\tau /2)$, which is the strike level
where the absolute magnitude of put and call deltas are identical.} $\sigma
_{ATM}$, where $K=F(t)$), whereas $\Sigma (t;\tau ,0.25)$ and $\Sigma
(t;\tau ,0.75)$ must be constructed from quotes for \emph{risk-reversals}
(RRs) and \emph{butterflies} (BFs) (also known as strangles). The relevant
formulas\footnote{%
We are omitting some complications in the definition of risk reversals and
butterflies. Besides being imprecise about the correct definitions of delta
and the ATM strike, we have chosen to define the butterflies as so-called 
\emph{smile stranges}, rather than the more common \emph{market strangles}.
The latter definition can be found in \cite{clark} and, unfortunately, does
not allow one to uniquely extract $\Sigma $.} are%
\begin{align*}
\sigma _{ATM}(t;\tau )& =\Sigma (t;\tau ,0.5), \\
RR(t;\tau )& =\Sigma (t;\tau ,0.75)-\Sigma (t;\tau ,0.25), \\
BF(t;\tau )& =\frac{1}{2}\left( \Sigma (t;\tau ,0.75)+\Sigma (t;\tau
,0.25)-2\sigma _{ATM}(t;\tau )\right) ,
\end{align*}%
which trivially allows us to construct $\Sigma (t;\tau ,0.75)$ and $\Sigma
(t;\tau ,0.25)$ from knowledge of $\sigma _{ATM}$, $RR$, and $BF$. It is
clear that the RRs and BFs are closely related to the slope and convexity,
respectively, of $\Sigma (t;\tau ,\Delta )$ around $\Delta =0.5$.
Specifically, we can write (omitting arguments) 
\begin{equation}
RR\approx \frac{1}{2}\frac{\partial \Sigma }{\partial \Delta };\quad
BF\approx \frac{1}{32}\frac{\partial ^{2}\Sigma }{\partial \Delta ^{2}},
\label{eq:approx}
\end{equation}%
where the derivatives are evaluated around $\Delta =0.5$.

In Table \ref{tab:inputs} below, we show some sample market quotes for $%
\sigma _{ATM}\left( \tau \right) $, $RR\left( \tau \right) $, and $BF\left(
\tau \right) $ for the USD/JPY currency pair. We highlight that the figure
suggests the existence of non-zero finite limits for all three quotes $%
\sigma _{ATM},$ $RR$, and $BF$ as $\tau $ approaches zero. From (\ref%
{eq:approx}), this essentially translates into non-zero finite limits for
both $\partial \Sigma /\partial \Delta $ and $\partial ^{2}\Sigma /\partial
\Delta ^{2}$ around the ATM point.

\begin{table}[h]
\centering\noindent 
\makebox[\textwidth]{
    \begin{tabular}{ |c||c|c|c|c|c|c|c|c|c|c|c|c|}
        \hline
          & 1d & 1w & 2w & 1m & 2m & 3m & 6m & 1y & 2y & 3y & 4y & 5y \\
        \hline\hline
        ATM &  9.87 & 12.15 & 11.27 & 10.35 & 10.34 & 10.52 & 11.04 & 11.80 & 12.90 & 13.80 & 14.30 & 15.05 \\
        RR25 & -1.25 & -1.00 & -0.46 & -0.30 & -0.16 & -0.10 & -0.05 & -0.02 & -0.35 & -0.78 & -1.20 & -1.48 \\
        BF25 & 0.30 & 0.30 & 0.30 & 0.30 & 0.32 & 0.33 & 0.46 & 0.59 & 0.60 & 0.58 & 0.53 & 0.46 \\
        RR10 & -2.22 & -1.77 & -0.79 & -0.49 & -0.22 & -0.09 & 0.07 & 0.18 & -0.50 & -1.41 & -2.30 & -2.86 \\
        BF10 & 1.11 & 1.08 & 1.03 & 1.02 & 1.08 & 1.12 & 1.54 & 1.98 & 2.11 & 2.22 & 2.28 & 2.29 \\
        \hline
  \end{tabular}
  }
\caption{The behavior of $\protect\sigma _{ATM}\left( \protect\tau \right) ,$
$RR\left( \protect\tau \right) $, and $BF\left( \protect\tau \right) $
(expressed in per cent) for USD/JPY on March 30th, 2012. $RR25$ and $RR10$
are the risk-reversals at $\Delta =0.25$ and $\Delta =0.10$, respectively
(and similar for $BF25$, $BF10$).}
\label{tab:inputs}
\end{table}

To understand the implications of Table \ref{tab:inputs} for the volatility
smile in strike space, recall our definition of \emph{log-moneyness} $k=\ln
(K/F(t))$ and notice that 
\begin{equation}
\frac{\partial \Sigma }{\partial \Delta }=-\frac{\partial \sigma _{imp}}{%
\partial k}\sqrt{2\pi \tau },\quad \frac{\partial ^{2}\Sigma }{\partial
\Delta ^{2}}=\left( \frac{\partial ^{2}\sigma _{imp}}{\partial k^{2}}-\frac{1%
}{2}\frac{\partial \sigma _{imp}}{\partial k}\right) 2\pi \tau ,
\label{eq:sigma_der}
\end{equation}%
where the derivatives are taken around $\Delta =0.5$ or, equivalently, $k=0$%
. To match non-zero limits of $\partial \Sigma /\partial \Delta $ and $%
\partial ^{2}\Sigma /\partial \Delta ^{2}$ for $\tau \rightarrow 0$, we
evidently need both the smile skew $\partial \sigma _{imp}/\partial k$ and
the smile convexity $\partial ^{2}\sigma _{imp}/\partial k^{2}$ to approach
infinity at $k=0$ for small $\tau $, at rates of $\tau ^{-1/2}$ and $\tau
^{-1}$, respectively. Such a requirement, however, would rule out that FX
dynamics are driven by a pure diffusion process (such as LSV models), since
it is known (see Section \ref{LVPsS}) that such processes always result in a
finite limit of $\partial \sigma _{imp}/\partial k$ at $k=0$. Motivated by
this observation, we proceed below to introduce the class of L\'{e}vy
processes.

\begin{figure}[h]
\includegraphics[width=1.00\textwidth, angle=0]
{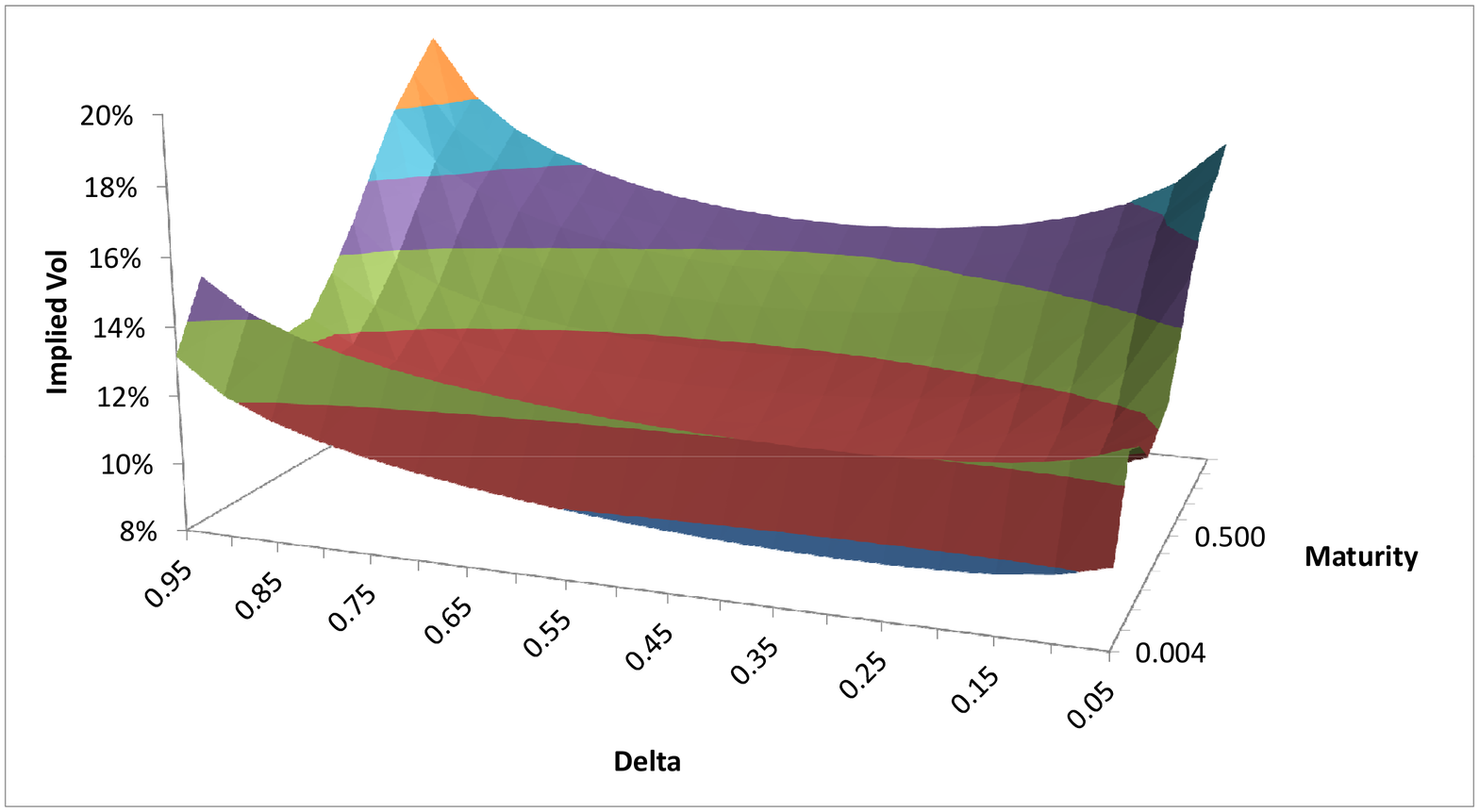}
\caption{{}USDJPY implied volatility surface. The quotes are for March 30th,
2012.}
\label{fig:USDJPYQuotes1}
\end{figure}

The data in Table \ref{tab:inputs} can be converted (with the aid of
interpolation and extrapolation) into the function $\Sigma \left( \tau
,\Delta \right) $. The result is shown in Figure \ref{fig:USDJPYQuotes1}. It
is clear that the FX options exhibit strong smile for short and medium
maturities.

\section{Background on Exponential L\'{e}vy Processes\label{ELPs}}

\subsection{Basic Setup\label{Basics}}

In this section we consider exponential L\'{e}vy processes (ELPs). General
properties of these processes are discussed in detail by numerous
researchers, see, e.g., \cite{bertoin}, \cite{sato}, \cite{appelbaum2},
among many others. The realization that these processes have important
applications to finance can be traced back to \cite{mandelbrot} (and, in
fact, to much earlier work); these applications are discussed in many books
and papers, see, e.g., \cite{eberlein}, \cite{benhamou}, \cite{raible}, \cite%
{barndorff-2}, \cite{boyar-2}, \cite{schoutens}, \cite{appelbaum}, \cite%
{cont-tankov}, to mention just a few.

Let $X(t)$ be a L\'{e}vy process, i.e. a cadlag process with stationary and
independent increments, satisfying $X(0)=0.$ It is known that every L\'{e}vy
process is characterized by a triplet $(\bar{\gamma},\sigma ,\nu )$, where $%
\bar{\gamma}$ and $\sigma $ are constants, and $\nu $ is a (possibly
infinite) Radon measure, known as the \emph{L\'{e}vy measure}. The L\'{e}vy
measure must always satisfy 
\begin{equation}
\int_{\mathbb{R}}\min \left( x^{2},1\right) \nu (dx)<\infty .
\label{eq:min_levy}
\end{equation}%
To characterize the infinitesimal generator of a L\'{e}vy process, let $%
\mathrm{E}$ be the expectation operator, and define 
\begin{equation*}
V(t,x)=\mathrm{E}\left( V_{T}(X(T))|X(t)=x\right)
\end{equation*}%
for some suitably regular function $V_{T}(\cdot )$. It can be shown that $V$
solves a partial integro-differential equation (PIDE) of the form%
\begin{equation*}
V_{t}+\bar{\gamma}V_{x}+\frac{1}{2}\sigma ^{2}V_{xx}+\int_{\mathbb{R}}\left(
V(t,x+y)-V\left( t,x\right) -y1_{|y|\leq 1}V_{x}\left( t,x\right) \right)
\,\nu (dy)=0,
\end{equation*}%
subject to the terminal condition $V\left( T,x\right) =V_{T}\left( x\right) $%
. More conveniently, we write this PIDE as%
\begin{equation}
V_{t}+\gamma V_{x}+\frac{1}{2}\sigma ^{2}\left( V_{xx}-V_{x}\right) +\int_{%
\mathbb{R}}\left( V(t,x+y)-V\left( t,x\right) -y1_{|y|\leq 1}V_{x}\left(
t,x\right) \right) \,\nu (dy)=0,  \label{eq:PIDE}
\end{equation}%
where $\gamma =\bar{\gamma}+\sigma ^{2}/2$. Loosely speaking, (\ref{eq:PIDE}%
) demonstrates that any L\'{e}vy process $X(t)$ is the sum of a
deterministic drift term $\gamma t$, a scaled drift-adjusted Brownian motion 
$\sigma W(t)-\sigma ^{2}t/2$, and a pure jump process.

By choosing $V_{T}\left( x\right) =\exp (iux)$ and solving (\ref{eq:PIDE})
through an affine ansatz 
\begin{equation*}
V(t,x)=\exp (\psi (u)(T-t)+iux),
\end{equation*}%
one arrives at the famous\emph{\ }\textit{L\'{e}vy-Khinchine formula,}%
\begin{equation*}
\phi (t,u)\triangleq \mathrm{E}\left( e^{iuX(t)}\right) =\exp \left( \psi
(u)t\right) ,
\end{equation*}%
\begin{equation}
\psi (u)=\gamma iu-\frac{1}{2}\sigma ^{2}u\left( u+i\right) +\int_{\mathbb{R}%
}\left( e^{iux}-1-iux1_{|x|\leq 1}\right) \,\nu (dx),\quad u\in \mathbb{R},
\label{eq:levy_bas}
\end{equation}%
where $\psi $ is the so-called \emph{L\'{e}vy exponent}. In practice, a L%
\'{e}vy process may be specified either by its exponent $\psi $ or by its L%
\'{e}vy measure $\nu $. It is frequently convenient to split $\psi (u)$ into
two parts as follows%
\begin{equation}
\psi (u)=-\frac{1}{2}\sigma ^{2}u(u+i)+\psi _{0}(u),  \label{eq:psi_0}
\end{equation}%
where $\psi _{0}(u)$ is the pure-jump and drift component of $\psi (u)$.

Let $F(t)>0$ denote the time $t$ price of a financial asset, and consider
modeling its evolution as an exponential L\'{e}vy process of the type (\ref%
{eq:S}), where we wish for $F(t)$ to be a martingale in some pricing
measure. For this, we impose that, for any $t>0$, 
\begin{equation}
\mathrm{E}\left( e^{X(t)}\right) =1,  \label{eq:mart}
\end{equation}%
which requires that the first exponential moment of $X(t)$ exists in the
first place, i.e. that large positive jumps be suitably bounded:%
\begin{equation}
\int_{|x|>1}e^{x}\nu (dx)<\infty .  \label{eq:reg}
\end{equation}%
Equivalently, we require that $\phi (t,z)$ exists in the complex plane strip 
\begin{equation*}
\mathcal{S}=\{z\in \mathbb{C}|\mathrm{Im}(z)\in \lbrack -1,0]\},
\end{equation*}%
and that 
\begin{equation*}
\phi (t,-i)=\exp \left( \gamma t+t\int_{\mathbb{R}}\left(
e^{x}-1-x1_{|x|\leq 1}\right) \,\nu (dx)\right) =1,
\end{equation*}%
in order to satisfy condition (\ref{eq:mart}). This implies the fundamental
martingale constraint 
\begin{equation}
\gamma =-\int_{\mathbb{R}}\left( e^{x}-1-x1_{|x|\leq 1}\right) \,\nu (dx).
\label{eq:gamma_mart}
\end{equation}

Using (\ref{eq:gamma_mart}) to eliminate $\gamma $ from (\ref{eq:levy_bas}),
the martingale restriction on $F$ allows us to write the L\'{e}vy exponent
in the form 
\begin{equation}
\psi (u)=-\frac{1}{2}\sigma ^{2}u(u+i)+\int_{\mathbb{R}}\left(
e^{iux}-1-iu\left( e^{x}-1\right) \right) \,\nu (dx).  \label{eq:LK}
\end{equation}%
Similarly, we may write the PIDE (\ref{eq:PIDE}) as 
\begin{equation}
V_{t}+\frac{1}{2}\sigma ^{2}\left( V_{xx}-V_{x}\right) +\int_{\mathbb{R}%
}\left( V(t,x+y)-V\left( t,x\right) -\left( e^{y}-1\right) V_{x}\left(
t,x\right) \right) \,\nu (dy)=0.  \label{eq:PIDE2}
\end{equation}

Whenever possible (see Section \ref{sec:class}), it is often more convenient
to use the following forms of the PIDE (\ref{eq:PIDE}), namely, 
\begin{equation}
V_{t}+\gamma ^{\prime }V_{x}+\frac{1}{2}\sigma ^{2}\left(
V_{xx}-V_{x}\right) +\int_{-\infty }^{\infty }\left( V\left( t,x+y\right)
-V\left( t,x\right) \right) \nu (dy)=0,  \label{eq:PIDE_1}
\end{equation}%
\begin{equation}
V_{t}+\gamma ^{\prime \prime }V_{x}+\frac{1}{2}\sigma ^{2}\left(
V_{xx}-V_{x}\right) +\int_{-\infty }^{\infty }\left( V\left( t,x+y\right)
-V\left( t,x\right) -yV_{x}\left( t,x\right) \right) \nu (dy)=0,
\label{eq:PIDE_2}
\end{equation}%
where 
\begin{equation}
\gamma ^{\prime }=-\int_{\mathbb{R}}\left( e^{x}-1\right) \,\nu (dx)=\gamma
-\int_{|x|\leq 1}x\nu (dx),  \label{eq:gamma_mart_1}
\end{equation}%
\begin{equation}
\gamma ^{\prime \prime }=-\int_{\mathbb{R}}\left( e^{x}-1-x\right) \,\nu
(dx)=\gamma +\int_{|x|>1}x\nu (dx).  \label{eq:gamma_mart_2}
\end{equation}

\subsection{Classification of Exponential L\'{e}vy Processes\label{sec:class}%
}

Depending on how \textquotedblleft singular\textquotedblright\ the L\'{e}vy
measure $\nu $ is, we can define various sub-groups of L\'{e}vy processes.
Each group allow us to decompose equations (\ref{eq:levy_bas}) and (\ref%
{eq:PIDE}) in slightly different ways.

\subsubsection{Finite Activity\label{FiniteActivity}}

First, if the L\'{e}vy measure is finite (i.e., the jump component of the
process has \emph{finite activity}), the resulting process for $X(t)$ is a
combination of a Brownian motion and an ordinary compound Poisson
jump-process. We may then replace $\nu (dx)$ with 
\begin{equation}
\nu (dy)=\lambda \,j(dy),\quad \lambda \triangleq \int_{\mathbb{R}}\nu
(dx)<\infty ,  \label{eq:fa}
\end{equation}%
where $j(dx)=\nu (dx)/\lambda $ is now a properly normed probability measure
for the distribution of jump sizes in $X$, and $\lambda $ is the (Poisson)
arrival intensity of jumps. In this case, 
\begin{equation*}
\psi (u)=\gamma ^{\prime }iu-\frac{1}{2}\sigma ^{2}u\left( u+i\right)
+\lambda \left( \int_{\mathbb{R}}e^{iux}j(\,dx)-1\right) ,
\end{equation*}%
where the martingale restriction requires that $\gamma ^{\prime }$ satisfies
(\ref{eq:gamma_mart_1}). Notice that if we define a random variable $J$ with
density $j(dx)$, then we can, in the finite activity case, interpret 
\begin{equation}
\psi (u)=\gamma ^{\prime }iu-\frac{1}{2}\sigma ^{2}u\left( u+i\right)
+\lambda \left( \phi _{J}(u)-1\right) ,  \label{eq:fin}
\end{equation}%
where $\phi _{J}(\cdot )$ is the characteristic function of the jump size%
\footnote{%
Specifically, if a jump of size $J$ takes place at time $t$, $X(t-)$ goes to 
$X(t+)=X(t-)+J$ and $S(t-)$ goes to $S(t+)=S(t-)e^{J}$.} $J$.

As for the PIDE (\ref{eq:PIDE}), for finite activity processes it simplifies
in an analogous way to the simpler form (\ref{eq:PIDE_1}): 
\begin{equation*}
V_{t}+\gamma ^{\prime }V_{x}+\frac{1}{2}\sigma ^{2}\left(
V_{xx}-V_{x}\right) +\lambda \int_{\mathbb{R}}V(t,x+y)\,\,j(dy)-\lambda V=0.
\end{equation*}

\subsubsection{Finite Variation\label{FiniteVariation}}

The jump component of a L\'{e}vy process is said to have \emph{finite
variation} if 
\begin{equation}
\int_{|x|\leq 1}|x|\nu (dx)<\infty .  \label{eq:fin_var}
\end{equation}%
Under this condition, truncation of the L\'{e}vy exponent $\psi $ around the
origin is not necessary, and we may write 
\begin{equation}
\psi (u)=\gamma ^{\prime }iu-\frac{1}{2}\sigma ^{2}u\left( u+i\right) +\int_{%
\mathbb{R}}\left( e^{iux}-1\right) \nu (\,dx),\ \ \ \gamma ^{\prime }=\gamma
-\int_{|x|\leq 1}x\nu (dx).  \label{eq:fv}
\end{equation}%
For the PIDE (\ref{eq:PIDE}), we then get the simpler form (\ref{eq:PIDE_1}).

\subsubsection{Finite First Moment\label{FiniteFM}}

Finally, for the case where the first moment exists, 
\begin{equation}
\int_{\mathbb{R}}|x|1_{|x|>1}\nu (\,dx)<\infty ,  \label{eq:fin_first}
\end{equation}%
we may write the L\'{e}vy exponent in purely compensated form: 
\begin{equation*}
\psi (u)=\gamma ^{\prime \prime }iu-\frac{1}{2}\sigma ^{2}u\left( u+i\right)
+\int_{\mathbb{R}}\left( e^{iux}-1-iux\right) \nu (\,dx),\ \ \ \gamma
^{\prime \prime }=\gamma +\int_{|x|>1}x\nu (dx).
\end{equation*}%
The corresponding PIDE can then be written in the simpler form (\ref%
{eq:PIDE_2}).

Going forward we shall often omit primes on $\gamma $ and simply use $\gamma 
$ to denote the drift term of the PIDE, implicitly choosing the right one.

\section{Examples of Exponential L\'{e}vy Processes\label{ExamplesELPs}}

\subsection{Tempered Stable Processes\label{sec:TSP}}

\subsubsection{Definitions and Basic Facts\label{DefinitionFact}}

Establishing short-time ATM volatility smile asymptotics for the completely
generic class of exponential L\'{e}vy processes appears to be a difficult
problem, so we narrow our focus to classes of processes important in
applications. Of primary importance to us are processes characterized by L%
\'{e}vy measures of the form

\begin{equation}
\nu (dx)=\left( \frac{c_{+}}{x^{\alpha +1}}e^{-\kappa _{+}x}1_{x>0}+\frac{%
c_{-}}{|x|^{\alpha +1}}e^{-\kappa _{-}|x|}1_{x<0}\right) dx,
\label{eq:LevyMes1}
\end{equation}%
where we require\footnote{%
The condition $\kappa _{+}\geq 1$ is required to satisfy (\ref{eq:reg}). In
most literature, the condition is the less tight $\kappa _{+}\geq 0.$} that $%
\kappa _{+}\geq 1$, $\kappa _{-}\geq 0,$ $c_{+}\geq 0$, $c_{-}\geq 0$, and $%
\alpha <2.$ The resulting class of processes\footnote{%
Only for $\alpha >0$ does the class behave like a stable process, but we
here allow for $\alpha \leq 0$ to include compound Poisson processes.} is
known as \emph{tempered }$\alpha $-\emph{stable L\'{e}vy processes (TSPs)},
see, e.g., \cite{koponen}, \cite{matacz}, \cite{boyar-1}, \cite{boyar-2}, 
\cite{carr et al}, \cite{cont-tankov}, \cite{rosinski}. Occasionally, a
different parametrization of the L\'{e}vy measure is used%
\begin{equation}
\nu (dx)=-\frac{\sec \left( \frac{\alpha \pi }{2}\right) }{\Gamma (-\alpha )}%
\left( \frac{\vartheta _{+}^{\alpha }}{x^{\alpha +1}}e^{-\kappa
_{+}x}1_{x>0}+\frac{\vartheta _{-}^{\alpha }}{|x|^{\alpha +1}}e^{-\kappa
_{-}|x|}1_{x<0}\right) dx,  \label{eq:LevyMes2}
\end{equation}%
$\vartheta _{+}\geq 0$, $\vartheta _{-}\geq 0$, and $\sec \left( \pi
/2\right) /\Gamma (-1)=-2/\pi $. This parametrization is particularly useful
when one analyzes changes occurring when $\alpha $ crosses unity. Asymmetry
of the TSP is often characterized by the non-dimensional number $\beta $,%
\begin{equation}
\beta =\frac{c_{+}-c_{-}}{c_{+}+c_{-}}.  \label{eq:beta}
\end{equation}%
It is clear that TSPs are natural extensions of the classical $\alpha $-%
\emph{stable L\'{e}vy processes (SPs),} where $\kappa _{\pm }=0$, see, e.g., 
\cite{zolotarev}, \cite{samorodinsky}, \cite{nolan}.

The overall behavior of the TS class is closely tied to the selection of the
power $\alpha $, as demonstrated in Table \ref{tab:classification}.

\begin{table}[h]
\centering\noindent 
\makebox[\textwidth]{
    \begin{tabular}[!h]{ |c|c|c|}
        \hline
         $\alpha $ & Activity & Variation  \\
        \hline\hline
        $<0$ & Finite & Finite\\
        $(0,1)$ & Infinite & Finite \\
        $\lbrack 1,2)$ & Infinite & Infinite \\
         \hline
  \end{tabular}
  }
\caption{The behavior of the TS L\'{e}vy class as a function of $\protect%
\alpha $.}
\label{tab:classification}
\end{table}

Some important special cases of the TS class include:

\begin{itemize}
\item The KoBoL (CGMY) model, where $c_{+}=c_{-}$, see \cite{koponen}, \cite%
{matacz}, \cite{boyar-1}, \cite{carr et al};

\item The exponential jump model, where $\alpha =-1$, see \cite{kou}, \cite%
{lipton-4};

\item The Gamma process, where $\alpha =0$ and either $c_{-}=0$ or $c_{+}=0$;

\item The Variance Gamma model, where $\alpha =0$, see \cite{madanseneta};

\item The Inverse Gaussian process, where $\alpha =1/2$ and either $c_{-}=0$
or $c_{+}=0$, see \cite{barndorff-1}, \cite{barndorff-2}.
\end{itemize}

For some of the special cases above, explicit formulas exist for the density
of $X(t)$ and for European call options. Section \ref{SkewedSP} lists such
formulas for the Inverse Gaussian process, which we use for testing various
results later on.

When $\alpha \neq 0$ and $\alpha \neq 1,$ the characteristic function for
the TS L\'{e}vy process can easily be shown to be, (see, e.g., \cite%
{cont-tankov},) 
\begin{align}
\psi (u)& =\sum\limits_{s=\pm }\Gamma \left( -\alpha \right) c_{s}\left(
\left( \kappa _{s}-siu\right) ^{\alpha }-iu\left( \kappa _{s}-s1\right)
^{\alpha }+\left( iu-1\right) \kappa _{s}^{\alpha }\right)  \label{eq:psi_TS}
\\
& =\sum_{s=\pm }a_{s}\left( \kappa _{s}-siu\right) ^{\alpha }+\gamma
iu+\delta ,  \notag
\end{align}%
where 
\begin{align}
a_{s}& =\Gamma \left( -\alpha \right) c_{s}=-\sec \left( \frac{\alpha \pi }{2%
}\right) \vartheta _{s}^{\alpha },\ \ \ \mathsf{\zeta }_{s}=\left( \kappa
_{s}^{\alpha }-\left( \kappa _{s}-s\right) ^{\alpha }\right) ,
\label{eq:defs} \\
\mathsf{\eta }_{s}& =-\kappa _{s}^{\alpha },\ \ \ \gamma =a_{+}\mathsf{\zeta 
}_{+}+a_{-}\mathsf{\zeta }_{-},\ \ \ \delta =\left( a_{+}\mathsf{\eta }%
_{+}+a_{-}\mathsf{\eta }_{-}\right) .  \notag
\end{align}%
We have imposed the martingale condition (\ref{eq:gamma_mart}) to express $%
\gamma $ as function of other parameters. In (\ref{eq:psi_TS}) the complex
power functions 
\begin{equation*}
\left( \kappa _{+}-iu\right) ^{\alpha },\ \ \ \left( \kappa _{-}+iu\right)
^{\alpha },
\end{equation*}%
are here (recall that $\alpha \neq 0,1,2)$ multi-valued functions, and an
appropriate branch cut is required. We need, as a minimum, that $\psi (u)$
is regular for the strip$\ \mathcal{S}$, i.e. when $u=u^{\prime }+iu^{\prime
\prime }$, $u^{\prime }\in \mathbb{R},$ $u^{\prime \prime }\in \lbrack -1,0]$%
. In this strip, the two power functions evaluate to 
\begin{equation*}
\left( \kappa _{+}+u^{\prime \prime }-iu^{\prime }\right) ^{\alpha },\ \ \
\left( \kappa _{-}-u^{\prime \prime }+iu^{\prime }\right) ^{\alpha }.
\end{equation*}%
As the real part of the argument of the power function is strictly positive
in $\mathcal{S}$, a branch cut in the left half-plane (e.g., the usual
principal value branch cut for the logarithm) will therefore suffice.

Equation (\ref{eq:psi_TS}) holds only for $\alpha \neq 1$ and $\alpha \neq 0$
(and $\alpha <2$, of course). When $\alpha =0,$ we get 
\begin{equation}
\psi (u)=\sum_{s=\pm }c_{s}\left( \log \left( \frac{\kappa _{s}}{\kappa
_{s}-siu}\right) -\log \left( \frac{\kappa _{s}}{\kappa _{s}-s}\right)
iu\right) .  \label{eq:psi_TS_0}
\end{equation}%
Proceeding as above, we can easily show that the arguments to the $\log $%
-functions in this expression are entirely in the right half-plane when $%
\mathrm{Im}u\in \lbrack -1,0]$; the principal value of the logarithm will
suffice. Finally, for the case $\alpha =1$, we have 
\begin{equation*}
\psi (u)=\sum_{s=\pm }c_{s}\left( \kappa _{s}-siu\right) \log \left( \frac{%
\kappa _{s}-siu}{\kappa _{s}}\right) -\left( \kappa _{s}-s\right) \log
\left( \frac{\kappa _{s}-s}{\kappa _{s}}\right) iu.
\end{equation*}%
Again, we may interpret $\log $ as the principal value of the logarithm.

By using (\ref{eq:psi_TS}) it is easy to show that the annualized standard
deviation of a TSP has the form%
\begin{equation}
stdev\left( X\left( 1\right) \right) =\left( \Gamma \left( 2-\alpha \right)
\left( \frac{c_{+}}{\kappa _{+}^{2-\alpha }}+\frac{c_{-}}{\kappa
_{-}^{2-\alpha }}\right) \right) ^{1/2}.  \label{eq:stdev_TSP}
\end{equation}%
Equation (\ref{eq:stdev_TSP}) allows us to get an idea of the magnitude of
the implied volatility of ATM\ options on TSPs.

\subsubsection{PIDEs and their Fractional Derivative Interpretation\label%
{PIDE}}

Consider now the TS L\'{e}vy class with an added Brownian motion with
volatility $\sigma $. For $\alpha \in (0,1)$ the PIDE (\ref{eq:PIDE_1})
applies, and has the form 
\begin{equation}
V_{t}+\gamma ^{\prime }V_{x}+\frac{1}{2}\sigma ^{2}\left(
V_{xx}-V_{x}\right) +\sum\limits_{s=\pm }c_{s}\int_{0}^{\infty }\left(
V\left( x+sy\right) -V\left( x\right) \right) \frac{e^{-\kappa _{s}y}dy}{%
y^{1+\alpha }}=0.  \label{eq:PIDE_alpha_3}
\end{equation}%
For $\alpha \in (1,2)$ the PIDE (\ref{eq:PIDE_2}) can be used, 
\begin{equation}
V_{t}+\gamma ^{\prime \prime }V_{x}+\frac{1}{2}\sigma ^{2}\left(
V_{xx}-V_{x}\right) +\sum\limits_{s=\pm }c_{s}\int_{0}^{\infty }\left(
V\left( x+sy\right) -V\left( x\right) -syV_{x}\left( x\right) \right) \frac{%
e^{-\kappa _{s}y}dy}{y^{1+\alpha }}=0.  \label{eq:PIDE_alpha_4}
\end{equation}

Interestingly, it is possible to rewrite both (\ref{eq:PIDE_alpha_3}) and (%
\ref{eq:PIDE_alpha_4}) in terms of so-called \emph{fractional derivatives}
(see, e.g., \cite{miller}, \cite{podlubny} for a survey, and \cite{zolotarev}
for a connection to regular, non-tempered stable L\'{e}vy processes). The
development of fractional derivatives originated in the nineteenth century
with Riemann, Liouville and Marchaud among others (see, e.g., \cite{marchaud}%
), and traditionally starts with the well-known Riemann-Liouville formula
which allows one to reduce the calculation of an $n$-fold integral to the
calculation of a single convolution integral. This formula can be extended
in a natural way for $\alpha $-fold integrals for any $\alpha >0$. Its
extension for negative $\alpha $, which can be viewed as $\alpha $-fold
differentiation, can be done in several different ways; we find that the
so-called \emph{Caputo definition} is the most convenient for our purposes.
For integer values of $\alpha ,$ the Caputo derivative coincides with the
regular derivative of order $\alpha .$ For non-integer values of $\alpha $,
consider first $\alpha \in \left( 0,1\right) $, and define left ($s=-1$) and
right ($s=1$) fractional derivatives of order $\alpha $ as follows

\begin{equation*}
\mathfrak{D}_{s}^{\alpha }V\left( x\right) =\frac{\left( -s\right) ^{\alpha }%
}{\Gamma \left( -\alpha \right) }\int_{0}^{\infty }\left( V\left(
x+sy\right) -V\left( x\right) \right) \frac{dy}{y^{1+\alpha }},\ \ \ \ \
s=\pm .
\end{equation*}%
Under some mild regularity assumptions we can perform integration by parts
and write 
\begin{equation*}
\mathfrak{D}_{s}^{\alpha }V\left( x\right) =\frac{\left( -s\right) ^{\alpha
-1}}{\Gamma \left( 1-\alpha \right) }\int_{0}^{\infty }\frac{dV\left(
x+sy\right) }{dy}\frac{dy}{y^{\alpha }}.
\end{equation*}%
For all non-integer values of $\alpha \in (1,\infty )$ we may then define%
\begin{equation*}
\mathfrak{D}_{s}^{\alpha }V\left( x\right) =\frac{\left( -s\right) ^{\alpha
-1-\left\lfloor \alpha \right\rfloor }}{\Gamma \left( 1+\left\lfloor \alpha
\right\rfloor -\alpha \right) }\int_{0}^{\infty }\frac{d^{\left\lfloor
\alpha \right\rfloor +1}V\left( x+sy\right) }{dy^{\left\lfloor \alpha
\right\rfloor +1}}\frac{dy}{y^{\alpha -\left\lfloor \alpha \right\rfloor }},
\end{equation*}%
where $\left\lfloor \alpha \right\rfloor $ is the floor function, i.e., the
largest integer such that $\left\lfloor \alpha \right\rfloor <\alpha $.
Notice that in general $\mathfrak{D}_{+}^{\alpha }V\left( x\right) $ is
complex-valued even when $V\left( x\right) $ is real-valued.

We emphasize that with this definition, irrespective of $s$, 
\begin{equation*}
\mathfrak{D}_{s}^{\alpha }e^{iux}=(iu)^{\alpha }e^{iux},
\end{equation*}%
consistent with what one would expect from a generalization of a regular
derivative.

\begin{lemma}
For $\alpha \in (0,2)$, $\alpha \neq 1$, the PIDEs (\ref{eq:PIDE_alpha_3}), (%
\ref{eq:PIDE_alpha_4}) may be written 
\begin{equation}
V_{t}+\gamma V_{x}+\frac{1}{2}\sigma ^{2}\left( V_{xx}-V_{x}\right)
+\sum\limits_{s=\pm }\left( -s\right) ^{\alpha }a_{s}e^{s\kappa _{s}x}%
\mathfrak{D}_{s}^{\alpha }\left( e^{-s\kappa _{s}x}V\right) +\delta V=0,
\label{eq: frac der 1}
\end{equation}%
where $a_{s},\gamma ,\delta $ are given by expression (\ref{eq:defs}).

\begin{proof}
See Appendix \ref{FracPIDE}.
\end{proof}
\end{lemma}

\begin{corollary}
In particular, for non-tempered stable processes with $\kappa _{\pm }=0$,
the corresponding PIDE has the form%
\begin{equation*}
V_{t}+\frac{1}{2}\sigma ^{2}\left( V_{xx}-V_{x}\right) +\sum\limits_{s=\pm
}\left( -s\right) ^{\alpha }a_{s}\left( \mathfrak{D}_{s}^{\alpha
}V-V_{x}\right) =0.
\end{equation*}
\end{corollary}

\subsubsection{Numerical Methods Based on Fractional Derivatives\label%
{Numerics}}

The advantage of restating the original pricing PIDEs as fractional
differential equations is that we may lean on a large body of literature on
numerical methods for such equations. These methods have been developed over
the last twenty years due to the fact that fractional differential equations
have numerous physical applications, especially to the so-called \emph{%
anomalous diffusions}, see, e.g., \cite{chaves}, \cite{sokolov} among many
others. In this section we present some extensions of these methods for our
setting where both left $(s=-1)$ and right $(s=+1)$ derivatives must be
considered \emph{simultaneously}.

Before proceeding, let us remind the reader that an $N\times N$ matrix $%
\mathcal{H}$ such that $h_{ij}=0$ when $j\leq i-2$ ($h_{ij}=0$ when $j\geq
i+2$) is called an upper (lower) \emph{Hessenberg matrix}. Such a matrix can
be viewed as a generalization of a tri-diagonal matrix. An equation of the
form 
\begin{equation}
\mathcal{H}\vec{p}=\vec{q},  \label{eq: HesMat}
\end{equation}%
can be solved for the vector $\vec{p}$ via an appropriate extension of the
Thomas algorithm for tri-diagonal matrices at the cost of $O\left(
N^{2}\right) $ operations. The advantages of using Hessenberg structure of
the problem are manyfold (see \cite{stewart}), the most obvious being the
ability to rely on highly parallelizable solvers. Hessenberg matrices
naturally arise when one wants to solve pricing equations with fractional
derivatives. When such derivatives are one-sided (or, equivalently, jumps
are one-sided) we can naturally represent them via Hessenberg matrices on a
grid by virtue of an appropriately discretized $\mathfrak{D}_{s}^{\alpha }$.
When they are two-sided, we can split the problem into two and consider
positive and negative jumps in turn at each step. Diffusion can be added as
needed. The Gr\"{u}nwald-Letnikov formula is often used for discretizing
fractional derivatives, see, e.g., \cite{meerschaert}, \cite{diethelm}, \cite%
{tadjeran}. However, it is more convenient to discretize $\mathfrak{D}%
_{s}^{\alpha }$ directly, see, e.g., \cite{sousa} for a discretization
scheme that guarantees that the resulting finite difference scheme is
stable. In this discretization scheme, fractional derivatives $\mathfrak{D}%
_{s}^{\alpha }$ in (\ref{eq: frac der 1}) turn into Hessenberg matrices $%
\mathcal{H}_{s}^{\alpha }$; while the diffusion-advection term turns into a
standard tri-diagonal matrix $\mathcal{D}$. In order to take full advantage
of the nature of the matrices $\mathcal{H}_{s}^{\alpha },\mathcal{D}$, we
solve a generic evolution equation of the form (\ref{eq: frac der 1}) by
splitting a typical time step $t^{m+1}\rightarrow t^{m},$ $\vec{V}%
^{m+1}\rightarrow \vec{V}^{m}$ into three, $\vec{V}^{m+1}\rightarrow \vec{V}%
^{\ast }$, $\vec{V}^{\ast }\rightarrow \vec{V}^{\ast \ast }$, $\vec{V}^{\ast
\ast }\rightarrow \vec{V}^{m}$ as follows:%
\begin{align*}
\left( I-\frac{1}{2}\Delta t\mathcal{H}_{+}^{\alpha }\right) \vec{V}^{\ast
}& =\left( I+\frac{1}{2}\Delta t\mathcal{H}_{+}^{\alpha }\right) \vec{V}%
^{m+1}, \\
\left( I-\frac{1}{2}\Delta t\mathcal{H}_{-}^{\alpha }\right) \vec{V}^{\ast
\ast }& =\left( I+\frac{1}{2}\Delta t\mathcal{H}_{-}^{\alpha }\right) \vec{V}%
^{\ast }, \\
\left( I-\frac{1}{2}\Delta t\mathcal{D}\right) \vec{V}^{m}& =\left( I+\frac{1%
}{2}\Delta t\mathcal{D}\right) \vec{V}^{\ast \ast }.
\end{align*}%
It is clear that at the first and second intermediate steps we have to solve
a Hessenberg system of equations, while at the third step we need to solve a
tri-diagonal system of equations. We illustrate the scheme by numerically
constructing the probability density function (PDF) of a typical TSP in two
different ways, namely by solving the corresponding PIDEs and by numerically
calculating its integral representation. Our results are shown in Figure \ref%
{fig:Hessenberg1}, and it is clear that our method reproduces the
corresponding PDF very accurately.

\begin{figure}[h]
\includegraphics[width=1.00\textwidth, angle=0]
{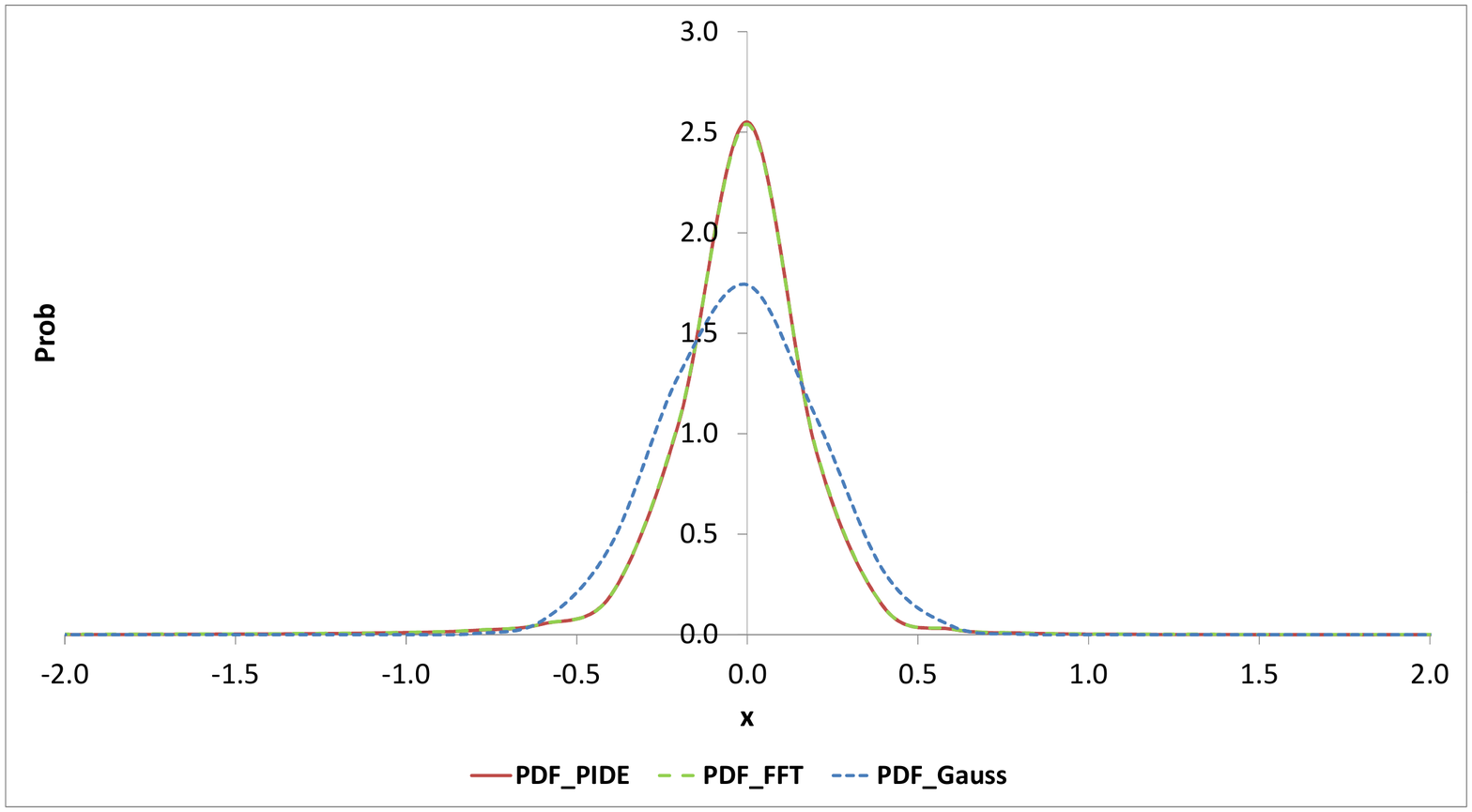}
\caption{{}Numerical and analytical PDFs for a TSP. For comparison, PDFs for
the moment-matched Gaussian process is shown as well, see (\protect\ref%
{eq:stdev_TSP}). The corresponding parameters are $\protect\alpha =1.50$, $%
\protect\sigma =0.01\%$, $c_{+}=0.0069$, $c_{-}=0.0063$, $\protect\kappa %
_{+}=1.9320$, $\protect\kappa _{-}=0.4087$. The choice of parameters is
justified in Section \protect\ref{Calibration}.}
\label{fig:Hessenberg1}
\end{figure}

There are several well-known Fourier-transform-based and approximation-based
approaches to solving the corresponding pricing equation in the spatially
homogeneous case, see, e.g., \cite{and-and}, \cite{asmusen}, \cite{dhalluin}%
, \cite{chen}, \cite{wang}, \cite{itkin}, \cite{lipton-5}, so, if this was
the only case one is interested in, it would not be worthwhile adding yet
another method to this set. However, to the best of our knowledge, none of
these methods can be straightforwardly extended to the inhomogeneous case,
nor to the case when barriers are present. Our method on the other hand can
be extended to cover these cases almost verbatim.

\subsubsection{Maximally Skewed Stable Processes\label{SkewedSP}}

If we set $\kappa _{+}=\kappa _{-}=0$, the L\'{e}vy measure (\ref%
{eq:LevyMes1}) becomes that of a \emph{regular stable process}. This type of
process has limited applications in finance, as we only can satisfy the
martingale restriction (\ref{eq:mart}) if we mandate that $c_{+}=0$ ($\beta
=-1$, where $\beta $ is given by (\ref{eq:beta})), i.e. when all jumps are
negative. The resulting process is sometimes known as the \emph{maximally
negatively skewed stable process}, and has received some interest in the
literature, see \cite{carr-wu-2}. The characteristic exponent for such
processes is known in closed form. Using parametrization (\ref{eq:LevyMes2})
and assuming that the process is a martingale, $\psi (u)$ can be defined as
follows%
\begin{equation}
\psi (u)=\left\{ 
\begin{array}{cc}
\vartheta ^{\alpha }\sec \left( \alpha \pi /2\right) \left( iu-(iu)^{\alpha
}\right) , & \alpha \neq 1, \\ 
\frac{2}{\pi }\vartheta iu\ln (iu), & \alpha =1,%
\end{array}%
\right.  \label{eq:psi_stable}
\end{equation}%
in the complex strip $\mathcal{S}$ defined by $\mathrm{Im}\left( u\right)
\in \lbrack -1,0]$. Despite the fact that jumps can only \textquotedblleft
go down\textquotedblright\ when $\beta =-1$, the already present
drift-correction around the origin in the L\'{e}vy-Khinchine theorem makes
the resulting process a martingale for $\alpha =1$, so no additional drift
correction is necessary.

\begin{remark}
There are three classical cases when the PDF for the stable L\'{e}vy process
can be calculated explicitly: i)\ $\alpha =1/2,\beta =\pm 1$ (L\'{e}vy
distribution); ii)\ $\alpha =1,\beta =0$ (Cauchy distribution); iii)\ $%
\alpha =2,1\leq \beta \leq 1$ (Gaussian distribution). Of these three cases,
for option pricing purposes the only nontrivial case is $\alpha =1/2$ and $%
\beta =-1$, where 
\begin{equation*}
\nu (dx)=\frac{\sqrt{2\vartheta }}{2\sqrt{\pi }\left( -x\right) ^{3/2}}%
\mathbf{1}_{x<0}dx.
\end{equation*}%
Here, the distribution of $X(t)$ is%
\begin{equation}
\mathrm{P}\left( X(t)\in \lbrack x,x+dx]\right) =\frac{\iota }{2\sqrt{\pi }}%
\frac{\exp \left( -\frac{\iota ^{2}}{4(\iota -x)}\right) }{\left( \iota
-x\right) ^{3/2}}\mathbf{1}_{x<\iota }dx,  \label{eq:LGP}
\end{equation}%
where $\iota =\sqrt{2\vartheta }t$ and condition (\ref{eq:mart}) is clearly
satisfied, so that $\exp (X(t))$ is a martingale. Notice that in this case $%
X(t)$ is bounded from above by $\iota $, which may not be particularly
realistic. Still, for testing purposes it is worthwhile developing this case
further, so Proposition \ref{prop:GLCall} develops an analytic call option
formula for this model.
\end{remark}

In order to avoid possible confusion, we call the maximally skewed L\'{e}vy
process with $\alpha =1/2,\beta =\pm 1$, the L\'{e}vy-Gauss Process (LGP).
By doing so we emphasize the so-called \emph{duality} relation between the L%
\'{e}vy process with $\alpha =1/2,\beta =\pm 1$, and the standard Gaussian
process with $\alpha =2$, see, e.g., \cite{zolotarev}.

Maximally skewed stable processes possess are scaling invariant, which
allows one to write their Green's functions in a simple form. Interestingly,
a maximally skewed \emph{tempered} stable process ($c_{+}=0$ and $\kappa
_{-}>0$) can be transformed into a maximally skewed stable process ($c_{+}=0$
and $\kappa _{-}=0$). This result, which is listed in Appendix \ref{GreenFun}%
, allows us to represent the Green's function for maximally skewed tempered
stable process in closed form. It also allows us to derive closed-form call
option prices for certain TSPs, which shall be useful later on for testing
purposes. For instance, we have:

\begin{proposition}
\label{prop:GLCall}Consider the price of a $T$-maturity, $K$-strike call
option in an exponential tempered L\'{e}vy-Gauss process with $\alpha =1/2$
and $c_{+}=0$. Set $\iota =\sqrt{2\vartheta }\tau $, $\iota _{1}=\sqrt{%
\kappa _{-}}\iota $, $\iota _{2}=\sqrt{\kappa _{-}+1}\iota $, and $k=\ln
(K/F(t))$. Further, assume that $k<\iota _{2}-\iota _{1}$, and set 
\begin{equation*}
v=\frac{2\iota ^{2}}{\left( \iota _{2}-\iota _{1}-k\right) }.
\end{equation*}%
We then have%
\begin{align}
\frac{C\left( t;T,K\right) }{F(t)}& =\mathsf{C}^{GL}\left( \tau ,k\right) ,
\label{eq:GLP} \\
\mathsf{C}^{GL}\left( \tau ,k\right) & =e^{\iota _{2}}\mathsf{D}\left(
v,2\iota _{2}\right) -e^{k+\iota _{1}}\mathsf{D}\left( v,2\iota _{1}\right) ,
\notag
\end{align}%
where $\mathsf{D}\left( v,k\right) $ is the symmetrized version of the
Black-Scholes price given by 
\begin{equation}
\mathsf{D}\left( v,k\right) =e^{-\frac{k}{2}}\left( 1-\mathsf{C}^{BS}\left(
v,k\right) \right) =e^{-\frac{k}{2}}\Phi \left( -d_{+}\right) +e^{\frac{k}{2}%
}\Phi \left( d_{-}\right) .  \label{eq: Upsilon}
\end{equation}

\begin{proof}
See Appendix \ref{GreenFun}.
\end{proof}
\end{proposition}

Similar formulas, albeit in a somewhat different context are given in \cite%
{gerber-shiu}.

\begin{remark}
When $\alpha =1/2$, $c_{+}=0$, the price process $F(T)$ is always bounded
from above by 
\begin{equation*}
F_{\max }\left( t,T\right) =F(t)e^{\iota _{2}-\iota _{1}}.
\end{equation*}%
A similar bound exists for all $\alpha <1$. Paradoxically, when $\alpha \geq
1$ and $c_{+}=0$, $F(T)$ can take any value in $(0,\infty )$, despite all
jumps being strictly negative.
\end{remark}

\subsection{Normal Inverse Gaussian Processes\label{NIG}}

As we have seen earlier, in some cases L\'{e}vy processes (such as a
tempered L\'{e}vy-Gauss process) are localized on a semi-axis. Processes of
that nature can be used in order to create new interesting processes by
changing time for the regular Brownian motion. Below we discuss some
examples along these lines. A useful aspect of this approach is that it
generates L\'{e}vy processes with PDFs known in closed form. One of the
well-known examples is the so-called \textit{Variance Gamma} process (VGP),
which is a special case of a TSP with $\alpha =0$, see, e.g., \cite%
{madanseneta}. Another popular choice is the so-called \textit{Normal
Inverse Gaussian} process (NIGP), see, e.g., \cite{barndorff-1}, \cite%
{eriksson}, which we consider in some detail.

The \textit{Inverse Gaussian} (IG) process describes the density of the
hitting time $\tau $ of a level $\mathcal{\varkappa }_{1}t$ by a standard
Brownian motion with volatility $\sigma _{1}$ and drift $\varkappa _{1}$.
Our choice of parameters ensures that $\mathbb{E}\left( \tau \right) =t$. It
is easy to verify that IGP as a TSP with $\alpha =1/2$, $\beta =1$, $\kappa
_{+}=\varkappa _{1}^{2}/2\sigma _{1}^{2}$. It is clear that IGPs and TLGPs
are closely related and can be transformed into each other via shift and
reflection. IG process can be used as a subordinator in order to build the
NIGP out of a standard Brownian motion.

The NIGP can be obtained by time changing of a standard Brownian motion with
volatility $\sigma $ and drift $-\sigma ^{2}/2$. The drift is chosen in a
way ensuring that the corresponding NIG is a martingale. The corresponding
subordinator is distributed as a hitting time of a level $\mathcal{\varkappa 
}_{1}t$ by an independent BM with volatility $\sigma _{1}$ and drift $%
\mathcal{\varkappa }_{1}$. The convolution of these two processes yields%
\begin{equation}
f^{NIG}\left( t,x\right) =\frac{\bar{\omega}\mathcal{\bar{\varkappa}}\sigma
^{2}te^{-\frac{x}{2}+\mathcal{\bar{\varkappa}}^{2}\sigma ^{2}t}}{\pi \sqrt{%
x^{2}+\mathcal{\bar{\varkappa}}^{2}\sigma ^{4}t^{2}}}K_{1}\left( \bar{\omega}%
\sqrt{x^{2}+\mathcal{\bar{\varkappa}}^{2}\sigma ^{4}t^{2}}\right) ,
\label{eq:NIG_pdf_1}
\end{equation}%
where $K_{1}\left( .\right) $ is the modified Bessel function, and $\bar{%
\omega}=\sqrt{\mathcal{\bar{\varkappa}}^{2}+1/4}$, $\mathcal{\bar{\varkappa}%
=\varkappa }_{1}/\sigma _{1}\sigma $. The corresponding L\'{e}vy exponent
and density have the form%
\begin{align*}
\psi ^{NIG}\left( u\right) & =\sigma ^{2}\mathcal{\bar{\varkappa}}\left( 
\mathcal{\bar{\varkappa}}-\sqrt{\mathcal{\bar{\varkappa}}^{2}+u\left(
u+i\right) }\right) , \\
\nu ^{NIG}\left( dx\right) & =\left( \lim_{t\rightarrow 0}\frac{f\left(
t,x\right) }{t}\right) dx=\frac{\sigma ^{2}\bar{\omega}\mathcal{\bar{%
\varkappa}}e^{-\frac{x}{2}}}{\pi \left\vert x\right\vert }K_{1}\left( \bar{%
\omega}\left\vert x\right\vert \right) dx,\ \ \ \ \ x\neq 0.
\end{align*}

Although NIGPs are not a special case of TSPs, they are closely related to
TSPs with $\alpha =1/2$ and $\alpha =1$, due to the fact that%
\begin{equation*}
K_{1}\left( \bar{\omega}\left\vert x\right\vert \right) \underset{\left\vert
x\right\vert \rightarrow 0}{\sim }\frac{1}{\bar{\omega}\left\vert
x\right\vert },\ \ \ K_{1}\left( \bar{\omega}\left\vert x\right\vert \right) 
\underset{\left\vert x\right\vert \rightarrow \infty }{\sim }\sqrt{\frac{\pi 
}{2\bar{\omega}\left\vert x\right\vert }}\left( 1+\frac{3}{8\bar{\omega}%
\left\vert x\right\vert }\right) e^{-\bar{\omega}\left\vert x\right\vert },
\end{equation*}%
so that%
\begin{equation*}
\nu ^{NIG}\left( dx\right) \underset{\left\vert x\right\vert \rightarrow 0}{%
\sim }\frac{\sigma ^{2}\mathcal{\bar{\varkappa}}e^{-\frac{x}{2}}}{\pi
\left\vert x\right\vert ^{2}},\ \ \ \ \ \nu ^{NIG}\left( dx\right) \underset{%
\left\vert x\right\vert \rightarrow \infty }{\sim }\frac{\sigma ^{2}\sqrt{%
\bar{\omega}}\mathcal{\bar{\varkappa}}}{\sqrt{2\pi }\left\vert x\right\vert
^{3/2}}e^{-\left( \frac{x}{2}+\bar{\omega}\left\vert x\right\vert \right) }.
\end{equation*}

In the spirit of formula (\ref{eq: BS non-dim}), the price of a call option
with log-strike $k$ can be written in the form%
\begin{equation}
\frac{C^{NIG}\left( t;T,K;\sigma ,\mathcal{\varkappa }_{1},\sigma
_{1}\right) }{F(t)}=\mathsf{C}^{NIG}\left( v,k;\mathcal{\bar{\varkappa}}%
\right) ,  \label{eq: call NIG dir}
\end{equation}%
where, as usual, $v=\sigma ^{2}\tau $, and%
\begin{equation}
\mathsf{C}^{NIG}\left( v,k;\mathcal{\bar{\varkappa}}\right) =\frac{\bar{%
\omega}\mathcal{\bar{\varkappa}}ve^{\mathcal{\bar{\varkappa}}^{2}v}}{\pi }%
\dint\limits_{k}^{\infty }\frac{\left( e^{\frac{x}{2}}-e^{k-\frac{x}{2}%
}\right) }{\sqrt{x^{2}+\mathcal{\bar{\varkappa}}^{2}v^{2}}}K_{1}\left( \bar{%
\omega}\sqrt{x^{2}+\mathcal{\bar{\varkappa}}^{2}v^{2}}\right) dx.
\label{eq: CNIG}
\end{equation}%
Clearly, the normalized call price for a NIGP $\mathsf{C}^{NIG}$ depends on
one non-dimensional parameter. (Recall that $\mathsf{C}^{BS}$ does not
depend on any parameters.) In Section \ref{LLFormula} we derive an
alternative integral representation for $\mathsf{C}^{NIG}$, given by the
Lewis-Lipton formula. We use one or the other of these two expressions as
convenient.

As before, it is easy to compute the annualized standard deviation of a
NIGP. The corresponding expression has the form%
\begin{equation}
stdev\left( X\left( 1\right) \right) =\frac{\bar{\omega}\sigma }{\mathcal{%
\bar{\varkappa}}}.  \label{eq:stdev_NIGP}
\end{equation}

\begin{remark}
We note in passing that the so-called \emph{Generalized Inverse Gaussian}
(GIG) processes can be viewed as a natural generalizations of IGPs. In
general, the density of such a process is not known analytically, while its L%
\'{e}vy density is given by a fairly complicated expression, so that GIG
process is not easy (or necessary) to use in practice. It is not in the TSP
class in any case.
\end{remark}

\subsection{Merton Processes\label{MP}}

One popular process that is not related to the tempered stable class is the
(pure-jump) Merton process (MP), with 
\begin{equation*}
\nu ^{M}(dx)=\frac{\lambda }{\sqrt{2\pi }\eta }e^{-\frac{(x-\mu )^{2}}{2\eta
^{2}}}dx.
\end{equation*}%
This process has finite activity, so we may use (\ref{eq:fin}) to establish $%
\psi (u)$. We get 
\begin{equation*}
\int_{\mathbb{R}}\left[ e^{iux}-1\right] \nu (dx)=\lambda \left( e^{iu\mu -%
\frac{1}{2}\eta ^{2}u^{2}}-1\right) ,
\end{equation*}%
so, for a pure-jump MP, 
\begin{equation}
\psi (u)=\lambda \left( e^{iu\mu -\frac{1}{2}\eta ^{2}u^{2}}-1\right)
+\lambda \left( 1-e^{\mu +\frac{1}{2}\eta ^{2}}\right) iu=\lambda \left(
e^{iu\mu -\frac{1}{2}\eta ^{2}u^{2}}-1\right) +\gamma iu,  \label{eq: MP psi}
\end{equation}%
where $\gamma =\lambda \left( 1-e^{q}\right) ,\ \ \ q=\mu +\frac{1}{2}\eta
^{2}$. The corresponding PDF can be calculated explicitly, 
\begin{align*}
f\left( t,x\right) & =e^{-\lambda t}\sum_{n=0}^{\infty }\frac{\left( \lambda
t\right) ^{n}}{n!}f_{n}\left( t,x\right) , \\
f_{n}\left( t,x\right) & =\left\{ 
\begin{array}{cc}
\delta \left( x-\gamma t\right) , & n=0, \\ 
\frac{1}{\sqrt{2\pi n}\eta }e^{-\frac{(x-\gamma t-n\mu )^{2}}{2n\eta ^{2}}},
& n>0,%
\end{array}%
\right.
\end{align*}%
It can be shown easily that 
\begin{equation*}
\nu ^{M}\left( dx\right) =\left( \lim_{t\rightarrow 0}\frac{f\left(
t,x\right) }{t}\right) dx=\lambda f_{1}\left( 0,x\right) =\frac{1}{\sqrt{%
2\pi }\eta }e^{-\frac{(x-\mu )^{2}}{2\eta ^{2}}},\ \ \ x\neq 0.
\end{equation*}%
A simple calculation yields the following expression from \cite{merton-2}
for the price of a call option%
\begin{align}
\frac{C^{M}\left( t;T,K;\lambda ,\mu ,\eta \right) }{F(t)}& =\mathsf{C}%
^{M}\left( v,k;\lambda ,\mu ,\eta \right) ,  \label{eq: call MP dir} \\
\mathsf{C}^{M}\left( v,k;\mu ,\eta \right) & =\sum_{n=0}^{\infty }c_{n}%
\mathsf{C}^{BS}\left( n\eta ^{2},l-nq\right) ,  \notag
\end{align}%
where 
\begin{equation*}
c_{n}=e^{-e^{q}v}\left( e^{q}v\right) ^{n}/n!,\ \ \ \ \ l=k-\left(
1-e^{q}\right) v,\ \ \ \ \ v=\lambda \tau .
\end{equation*}

Analysis of MPs with diffusion component is very similar to the one
performed above. The corresponding characteristic function has the form%
\begin{equation*}
\psi (u)=-\frac{1}{2}\sigma ^{2}u\left( u+i\right) +\lambda \left( e^{iu\mu -%
\frac{1}{2}\eta ^{2}u^{2}}-1\right) +\gamma iu,
\end{equation*}%
while the price of a call option can be written as follows%
\begin{align*}
\frac{C^{M}\left( t;T,K;\lambda ,\mu ,\eta \right) }{F(t)}& =\mathsf{C}%
^{M}\left( v,k;\hat{\sigma},\mu ,\eta \right) , \\
\mathsf{C}^{M}\left( v,k;\hat{\sigma},\mu ,\eta \right) &
=\sum_{n=0}^{\infty }c_{n}\mathsf{C}^{BS}\left( \hat{\sigma}^{2}v+n\eta
^{2},l-nq\right) ,
\end{align*}%
where $c_{n},l$ have the same form as before, and $\hat{\sigma}=\sigma /%
\sqrt{\lambda }$.

The annualized standard deviation of a MP has the form%
\begin{equation}
stdev\left( X\left( 1\right) \right) =\left( \sigma ^{2}+\lambda \left( \mu
^{2}+\eta ^{2}\right) \right) ^{1/2}.  \label{eq:stdev_MP}
\end{equation}%
Once again, this formula allows us to get a rough estimate of the magnitude
of the implied volatility of ATM\ options on MPs.

\section{The Lewis-Lipton Option Price Formula and its Implications\label%
{LLFormula}}

\subsection{Exponential L\'{e}vy Processes\label{ELPsLL}}

The key formula allowing one to analyze option prices for L\'{e}vy processes
is known as the Lewis-Lipton (LL) formula; it has been independently
proposed by Lewis \cite{lewis-2} and Lipton \cite{lipton-2}. This formula is
based on the Fourier transform of an appropriately modified payoff of the
call option. A complementary method is proposed in \cite{carrmadan};
additional information can be found in \cite{lee1}.

\begin{proposition}
\label{Prop:LLFormula} The normalized price of a call option written on an
underlying driven by an exponential L\'{e}vy process with L\'{e}vy-Khinchine
exponent $\psi $ has the form%
\begin{equation}
\mathsf{C}\left( \tau ,k\right) =1-\frac{1}{2\pi }\int_{-\infty }^{\infty }%
\frac{E\left( \tau ,u\right) }{Q\left( u\right) }e^{-k\left( iu-\frac{1}{2}%
\right) }du.  \label{eq: LL1}
\end{equation}%
Here%
\begin{equation}
E\left( \tau ,u\right) =\exp \left( \tau \upsilon \left( u\right) \right)
=\exp \left( \tau \left( \upsilon _{0}\left( u\right) -\frac{1}{2}\sigma
^{2}Q\left( u\right) \right) \right) ,  \label{eq: LLE}
\end{equation}%
\begin{equation}
\upsilon \left( u\right) =\psi \left( u-\frac{i}{2}\right) ,\ \ \ \ \
\upsilon _{0}\left( u\right) =\psi _{0}\left( u-\frac{i}{2}\right) ,\ \ \ \
\ Q\left( u\right) =u^{2}+\frac{1}{4}.  \label{eq: LLpsi}
\end{equation}%
and $\psi _{0}$ is given by (\ref{eq:psi_0}).

\begin{proof}
See \cite{lewis-1}, \cite{lewis-2}, \cite{lipton-book}, \cite{lipton-2}.
\end{proof}
\end{proposition}

As a corollary, we have:

\begin{corollary}
\label{Coro:LLFormula}The derivatives with respect to $k$ of the normalized
price of a call option are%
\begin{align}
\mathsf{C}_{k}\left( \tau ,k\right) & =-\frac{1}{2\pi }\int_{-\infty
}^{\infty }\frac{E\left( \tau ,u\right) }{Q\left( u\right) }e^{-k\left( iu-%
\frac{1}{2}\right) }\left( -iu+\frac{1}{2}\right) du,  \label{eq: LL2} \\
\mathsf{C}_{kk}\left( \tau ,k\right) & =-\frac{1}{2\pi }\int_{-\infty
}^{\infty }\frac{E\left( \tau ,u\right) }{Q\left( u\right) }e^{-k\left( iu-%
\frac{1}{2}\right) }\left( -iu+\frac{1}{2}\right) ^{2}du.  \label{eq: LL3}
\end{align}
\end{corollary}

Applying this result to the processes introduced in Section \ref%
{ExamplesELPs} we easily get the following lemma.

\begin{lemma}
For TSPs, NIGPs, and MPs, the corresponding $E\left( \tau ,u\right) $ have
the form%
\begin{align*}
E^{TS}\left( \tau ,u\right) & =\exp \left( \tau \left( -\frac{1}{2}\sigma
^{2}Q\left( u\right) +\sum_{s=\pm }a_{s}\left( \kappa _{s}-s\left( iu+\frac{1%
}{2}\right) \right) ^{\alpha }+\gamma \left( iu+\frac{1}{2}\right) +\delta
\right) \right) , \\
E^{NIG}\left( \tau ,u\right) & =\exp \left( \tau \sigma ^{2}\mathcal{\bar{%
\varkappa}}\left( \mathcal{\bar{\varkappa}}-\sqrt{\mathcal{\bar{\omega}}%
^{2}+u^{2}}\right) \right) , \\
E^{M}\left( \tau ,u\right) & =\exp \left( \tau \left( -\frac{1}{2}\sigma
^{2}Q\left( u\right) +\lambda \left( e^{q\left( iu+\frac{1}{2}\right) -\frac{%
\eta ^{2}}{2}Q\left( u\right) }-1+\left( 1-e^{q}\right) \left( iu+\frac{1}{2}%
\right) \right) \right) \right) .
\end{align*}
\end{lemma}

\begin{remark}
It is worth mentioning that $E^{TS}\left( \tau ,u\right) $ and $%
E^{NIG}\left( \tau ,u\right) $ rapidly decay when $\left\vert u\right\vert
\rightarrow \infty $, so that computation of the integrals (\ref{eq: LL1}), (%
\ref{eq: LL2}), and (\ref{eq: LL3}) is straightforward. When $\sigma \neq 0$%
, $\ E^{M}\left( \tau ,u\right) $ is rapidly decaying as well. However, when 
$\sigma =0$, the situation is more difficult. The corresponding integrals (%
\ref{eq: LL1}), (\ref{eq: LL2}) for this case can be computed efficiently by
using 
\begin{equation*}
\frac{E^{M,\infty }\left( \tau ,u\right) }{Q\left( u\right) }=\frac{\exp
\left( \lambda \tau \left( -1+\left( 1-e^{q}\right) \left( iu+\frac{1}{2}%
\right) \right) \right) }{Q\left( u\right) },
\end{equation*}%
as a control variate.
\end{remark}

\begin{remark}
In the case of a BSM diffusion process, (\ref{eq: LL1}) yields%
\begin{equation*}
\mathsf{C}^{BS}\left( v,k\right) =1-\frac{1}{2\pi }\int_{-\infty }^{\infty }%
\frac{e^{-\frac{1}{2}vQ\left( u\right) -k\left( iu-\frac{1}{2}\right) }}{%
Q\left( u\right) }du,
\end{equation*}%
so that%
\begin{equation}
\frac{1}{2\pi }\int_{-\infty }^{\infty }\frac{e^{-\frac{1}{2}vQ\left(
u\right) -k\left( iu-\frac{1}{2}\right) }}{Q\left( u\right) }du=\Phi \left( 
\frac{k}{\sqrt{v}}-\frac{1}{2}\sqrt{v}\right) +e^{k}\Phi \left( -\frac{k}{%
\sqrt{v}}-\frac{1}{2}\sqrt{v}\right) .  \label{eq: BSLL1}
\end{equation}%
In the limiting case $v=0$, we get%
\begin{equation}
\frac{1}{2\pi }\int_{-\infty }^{\infty }\frac{e^{-k\left( iu-\frac{1}{2}%
\right) }}{Q\left( u\right) }du=e^{k^{-}}.  \label{eq: BSLL}
\end{equation}%
These useful formulas shall be used repeatedly in what follows.
\end{remark}

The usefulness of the LL formula becomes apparent when one wants to study
the asymptotic behavior of the call price and its derivatives (the Greeks)
in the limiting cases of $\tau \rightarrow 0$, $\tau \rightarrow \infty $, $%
\left\vert k\right\vert \rightarrow \infty $, since it allows one to use
enormous body of work dedicated to the asymptotics of integrals depending on
large and small parameters. This is done in the remainder of the paper where
it is shown how to apply the saddlepoint method, the high frequency Fourier
integrals estimates, and other tricks to the problem at hand. We return to
these asymptotics in Sections \ref{AsymptoticsLong}, \ref{AsymptoticsShort}, %
\ref{AsymptoticsWing}. While this is not the focus of this paper, the LL
formula can also be used to study the small jump asymptotics, as is briefly
shown in the next section.

\subsection{Small jumps asymptotics\label{small jumps}}

When the jump component of a L\'{e}vy process is small compared to its
diffusion component, by expanding $E\left( \tau ,u\right) $ in (\ref{eq: LLE}%
), the call price can be written in the form%
\begin{align}
\mathsf{C}\left( \tau ,k\right) & =1-\frac{1}{2\pi }\int_{-\infty }^{\infty }%
\frac{e^{-\frac{1}{2}\sigma ^{2}\tau Q\left( u\right) -k\left( iu-\frac{1}{2}%
\right) }}{Q\left( u\right) }du  \label{eq: LL small} \\
& -\tau \frac{1}{2\pi }\int_{-\infty }^{\infty }\frac{e^{-\frac{1}{2}\sigma
^{2}\tau Q\left( u\right) -k\left( iu-\frac{1}{2}\right) }}{Q\left( u\right) 
}\upsilon _{0}\left( u\right) du+...,  \notag
\end{align}%
where $\upsilon _{0}\left( u\right) $ is given in (\ref{eq: LLpsi}),
provided that the second integral converges. This expression is particularly
useful for the qualitative study of perturbations of the flat volatility
surface caused by jumps. Indeed, by comparing (\ref{eq: LL small}) with the
expansion of the BS formula around $\sigma _{imp}=\sigma $ and matching
terms, the implied volatility surface can be represented in the form%
\begin{equation*}
\sigma _{imp}\left( \tau ,k\right) =\sigma +\sigma _{1}\left( \tau ,k\right)
+...,
\end{equation*}%
where $\sigma _{1}$ is of the same order of magnitude as $\upsilon _{0}$,
and is given by the following expression%
\begin{align*}
\sigma _{1}\left( \tau ,k\right) & =\tau ^{1/2}\frac{e^{\frac{k^{2}}{2\sigma
^{2}\tau }}}{\sqrt{2\pi }}\int_{-\infty }^{\infty }\frac{e^{-\frac{1}{2}%
\sigma ^{2}\tau u^{2}-iku}}{Q\left( u\right) }\upsilon _{0}\left( u\right) du
\\
& =\tau ^{1/2}\frac{1}{\sqrt{2\pi }}\int_{-\infty }^{\infty }\frac{e^{-\frac{%
1}{2}\left( \sigma \tau ^{1/2}u+\frac{ik}{\sigma \tau ^{1/2}}\right) ^{2}}}{%
Q\left( u\right) }\upsilon _{0}\left( u\right) du.
\end{align*}%
A similar formula, albeit derived in a much more complicated way, is given
in \cite{matytsin}.

\subsection{Quadratic Volatility Processes\label{QVPsLL}}

On rare occasions, formulas similar to the LL formula can be used for
spatially inhomogeneous processes as well. The best-known example are the
so-called \emph{quadratic volatility processes} (QVPs), which we shall
briefly describe below. The reader is referred to \cite{lipton-2} and \cite%
{andersen} for further details. The reason why these processes are
considered alongside ELPs is due to the fact that after appropriate
transforms they can be made translationally invariant, see, e.g., \cite%
{lipton-book}, \cite{lipton-2}, \cite{carrlipton}.

When the volatility is quadratic, (including the limiting case when it is
linear), the dynamics of the corresponding underlier is driven by the
following local volatility SDE%
\begin{equation*}
dF\left( t\right) =\sigma _{loc}^{N}\left( F\left( t\right) \right) dW\left(
t\right) ,\ \ \ \ \ F\left( 0\right) =F_{0},
\end{equation*}%
where%
\begin{equation*}
\sigma _{loc}^{N}\left( F\right) =\mathfrak{a}F^{2}+\mathfrak{b}F+\mathfrak{c%
},\ \ \ \ \ \mathfrak{a}>0.
\end{equation*}%
Consider the following quadratic equation 
\begin{equation}
\sigma _{loc}^{N}\left( F\right) =0.  \label{eq:quad_roots}
\end{equation}%
If we want to ensure that $\sigma _{loc}^{N}\left( F\right) >0$ for $%
0<F<\infty $, we have to consider two possibilities: (A)\ (\ref%
{eq:quad_roots}) negative real has roots; (B) (\ref{eq:quad_roots}) has
complex roots. For brevity, we concentrate on the second possibility, which
is most often what real market data dictates, and write%
\begin{equation*}
\sigma _{loc}^{N}\left( F\right) =\mathfrak{a}\left( \left( F-\mathfrak{p}%
\right) ^{2}+\mathfrak{q}^{2}\right) ,\ \ \ \ \ \mathfrak{q>}0.
\end{equation*}%
It turns out that in the case in question the following proposition holds.

\begin{proposition}
\label{Prop:QuadVol} In the case when volatility is quadratic with complex
roots, we can represent the price of a call option using an eigenfunction
expansion representation in the form%
\begin{align}
\frac{C\left( t;\tau ,K\right) }{F_{t}}& =1-\frac{1}{F_{t}\left( \mathfrak{p}%
\sin \left( X_{F_{t}}\right) +\mathfrak{q}\cos \left( X_{F_{t}}\right)
\right) }  \label{eq: LLQV} \\
& \times \sum\limits_{l=1}^{\infty }\frac{e^{-\frac{1}{2}vR\left(
k_{l}\right) }}{R\left( k_{l}\right) }\left( \mathfrak{\zeta }^{s}\sin
\left( 2k_{l}X_{K}\right) -2k_{l}\mathfrak{\zeta }^{c}\cos \left(
2k_{l}X_{K}\right) \right) \sin \left( 2k_{l}X_{F}\right) ,  \notag
\end{align}%
where $F_{t}=F\left( t\right) $, $v=4\mathfrak{a}^{2}\mathfrak{q}^{2}\tau $, 
$k_{l}=\pi l/2X_{\infty }$, and%
\begin{align*}
X_{F_{t}}& =\left( \arctan \left( \frac{F_{t}-\mathfrak{p}}{\mathfrak{q}}%
\right) +\arctan \left( \frac{\mathfrak{p}}{\mathfrak{q}}\right) \right) , \\
X_{K}& =\left( \arctan \left( \frac{K-\mathfrak{p}}{\mathfrak{q}}\right)
+\arctan \left( \frac{\mathfrak{p}}{\mathfrak{q}}\right) \right) , \\
X_{\infty }& =\left( \frac{\pi }{2}+\arctan \left( \frac{\mathfrak{p}}{%
\mathfrak{q}}\right) \right) , \\
\mathfrak{\zeta }^{s}& =\frac{1}{2X_{\infty }}\left( \left( \mathfrak{p}^{2}+%
\mathfrak{q}^{2}-Kp\right) \cos \left( X_{K}\right) +Kq\sin \left(
X_{K}\right) \right) , \\
\mathfrak{\zeta }^{c}& =\frac{1}{2X_{\infty }}\left( \left( \mathfrak{p}^{2}+%
\mathfrak{q}^{2}-Kp\right) \sin \left( X_{K}\right) -Kq\cos \left(
X_{K}\right) \right) .
\end{align*}

\begin{proof}
See \cite{lipton-2}, \cite{andersen}.
\end{proof}

\begin{remark}
Expression (\ref{eq: LLQV}) is more compact, but equivalent, to the one
given in \cite{andersen}. As in \cite{andersen}, solution based on the
method of images is also possible. As usual, the eigenfunction expansion
based solution and the method of images based solution should be used when $%
\tau \rightarrow \infty $ and $\tau \rightarrow 0$, respectively.
\end{remark}

\begin{remark}
It is worth noting that drift-free processes with quadratic volatility are
not martingales, but rather supermatingales, see., e.g., \cite{andersen}. As
a consequence, the solution of the corresponding pricing problem is not
unique and a proper one has to be chosen. Such a choice is implicitly done
above. Non-uniqueness also means that there are (many) non-zero solutions of
the pricing problem with zero initial and boundary conditions. To
demonstrate, assume for brevity that $\mathfrak{p}=\mathfrak{q}=0$, so that 
\begin{equation*}
\sigma _{loc}^{N}\left( F\right) =\mathfrak{a}F^{2},\ \ \ \ \ \mathfrak{a}>0.
\end{equation*}%
The corresponding homogeneous pricing problem can be represented as follows:%
\begin{equation*}
V_{\tau }-\frac{1}{2}\mathfrak{a}^{2}F^{4}V_{FF}=0,
\end{equation*}%
\begin{equation*}
V\left( 0,F\right) =0,\ \ \ V\left( \tau ,0\right) =0.
\end{equation*}%
It can be shown that the generic non-trivial solution of the above problem
has the form%
\begin{equation*}
V\left( \tau ,F;h\right) =\frac{1}{\mathfrak{a}}\int_{0}^{\tau }\frac{%
e^{-\left( 2\mathfrak{a}^{2}F^{2}\left( \tau -\tau ^{\prime }\right) \right)
^{-1}}}{\sqrt{2\pi \left( \tau -\tau ^{\prime }\right) ^{3}}}h\left( \tau
^{\prime }\right) d\tau ^{\prime }.
\end{equation*}%
In particular, when $h\left( \tau \right) =1$, we have%
\begin{equation*}
V\left( \tau ,F;1\right) =2F\Phi \left( -\frac{1}{\mathfrak{a}F\sqrt{\tau }}%
\right) .
\end{equation*}
\end{remark}
\end{proposition}

\subsection{Heston Stochastic Volatility Processes\label{HSVPsLL}}

Another important case when the LL formula can be used successfully is the
so-called \emph{Heston model}\footnote{%
We note in passing that many of our results are applicable for other cases,
for instance, for the Stein-Stein stochastic volatility processes.}. Heston
stochastic volatility processes (HSVPs) are governed by a system of SDEs of
the form%
\begin{align*}
dF\left( t\right) & =\sqrt{\varpi \left( t\right) }F\left( t\right)
dW^{\left( F\right) }\left( t\right) ,\ \ \ \ \ F\left( 0\right) =F_{0}, \\
d\varpi \left( t\right) & =\kappa \left( \theta -\varpi \left( t\right)
\right) dt+\varepsilon \sqrt{\varpi \left( t\right) }dW^{\left( \varpi
\right) }\left( t\right) ,\ \ \ \ \ \varpi \left( 0\right) =\varpi _{0},
\end{align*}%
where $dW^{\left( F\right) }\left( t\right) W^{\left( \varpi \right) }\left(
t\right) =\rho dt$, see \cite{heston}. According to \cite{lipton-book}, \cite%
{lipton-2} for the Heston model we can represent the price of a call option
as follows.

\begin{proposition}
\label{Prop:LLHFormula} The normalized price of a call written on an
underlying driven by a square-root stochastic volatility process has the form%
\begin{equation}
\mathsf{C}\left( \tau ,k\right) =1-\frac{1}{2\pi }\int_{-\infty }^{\infty }%
\frac{E\left( \tau ,u\right) }{Q\left( u\right) }e^{-k\left( iu-\frac{1}{2}%
\right) }du,  \label{eq: LLH}
\end{equation}%
where%
\begin{align}
E\left( \tau ,u\right) & =e^{\mathcal{A}\left( \tau ,u\right) -\mathcal{B}%
\left( \tau ,u\right) \varpi _{0}Q\left( u\right) }\equiv e^{\mathcal{C}%
\left( \tau ,u\right) },  \label{eq: Hest Form} \\
\mathcal{A}\left( \tau ,u\right) & =-\frac{\kappa \theta }{\varepsilon ^{2}}%
\left( \mathcal{F}_{+}\left( u\right) \tau +2\ln \left( \frac{\mathcal{F}%
_{-}\left( u\right) +\mathcal{F}_{+}\left( u\right) \exp \left( -\mathcal{Z}%
\left( u\right) \tau \right) }{2\mathcal{Z}\left( u\right) }\right) \right) ,
\notag \\
\mathcal{B}\left( \tau ,u\right) & =\frac{1-\exp \left( -\mathcal{Z}\left(
u\right) \tau \right) }{\mathcal{F}_{-}\left( u\right) +\mathcal{F}%
_{+}\left( u\right) \exp \left( -\mathcal{Z}\left( u\right) \tau \right) }, 
\notag \\
\mathcal{C}\left( \tau ,u\right) & =\mathcal{A}\left( \tau ,u\right) -%
\mathcal{B}\left( \tau ,u\right) \varpi _{0}Q\left( u\right) ,  \notag
\end{align}%
and%
\begin{align}
\mathcal{F}_{\pm }\left( u\right) & =\pm \left( \rho \varepsilon \left( iu+%
\frac{1}{2}\right) -\kappa \right) +\mathcal{Z}\left( u\right) ,
\label{eq: Hest Term} \\
\mathcal{Z}\left( u\right) & =\sqrt{\left( \rho \varepsilon \left( iu+\frac{1%
}{2}\right) -\kappa \right) ^{2}+\varepsilon ^{2}Q\left( u\right) },  \notag
\end{align}%
or, equivalently,%
\begin{align*}
\mathcal{F}_{\pm }\left( u\right) & =\pm \left( \rho \varepsilon iu-\hat{%
\kappa}\right) +\mathcal{Z}\left( u\right) , \\
\mathcal{Z}\left( u\right) & =\sqrt{\left( \rho \varepsilon iu-\hat{\kappa}%
\right) ^{2}+\varepsilon ^{2}Q\left( u\right) },
\end{align*}%
where $\hat{\kappa}=\kappa -\rho \varepsilon /2.$

\begin{remark}
The martingale condition, which is easy to verify, reads%
\begin{equation*}
\mathcal{A}\left( \tau ,-\frac{i}{2}\right) =0.
\end{equation*}
\end{remark}

\begin{remark}
We note in passing that $\mathcal{C}\left( \tau ,u\right) $ can be
represented in the form 
\begin{equation*}
\mathcal{C}\left( \tau ,u\right) =\theta \mathcal{C}_{1}\left( \tau
,u\right) +\varpi _{0}\mathcal{C}_{2}\left( \tau ,u\right) ,
\end{equation*}%
which emphasizes the contributions of average and instantaneous variance,
respectively.
\end{remark}
\end{proposition}

It is clear that $\mathcal{C}\left( \tau ,u\right) $ in Proposition \ref%
{Prop:LLHFormula} is not a linear function of $\tau $, so that the Heston
process is not a L\'{e}vy process. However, in the limits of $\tau
\rightarrow \infty $ and $\tau \rightarrow 0$ it can be viewed as such.

\begin{proposition}
For $\tau \rightarrow \infty $ we can represent $\mathcal{C}\left( \tau
,u\right) $ as follows%
\begin{equation}
\mathcal{C}\left( \tau ,u\right) =-\frac{\kappa \theta \mathcal{F}_{+}\left(
u\right) \tau }{\varepsilon ^{2}}-\frac{2\kappa \theta }{\varepsilon ^{2}}%
\ln \left( 1-\frac{\mathcal{F}_{+}\left( u\right) }{2\mathcal{Z}\left(
u\right) }\right) -\frac{\varpi _{0}}{\varepsilon ^{2}}\mathcal{F}_{+}\left(
u\right) +O\left( \frac{1}{\tau }\right) .  \label{eq: gammal}
\end{equation}%
For $\tau \rightarrow 0$ we have%
\begin{align}
\mathcal{C}\left( \tau ,\frac{v}{\tau }\right) & =-\frac{i\varpi _{0}v\sinh
\left( \frac{\bar{\rho}\varepsilon v}{2}\right) }{\tau \varepsilon \sinh
\left( \frac{\bar{\rho}\varepsilon v}{2}+i\phi \right) }-\frac{\kappa \theta 
}{\varepsilon ^{2}}\left( i\rho \varepsilon v+2\ln \left( -\frac{i\sinh
\left( \frac{\bar{\rho}\varepsilon v}{2}+i\phi \right) }{\bar{\rho}}\right)
\right)  \label{eq: gammas} \\
& +\frac{i\hat{\kappa}\varpi _{0}v\left( -\frac{\rho \bar{\rho}\varepsilon v%
}{2}+\sinh \left( \frac{\bar{\rho}\varepsilon v}{2}\right) \cosh \left( 
\frac{\bar{\rho}\varepsilon v}{2}+i\phi \right) \right) }{\bar{\rho}%
\varepsilon ^{2}\sinh ^{2}\left( \frac{\bar{\rho}\varepsilon v}{2}+i\phi
\right) ^{2}}+O\left( \tau \right) ,  \notag
\end{align}%
where $\bar{\rho}=\sqrt{1-\rho ^{2}}$, and $\phi =\arctan \left( \bar{\rho}%
/\rho \right) $.

\begin{proof}
Straightforward but tedious calculation leads to (\ref{eq: gammal}). To
derive (\ref{eq: gammas}), we use for inspiration the well-known duality
between the Brownian motions $W\left( \tau \right) $ and $\tau W\left(
1/\tau \right) $, introduce a new variable $v=\tau u$ and then the formula
follows.
\end{proof}
\end{proposition}

It is easy to see from formula (\ref{eq: gammal}) that a Heston process with
zero correlation is asymptotically equivalent to an NIGP. Naturally, in the
long-time limit, to the leading order $\mathcal{C}$ does not depend on $%
\varpi _{0}$. Thus, to the leading order a Heston process can be viewed as a
Levy process but with time inverted. Also notice that in the short-time
limit $\mathcal{C}$ does not depend on $\theta $ to the leading order.

A useful survey of modern approaches to option pricing in the Heston
framework is given in \cite{zeliade}.

\subsection{Calibration to Market Data\label{Calibration}}

For later tests we shall need concrete parametrizations of the various
processes introduced in this paper. For this purpose, let us attempt to
calibrate to the set of market data given in Table \ref{tab:inputs}. Since
we are only interested in the case when parameters are constant in time, it
is not possible to calibrate any of the processes of interest to the \emph{%
entire} set of market quotes. Instead, we shall choose a representative
maturity, $\tau =2y$, say, and calibrate parameters to the set of five
option volatilities. This will allow us to identify proper order of
magnitude for these parameters. We consider six representative ELPs and
present the corresponding calibrated parameters in Table \ref{tab:
calibparam}.

\begin{table}[h]
\centering\noindent 
\makebox[\textwidth]{
    \begin{tabular}{ |c|c|c|c|c|c|c|}
        \hline
        TSP1 & $\alpha=0.66$ & $\sigma =0.07\%$ & $c_{+}=0.1305$ & $c_{-}=0.0615$ & $\kappa _{+}=6.5022$ & $\kappa _{-}=3.0888$ \\
        TSP2 & $\alpha=1.50$ & $\sigma =0.01\%$ & $c_{+}=0.0069$ & $c_{-}=0.0063$ & $\kappa _{+}=1.9320$ & $\kappa _{-}=0.4087$ \\
        MP   & $\lambda=35.33\%$ & $\mu =-0.0318$ & $\eta =0.2023$ & & & \\
	HSVP & $\kappa =2.2707$ & $\theta =0.0225$ & $\varepsilon=0.6200$ & $\rho =-0.0541$ & $\varpi _{0}=0.01374$ & \\
	NIGP & $\sigma =14.90\%$ & $\bar{\chi}=3.20$ &  &  &  & \\
        QVP  & $\mathfrak{a}=1.322$ & $\mathfrak{p}=0.967$ & $\mathfrak{q}=0.301$ &  &  & \\
        \hline
  \end{tabular}
  }
\caption{Parameters for six representative processes calibrated to market
quotes at $\protect\tau =2y$.}
\label{tab: calibparam}
\end{table}

\begin{figure}[h]
\subfigure[TSP,MP,HSVP] {\includegraphics[width=1.0\textwidth, angle=0]
{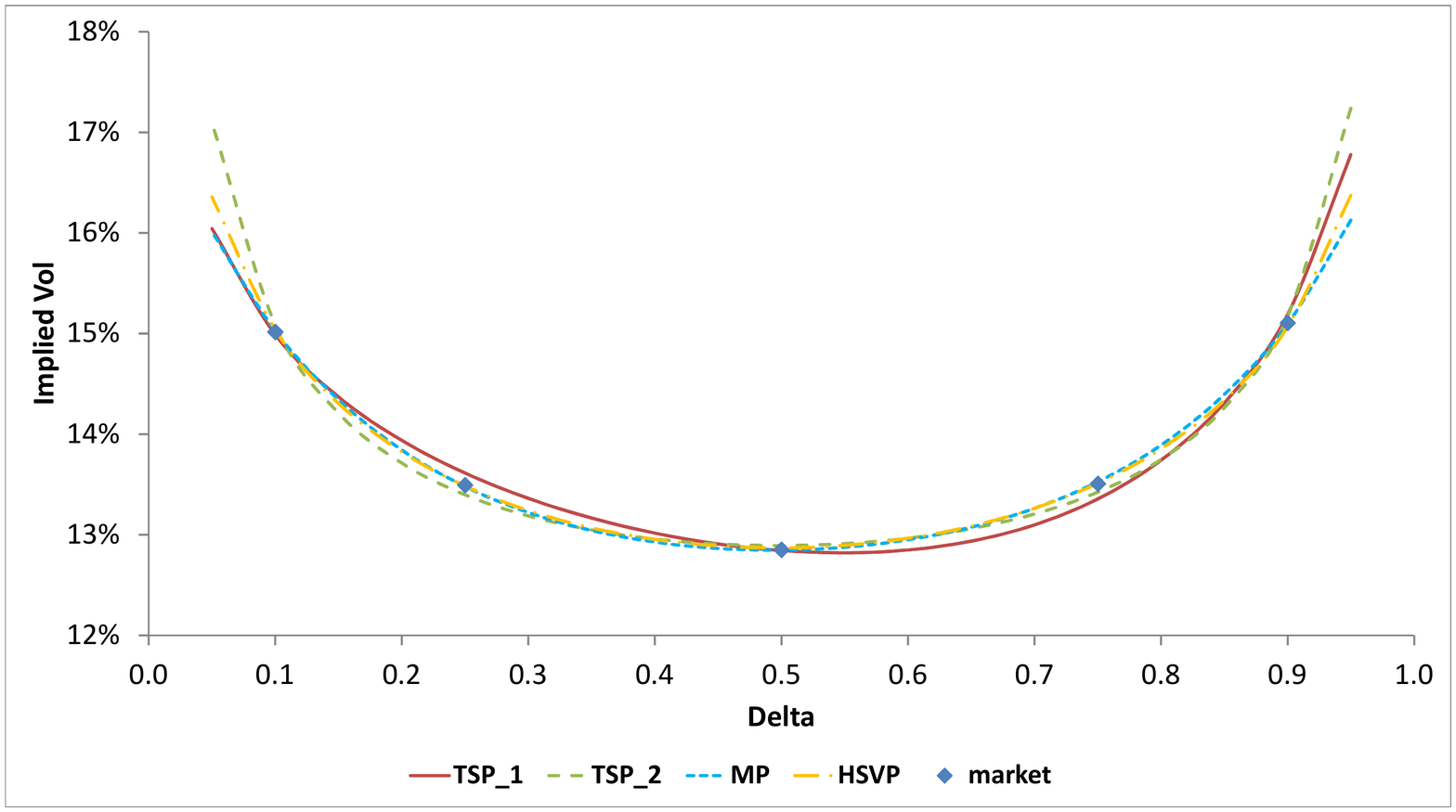}} 
\subfigure[NIGP,QVP] {\includegraphics[width=1.0\textwidth, angle=0]
{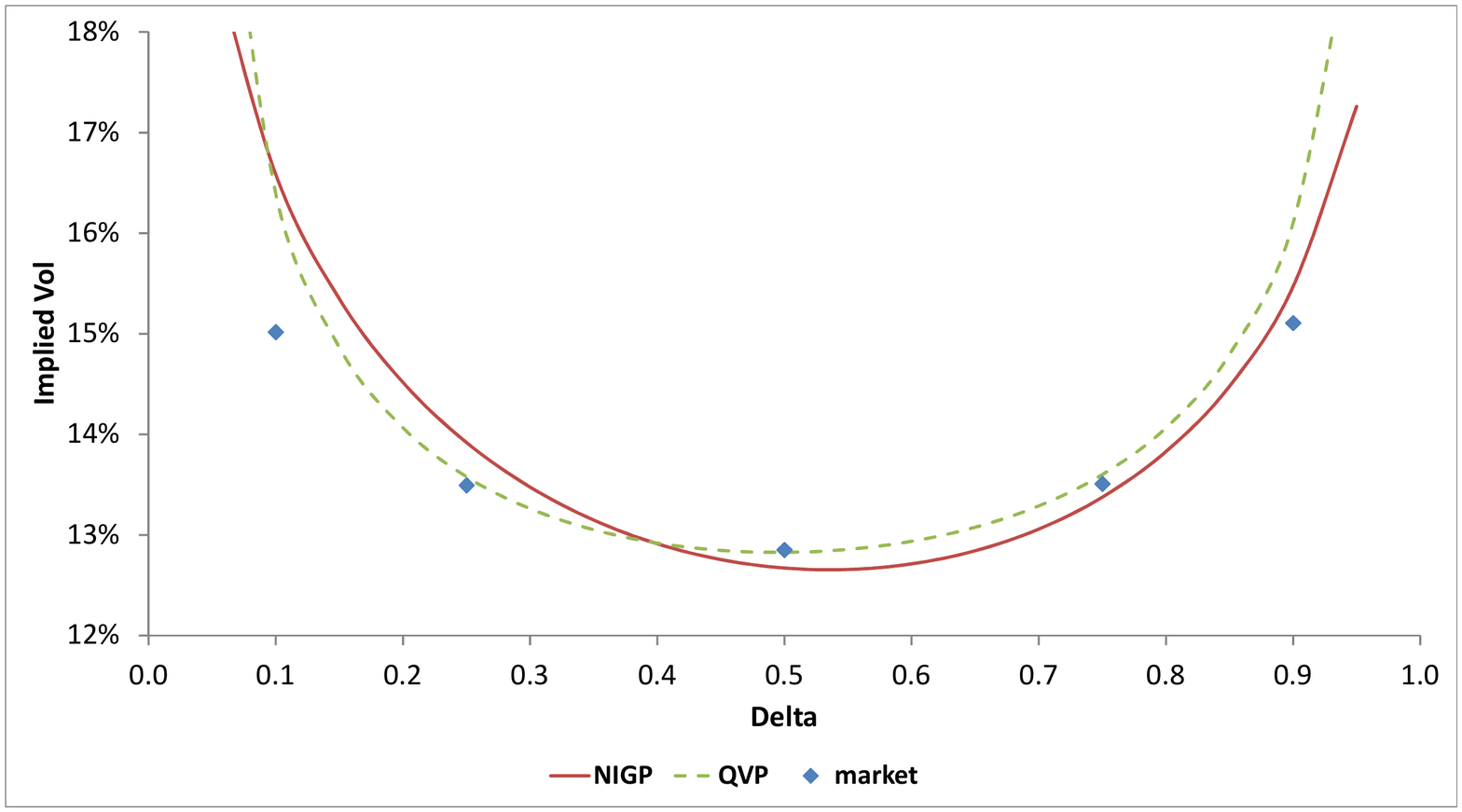}}
\caption{{}$\protect\sigma \left( \Delta \right) $ calibrated to five market
quotes (see Table \protect\ref{tab:inputs}) for four representative
processes. The corresponding maturity is 2y.}
\label{fig: calibration}
\end{figure}

The quality of the calibration for two TSPs, MPs and HSVPs, which is fairly
high, is shown in Figure \ref{fig: calibration}(a). Although for NIGPs and
QVPs do not have enough parameters to ensure successful calibration to the
market, our choice of parameters generates satisfactory (but not perfect)
fits shown in Figure \ref{fig: calibration}(b). 

\clearpage

\section{Asymptotics of the Option Price in the Black-Scholes-Merton
Framework \label{BSAsymptotics}}

In Sections \ref{AsymptoticsLong}, \ref{AsymptoticsShort}, \ref%
{AsymptoticsWing} we shall use the LL formula of Section \ref{LLFormula} to
establish a variety of asymptotic expressions for option prices and implied
volatilities. Before this, however, we take a brief detour into the analysis
of the limiting behavior of BS\ prices when variances $v$ and strikes $k$
are either large or small. This analysis is required later in order to
convert option price asymptotics into implied volatility asymptotics.

As was mentioned earlier, the undiscounted price of a call option can be
written in the form $\mathsf{C}^{BS}\left( v,k\right) $, where $v$ is the
annualized variance, and $k$ is the log strike. The well-known relations 
\begin{equation}
\Phi \left( x\right) =1-\Phi \left( -x\right) ,\ \ \ \ \ \Phi \left(
x\right) \underset{x\rightarrow -\infty }{\sim }-\frac{\phi \left( x\right) 
}{x}\left( 1-\frac{1}{x^{2}}\right) ,  \label{eq:cum_norm_as}
\end{equation}%
allow us to simplify the above formula in various asymptotic regimes.
Specifically, we are interested in the following three cases: A) $%
v\rightarrow \infty $, $k=\hat{k}v$, $\hat{k}$ is fixed (long-time
asymptotics); (B) $v\rightarrow 0$, $k$ is fixed (short-time asymptotics);
(C) $v=\hat{v}\left\vert k\right\vert $, $\left\vert k\right\vert
\rightarrow \infty $, $0\leq \hat{v}\leq 2$ (wing asymptotics). Specifically
we have the following proposition:

\begin{proposition}
\label{BS_asymptotics}In case (A) we have%
\begin{equation}
\mathsf{C}^{BS}\left( v,\hat{k}v\right) =\left\{ 
\begin{array}{ll}
\begin{array}{l}
\left( 1-e^{\hat{k}v}\right) \mathbf{1}_{\left\{ \hat{k}<-\frac{1}{2}%
\right\} }+\mathbf{1}_{\left\{ -\frac{1}{2}<\hat{k}<\frac{1}{2}\right\} } \\ 
+\frac{\phi \left( \left( \hat{k}-\frac{1}{2}\right) \sqrt{v}\right) }{%
R\left( \hat{k}\right) \sqrt{v}}\left( 1-\frac{1}{v}\frac{1}{R\left( \hat{k}%
\right) }\left( 3+\frac{1}{R\left( \hat{k}\right) }\right) \right) +...,%
\end{array}
& \hat{k}\neq \pm \frac{1}{2}, \\ 
\begin{array}{l}
\frac{1}{2}\mathbf{1}_{\left\{ \hat{k}=\frac{1}{2}\right\} }+\left( 1-\frac{1%
}{2}e^{-v/2}\right) \mathbf{1}_{\left\{ \hat{k}=-\frac{1}{2}\right\} } \\ 
-\frac{e^{\left( \hat{k}-1/2\right) v/2}}{\sqrt{v}}\left( 1-\frac{1}{v}%
\right) +...,%
\end{array}
& \hat{k}=\pm \frac{1}{2},%
\end{array}%
\right.  \label{eq:BS_asym_A}
\end{equation}%
where 
\begin{equation}
R\left( \hat{k}\right) =-Q\left( i\hat{k}\right) =\hat{k}^{2}-\frac{1}{4}.
\label{eq: R}
\end{equation}%
In case (B) we have%
\begin{equation}
\mathsf{C}^{BS}\left( v,k\right) =\left( 1-e^{k}\right) ^{+}+\phi \left( 
\frac{k}{\sqrt{v}}\right) \frac{e^{\frac{k}{2}}v^{3/2}}{k^{2}}\left( 1-\frac{%
1}{8}v\right) +....  \label{eq:BS_asym_B}
\end{equation}%
Finally, in case (C) we have%
\begin{equation}
\mathsf{C}^{BS}\left( \hat{v}\left\vert k\right\vert ,k\right) =\left(
1-e^{k}\right) ^{+}+\frac{\phi \left( \left( 1/\sqrt{\hat{v}}-\sqrt{\hat{v}}%
/2\right) \sqrt{\left\vert k\right\vert }\right) }{R\left( 1/\hat{v}\right) 
\sqrt{\hat{v}\left\vert k\right\vert }}\left( 1-\frac{3/\hat{v}^{2}+4}{%
\left( R\left( 1/\hat{v}\right) \right) ^{2}\hat{v}\left\vert k\right\vert }%
\right) +....  \label{eq:BS_asym_C}
\end{equation}

\begin{proof}
Simple but slightly tedious application of (\ref{eq:cum_norm_as}) to BS
formula (\ref{eq: BS non-dim}), see, e.g., \cite{tehranchi}, \cite%
{lopez-forde-jacquier}, \cite{gao}.
\end{proof}
\end{proposition}

\begin{remark}
For case (A), it is clear that cases $\hat{k}=\pm 1/2$ require separate
treatment, with $\hat{k}=1/2$ being particularly important since in order to
get $\Delta \sim 1/2$, we have to have $\hat{k}-1/2\sim 1/\sqrt{v}$, in
which case%
\begin{equation*}
\Delta \sim \Phi \left( \left( \hat{k}-\frac{1}{2}\right) \sqrt{v}\right)
\sim \frac{1}{2}.
\end{equation*}
\end{remark}

\begin{remark}
For future reference, we wish to generalize formulas (\ref{eq:BS_asym_A})
and (\ref{eq:BS_asym_B}) and analyze the asymptotics of $\mathsf{C}%
^{BS}\left( \mathsf{g}\left( v\right) v,\hat{k}v\right) $, with $\mathsf{g}%
\left( v\right) =\left( 1+\alpha _{1}/v+\alpha _{2}/v^{2}\right) $, $\hat{k}$
is fixed and $v\rightarrow \infty $, and $\mathsf{C}^{BS}\left( \mathsf{f}%
\left( v\right) v,k\right) $ with $\mathsf{f}\left( v\right) =\left(
1+\alpha _{1}v+\alpha _{2}v^{2}\right) $, $k\neq 0$ is fixed and $%
v\rightarrow 0$. Straightforward computation yields%
\begin{align}
& \mathsf{C}^{BS}\left( \mathsf{g}\left( v\right) v,\hat{k}v\right) =\mathbf{%
1}_{\left\{ \hat{k}<-\frac{1}{2}\right\} }\left( 1-e^{\hat{k}v}\right) +%
\mathbf{1}_{\left\{ -\frac{1}{2}<\hat{k}<\frac{1}{2}\right\} }+e^{\frac{1}{2}%
\alpha _{1}R\left( \hat{k}\right) }\frac{\phi \left( \left( -\hat{k}+\frac{1%
}{2}\right) \sqrt{v}\right) }{R\left( \hat{k}\right) \sqrt{v}}
\label{eq: BS ass 1} \\
& \ \ \ \ \ \ \times \left( 1-\frac{1}{v}\left( \left( \frac{1}{R\left( \hat{%
k}\right) }-\frac{\alpha _{1}}{2}\right) \left( 3+\frac{1}{R\left( \hat{k}%
\right) }\right) +\frac{\alpha _{1}^{2}}{8}+\frac{R\left( \hat{k}\right) }{2}%
\left( \alpha _{1}^{2}-\alpha _{2}\right) \right) \right) ,  \notag
\end{align}%
and%
\begin{align}
\mathsf{C}^{BS}\left( \mathsf{f}\left( v\right) v,k\right) & =\left(
1-e^{k}\right) ^{+}+\phi \left( \frac{k}{\sqrt{v}}\right) \frac{e^{\frac{k}{2%
}+\frac{\alpha _{1}k^{2}}{2}}v^{3/2}}{k^{2}}  \label{eq: BS ass 2} \\
& \times \left( 1-\frac{1}{8}\left( 1-12\alpha _{1}+4\left( \alpha
_{1}^{2}-\alpha _{2}\right) k^{2}\right) v\right) +....  \notag
\end{align}%
These formulas are used below for studying the asymptotic behavior of the
implied volatility in the long-time and short-time limits, respectively. A
more complicated expression, which is equivalent to (\ref{eq: BS ass 1}) is
given by \cite{lopez-forde-jacquier}.
\end{remark}

\section{Long-time Asymptotics\label{AsymptoticsLong}}

\subsection{General Remarks\label{GeneralRemaksLong}}

For the purpose of establishing long-time implied volatility asymptotics via
the LL formula, let us briefly introduce the so-called \textit{saddlepoint
method}.

Saddlepoint approximation is a method for computing integrals of the form 
\begin{equation}
g\left( \tau \right) =\frac{1}{2\pi }\int\limits_{\mathfrak{C}}f\left(
z\right) e^{\tau S\left( z\right) }dz  \label{eq:int_rep}
\end{equation}%
when $\tau \rightarrow +\infty $. Here $\mathfrak{C}$ is a contour in the
complex plane, and the amplitude and phase functions $f\left( z\right)
,S\left( z\right) $ are analytic is a domain $\mathcal{D}$ containing $%
\mathfrak{C}$. The extremal points of $S$, i.e. zeroes of $S^{\prime }$, are
called the \textit{saddlepoints} of $S$. Under reasonable conditions, the
contribution to $g\left( \tau \right) $ from a vicinity of a saddlepoint $%
z_{0}$, where $S^{\prime \prime }\left( z_{0}\right) \neq 0$, is given by 
\begin{equation}
g_{z_{0},0}\left( \tau \right) =\frac{1}{\sqrt{-2\pi \tau S^{\prime \prime
}\left( z_{0}\right) }}e^{\tau S\left( z_{0}\right) }f\left( z_{0}\right)
\left( 1+O\left( \tau ^{-1}\right) \right) ,  \label{eq:saddle_0}
\end{equation}%
see, e.g., \cite{fedoryuk}, \cite{luggannani}, \cite{jensen}, It is clear
that the main contribution to the integral comes from the saddlepoints where 
$\func{Re}\left[ S\right] $ attains its absolute maximum. The second-order
approximation has the form%
\begin{equation}
g_{z_{0},1}\left( \tau \right) =g_{z_{0},0}\left( \tau \right) \left( 1-%
\frac{1}{2\tau }\left( \frac{\left( \frac{f^{\prime }\left( z_{0}\right) }{%
S^{\prime \prime }\left( z_{0}\right) }\right) ^{\prime }}{f\left(
z_{0}\right) }+\frac{5\left( S^{^{\prime \prime \prime }}\left( z_{0}\right)
\right) ^{2}}{12\left( S^{\prime \prime }\left( z_{0}\right) \right) ^{3}}-%
\frac{S^{^{\prime \prime \prime \prime }}\left( z_{0}\right) }{4\left(
S^{\prime \prime }\left( z_{0}\right) \right) ^{2}}\right) +O\left( \tau
^{-2}\right) \right) .  \label{eq:saddle_1}
\end{equation}

Saddlepoint approximation has been successfully used by many authors for a
variety of financial applications, see, e.g., \cite{rogers-2}, \cite{lewis-1}%
, \cite{carrmadan2}, \cite{martin}, among others.

\subsection{BS Asymptotics via the Saddlepoint Approximation\label{BSSaddle}}

To motivate subsequent developments, let us briefly derive some large $v$
asymptotic results for the BS case. We are specifically interested in the
case when $v\rightarrow \infty $, $k=\hat{k}v$, $\hat{k}$ is fixed. We have%
\begin{equation*}
\mathsf{C}^{BS}\left( v,\hat{k}v\right) =1-\frac{1}{2\pi }\int_{-\infty
}^{\infty }\frac{e^{-v\left( \frac{1}{2}Q\left( u\right) +\hat{k}\left( iu-%
\frac{1}{2}\right) \right) }}{Q\left( u\right) }du=1-\frac{1}{2\pi }%
\int_{-\infty }^{\infty }e^{vS\left( u,\hat{k}\right) }f\left( u\right) du,
\end{equation*}%
where%
\begin{equation*}
S\left( u,\hat{k}\right) =-\frac{1}{2}Q\left( u\right) -\hat{k}\left( iu-%
\frac{1}{2}\right) ,\ \ \ \ \ f\left( u\right) =\frac{1}{Q\left( u\right) }.
\end{equation*}%
The integrand is meromorphic in the entire complex plain where it has two
simple poles located at the points%
\begin{equation*}
u_{\pm }=\pm \frac{i}{2},\ \ \ S\left( \frac{i}{2},\hat{k}\right) =\hat{k},\
\ \ S\left( -\frac{i}{2},\hat{k}\right) =0.
\end{equation*}%
We calculate the location of the saddlepoint $u_{\ast }$ by solving the
equation%
\begin{equation*}
S^{\prime }\left( u,\hat{k}\right) =-u-i\hat{k}=0,
\end{equation*}%
so that%
\begin{equation*}
u_{\ast }=-i\hat{k},\ \ \ S^{\prime \prime }\left( u_{\ast }\right) =-1,\ \
\ S\left( u_{\ast }\right) =-\frac{1}{2}\left( \hat{k}-\frac{1}{2}\right)
^{2},\ \ \ f\left( u_{\ast }\right) =-\frac{1}{R\left( \hat{k}\right) }.
\end{equation*}%
and $R(\hat{k})$ is given by (\ref{eq: R}). In order to apply the
saddlepoint approximation, we need to transform the contour of integration
in such a way that it passes through the saddlepoint, so that its immediate
vicinity dominates the entire integral. We can achieve this goal by parallel
shift of the contour of integration from the real axis to the contour given
by 
\begin{equation*}
\func{Im}\left( u\right) =-\hat{k}.
\end{equation*}%
In the process of doing so, three possibilities might occur: (A) the lower
pole $u_{-}$ is crossed (OTM case); (B) no poles are crossed (near-ATM
case); (C) the upper pole $u_{-}$ is crossed (ITM case). When poles are
crossed, their contributions have to be computed via the Cauchy formula:%
\begin{equation*}
\Pi _{+}=e^{\hat{k}v},\ \ \ \ \ \Pi _{-}=1.
\end{equation*}%
The contribution of the saddlepoint has the form (\ref{eq:saddle_1}).
Summing up all these contributions, we obtain formula (\ref{eq:BS_asym_A})
of Section \ref{BSAsymptotics}, where it is derived by elementary means.

\subsection{Exponential L\'{e}vy Processes\label{ELPsLong}}

We can now consider more general LPs.

\subsubsection{Generic Exponential L\'{e}vy Processes\label{GenericELPsLong}}

For a generic ELP the\ LL formula yields 
\begin{equation*}
C\left( \tau ,k\right) =1-\frac{1}{2\pi }\int_{-\infty }^{\infty }e^{\tau
S\left( u\right) }f\left( u,k\right) du,
\end{equation*}%
where 
\begin{equation*}
S\left( u\right) =\upsilon _{0}\left( u\right) -\frac{1}{2}\sigma
^{2}Q\left( u\right) ,\ \ \ \ \ f\left( u,k\right) =\frac{e^{-k\left( iu-%
\frac{1}{2}\right) }}{Q\left( u\right) },
\end{equation*}%
where $\upsilon _{0}\left( u\right) $ is given by (\ref{eq: LLpsi}). Thus,
for large $\tau $ the integral of interest has the form (\ref{eq:int_rep}),
and can be analyzed via formulas (\ref{eq:saddle_0}) or (\ref{eq:saddle_1}).
However, it is more natural to assume that $k\sim \bar{k}\tau $, so that the
strike moves deeper in or out of the money when maturity increases (this
case is necessary to consider in order to study the asymptotic behavior of
volatility as a function of delta). Under this assumption we have%
\begin{equation*}
\mathsf{C}\left( \tau ,k\right) =1-\frac{1}{2\pi }\int_{-\infty }^{\infty
}e^{\tau S\left( u,\bar{k}\right) }f\left( u\right) du,
\end{equation*}%
where%
\begin{equation*}
S\left( u,\bar{k}\right) =\upsilon _{0}\left( u\right) -\frac{1}{2}\sigma
^{2}Q\left( u\right) -\bar{k}\left( iu-\frac{1}{2}\right) ,\ \ \ \ \ f\left(
u\right) =\frac{1}{Q\left( u\right) }.
\end{equation*}

Consider now the general case of ELPs. Since we are dealing with shifted
characteristic functions, we can use symmetry and prove that the saddlepoint
of interest is located on the imaginary axis. Accordingly, on the interval
of analyticity of $S\left( u\right) $, given by the inequalities $\mathsf{Y}%
_{-}<\func{Im}\left( u\right) <\mathsf{Y}_{+}$, we can define a real-valued
function $\Xi _{01}\left( y,\bar{k}\right) $ of real arguments $y,\bar{k}$
(since on the imaginary axis the value of $S\left( u,\bar{k}\right) $ is
real, see, e.g., \cite{lewis-1}, \cite{lukacs}) as follows%
\begin{equation*}
\Xi _{01}\left( y,\bar{k}\right) =\Xi _{0}\left( y\right) +\bar{k}\left( y+%
\frac{1}{2}\right) ,
\end{equation*}%
where%
\begin{equation}
\Xi _{0}\left( y\right) =\upsilon _{0}\left( iy\right) +\frac{1}{2}\sigma
^{2}R\left( y\right) ,  \label{eq_Xi_0}
\end{equation}%
and $R\left( y\right) $ is given by (\ref{eq: R}). It is clear that we can
find the location of the saddlepoint by solving the following equation%
\begin{equation}
\Xi _{0}^{\prime }\left( y\right) +\bar{k}=0.  \label{eq: saddle_tsp}
\end{equation}%
Since, in general, this equation cannot be solved analytically, we prefer a
parametric approach, by expressing $\bar{k}$ in terms of $y$, rather than
the other way around. This approach leads to the following proposition.

\begin{proposition}
\label{GeneralLongProp}Consider $y\in \left( -\mathsf{Y}_{+},\mathsf{Y}%
_{-}\right) $ and define $\Xi _{0}\left( y\right) ,\ \Xi _{1}\left( y\right) 
$ and $\Xi _{01,\pm }\left( y\right) $ as follows%
\begin{align}
\Xi _{0}\left( y\right) & =\upsilon _{0}\left( iy\right) +\frac{1}{2}\sigma
^{2}R\left( y\right) ,\ \ \ \ \ \Xi _{1}\left( y\right) =\Xi _{0}^{\prime
}\left( y\right) ,  \label{eq:xi0} \\
\Xi _{01,\pm }\left( y\right) & =\Xi _{0}\left( y\right) -\Xi _{1}\left(
y\right) \left( y\pm \frac{1}{2}\right) .  \notag
\end{align}%
Then for $\bar{k}=-\Xi _{1}\left( y\right) $ the corresponding $\sigma
_{imp}\left( \tau ,\tau \bar{k}\right) $ can be written in the form%
\begin{equation}
\sigma _{imp}\left( \tau ,\tau \bar{k}\right) =\left( a_{0}\left( y\right) +%
\frac{a_{1}\left( y\right) }{\tau }+\frac{a_{2}\left( y\right) }{\tau ^{2}}%
+...\right) ^{1/2},  \label{eq: sigma imp}
\end{equation}%
so that $\sigma _{imp}\left( \tau ,\tau \bar{k}\right) $ can be represented
as follows%
\begin{equation}
\sigma _{imp}\left( \tau ,\tau \bar{k}\right) =b_{0}\left( y\right) +\frac{%
b_{1}\left( y\right) }{\tau }+\frac{b_{2}\left( y\right) }{\tau ^{2}}+....
\label{eq: sigma imp series}
\end{equation}%
where%
\begin{align}
a_{0}\left( y\right) & =\left( \mathrm{sign}\left( y+\frac{1}{2}\right) 
\sqrt{-2\Xi _{01,+}\left( y\right) }-\mathrm{sign}\left( y-\frac{1}{2}%
\right) \sqrt{-2\Xi _{01,-}\left( y\right) }\right) ^{2},  \label{eq: a_0} \\
b_{0}\left( y\right) & =a_{0}^{1/2}\left( y\right) ,  \label{eq: b_0}
\end{align}%
and higher order coefficients $a_{i}\left( y\right) $ and $b_{i}\left(
y\right) $ have the form (\ref{eq: a_i}), (\ref{eq: b_i}).

\begin{proof}
See Appendix \ref{ELPsLA}.
\end{proof}
\end{proposition}

Fundamentally, Proposition \ref{GeneralLongProp} is derived by comparing the
asymptotic expansion obtained via the saddlepoint method with the asymptotic
expansion for BS price given by (\ref{eq: BS ass 1}), and balancing terms.
Below we use the following notation 
\begin{equation}
\sigma _{imp,i}\left( y\right) =\dsum\limits_{j=0}^{i}\frac{b_{j}\left(
y\right) }{\tau ^{j}},  \label{eq:sigma_i_long}
\end{equation}%
for partial sums of an asymptotic series (\ref{eq: sigma imp series}), and
similarly in other instances.

\begin{remark}
While in our approach Proposition \ref{GeneralLongProp} follows from
asymptotic formulas for complex integrals, it can be obtained via the large
deviation principle as well, see, e.g., \cite{varadan1}, \cite{varadan}, 
\cite{dembo}.
\end{remark}

\subsubsection{Specific Exponential L\'{e}vy Processes\label%
{ConcreteELPsLong}}

For specific ELPs, such as TSPs, NIGPs, MPs, etc., the\ corresponding
formulas can be made explicit.

\begin{proposition}
For TSPs, NIGPs, and MPs Proposition \ref{GeneralLongProp} holds provided
that the corresponding $\Xi _{0}\left( y\right) $ are defined as follows:%
\begin{align}
\Xi _{0}^{TS}\left( y\right) & =\frac{1}{2}\sigma ^{2}R\left( y\right)
+\sum_{s=\pm }a_{s}\left( \kappa _{s}+s\left( y-\frac{1}{2}\right) \right)
^{\alpha }-\gamma \left( y-\frac{1}{2}\right) +\delta ,  \label{eq: xi0} \\
\Xi _{0}^{NIG}\left( y\right) & =\sigma ^{2}\mathcal{\bar{\varkappa}}\left( 
\mathcal{\bar{\varkappa}}-\sqrt{\mathcal{\bar{\omega}}^{2}-y^{2}}\right) , 
\notag \\
\Xi _{0}^{M}\left( y\right) & =\frac{1}{2}\sigma ^{2}R\left( y\right)
+\lambda \left( e^{-q\left( y-\frac{1}{2}\right) +\frac{\eta ^{2}}{2}R\left(
y\right) }-1-\left( 1-e^{q}\right) \left( y-\frac{1}{2}\right) \right) . 
\notag
\end{align}%
For TSPs, $y\in \left( -\kappa _{+}+\frac{1}{2},\kappa _{-}+\frac{1}{2}%
\right) $; for NIGPs, $y\in \left( -\bar{\omega},\bar{\omega}\right) $; for
MPs, $y\in \left( -\infty ,\infty \right) $.

\begin{proof}
Proof is straightforward. For further details, see Appendices \ref{TSPsLA}, %
\ref{NIGPsLA}, \ref{MPsLApp}, respectively.
\end{proof}
\end{proposition}

\begin{figure}[h]
\subfigure[T=10y] {\includegraphics[width=1.0\textwidth, angle=0]
{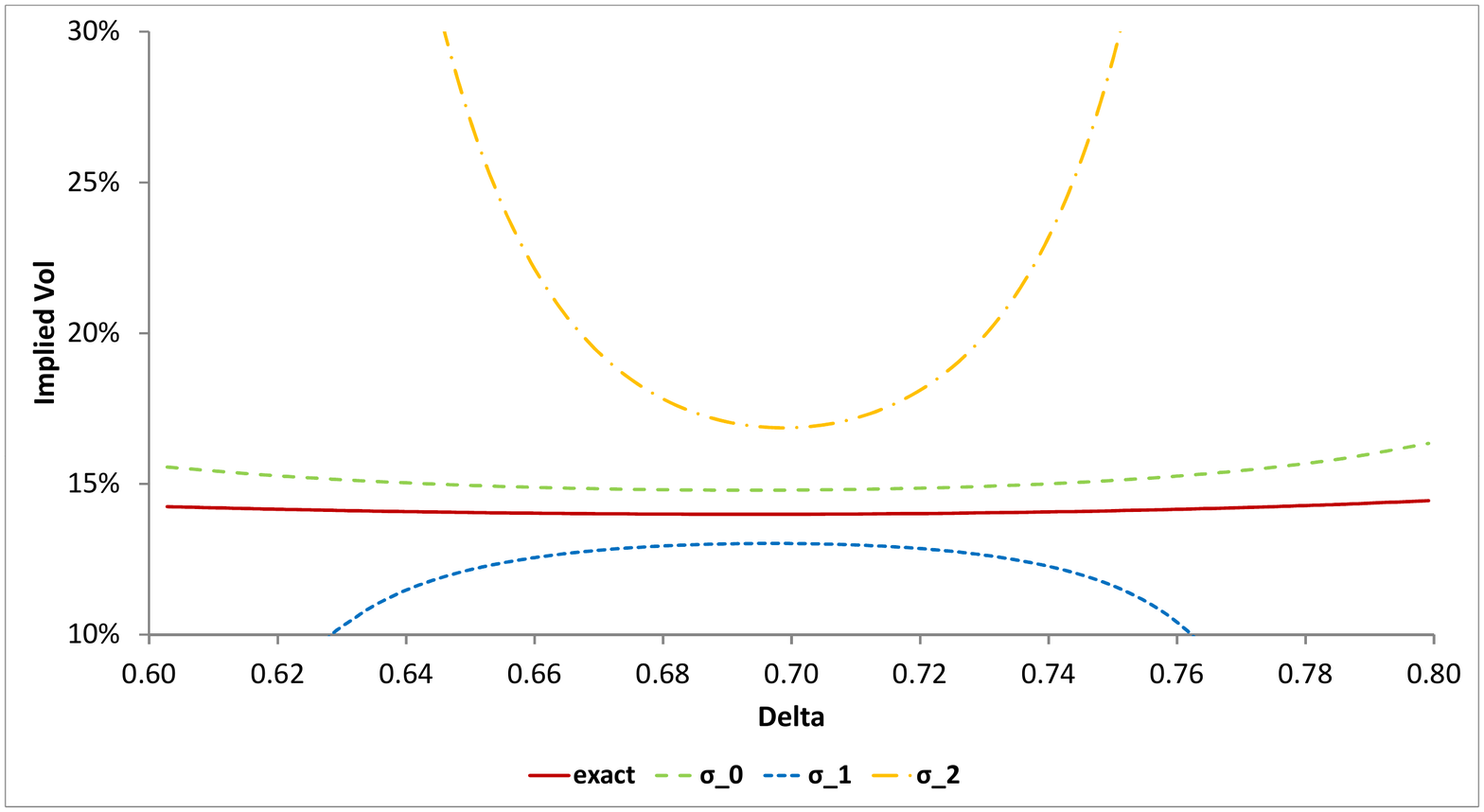}} 
\subfigure[T=50y] {\includegraphics[width=1.0\textwidth, angle=0]
{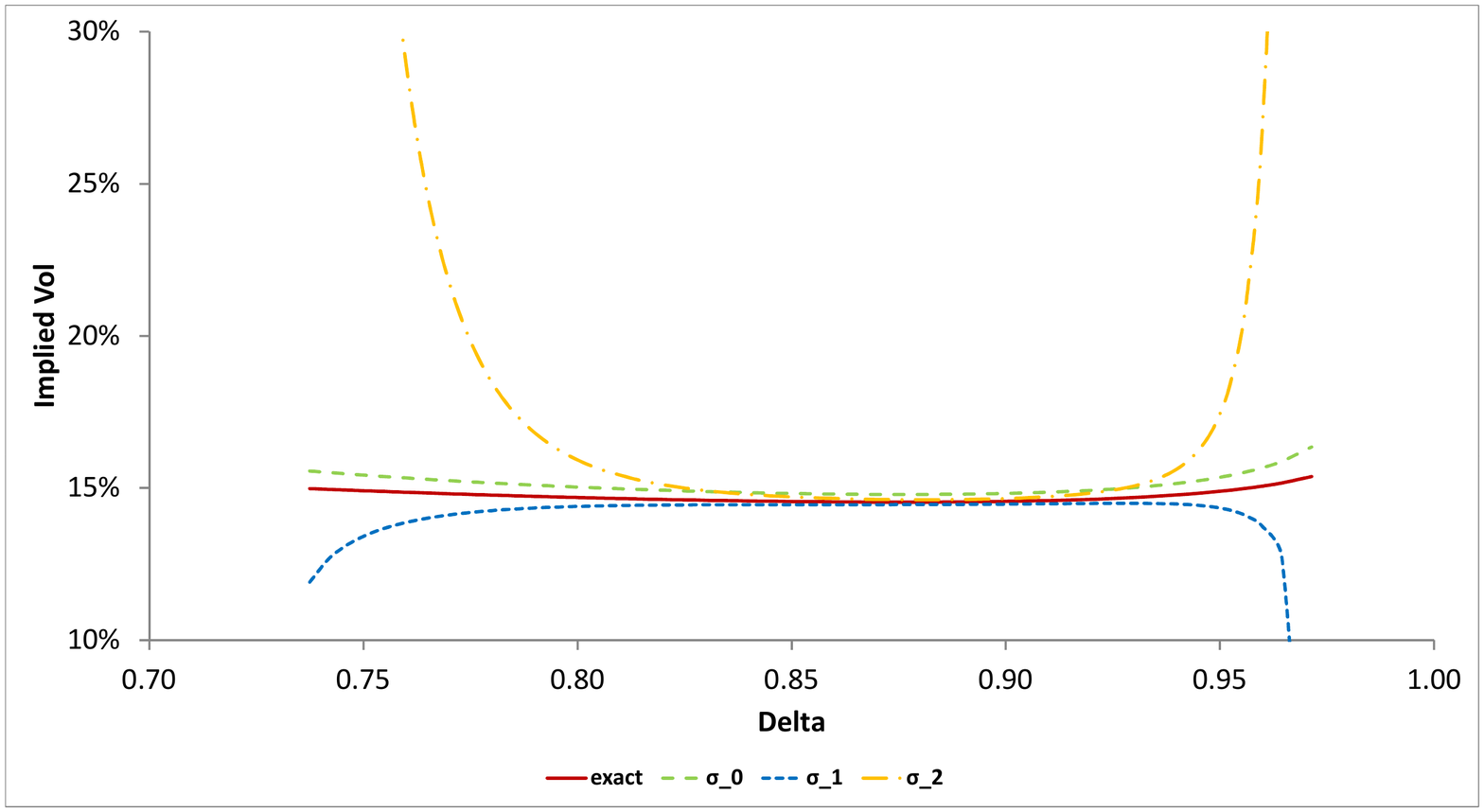}}
\caption{Comparison of the exact and asymptotic expressions for $\protect%
\sigma _{imp}$ for the calibrated TSP for different maturities, $\protect%
\alpha =3/2$.}
\label{fig:TSPsLong_150}
\end{figure}

\begin{figure}[h]
\subfigure[T=2y] {\includegraphics[width=1.0\textwidth, angle=0]
{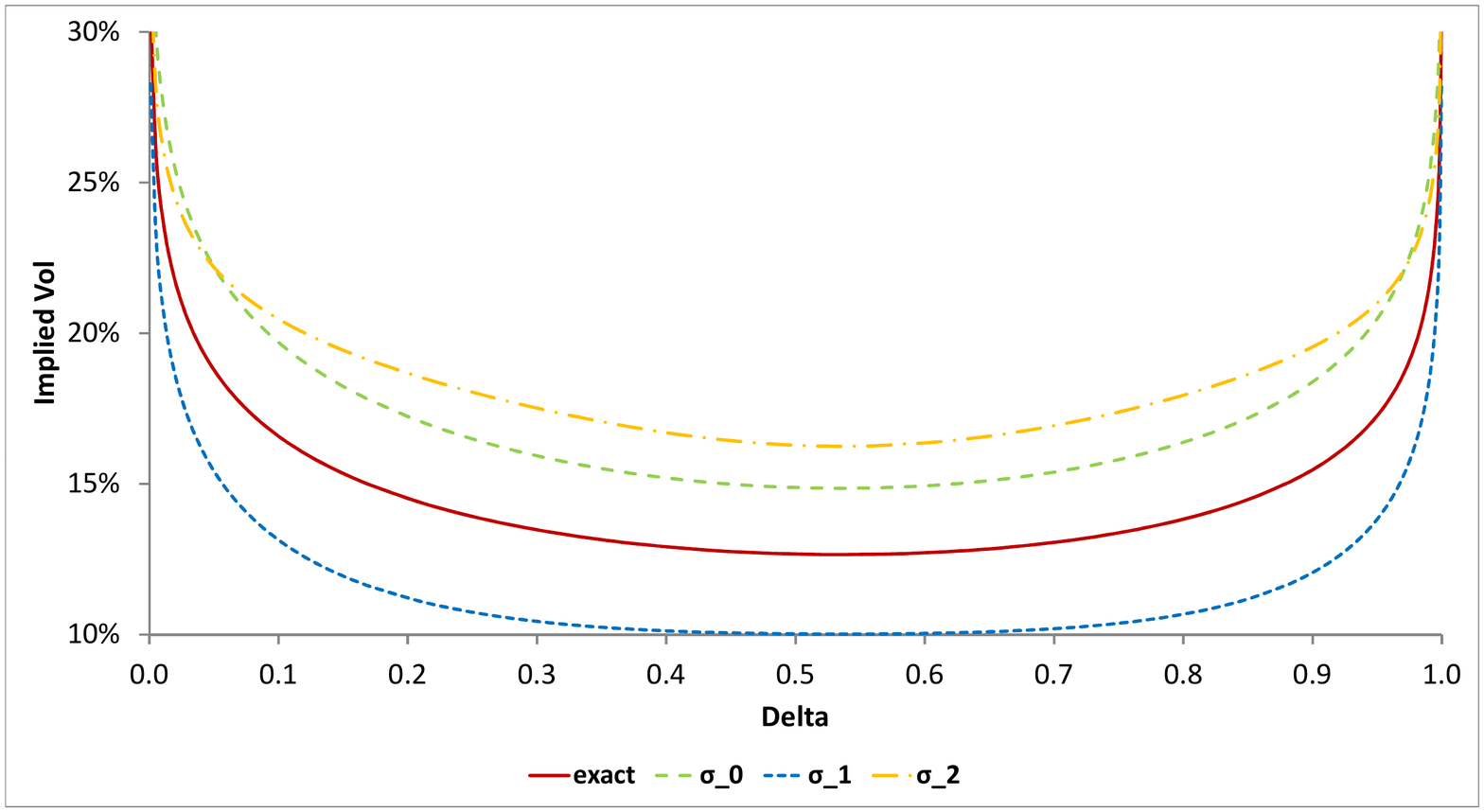}} 
\subfigure[T=10y] {\includegraphics[width=1.0\textwidth, angle=0]
{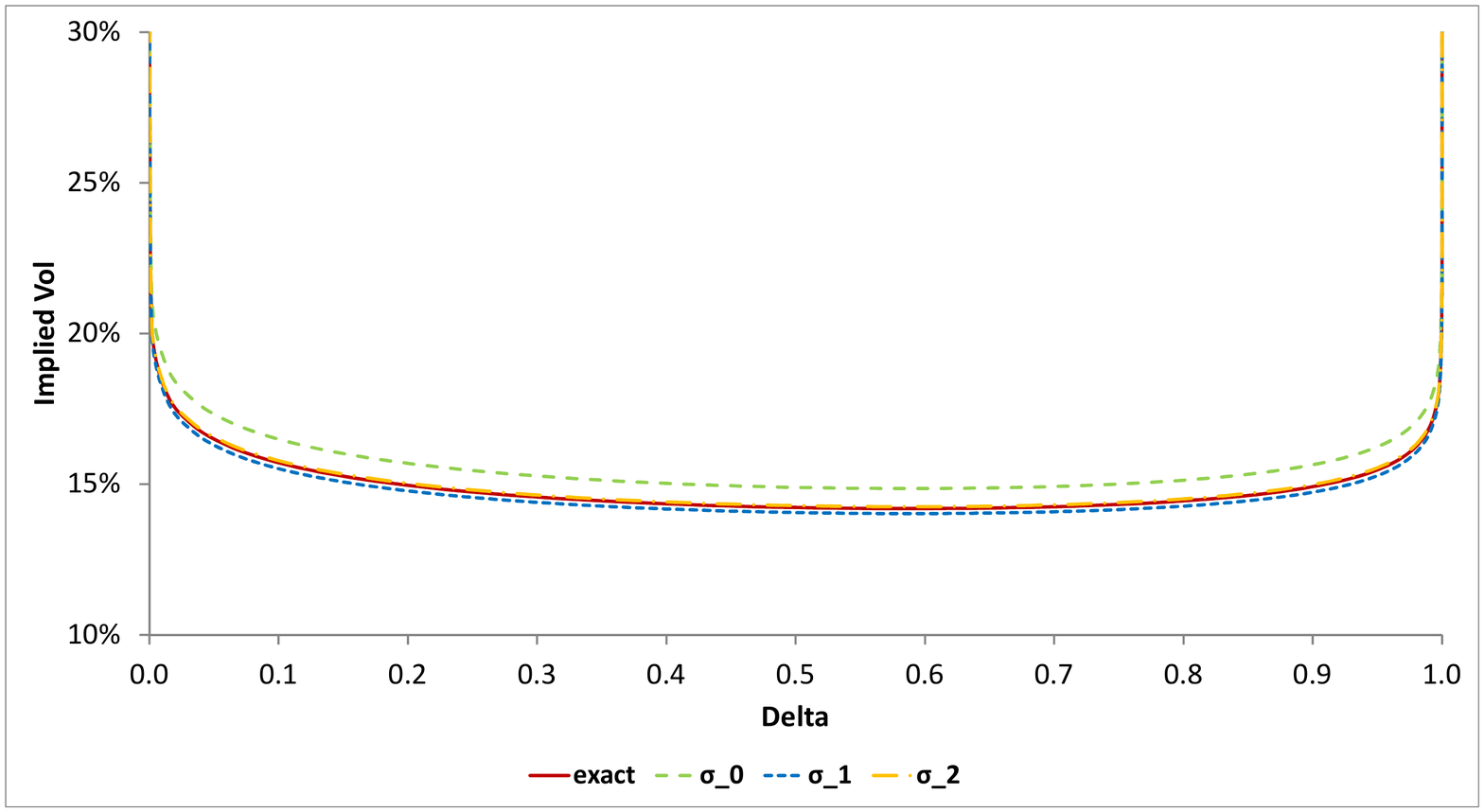}}
\caption{{}Comparison of the exact and asymptotic expressions for $\protect%
\sigma _{imp}$ for the calibrated NIGP for different maturities.}
\label{fig:NIGPsLong}
\end{figure}

The quality of the saddlepoint approximation in the limit of infinite
maturity for representative TSP, NIGP, and MP is illustrated in Figures \ref%
{fig:TSPsLong_066}, \ref{fig:TSPsLong_150}, \ref{fig:NIGPsLong}, and \ref%
{fig:MPsLong}, respectively. The relevant parameters are calibrated to the
market. It is clear that for the calibrated TSP with $\alpha =2/3$, NIGP,
and MP saddlepoint approximation is very good, while for the calibrated TSP
with $\alpha =3/2$ it fails.

\begin{figure}[h]
\subfigure[T=2y] {\includegraphics[width=1.0\textwidth, angle=0]
{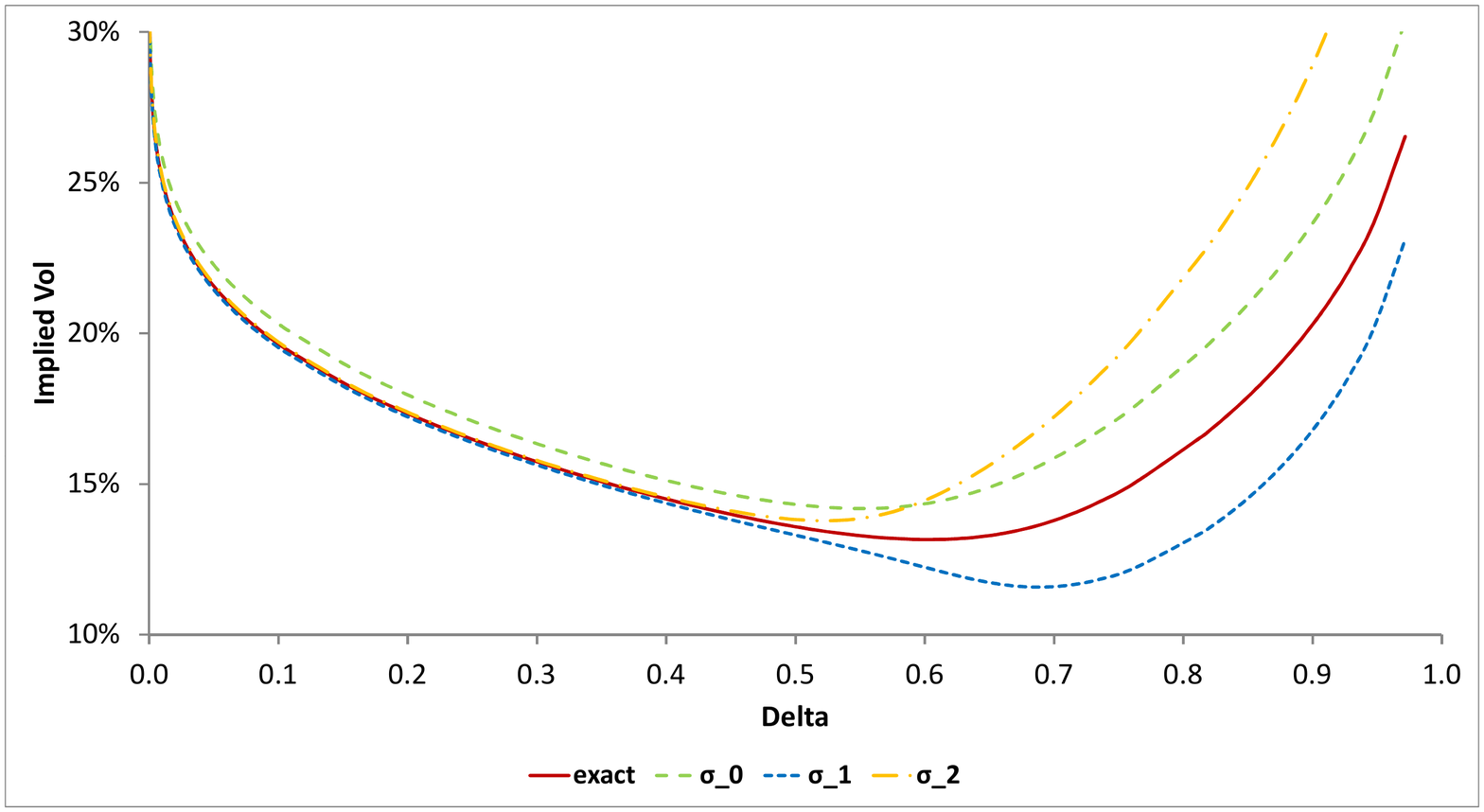}} 
\subfigure[T=10y] {\includegraphics[width=1.0\textwidth, angle=0]
{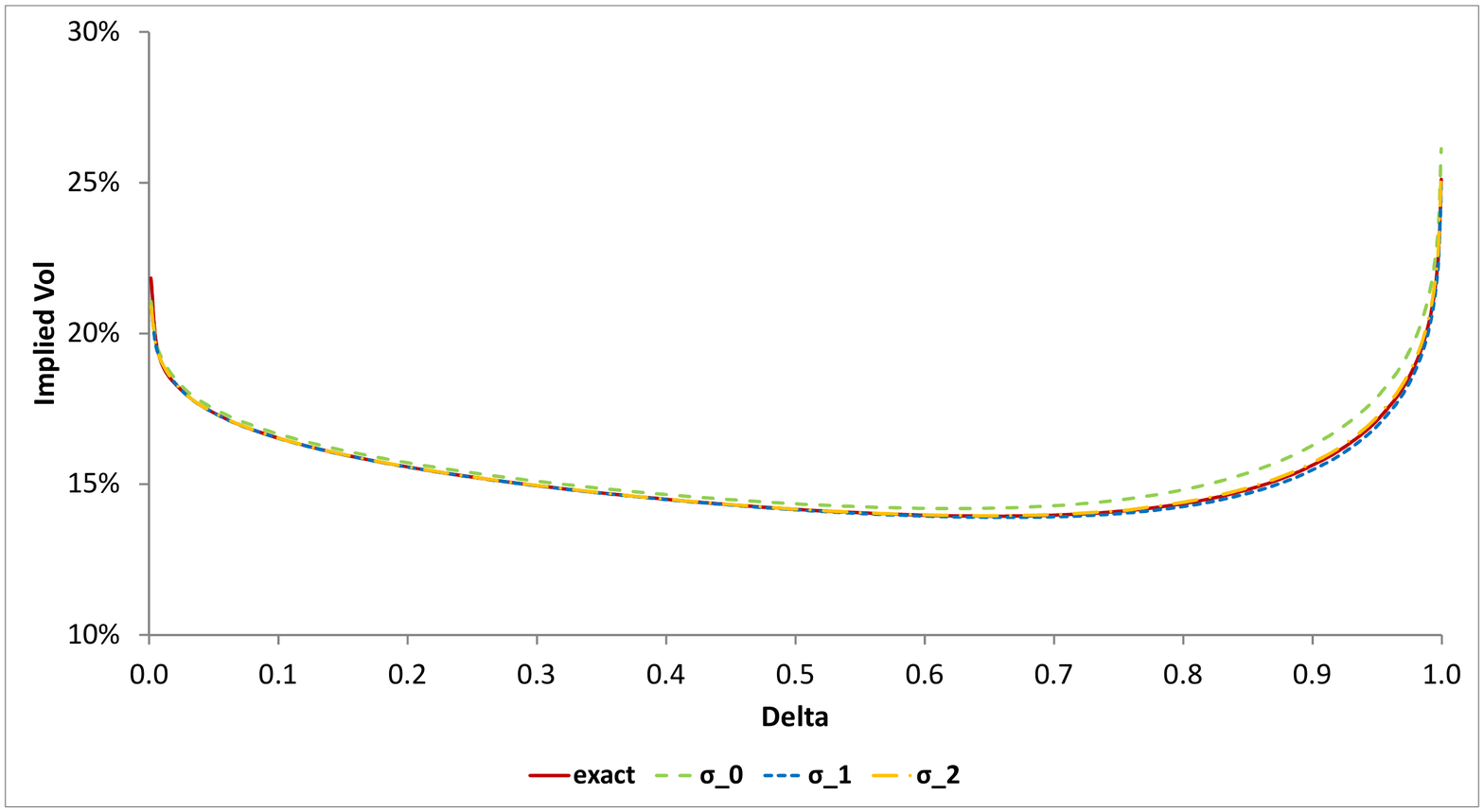}}
\caption{Comparison of the exact and asymptotic expressions for $\protect%
\sigma _{imp}$ for the calibrated TSP for different maturities, $\protect%
\alpha =2/3$. Here and in Figures \protect\ref{fig:TSPsLong_150}, \protect
\ref{fig:NIGPsLong}, and \protect\ref{fig:MPsLong}, "exact" denotes the
implied volatility calculated by virtue of the LL formula, while $\protect%
\sigma \_0$, $\protect\sigma \_1$, $\protect\sigma \_2$ are given by (%
\protect\ref{eq:sigma_i_long}).}
\label{fig:TSPsLong_066}
\end{figure}

\clearpage

\clearpage

\clearpage

\begin{figure}[h]
\subfigure[T=2y] {\includegraphics[width=1.0\textwidth, angle=0]
{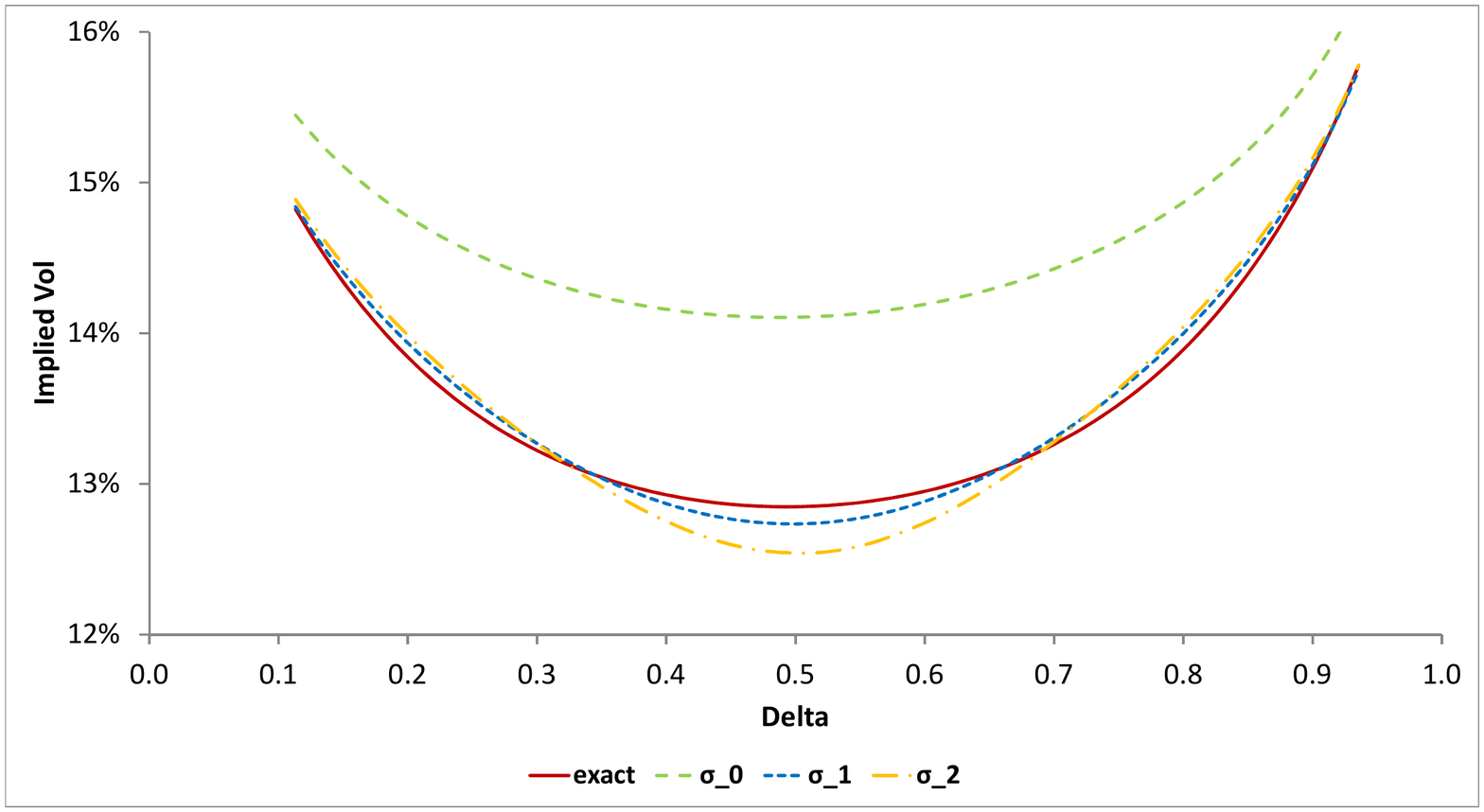}} 
\subfigure[T=10y] {\includegraphics[width=1.0\textwidth, angle=0]
{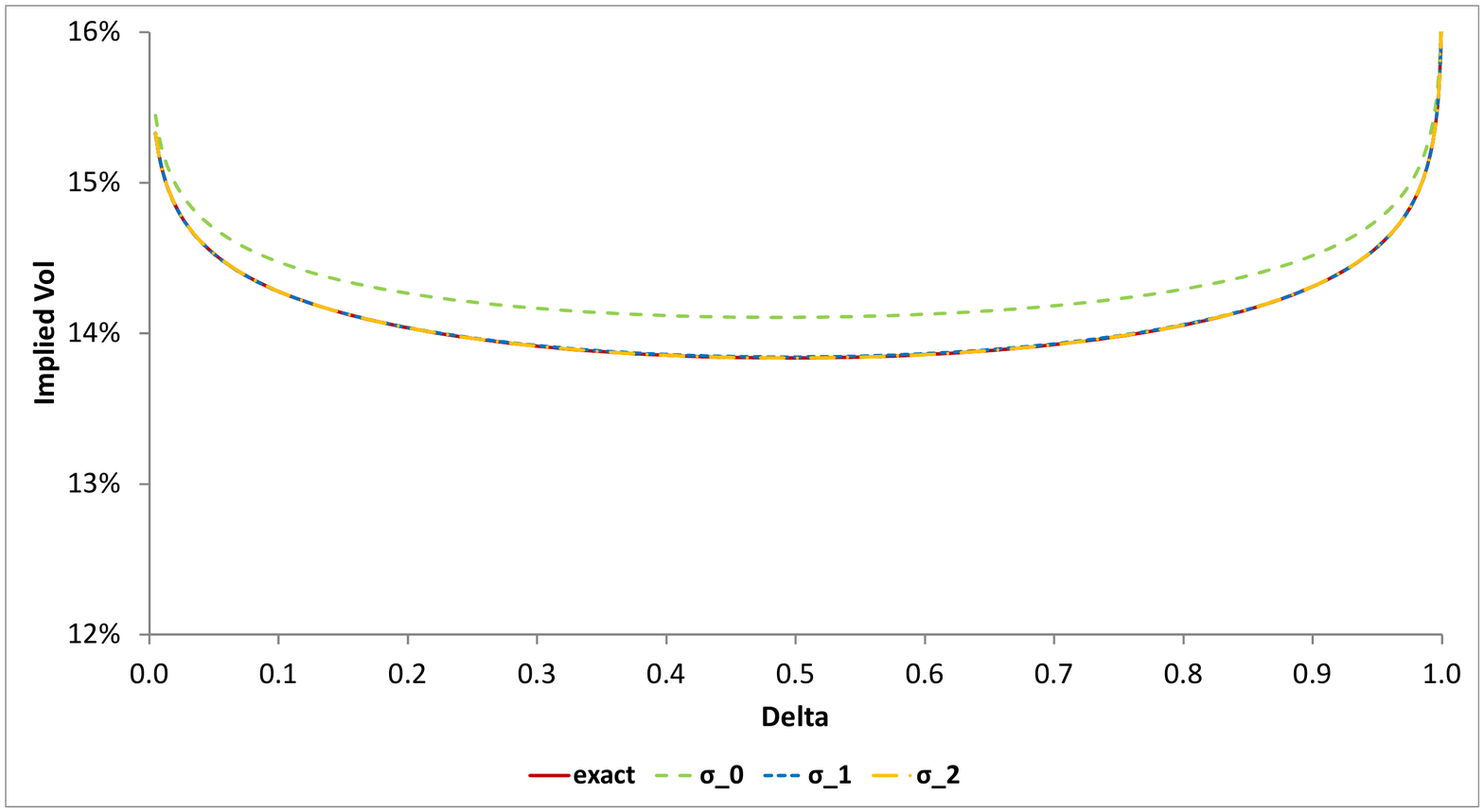}}
\caption{{}Comparison of the exact and asymptotic expressions for $\protect%
\sigma _{imp}$ for the calibrated MP for different maturities. }
\label{fig:MPsLong}
\end{figure}

\clearpage

\begin{remark}
For some special cases the above formulae can be made more specific. For the
general maximally skewed TSP we can solve (\ref{eq: saddle_tsp}) explicitly
and avoid using the parametric representation. Specifically, we have 
\begin{align*}
\Xi _{0}\left( y\right) & =a\left( \left( \kappa -\left( y-\frac{1}{2}%
\right) \right) ^{\alpha }-\mathsf{\zeta }\left( y-\frac{1}{2}\right) +%
\mathsf{\eta }\right) , \\
\Xi _{1}\left( y\right) & =a\left( -\alpha \left( \kappa -\left( y-\frac{1}{2%
}\right) \right) ^{\alpha -1}-\mathsf{\zeta }\right) =-\bar{k},
\end{align*}%
where $\mathsf{\zeta }$, $\mathsf{\eta }$ are given by (\ref{eq:defs}), so
that%
\begin{align*}
y_{\ast }\left( \bar{k}\right) & =\kappa +\frac{1}{2}-\left( \frac{1}{\alpha 
}\left( \frac{\bar{k}}{a}-\mathsf{\zeta }\right) \right) ^{\frac{1}{\alpha -1%
}}, \\
\Xi _{01,\pm }\left( \bar{k}\right) & =a\left( \left( 1-\alpha \right)
\left( \frac{1}{\alpha }\left( \frac{\bar{k}}{a}-\mathsf{\zeta }\right)
\right) ^{\frac{\alpha }{\alpha -1}}+\kappa \left( \frac{\bar{k}}{a}-\mathsf{%
\zeta }\right) +\mathsf{\eta }\right) +\frac{1}{2}\left( \bar{k}\pm \bar{k}%
\right) , \\
\bar{k}_{\pm }& =a\left( \alpha \left( \kappa +\frac{1}{2}\mp \frac{1}{2}%
\right) ^{\alpha -1}+\mathsf{\zeta }\right) , \\
\sigma _{imp,\infty }^{\left( \alpha ,\kappa \right) }\left( \bar{k}\right)
& =\mathrm{sign}\left( \bar{k}_{-}-\bar{k}\right) \sqrt{-2\Xi _{01,+}\left( 
\bar{k}\right) }-\mathrm{sign}\left( \bar{k}_{+}-\bar{k}\right) \sqrt{-2\Xi
_{01,-}\left( \bar{k}\right) }.
\end{align*}%
These formulas show that for TLGP the ATM implied volatility approaches from
below the following level 
\begin{equation*}
\sigma _{imp,\infty }^{\left( 1/2,\kappa \right) }\left( 0\right)
=2^{3/4}\vartheta ^{1/4}\left( \sqrt{\kappa +1}-\sqrt{\kappa }\right) ^{3/2}.
\end{equation*}
\end{remark}

We compare asymptotic and exact implied volatilities for TLGP in Figures \ref%
{fig:GLPLong} (a), (b) below. These Figures show that, similarly to the
general case of TSPs with $0<\alpha <1$, for TLGPs the saddlepoint
approximation is acceptable.

\begin{figure}[h]
\subfigure[T=5y] {\includegraphics[width=1.0\textwidth, angle=0]
{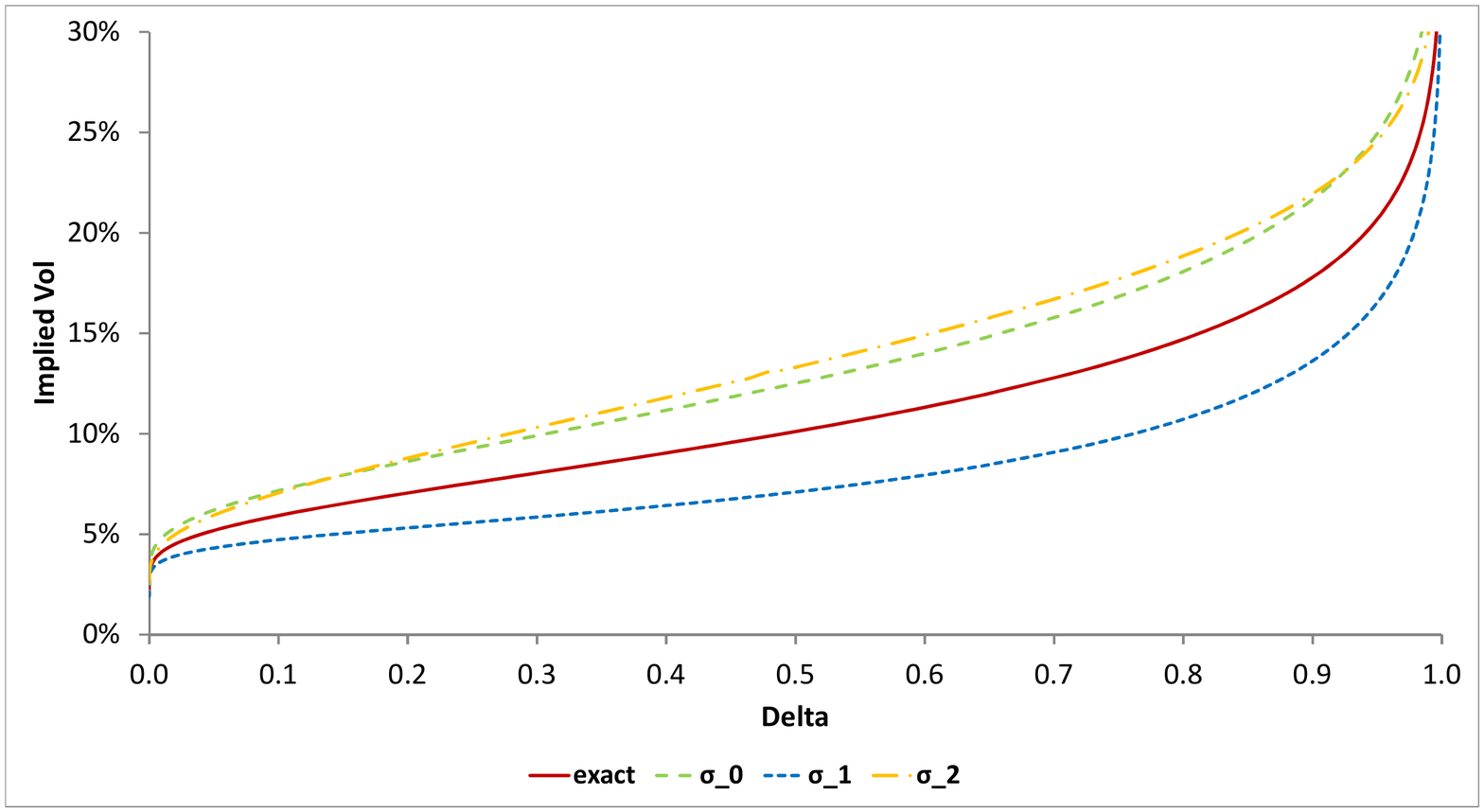}} 
\subfigure[T=15y] {\includegraphics[width=1.0\textwidth, angle=0]
{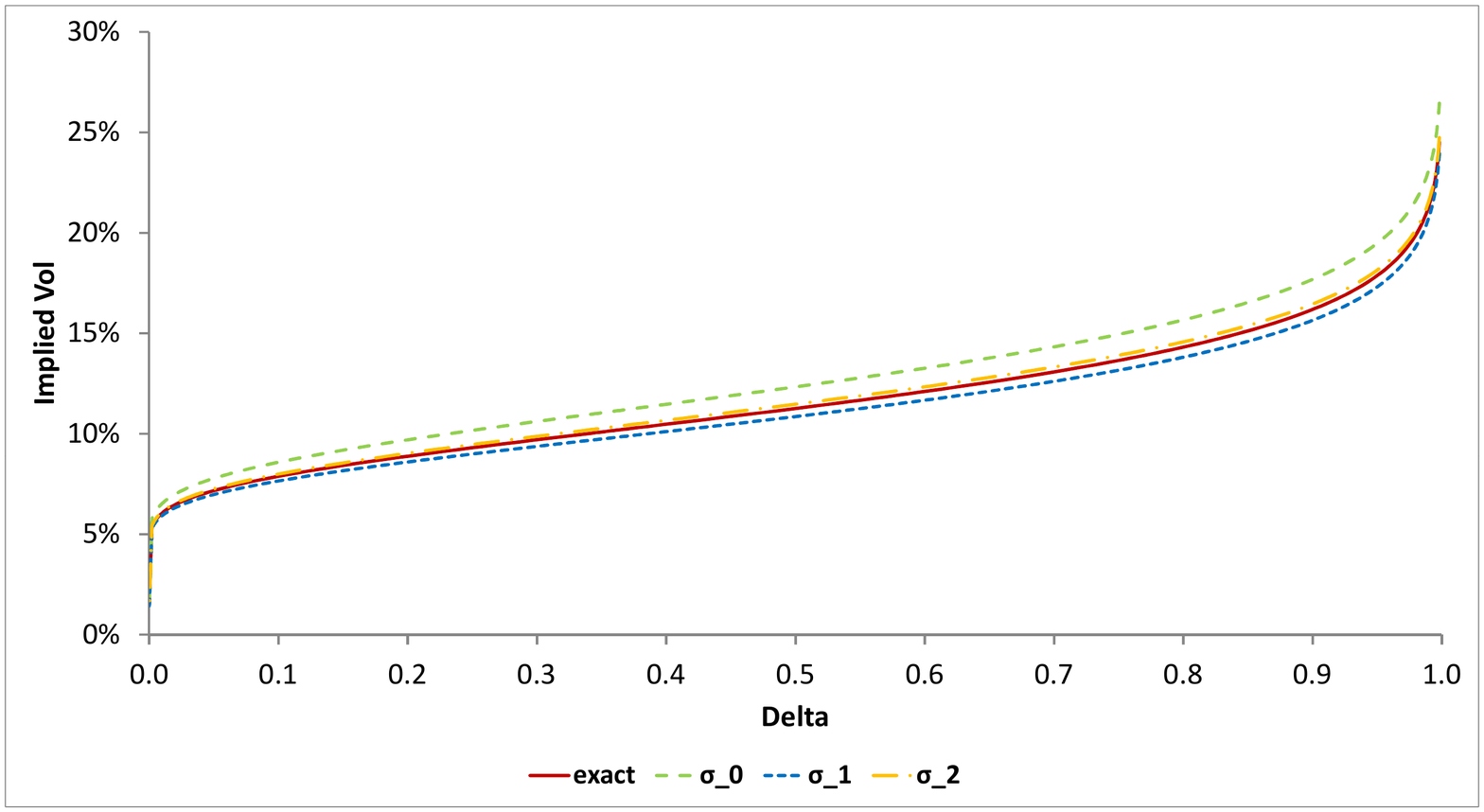}}
\caption{{}Comparison of the exact and asymptotic expressions for $\protect%
\sigma _{imp}$ for the GLP with $\protect\vartheta =0.0075$, $\protect\kappa %
=1.0$ for different maturities. Here $\protect\sigma _{imp,exact}$ denotes
the exact volatility calculated by virtue of (\protect\ref{eq:GLP}), while $%
\protect\sigma _{imp,i}$ are given by (\protect\ref{eq:sigma_i_long}).}
\label{fig:GLPLong}
\end{figure}

\clearpage

\begin{remark}
For NIGP we can easily invert the equation $\Xi _{0}^{\prime }\left(
y\right) +\bar{k}=0$, and avoid using a parametric representation. We have%
\begin{align*}
y_{0}\left( \bar{k}\right) & =-\frac{\bar{\omega}\bar{k}}{\sqrt{\sigma ^{4}%
\mathcal{\bar{\varkappa}}^{2}+\bar{k}^{2}}},\ \ \ \Xi _{01,\pm }\left( \bar{k%
}\right) =\sigma ^{2}\mathcal{\bar{\varkappa}}^{2}-\bar{\omega}\sqrt{\sigma
^{4}\mathcal{\bar{\varkappa}}^{2}+\bar{k}^{2}}\pm \frac{1}{2}\bar{k},\ \ \ 
\bar{k}_{\pm }=\mp \frac{\sigma ^{2}}{2}, \\
\sigma _{imp,\infty }^{NIG}\left( \bar{k}\right) & =\mathrm{sign}\left( \bar{%
k}_{-}-\bar{k}\right) \sqrt{-2\Xi _{01,+}\left( \bar{k}\right) }-\mathrm{sign%
}\left( \bar{k}_{+}-\bar{k}\right) \sqrt{-2\Xi _{01,-}\left( \bar{k}\right) }%
.
\end{align*}
\end{remark}

\begin{remark}
By using (\ref{eq: CNIG}) which provides the call price for a NIGP
explicitly, we can calculate long-time asymptotics for call prices directly,
see Appendix \ref{NIGPsLA} for details. The asymptotics obtained via two
complementary methods naturally agree.
\end{remark}

\subsection{Heston Stochastic Volatility Processes\label{HSVPsLong}}

Equation (\ref{eq: Hest Form}) shows that for the Heston model the LL
exponent is not proportional to time. However, it is proportional to time to
the leading order when $\tau \rightarrow \infty $. Formal expansion in
powers of $\exp \left( -\tau \mathcal{Z}\left( u\right) \right) $ yields%
\begin{equation*}
\frac{E\left( \tau ,u,k\right) }{Q\left( u\right) }=\frac{e^{\mathcal{A}%
\left( \tau ,u\right) -\mathcal{B}\left( \tau ,u\right) \varpi _{0}Q\left(
u\right) -\tau \bar{k}\left( iu-\frac{1}{2}\right) }}{Q\left( u\right) }\sim
e^{\tau S\left( u,k\right) }f\left( u\right) ,
\end{equation*}%
where 
\begin{align}
S\left( u,\bar{k}\right) & =-\frac{\kappa \theta }{\varepsilon ^{2}}\mathcal{%
F}_{+}\left( u\right) -\bar{k}\left( iu-\frac{1}{2}\right) ,
\label{eq:S_HSVP} \\
f\left( u\right) & =\frac{e^{-\varpi _{0}\mathcal{F}_{+}\left( u\right)
/\varepsilon ^{2}}}{Q\left( u\right) \left( \frac{\mathcal{F}_{-}\left(
u\right) }{2\mathcal{Z}\left( u\right) }\right) ^{2\kappa \theta
/\varepsilon ^{2}}},  \notag
\end{align}%
and $\mathcal{F}_{+}\left( u\right) ,\mathcal{Z}\left( u\right) $ are given
by (\ref{eq: Hest Term}). As before, we can use saddlepoint method to obtain
the asymptotic of the LL\ integral. It is easy to check that the
corresponding saddlepoint has to be purely imaginary, so that we can proceed
as before.

\begin{proposition}
Consider $y\in \left( -\mathsf{Y}_{+},\mathsf{Y}_{-}\right) $, where%
\begin{equation}
\mathsf{Y}_{\pm }=\frac{\left( \mp \rho \check{\kappa}+\sqrt{\check{\kappa}%
^{2}+\frac{1}{4}\bar{\rho}^{2}}\right) }{\bar{\rho}^{2}},  \label{eq:Y_pm}
\end{equation}%
and $\check{\kappa}=\hat{\kappa}/\varepsilon $, and define%
\begin{align}
\Xi _{0}\left( y\right) & =\frac{\kappa \theta }{\varepsilon }\left( \rho y+%
\check{\kappa}-\varsigma \left( y\right) \right) ,  \label{eq: xi_i HSVP} \\
\Xi _{1}\left( y\right) & =\frac{\kappa \theta }{\varepsilon }\left( \rho -%
\frac{-\bar{\rho}^{2}y+\rho \check{\kappa}}{\varsigma \left( y\right) }%
\right) ,  \notag \\
\Xi _{01,\pm }\left( y\right) & =\Xi _{0}\left( y\right) -\Xi _{1}\left(
y\right) \left( y\pm \frac{1}{2}\right) ,  \notag
\end{align}%
where 
\begin{equation}
\varsigma \left( y\right) =\sqrt{-\bar{\rho}^{2}y^{2}+2\rho \check{\kappa}y+%
\check{\kappa}^{2}+1/4}.  \label{eq:zeta(y)}
\end{equation}%
Then for $\bar{k}=-\Xi _{1}\left( y\right) $ the corresponding $\sigma
_{imp}\left( \tau ,\tau \bar{k}\right) $ can be written in the form%
\begin{equation*}
\sigma _{imp}\left( \tau ,\tau \bar{k}\right) =\left( a_{0}\left( y\right) +%
\frac{a_{1}\left( y\right) }{\tau }+...\right) ^{1/2}=b_{0}\left( y\right) +%
\frac{b_{1}\left( y\right) }{\tau }+...,
\end{equation*}%
where the leading order coefficients $a_{0}\left( y\right) $ and $%
b_{0}\left( y\right) $ have the form (\ref{eq: a_0}), (\ref{eq: b_0}), and
higher order coefficients $a_{1}\left( y\right) $ and $b_{1}\left( y\right) $
have the form (\ref{eq: a_i}), (\ref{eq: b_i}).
\end{proposition}

\begin{proof}
The proof is similar to the one of Proposition \ref{GeneralLongProp} and is
omitted for brevity.
\end{proof}

\begin{figure}[h]
\subfigure[T=2y] {\includegraphics[width=0.75\textwidth, angle=0]
{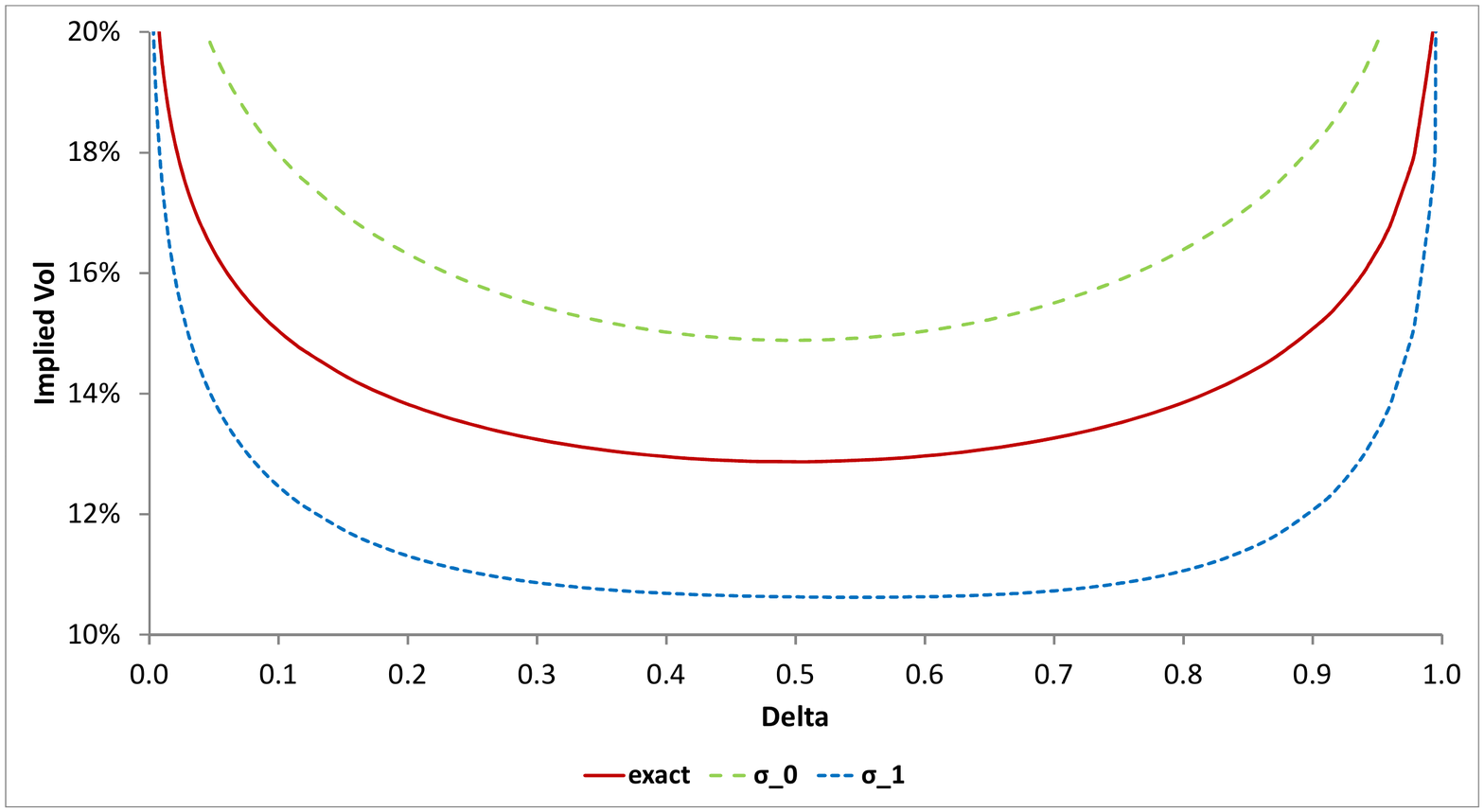}} 
\subfigure[T=10y] {\includegraphics[width=0.75\textwidth, angle=0]
{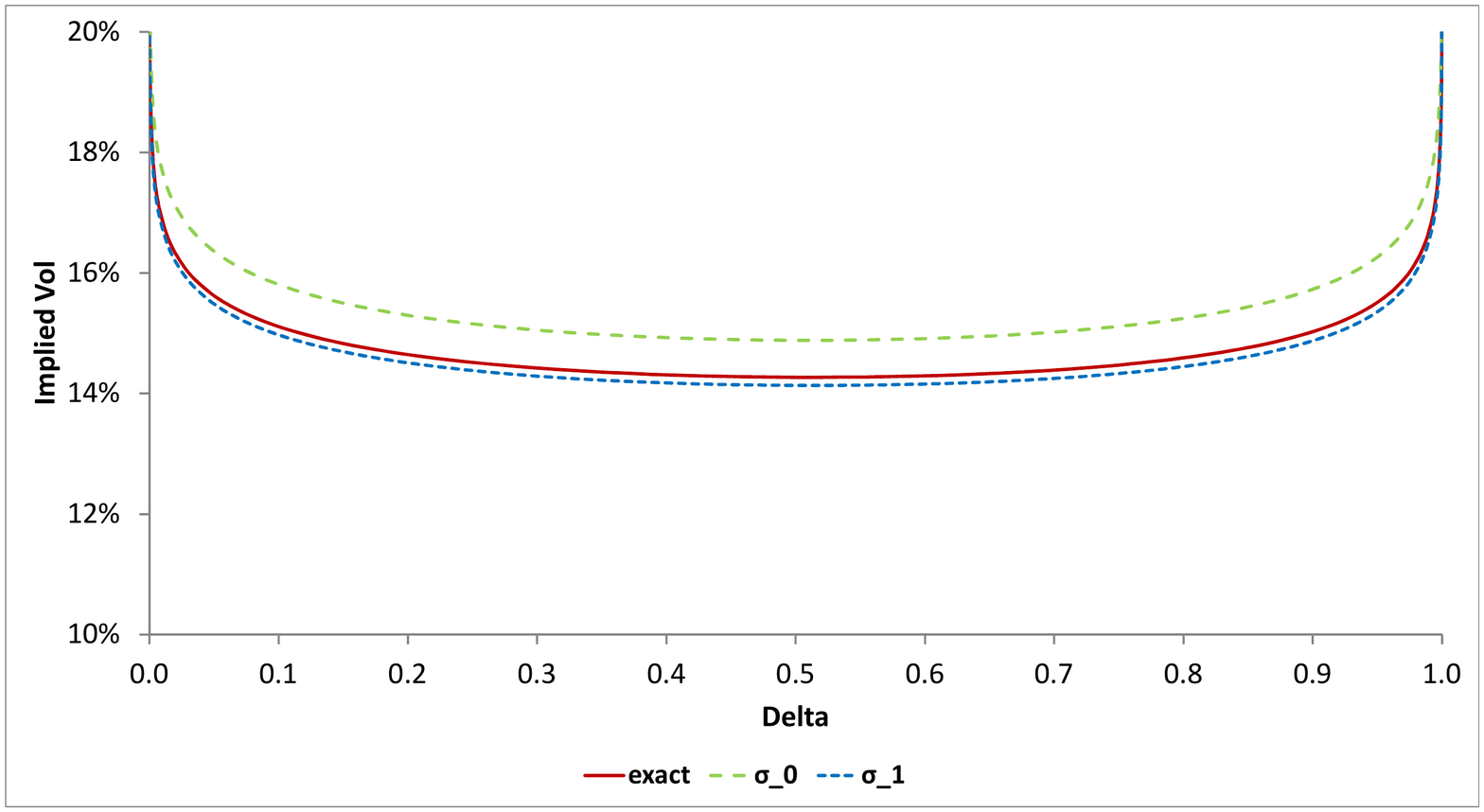}}
\caption{{}Comparison of the exact and asymptotic expressions for $\protect%
\sigma _{imp}$ for the calibrated HSVP for different maturities. Here
"exact" denotes the implied volatility calculated by virtue of the LL
formula, while $\protect\sigma \_0$, $\protect\sigma \_1$ (\protect\ref%
{eq:sigma_i_long}).}
\label{fig:HSVPsLong}
\end{figure}

As expected, the leading-order term in the volatility expansion does not
depend on $\varpi _{0}$. The quality of the above approximation is
illustrated in Figures \ref{fig:HSVPsLong} (a), (b). These Figures show that
for HSVPs the saddlepoint approximation works well.

\clearpage

\begin{remark}
For HSVPs we can easily invert the equation $\Xi _{0}^{\prime }\left(
y\right) +\bar{k}=0$, and avoid using the parametric representation. A
simple calculation performed in Appendix \ref{HSVPsLA} yields%
\begin{align}
\sigma _{imp,0}\left( \bar{k}\right) & =\mathrm{sign}\left( y^{\ast }\left( 
\bar{k}\right) +\frac{1}{2}\right) \sqrt{-2\Xi _{01,+}\left( \bar{k}\right) }%
-\mathrm{sign}\left( y^{\ast }\left( \bar{k}\right) -\frac{1}{2}\right) 
\sqrt{-2\Xi _{01,-}\left( \bar{k}\right) },  \label{eq: sigma_0 HSVP} \\
y^{\ast }\left( \bar{k}\right) & =\frac{\rho \check{\kappa}-\mathrm{sign}%
\left( l\right) \sqrt{\check{\kappa}^{2}-\bar{\rho}^{2}\left( \check{\kappa}%
^{2}-\frac{1}{4}l^{2}\right) /\left( \bar{\rho}^{2}+l^{2}\right) }}{\bar{\rho%
}^{2}},  \notag \\
\Xi _{01,\pm }\left( \bar{k}\right) & =\frac{\kappa \theta }{\varepsilon l}%
\left( \left( \bar{\rho}^{2}+\rho l\right) y^{\ast }\left( \bar{k}\right)
+\left( l-\rho \right) \check{\kappa}\right) \pm \frac{1}{2}\bar{k},  \notag
\end{align}%
where $l=\rho +\varepsilon \bar{k}/\kappa \theta $.
\end{remark}

Additional information is given in \cite{forde3}, \cite{jacquer}.

\section{Short-time Asymptotics\label{AsymptoticsShort}}

\subsection{General Remarks\label{GeneralRemarksShort}}

From the results in Proposition \ref{Prop:LLFormula}, and Corollary \ref%
{Coro:LLFormula}, we see that in order to analyze the asymptotic behavior of
the ATM\ option price ($k=0$) and its derivatives for $\tau \rightarrow 0$,
we need to study the following integrals%
\begin{align}
\mathfrak{l}\left( \tau \right) & =\frac{1}{2\pi }\int_{-\infty }^{\infty }%
\frac{\left( 1-E\left( \tau ,u\right) \right) }{Q\left( u\right) }du=\frac{1%
}{\pi }\func{Re}\left\{ \int_{0}^{\infty }\frac{\left( 1-E\left( \tau
,u\right) \right) }{Q\left( u\right) }du\right\} ,
\label{eq: Main Integrals} \\
\mathfrak{m}\left( \tau \right) & =\frac{1}{2\pi }\int_{-\infty }^{\infty }%
\frac{E\left( \tau ,u\right) }{Q\left( u\right) }iudu=-\frac{1}{\pi }\func{Im%
}\left\{ \int_{0}^{\infty }\frac{E\left( \tau ,u\right) }{Q\left( u\right) }%
udu\right\} ,  \notag \\
\mathfrak{n}\left( \tau \right) & =\frac{1}{2\pi }\int_{-\infty }^{\infty
}E\left( \tau ,u\right) du=\frac{1}{\pi }\func{Re}\left\{ \int_{0}^{\infty
}E\left( \tau ,u\right) du\right\} ,  \notag
\end{align}%
where%
\begin{equation*}
E\left( \tau ,u\right) =\exp \left( \tau \upsilon _{0}\left( u\right) -\frac{%
\sigma ^{2}\tau }{2}Q\left( u\right) \right) .
\end{equation*}

The relation between $\mathsf{C}\left( \tau ,0\right) ,\mathsf{C}_{k}\left(
\tau ,0\right) ,\mathsf{C}_{kk}\left( \tau ,0\right) $ and $\mathfrak{l}%
\left( \tau \right) ,\mathfrak{m}\left( \tau \right) ,\mathfrak{n}\left(
\tau \right) $ is straightforward.

\begin{proposition}
$\mathsf{C}\left( \tau ,0\right) ,\mathsf{C}_{k}\left( \tau ,0\right) ,%
\mathsf{C}_{kk}\left( \tau ,0\right) $ can be expressed in terms of $%
\mathfrak{l}\left( \tau \right) ,\mathfrak{m}\left( \tau \right) ,\mathfrak{n%
}\left( \tau \right) $ as follows%
\begin{align}
\mathsf{C}\left( \tau ,0\right) & =\mathfrak{l}\left( \tau \right) ,
\label{eq: cl} \\
\mathsf{C}_{k}\left( \tau ,0\right) & =\mathfrak{m}\left( \tau \right) +%
\frac{1}{2}\left( \mathfrak{l}\left( \tau \right) -1\right) ,  \notag \\
\mathsf{C}_{kk}\left( \tau ,0\right) & =\mathfrak{n}\left( \tau \right) +%
\mathfrak{m}\left( \tau \right) +\frac{1}{2}\left( \mathfrak{l}\left( \tau
\right) -1\right) .  \notag
\end{align}

\begin{proof}
Straightforward computation.
\end{proof}
\end{proposition}

We assume that the implied volatility can be expanded in powers of $k$:%
\begin{equation}
\sigma _{imp}\left( \tau ,k\right) =\sum\limits_{n\geq 0}\chi _{n}\left(
\tau \right) \frac{k^{n}}{n!},  \label{eq:sigma_imp}
\end{equation}%
and demonstrate how to express $\chi _{0},\chi _{1},\chi _{2}$ in terms of $%
\mathfrak{l},\mathfrak{m},\mathfrak{n}$.

\begin{proposition}
\label{price to vol general} In the LL framework we get%
\begin{align*}
\chi _{0}\left( \tau \right) & \sim \mathfrak{\hat{l}}\left( \tau \right)
\left( 1+\frac{\mathfrak{\hat{l}}^{2}\left( \tau \right) }{24}\right) \tau
^{-1/2}, \\
\chi _{1}\left( \tau \right) & \sim \mathfrak{\hat{m}}\left( \tau \right)
\left( 1+\frac{\mathfrak{\hat{l}}^{2}\left( \tau \right) }{8}\right) \tau
^{-1/2}, \\
\chi _{2}\left( \tau \right) & \sim \left( -\frac{1}{\mathfrak{\hat{l}}%
\left( \tau \right) }+\frac{\mathfrak{\hat{l}}\left( \tau \right) }{24}+%
\frac{1}{4}\mathfrak{\hat{l}}\left( \tau \right) \mathfrak{\hat{m}}%
^{2}\left( \tau \right) +\mathfrak{\hat{n}}\left( \tau \right) \left( 1+%
\frac{\mathfrak{\hat{l}}^{2}\left( \tau \right) }{8}\right) \right) \tau
^{-1/2},
\end{align*}%
where%
\begin{equation}
\mathfrak{\hat{l}}\left( \tau \right) =\sqrt{2\pi }\mathfrak{l}\left( \tau
\right) ,\ \ \ \ \ \mathfrak{\hat{m}}\left( \tau \right) =\sqrt{2\pi }%
\mathfrak{m}\left( \tau \right) ,\ \ \ \ \ \mathfrak{\hat{n}}\left( \tau
\right) =\sqrt{2\pi }\mathfrak{n}\left( \tau \right) .  \label{eq:lmn_hat}
\end{equation}

\begin{proof}
See Appendix \ref{ImpVolExp}.
\end{proof}
\end{proposition}

\subsection{Exponential L\'{e}vy Processes\label{ELPsS}}

\subsubsection{Non-ATM\ Options on Tempered Stable Processes, Simple
Heuristics\label{non-ATM on TSPs}}

Using the LL formula, it is straightforward to develop small-time
asymptotics for non-ATM options on TSPs. The argument goes as follows (see,
e.g., \cite{levendorskii-1} among others).

\begin{proposition}
Assume that $\sigma =0$ and consider 
\begin{equation*}
\mathsf{C}\left( \tau ,k\right) =1-\frac{1}{2\pi }\int_{-\infty }^{\infty }%
\frac{e^{\tau \upsilon _{0}\left( u\right) -k\left( iu-\frac{1}{2}\right) }}{%
Q\left( u\right) }du,\ \ \ \ \ k\neq 0,
\end{equation*}%
where $\upsilon _{0}\left( u\right) $ is given by (\ref{eq: LLpsi}). If $%
\alpha \in \left( 0,1\right) $, then%
\begin{equation*}
\frac{\partial }{\partial \tau }\mathsf{C}\left( 0,k\right) \sim -\frac{1}{%
2\pi }\int_{-\infty }^{\infty }\frac{e^{-k\left( iu-\frac{1}{2}\right) }}{%
Q\left( u\right) }\upsilon _{0}\left( u\right) du,
\end{equation*}%
where the integral clearly converges. If $\alpha \in \left( 1,2\right) $,
then integration by parts (which is possible because $k\neq 0$) shows that:%
\begin{equation*}
\frac{\partial }{\partial \tau }\mathsf{C}\left( \tau ,k\right) \sim -\frac{1%
}{2\pi ik}\int_{-\infty }^{\infty }e^{-k\left( iu-\frac{1}{2}\right) }\left( 
\frac{\upsilon _{0}\left( u\right) }{Q\left( u\right) }\right) ^{\prime }du,
\end{equation*}%
where the integral clearly converges. Accordingly, in both cases, the time
value of the call option $\mathsf{\delta C}\left( \tau ,k\right) $ is linear
in time,%
\begin{equation*}
\mathsf{\delta C}\left( \tau ,k\right) =\mathsf{c}\left( k\right) \tau
+o\left( \tau \right) .
\end{equation*}

\begin{proof}
See Appendix \ref{TSPsSA}.
\end{proof}
\end{proposition}

Broadly speaking, this result means the time decay of the time value $%
\mathsf{\delta C}\left( \tau ,k\right) $ is much slower that in the BS\
case, which is of order $\exp \left( -k^{2}/2\sigma ^{2}\tau \right) $, see (%
\ref{eq: BS non-dim}), (\ref{eq:cum_norm_as}). To compensate for that, the
corresponding non-ATM implied volatility must explode with order $O(1/\sqrt{%
\tau \ln (1/\tau )})$, as was already mentioned in Section \ref{Intro}.

\subsubsection{ATM\ Options on Tempered Stable Processes, Simple Heuristics 
\label{Heuristics}}

For ATM options we start with a simple heuristic argument. Consider a TSP
with $\sigma =0$ and $\alpha \in \left( 0,1\right) $ and calculate the
corresponding ATM\ asymptotics. From (\ref{eq:PIDE_1}), we have%
\begin{equation*}
\mathsf{C}_{\tau }\left( \tau ,x\right) =\gamma \mathsf{C}_{x}\left( \tau
,x\right) +\sum\limits_{s=\pm }c_{s}\int_{0}^{\infty }\frac{\left( \mathsf{C}%
\left( \tau ,x+sy\right) -\mathsf{C}\left( \tau ,x\right) \right) e^{-\kappa
_{s}y}}{y^{1+\alpha }}dy,
\end{equation*}%
\begin{equation*}
\mathsf{C}\left( 0,x\right) =\left( e^{x}-1\right) ^{+}.
\end{equation*}%
As we already know, the martingale condition yields 
\begin{equation*}
\gamma =-\sum\limits_{s=\pm }c_{s}\int_{0}^{\infty }\frac{\left(
e^{sy}-1\right) e^{-\kappa _{s}y}}{y^{1+\alpha }}dy=-p_{+}+p_{-},
\end{equation*}%
where 
\begin{equation*}
p_{s}=sc_{s}\int_{0}^{\infty }\frac{\left( e^{sy}-1\right) e^{-\kappa _{s}y}%
}{y^{1+\alpha }}dy>0,\ \ \ s=\pm .
\end{equation*}%
In order to calculate $\mathsf{C}\left( \tau ,0\right) $ we discretize $x$
on a grid $x_{n}=nh$, $-N\leq x\leq N$, $\mathsf{C}_{n}\left( \tau \right) =%
\mathsf{C}\left( \tau ,x_{n}\right) $, and use an explicit finite-difference
scheme. Since we have an advection term, we have to differentiate two cases:
(A)\ $\gamma >0$ $\left( p_{-}>p_{+}\right) $, and (B) $\gamma <0\ $ $\left(
p_{-}<p_{+}\right) $. In case (A) our finite-difference scheme yields%
\begin{equation*}
\frac{\mathsf{C}_{0}\left( \tau \right) }{\tau }=\gamma \frac{\left( \mathsf{%
C}_{1}\left( 0\right) -\mathsf{C}_{0}\left( 0\right) \right) }{h}%
+p_{+}=\gamma \frac{\left( e^{h}-1\right) }{h}+p_{+}\approx \gamma
+p_{+}=p_{-}.
\end{equation*}%
In case (B) our scheme yields%
\begin{equation*}
\frac{\mathsf{C}_{0}\left( \tau \right) }{\tau }=\gamma \frac{\left( \mathsf{%
C}_{0}\left( 0\right) -\mathsf{C}_{-1}\left( 0\right) \right) }{h}%
+p_{+}=p_{+}.
\end{equation*}%
Note, in both cases differentiation is performed upstream to ensure
stability of the scheme. We can combine the above formulae into one as
follows 
\begin{equation}
\mathsf{C}\left( \tau ,0\right) =\tau \max \left( p_{+},p_{-}\right) .
\label{eq:max_pp_pm}
\end{equation}%
We note that this formula coincides with (\ref{eq:tankov}) in Section \ref%
{Intro}. For finite variation TSPs, the ATM volatility therefore goes to
zero (to ensure that $\mathsf{C}\left( \tau ,0\right) $ decays linearly), in
marked contrast with non-ATM volatilities, which, as was noted in Section %
\ref{non-ATM on TSPs}, explode for $\tau \rightarrow 0$. We emphasize that
the technique above does not hold for $\alpha \geq 1$, as the $p_{\pm }$
integrals diverge. We also note that the result gives us no information
about the slope and convexity of the ATM implied volatility. To address
these issues, we now process with a more detailed analysis.

\subsubsection{Main Result for Tempered Stable Processes\label{Main Result}}

In view of the LL formula, in order to be able to evaluate the short-time
asymptotic behavior of $\sigma _{imp}\left( \tau ,0\right) $ and its $k$%
-derivatives $\left. \partial _{k}\sigma _{imp}\left( \tau ,k\right)
\right\vert _{k=0}$, $\left. \partial _{k}^{2}\sigma _{imp}\left( \tau
,k\right) \right\vert _{k=0}$, we need to evaluate the short-time limit of
the three integrals in (\ref{eq: Main Integrals}) with the shifted
characteristic exponent of the form 
\begin{equation*}
\upsilon _{0}\left( u\right) =\sum_{s=\pm }a_{s}\left( \kappa _{s}-s\left(
iu+\frac{1}{2}\right) \right) ^{\alpha }+\gamma \left( iu+\frac{1}{2}\right)
+\delta .
\end{equation*}

We have to distinguish four cases (A) $\alpha \in \left( 0,1\right) ,\sigma
=0$, (B) $\alpha \in \left( 1,2\right) ,\sigma =0$, (C) $\alpha \in \left(
0,1\right) ,\sigma >0$, (D) $\alpha \in \left( 1,2\right) ,\sigma >0$. We
calculate the corresponding integrals by scaling the independent variable $u$
as appropriate. Specifically, we use: $v=\tau u$ (case (A)); $v=\tau
^{1/\alpha }u$ (case (B)); $v=\tau ^{1/2}u$ (cases (C) and (D)). This
technique is similar to the approach in \cite{debruin}, for example; in case
(B), it has also been used by \cite{tankov}.

When considering the relevant integrals, we note that the dominated
convergence theorem (DCT), which is the principal tool for studying the
limiting behavior of integrals depending on parameters, cannot in all cases
be applied to establish the limits for $\tau \rightarrow 0$ directly, so
great case must be taken to avoid faulty conclusions. To demonstrate this,
let us just consider the simple example of establishing the limit of%
\begin{equation*}
\mathfrak{\tilde{l}}\left( \tau \right) =\frac{1}{2\pi }\dint\limits_{-%
\infty }^{\infty }\frac{1-e^{\tau \gamma \left( iu+\frac{1}{2}\right) }}{%
u^{2}+\frac{1}{4}}du,
\end{equation*}%
which is similar to $\mathfrak{l}\left( \tau \right) $. Formula (\ref{eq:
BSLL}) shows that%
\begin{equation*}
\mathfrak{\tilde{l}}\left( \tau \right) =1-e^{\tau \gamma ^{-}}\sim -\tau
\gamma ^{-}.
\end{equation*}%
Now, we wish to derive this formula asymptotically. We change variables $%
v=\tau u$, and represent $\mathfrak{\tilde{l}}\left( \tau \right) $ in the
form%
\begin{equation*}
\mathfrak{\tilde{l}}\left( \tau \right) =\tau \mathfrak{\tilde{L}}\left(
\tau \right) ,
\end{equation*}%
where%
\begin{equation}
\mathfrak{L}\left( \tau \right) =\frac{1}{2\pi }\int_{-\infty }^{\infty }%
\frac{1-e^{\frac{\tau \gamma }{2}+i\gamma v}}{\left( v^{2}+\frac{1}{4}\tau
^{2}\right) }dv=\frac{1}{\pi }\int_{0}^{\infty }\frac{1-e^{\frac{\tau \gamma 
}{2}}\cos \left( \gamma v\right) }{v^{2}+\frac{1}{4}\tau ^{2}}dv.
\label{eq: LMNE}
\end{equation}%
We notice that at $v=0$ the corresponding integrand is unbounded when $\tau
\rightarrow 0$, so that we cannot apply the DCT directly. However, we can
split the above integral into two parts as follows 
\begin{equation*}
\mathfrak{L}\left( \tau \right) =\mathfrak{L}_{0}^{\varepsilon }\left( \tau
\right) +\mathfrak{L}_{\varepsilon }^{\infty }\left( \tau \right) ,
\end{equation*}%
where 
\begin{equation*}
\mathfrak{L}_{v_{1}}^{v_{2}}\left( \tau \right) =\frac{1}{\pi }%
\int_{v_{1}}^{v_{2}}\frac{1-e^{\frac{\tau \gamma }{2}}\cos \left( \gamma
v\right) }{v^{2}+\frac{1}{4}\tau ^{2}}dv.
\end{equation*}%
In order to evaluate $\mathfrak{L}_{0}^{\varepsilon }\left( \tau \right) $
we expand the integrand around $v=0$ and get, to the leading order in $%
\varepsilon $,%
\begin{equation*}
\mathfrak{L}_{0}^{\varepsilon }\left( \tau \right) \sim -\frac{\tau \gamma }{%
2\pi }\int_{0}^{\varepsilon }\frac{1}{v^{2}+\frac{1}{4}\tau ^{2}}dv=-\frac{%
\gamma }{2\pi }\int_{0}^{\varepsilon /\tau }\frac{1}{Q\left( u\right) }%
du\sim -\frac{\gamma }{2}.
\end{equation*}%
In order to evaluate $\mathfrak{L}_{\varepsilon }^{\infty }\left( \tau
\right) $ we apply the DCT, which says that 
\begin{equation*}
\mathfrak{L}_{\varepsilon }^{\infty }\left( \tau \right) \underset{\tau
\rightarrow 0}{\rightarrow }\mathfrak{L}_{\varepsilon }^{\infty }\left(
0\right) ,
\end{equation*}%
since the corresponding integrands are uniformly bounded, yielding: 
\begin{equation*}
\mathfrak{L}_{\varepsilon }^{\infty }\left( \tau \right) \sim \frac{1}{\pi }%
\int_{\varepsilon }^{\infty }\frac{1-\cos \left( \gamma v\right) }{v^{2}}dv%
\underset{\varepsilon \rightarrow 0}{\rightarrow }\frac{\left\vert \gamma
\right\vert }{2}.
\end{equation*}%
Thus, 
\begin{equation*}
\mathfrak{L}\left( \tau \right) \sim \frac{-\gamma +\left\vert \gamma
\right\vert }{2}=-\gamma ^{-},
\end{equation*}%
which is the correct answer.

Judicious usage of the technique above, leads to the following important
result:

\begin{proposition}
\label{Main Result Integrals} Consider integrals $\mathfrak{l},\mathfrak{m},%
\mathfrak{n}$ defined by formulas (\ref{eq: Main Integrals}). Their
asymptotic behavior depends on $\alpha $ and $\sigma $. In cases (A)-(D) the
corresponding asymptotics has the form%
\begin{equation*}
\begin{tabular}{|c|c|c|c|}
\hline
& $\mathfrak{l}$ & $\mathfrak{m}$ & $\mathfrak{n}$ \\ \hline
$A$ & $C_{\mathfrak{L}}^{\left( A\right) }\tau $ & $C_{\mathfrak{M}}^{\left(
A\right) }$ & $C_{\mathcal{\mathfrak{N}}}^{\left( A\right) }\tau ^{-1}$ \\ 
\hline
$B$ & $C_{\mathfrak{L}}^{\left( B\right) }\tau ^{\alpha ^{\prime }}$ & $C_{%
\mathfrak{M}}^{\left( B\right) }$ & $C_{\mathcal{\mathfrak{N}}}^{\left(
B\right) }\tau ^{-\alpha ^{\prime }}$ \\ \hline
$C$ & $\frac{\sigma \tau ^{1/2}}{\sqrt{2\pi }}+C_{\mathfrak{L}}^{\left(
C\right) }\tau $ & $C_{\mathfrak{M}}^{\left( C\right) }\tau ^{1/2}$ & $\frac{%
1}{\sqrt{2\pi }\sigma \tau ^{1/2}}+C_{\mathcal{\mathfrak{N}}}^{\left(
C\right) }\tau ^{\left( 1-\alpha \right) /2}$ \\ \hline
$D$ & $\frac{\sigma \tau ^{1/2}}{\sqrt{2\pi }}+C_{\mathfrak{L}}^{\left(
D\right) }\tau ^{\left( 3-\alpha \right) /2}$ & $C_{\mathfrak{M}}^{\left(
D\right) }\tau ^{\left( 2-\alpha \right) /2}$ & $\frac{1}{\sqrt{2\pi }\sigma
\tau ^{1/2}}+C_{\mathcal{\mathfrak{N}}}^{\left( D\right) }\tau ^{\left(
1-\alpha \right) /2}$ \\ \hline
\end{tabular}%
,
\end{equation*}%
where%
\begin{equation*}
\begin{tabular}{|c|c|c|c|}
\hline
& $C_{\mathfrak{L}}$ & $C_{\mathfrak{M}}$ & $C_{\mathfrak{N}}$ \\ \hline
$A$ & $-\left( \left( \mathsf{\gamma }\right) ^{-}+\mathsf{\delta }+\mathsf{%
\varrho }\right) $ & $-\frac{1}{2}\mathrm{sign}\left( \mathsf{\gamma }%
\right) $ & $\delta _{D}\left( \mathsf{\gamma }\right) $ \\ \hline
$B$ & $\frac{1}{\pi }\Gamma \left( 1-\alpha ^{\prime }\right) r^{\alpha
^{\prime }}\cos \left( \alpha ^{\prime }\mathsf{\chi }\right) $ & $-\frac{1}{%
\pi }\alpha ^{\prime }\mathsf{\chi }$ & $\frac{1}{\pi }\Gamma \left(
1+\alpha ^{\prime }\right) r^{-\alpha ^{\prime }}\cos \left( \alpha ^{\prime
}\mathsf{\chi }\right) $ \\ \hline
$C$ & $-\left( \frac{1}{2}\mathsf{\gamma }+\mathsf{\delta }+\mathsf{\varrho }%
\right) $ & $-\frac{1}{\sqrt{2\pi }}\mathsf{\gamma }\sigma ^{-1/2}$ & $\frac{%
1}{\pi }2^{\left( \alpha -1\right) /2}\Gamma \left( \frac{\alpha +1}{2}%
\right) p\sigma ^{-\left( \alpha +1\right) }$ \\ \hline
$D$ & $-\frac{1}{\pi }2^{\left( \alpha -3\right) /2}\Gamma \left( \frac{%
\alpha -1}{2}\right) p\sigma ^{-\left( \alpha -1\right) }$ & $-\frac{1}{\pi }%
2^{\left( \alpha -2\right) /2}\Gamma \left( \frac{\alpha }{2}\right) q\sigma
^{-\alpha }$ & $\frac{1}{\pi }2^{\left( \alpha -1\right) /2}\Gamma \left( 
\frac{\alpha +1}{2}\right) p\sigma ^{-\left( \alpha +1\right) }$ \\ \hline
\end{tabular}%
.
\end{equation*}%
Here $\alpha ^{\prime }=1/\alpha $, $a_{\pm },\mathsf{\gamma ,\delta
,\varrho }$ are given by (\ref{eq:defs}), $\delta _{D}\left( \mathsf{.}%
\right) $ is the Dirac delta function, and 
\begin{equation*}
p=\left( a_{+}+a_{-}\right) \cos \left( \frac{\pi \alpha }{2}\right) <0,\ \
q=-\left( a_{+}-a_{-}\right) \sin \left( \frac{\pi \alpha }{2}\right) ,\ \ r=%
\sqrt{p^{2}+q^{2}},\ \ \mathsf{\chi }=\arctan \left( -\frac{q}{p}\right) .
\end{equation*}

\begin{proof}
See Appendix \ref{TSPsSA}.
\end{proof}
\end{proposition}

\begin{remark}
For $\mathfrak{m}^{\left( C\right) }$ a more accurate asymptotic expression
can be derived:%
\begin{equation*}
\mathfrak{m}^{\left( C\right) }\sim \left( C_{\mathfrak{M}}^{\left( C\right)
}\tau ^{1/2}+D_{\mathfrak{M}}^{\left( C\right) }\tau ^{\left( 2-\alpha
\right) /2}\right) ,
\end{equation*}%
where%
\begin{equation*}
D_{\mathfrak{M}}^{\left( C\right) }=-\frac{1}{\pi }2^{\alpha /2-1}\Gamma
\left( \frac{\alpha }{2}\right) q\sigma ^{-\alpha }.
\end{equation*}
\end{remark}

Given Proposition \ref{Main Result Integrals}, we can use Proposition \ref%
{price to vol general} to convert the price limits into the volatility
limits. The results are given in Proposition \ref{price to vol} below.

\begin{proposition}
\label{price to vol} In cases (A)-(D) we have the following expressions for $%
\chi _{i}$ in equation (\ref{eq:sigma_imp}):%
\begin{equation}
\begin{tabular}{|c|c|c|c|}
\hline
& $\chi _{0}-\sigma $ & $\chi _{1}$ & $\chi _{2}$ \\ \hline
$A$ & $\sqrt{2\pi }C_{\mathfrak{L}}^{\left( A\right) }\tau ^{1/2}$ & $\sqrt{%
2\pi }C_{\mathfrak{M}}^{\left( A\right) }\tau ^{-1/2}$ & $N/A$ \\ \hline
$B$ & $\sqrt{2\pi }C_{\mathfrak{L}}^{\left( B\right) }\tau ^{\alpha ^{\prime
}-1/2}$ & $\sqrt{2\pi }C_{\mathfrak{M}}^{\left( B\right) }\tau ^{-1/2}$ & $%
\left( -\frac{1}{\sqrt{2\pi }C_{\mathfrak{L}}^{\left( B\right) }}+\sqrt{2\pi 
}C_{\mathfrak{N}}^{\left( B\right) }\right) \tau ^{-\alpha ^{\prime }-1/2}$
\\ \hline
$C$ & $\sqrt{2\pi }C_{\mathfrak{L}}^{\left( C\right) }\tau ^{1/2}$ & $\sqrt{%
2\pi }C_{\mathfrak{M}}^{\left( C\right) }$ & $\frac{\sqrt{2\pi }C_{\mathfrak{%
L}}^{\left( C\right) }}{\sigma ^{2}}\tau ^{-1/2}$ \\ \hline
$D$ & $\sqrt{2\pi }C_{\mathfrak{L}}^{\left( D\right) }\tau ^{\left( 2-\alpha
\right) /2}$ & $\sqrt{2\pi }C_{\mathfrak{M}}^{\left( D\right) }\tau ^{\left(
1-\alpha \right) /2}$ & $\sqrt{2\pi }\left( \frac{C_{\mathfrak{L}}^{\left(
D\right) }}{\sigma ^{2}}+C_{\mathfrak{N}}^{\left( D\right) }\right) \tau
^{-\alpha /2}$ \\ \hline
\end{tabular}%
.  \label{eq: xi}
\end{equation}

\begin{proof}
Follows from Propositions \ref{price to vol general}, \ref{Main Result
Integrals}.
\end{proof}
\end{proposition}

\begin{remark}
For $\chi _{1}^{\left( C\right) },\chi _{2}^{\left( C\right) }$ more
accurate asymptotic expressions can be derived:%
\begin{align*}
\chi _{1}^{\left( C\right) }& \sim \sqrt{2\pi }\left( C_{\mathfrak{M}%
}^{\left( C\right) }+D_{\mathfrak{M}}^{\left( C\right) }\tau ^{\left(
1-\alpha \right) /2}\right) , \\
\chi _{2}^{\left( C\right) }& \sim \sqrt{2\pi }\left( \frac{C_{\mathfrak{L}%
}^{\left( C\right) }}{\sigma ^{2}}+C_{\mathfrak{N}}^{\left( C\right) }\tau
^{\left( 2-\alpha \right) /2}\right) \tau ^{-1/2}.
\end{align*}
\end{remark}

Motivated by our discussion in Section \ref{FXMarket}, it is of interest to
express these limits in terms of RRs and BFs.

\begin{proposition}
\label{RR and BF}In cases (A) - (D) we have the following expressions for
RRs and BFs:%
\begin{equation*}
\begin{tabular}{|c|c|c|}
\hline
& $RR$ & $BF$ \\ \hline
$A$ & $-\sqrt{2\pi ^{3}}C_{\mathfrak{L}}^{\left( A\right) }C_{\mathfrak{M}%
}^{\left( A\right) }\tau ^{1/2}$ & $N/A$ \\ \hline
$B$ & $-\sqrt{2\pi ^{3}}C_{\mathfrak{L}}^{\left( B\right) }C_{\mathfrak{M}%
}^{\left( B\right) }\tau ^{\alpha ^{\prime }-1/2}$ & $\frac{\sqrt{2\pi ^{3}}%
}{16}C_{\mathfrak{L}}^{\left( B\right) }\left( -1+2\pi C_{\mathfrak{L}%
}^{\left( B\right) }C_{\mathfrak{N}}^{\left( B\right) }+4\pi \left( C_{%
\mathfrak{M}}^{\left( B\right) }\right) ^{2}\right) \tau ^{\alpha ^{\prime
}-1/2}$ \\ \hline
$C$ & $-\pi C_{\mathfrak{M}}^{\left( C\right) }\tau ^{1/2}$ & $\frac{\sqrt{%
2\pi ^{3}}}{16}C_{\mathfrak{L}}^{\left( C\right) }\tau ^{1/2}$ \\ \hline
$D$ & $-\pi C_{\mathfrak{M}}^{\left( D\right) }\tau ^{\left( 2-\alpha
\right) /2}$ & $\frac{\sqrt{2\pi ^{3}}}{16}\left( C_{\mathfrak{L}}^{\left(
D\right) }+\sigma ^{2}C_{\mathfrak{N}}^{\left( D\right) }\right) \tau
^{\left( 2-\alpha \right) /2}$ \\ \hline
\end{tabular}%
.
\end{equation*}

\begin{proof}
Combining equations (\ref{eq:approx}) and (\ref{eq:sigma_der}) we easily get%
\begin{equation}
RR\left( \tau \right) \sim -\sqrt{\frac{\pi }{2}}\chi _{0}\chi _{1}\tau
^{1/2},  \label{eq: RR}
\end{equation}%
\begin{equation}
BF\left( \tau \right) \sim \frac{\pi }{32}\chi _{0}\left( 2\chi _{0}\chi
_{2}-\chi _{0}\chi _{1}+4\chi _{1}^{2}\right) \tau ,  \label{eq: BF}
\end{equation}%
The Proposition now follows directly from Proposition \ref{price to vol}.
\end{proof}
\end{proposition}

\begin{remark}
For $RR^{\left( C\right) }$, $RR^{\left( D\right) }$, $BF^{\left( B\right) }$%
, $BF^{\left( C\right) }$ more accurate asymptotic expressions can be
derived:%
\begin{align*}
RR^{\left( C\right) }& \sim -\sqrt{2\pi ^{3}}\left( \frac{1}{\sqrt{2\pi }}%
+C_{\mathfrak{L}}^{\left( C\right) }\tau ^{1/2}\right) C_{\mathfrak{M}%
}^{\left( C\right) }\tau ^{1/2}, \\
RR^{\left( D\right) }& \sim -\sqrt{2\pi ^{3}}\left( \frac{1}{\sqrt{2\pi }}%
+C_{\mathfrak{L}}^{\left( D\right) }\tau ^{\left( 2-\alpha \right)
/2}\right) C_{\mathfrak{M}}^{\left( D\right) }\tau ^{\left( 2-\alpha \right)
/2}, \\
BF^{\left( B\right) }& \sim \frac{\sqrt{2\pi ^{3}}}{16}C_{\mathfrak{L}%
}^{\left( B\right) }\left( -1+2\pi C_{\mathfrak{L}}^{\left( B\right) }C_{%
\mathfrak{N}}^{\left( B\right) }+4\pi \left( C_{\mathfrak{M}}^{\left(
B\right) }\right) ^{2}-\pi C_{\mathfrak{L}}^{\left( B\right) }C_{\mathfrak{M}%
}^{\left( B\right) }\tau ^{\alpha ^{\prime }}\right) \tau ^{\alpha ^{\prime
}-1/2}, \\
BF^{\left( C\right) }& \sim \frac{\sqrt{2\pi ^{3}}}{16}\left( C_{\mathfrak{L}%
}^{\left( C\right) }+\sigma ^{2}C_{\mathfrak{N}}^{\left( C\right) }\tau
^{\left( 1-\alpha \right) /2}\right) \tau ^{1/2}.
\end{align*}
\end{remark}

Examination of the table in Proposition \ref{RR and BF} shows that in all
cases (with the exception of the degenerate case (A)), RRs and BFs go to
zero for small $\tau $. In other words, TSPs cannot produce finite limits
for RRs and BFs for $\tau \rightarrow 0$. On the other hand, the rate of
decay is generally slower than for regular diffusions.

\subsubsection{Tests\label{Tests}}

In order to study the quality of the above asymptotic formulas in detail we
start with a representative LGP considered in Proposition \ref{prop:GLCall}.
As in Section \ref{ConcreteELPsLong}, we choose for illustrative purposes $%
\vartheta =0.0075$, $\kappa _{-}=1.00$. Straightforward calculation yields $%
C_{\mathfrak{L}}^{\left( A\right) }=0.0507$, $C_{\mathfrak{M}}^{\left(
A\right) }=-0.5$, $C_{\mathfrak{N}}^{\left( A\right) }=0.0$. The quality of
the corresponding asymptotic formulas vs. exact analytical expressions
calculated in Proposition \ref{prop:GLCall} is shown in Table \ref{tab:
comp_LGP}. Here and below $\bar{C}$ and $C$ stand for the values calculated
numerically and analytically, respectively, and similarly for other
quantities of interest.

\begin{table}[h]
\centering\noindent\ \noindent 
\makebox[\textwidth]{
    \begin{tabular}{ |c|c|c|c|}
        \hline
         $\log _{10}\left( \tau \right) $ & -2  & -4 & -6   \\
        \hline\hline
        $\bar{C}_{\mathfrak{L}}^{(A)}/C_{\mathfrak{L}}^{(A)}$ &0.941 & 0.994 & 0.999   \\
        $\bar{C}_{\mathfrak{M}}^{(A)}/C_{\mathfrak{M}}^{(A)}$ &0.941 & 0.994 & 0.999    \\
        $\bar{C}_{\mathfrak{N}}^{(A)}$ &0.304 & 0.030 & 0.003   \\
         \hline
  \end{tabular}}
\caption{Analytical vs. numerical values for $C_{\mathfrak{L}}^{\left(
A\right) }$, $C_{\mathfrak{M}}^{\left( A\right) }$, $C_{\mathfrak{N}%
}^{\left( A\right) }$.}
\label{tab: comp_LGP}
\end{table}

Next, we up the ante and consider realistic TSPs. Table \ref{tab: param}
lists their parameters as well as the values of $C_{\mathfrak{L}}$, $C_{%
\mathfrak{M}}$, $C_{\mathfrak{N}}$ as computed by Proposition \ref{Main
Result Integrals}. The quality of the corresponding asymptotic formulas is
summarized in Tables \ref{tab: comp}, \ref{tab: comp1}.

\begin{table}[h]
\centering\noindent 
\makebox[\textwidth]{
    \begin{tabular}{ |c|c|c|c|c|c|c|c|c|c|}
        \hline
          & $\alpha$ & $c_{+}$ & $c_{-}$ & $\kappa_{+}$ & $\kappa_{-}$ & $\sigma$ & $C_{\mathfrak{L}}$ & $C_{\mathfrak{M}}$ & $C_{\mathfrak{N}}$ \\
        \hline\hline       
        A & 0.66 &  0.1305 & 0.0615  & 6.5022 & 3.0888 & 0.00 & 0.1863 & 0.5000 & 0.0000\\
        B &1.50  & 0.0069  &  0.0063 &  1.9320 & 0.4087  & 0.00 & 0.0670 & 0.0096 & 3.6492\\       
        C & 0.66  & 0.0521  & 0.0245  & 6.5022 & 3.0888 & 0.10 & 0.0599  & 0.1353 & -2.2867 \\
        D &1.50  & 0.0028 &  0.0025 &   1.9320 & 0.4087 & 0.10 & 0.0192 & 0.0052 & -0.9610\\
         \hline
  \end{tabular}
  }
\caption{Parameters for four representative TSPs and analytical values for $%
C_{\mathfrak{L}}$, $C_{\mathfrak{M}}$, $C_{\mathfrak{N}}$. Parameters for
processes (A) and (B) are approximately calibrated to the market vols for 2y
options; parameters for processes (C) and (D) are chosen in such a way that
the market ATM vol for 2y options is recovered. In addition, $D_{\mathfrak{M}%
}^{\left( C\right) }=-0.2395$. }
\label{tab: param}
\end{table}

\begin{table}[h]
\centering\noindent 
\makebox[\textwidth]{
    \begin{tabular}{ |c|c|c|c|c|c|c|}
        \hline
           & $\log \left( \tau \right) $ & -2  & -4 & -6 & -8 & -10   \\
        \hline\hline
        A & $\bar{C}_{\mathfrak{L}}^{(A)}/C_{\mathfrak{L}}^{(A)}$ &0.74 & 0.94 & 0.99 & 1.00 & 1.00      \\
        A & $\bar{C}_{\mathfrak{M}}^{(A)}/C_{\mathfrak{M}}^{(A)}$ &0.60 & 0.92 & 0.81 & 0.76 & 1.28        \\
        \hline
        B & $\bar{C}_{\mathfrak{L}}^{(B)}/C_{\mathfrak{L}}^{(B)}$ &0.92 & 0.98 & 1.00 & 1.00 & 1.00     \\
        B & $\bar{C}_{\mathfrak{M}}^{(B)}/C_{\mathfrak{M}}^{(B)}$ &0.43 & 0.86 & 0.97 & 0.99 & 1.00      \\
        B & $\bar{C}_{\mathfrak{N}}^{(B)}/C_{\mathfrak{N}}^{(B)}$ &1.00 & 1.00 & 1.00 & 1.00 & 1.00    \\
        \hline
        C & $\bar{C}_{\mathfrak{L}}^{(C)}/C_{\mathfrak{L}}^{(C)}$ & 0.51 &0.76 & 0.87 & 0.95 & 0.98         \\
        C & $\bar{C}_{\mathfrak{M}}^{(C)}/C_{\mathfrak{M}}^{(C)}$ & 0.29 & 0.63 & 0.82 & 0.92 & 0.96         \\
        C & $\bar{D}_{\mathfrak{M}}^{(C)}/D_{\mathfrak{M}}^{(C)}$ & 0.86 & 0.96 & 0.99 & 1.00 & 1.00         \\
        C & $\bar{C}_{\mathfrak{N}}^{(C)}/C_{\mathfrak{N}}^{(C)}$ & 0.75 & 0.94 & 0.99 & 1.00 & 1.00         \\
        \hline
        D & $\bar{C}_{\mathfrak{L}}^{(D)}/C_{\mathfrak{L}}^{(D)}$ &0.83 & 0.94 & 0.98 & 0.99 & 1.00       \\
        D & $\bar{C}_{\mathfrak{M}}^{(D)}/C_{\mathfrak{M}}^{(D)}$ &0.27 & 0.72 & 0.91 & 0.97 & 0.99       \\
        D & $\bar{C}_{\mathfrak{N}}^{(D)}/C_{\mathfrak{N}}^{(D)}$ &0.91 & 0.97 & 0.99 & 1.00 & 1.00       \\
         \hline
  \end{tabular}
  }
\caption{Analytical vs. numerical values for $C_{\mathfrak{L}}$, $C_{%
\mathfrak{M}}$, $C_{\mathfrak{N}}$, and $D_{\mathfrak{M}}^{\left( C\right) }$%
.}
\label{tab: comp}
\end{table}

To numerically examine the performance of the short-time asymptotic
expansions, Table \ref{tab: comp1} compares the asymptotic results given by (%
\ref{eq: xi}) with the results for $\chi _{0}-\sigma $, $\chi _{1}$, and $%
\chi _{2}$ obtained by a direct computation of the LL integrals in Equations
(\ref{eq: LL1}), (\ref{eq: LL2}), (\ref{eq: LL3}). Note that high-precision
numerical computation of these integrals is a rather delicate affair; to
ensure stable results, we used adaptive Gauss-Kronrod quadrature combined
with both a judicious choice of integration region and a very high number of
integration nodes (often exceeding $10^{5}$ nodes). Table \ref{tab: comp1}
provides proof for the validity of the results given by (\ref{eq: xi}), as
the exact and asymptotic results converge to each other for sufficiently
small $\tau $. Unfortunately, this convergence is rather slow and only truly
satisfactory for $\tau $ around $10^{-4}$ years, i.e. for time-scales in the
order of minutes or seconds. While the asymptotic expressions do provide the
correct limiting behavior, it is therefore questionable how useful they are
for practical work.

\begin{table}[h]
\centering\noindent 
\makebox[\textwidth]{
    \begin{tabular}{ |c|c|c|c|c|c|c|c|}
        \hline
           & $\log\left( \tau \right) $ & 0 & -2  & -4 & -6 & -8 & -10   \\
        \hline\hline
        A & $\log\left(\chi_{0}-\sigma\right)$ & -0.92 &-1.46 & -2.36 & -3.34 & -4.33 & -5.33      \\
        A & $\log\left(\bar{\chi}_{0}-\sigma\right)$ & -0.33 & -1.33 & -2.33 & -3.33 & -4.33 & -5.33      \\
        A & $\log\left(\chi_{1}\right)$ & -1.34 &0.88 & 2.06 & 3.01 & 3.98 & 5.20      \\
        A & $\log\left(\bar{\chi}_{1}\right)$ & 0.10 & 1.10 & 2.10 & 3.10 & 4.10 & 5.10        \\
	  \hline
        B & $\log\left(\chi_{0}-\sigma\right)$ & -0.91 &-1.14 & -1.45 & -1.78 & -2.11 & -2.44    \\
        B & $\log\left(\bar{\chi}_{0}-\sigma\right)$ & -0.77 &-1.11 & -1.44 & -1.77 & -2.11 & -2.44    \\
        B & $\log\left(\chi_{1}\right)$ & -1.87 + $\pi i$ &-0.98 & 0.32 & 1.37 & 2.38 & 3.38     \\
        B & $\log\left(\bar{\chi}_{1}\right)$ & -1.62 &-0.62 & 0.38 & 1.38 & 2.38 & 3.38      \\
        B & $\log\left(\chi_{2}\right)$ & 0.23 & 2.78 & 5.16 & 7.50 & 9.84 & 12.17    \\
        B & $\log\left(\bar{\chi}_{2}\right)$ & 0.51 & 2.84 & 5.17 & 7.51 & 9.84 & 12.17    \\
        \hline
	  C & $\log\left(\chi_{0}-\sigma\right)$ & -1.57 &- 2.11 & -2.94 & -3.88 & -4.85 & --5.83         \\
        C & $\log\left(\bar{\chi}_{0}-\sigma\right)$ & -0.82 &-1.82 & -2.82 & -3.82 & -4.82 & -5.82         \\
        C & $\log\left(\chi_{1}\right)$ & -2.75 & -1.00 & -0.67 & -0.55 & -0.51 & -0.49        \\
        C & $\log\left(\bar{\chi}_{1}\right)$ & -0.47 & -0.47 &-0.47 &-0.47 & -0.47 & -0.47         \\
        C & $\log\left(\chi_{2}\right)$ & -0.25 & 1.71 & 3.00 & 4.10 & 5.14 & 6.16        \\
        C & $\log\left(\bar{\chi}_{2}\right)$ & 1.18 & 2.18 & 3.18 & 4.18 & 5.18 &6.18         \\
        \hline
	  D & $\log\left(\chi_{0}-\sigma\right)$ & -1.56 & -1.90& -2.34 & -2.83 & -3.32 & -3.82      \\
        D & $\log\left(\bar{\chi}_{0}-\sigma\right)$ & -1.32 &-1.82 & -2.32 &-2.82 & -3.32 &-3.82       \\
        D & $\log\left(\chi_{1}\right)$ & -2.42 + $\pi i$  &-1.95 & -1.03 & -0.43 & 0.10 & 0.61       \\
        D & $\log\left(\bar{\chi}_{1}\right)$ & -1.88 &-1.38 &-0.88 & -0.38 & 0.12 &0.62       \\
        D & $\log\left(\chi_{2}\right)$ & -0.36 &1.63 & 3.30 & 4.86 & 6.37 & 7.88       \\
        D & $\log\left(\bar{\chi}_{2}\right)$ & 0.38 & 1.88 & 3.38 & 4.88 & 6.38 & 7.88       \\
         \hline
  \end{tabular}
  }
\caption{Analytical and numerical values for $\log (\protect\chi _{0}-%
\protect\sigma )$, $\log \left( \protect\chi _{1}\right) $, and $\log \left( 
\protect\chi _{2}\right) $ as functions of $\log \left( 1/\protect\tau %
\right) $.}
\label{tab: comp1}
\end{table}

\begin{figure}[h]
\subfigure[$\chi_{0}-\sigma$] {\includegraphics[width=0.75\textwidth, angle=0]
{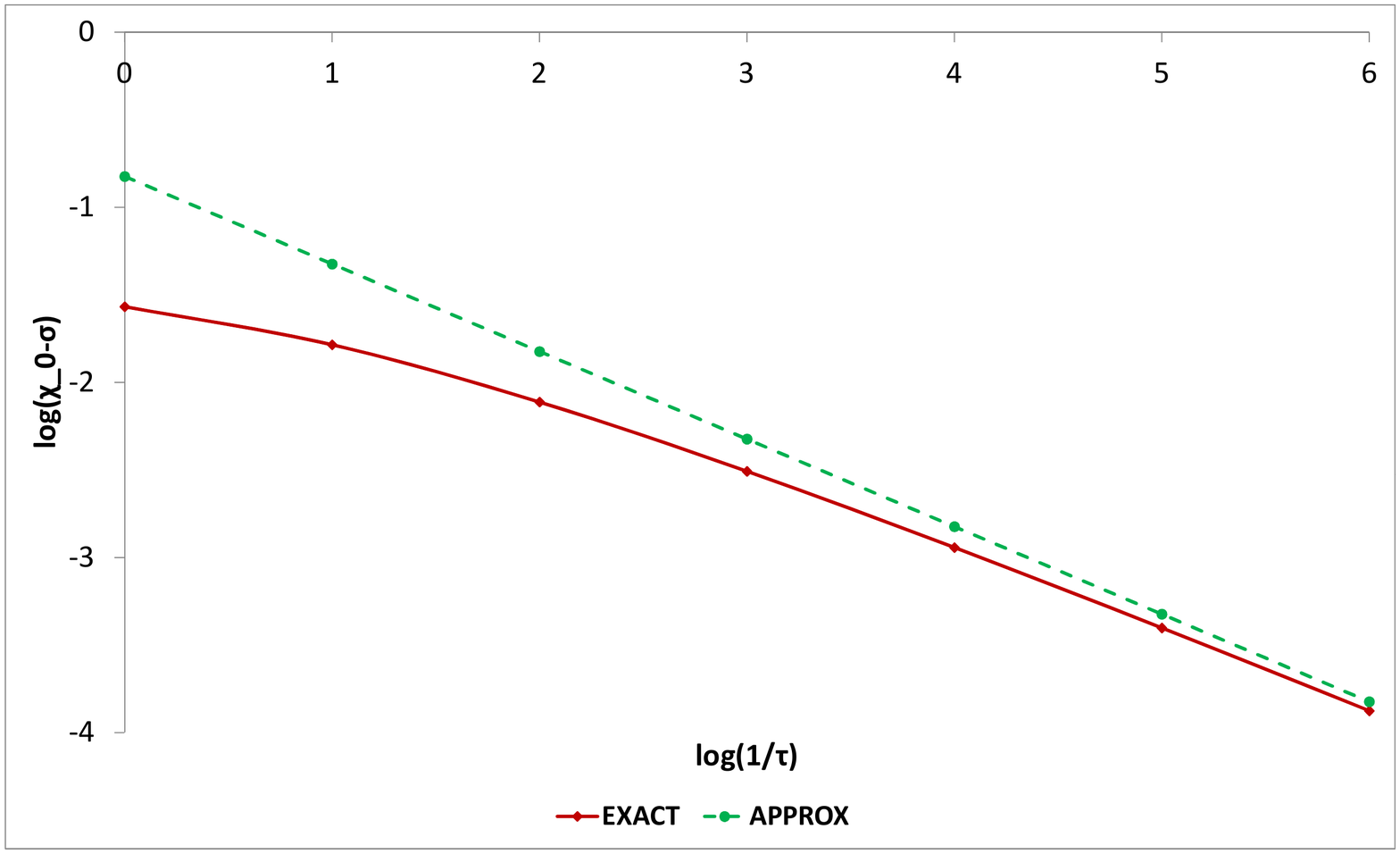}} 
\subfigure[$\chi_{1}$] {\includegraphics[width=0.75\textwidth, angle=0]
{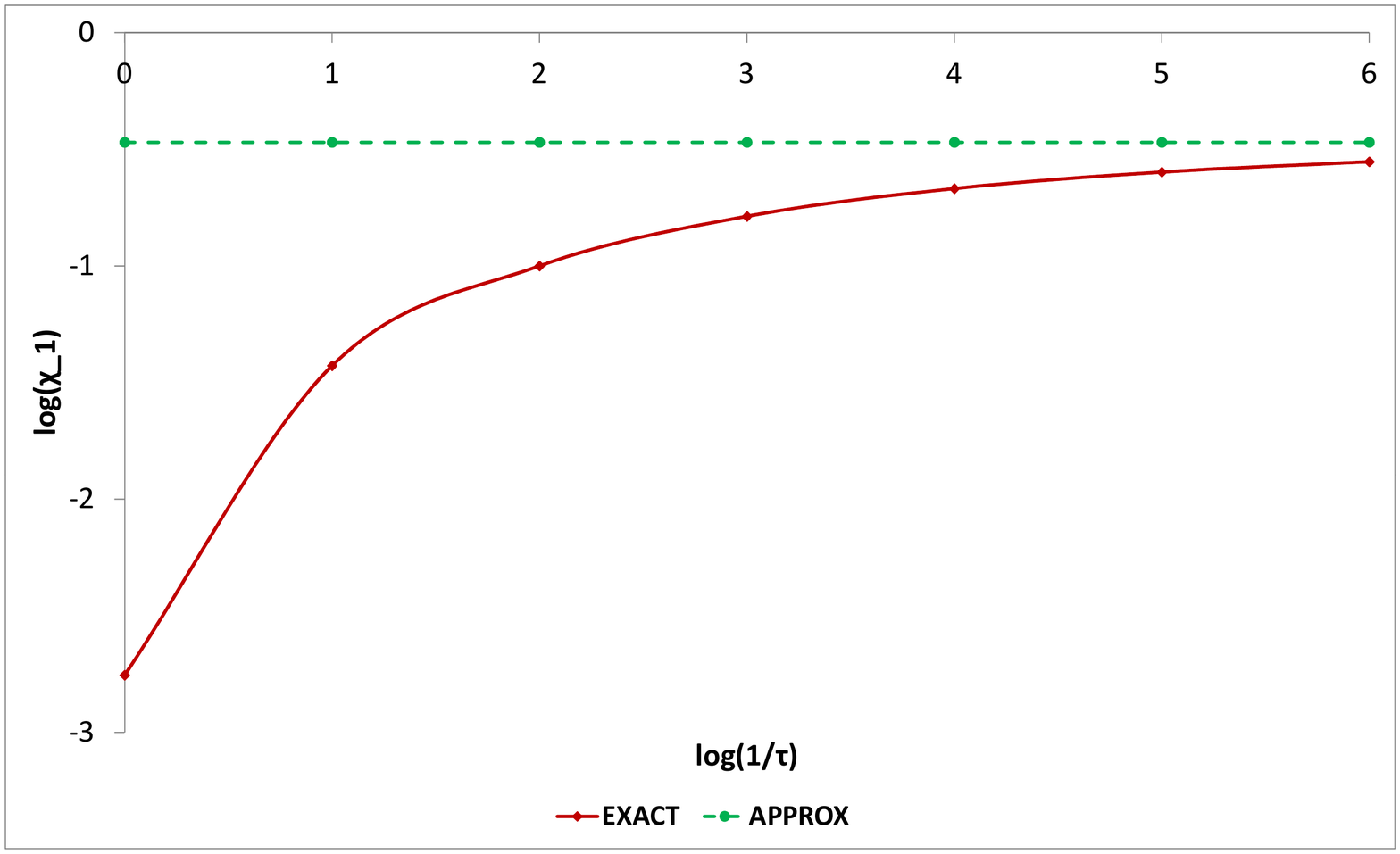}} 
\subfigure[$\chi_{2}$] {\includegraphics[width=0.75\textwidth, angle=0]
{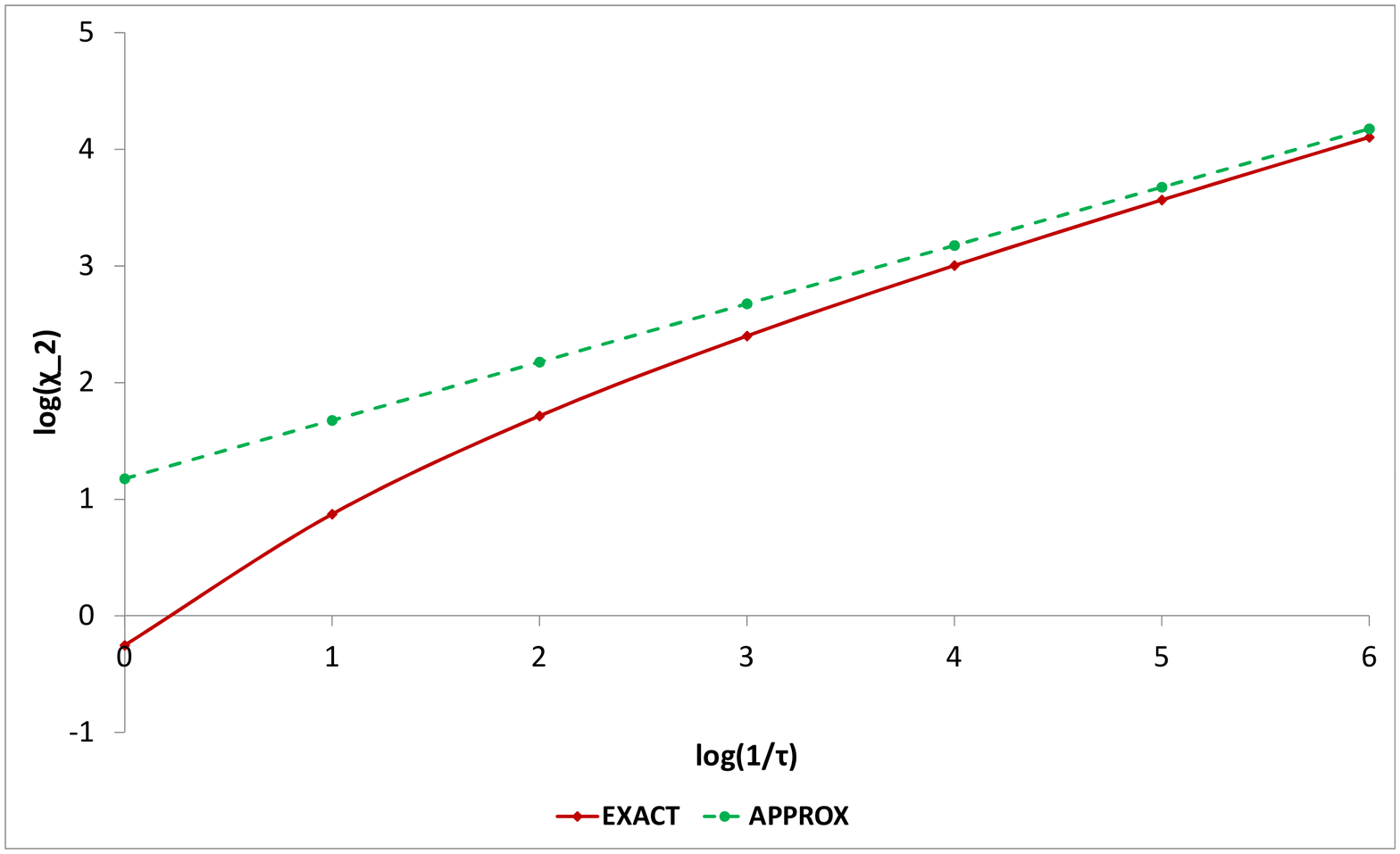}}
\caption{{}Comparison of the exact and asymptotic short-time behavior of $%
\protect\chi _{0}\left( \protect\tau \right) -\protect\sigma $, $\protect%
\chi _{1}\left( \protect\tau \right) $, and $\protect\chi _{2}\left( \protect%
\tau \right) $, for test case C.}
\label{fig: short-time TSP}
\end{figure}

To further illustrate the convergence of exact and asymptotic results,
Figure \ref{fig: short-time TSP} contains log-log plots of the short-time
behavior of $\chi _{0}-\sigma $, $\chi _{1}$, and $\chi _{2}$, for test case
C ($1<\alpha <2$, $\sigma >0$). It is evident that the straight-line
behavior predicted by (\ref{eq: xi}) is not realized for values of $\tau $
larger than, at most, a few hours.

\clearpage

\subsubsection{Options on Normal Inverse Gaussian Processes \label{non-ATM
and ATM on NIGPs}}

By using (\ref{eq:NIG_pdf_1}), we can calculate short-time asymptotics for
call prices both directly and via the LL formula.

\begin{proposition}
When $k\neq 0$ the asymptotic behavior of the time value of the call option $%
\mathsf{\delta C}^{NIG}\left( v,k\right) $ and its derivatives is linear in
time%
\begin{equation*}
\frac{\partial ^{p}\mathsf{\delta C}^{NIG}\left( v,k\right) }{\partial k^{p}}%
=\mathsf{c}_{p}^{NIG}\left( k\right) v+o\left( v\right) ,\ \ \ \ \ k\neq 0,
\end{equation*}%
where $v=\sigma ^{2}\tau $.

\begin{proof}
See Appendix \ref{NIGPsSA}.
\end{proof}
\end{proposition}

When $k=0$, we need to analyze the asymptotic behavior of of the integrals $%
\mathfrak{l},\mathfrak{m},\mathfrak{n}$ given by (\ref{eq: Main Integrals})
with $E^{NIG}\left( v,u\right) $ of the form%
\begin{equation}
E^{NIG}\left( v,u\right) =\exp \left( v\mathcal{\bar{\varkappa}}\left( 
\mathcal{\bar{\varkappa}}-\sqrt{\mathcal{\bar{\omega}}^{2}+u^{2}}\right)
\right) .  \label{eq: E NIGP}
\end{equation}

\begin{proposition}
Consider $\mathfrak{l},\mathfrak{m},\mathfrak{n}$ defined by formulas (\ref%
{eq: Main Integrals}) with $E^{NIG}\left( v,u\right) $ given by (\ref{eq: E
NIGP}) and the ATM call price $\mathsf{C}^{NIG}\left( v,0\right) $ and its
derivatives with respect to log-strike $k$. Their asymptotic behavior is
given by%
\begin{align*}
\mathfrak{l}\left( v\right) & \sim \mathsf{C}^{NIG}\left( v,0\right) \sim -%
\frac{\mathcal{\bar{\varkappa}}v\ln \left( v\right) }{\pi }, \\
\mathfrak{m}\left( v\right) & \sim \mathsf{C}_{k}^{NIG}\left( v,0\right)
\sim -\frac{1}{2}, \\
\mathfrak{n}\left( v\right) & \sim \mathsf{C}_{kk}^{NIG}\left( v,0\right)
\sim \frac{1}{\pi \mathcal{\bar{\varkappa}}v}.
\end{align*}

\begin{proof}
See Appendix \ref{NIGPsSA}.
\end{proof}
\end{proposition}

As for TSPs, the rate of convergence of the corresponding integrals for
NIGPs is slow.

\subsubsection{Options on Merton Processes\label{non-ATM and ATM on MPs}}

As was mentioned earlier, MPs do not belong to the TS class. Nevertheless,
they can be analyzed by the same methods. For brevity we assume that $\sigma
=0$. As before, we need to evaluate the integrals $\mathfrak{l},\mathfrak{m},%
\mathfrak{n}$ given by (\ref{eq: Main Integrals}) with $E^{M}\left(
v,u\right) $ of the form%
\begin{equation}
E^{M}\left( v,u\right) =\exp \left( v\left( \left( e^{q\left( iu+1/2\right)
-\eta ^{2}Q\left( u\right) /2}-1\right) +\left( 1-e^{q}\right) \left( iu+%
\frac{1}{2}\right) \right) \right) .  \label{eq: E Merton}
\end{equation}

\begin{proposition}
\label{prop: merton short}Consider $\mathfrak{l},\mathfrak{m},\mathfrak{n}$
defined by formulas (\ref{eq: Main Integrals}) with $E^{M}\left( \tau
,u\right) $ given by (\ref{eq: E Merton}) and the ATM call price $\mathsf{C}%
^{M}\left( v,0\right) $ and its derivatives with respect to log-strike $k$.
Their asymptotic behavior is given by%
\begin{align*}
\mathfrak{l}\left( v\right) & \sim \mathsf{C}^{M}\left( v,0\right) \sim
v\left( \left( 1-e^{q}\right) ^{+}+e^{q}\mathsf{C}^{BS}\left( \eta
,-q\right) \right) , \\
\mathfrak{m}\left( v\right) & \sim \mathsf{C}_{k}^{M}\left( v,0\right) \sim -%
\frac{1}{2}\mathrm{sign}\left( 1-e^{q}\right) , \\
\mathfrak{n}\left( v\right) & \sim \mathsf{C}_{kk}^{M}\left( v,0\right) \sim 
\frac{1}{v}\delta \left( 1-e^{q}\right) .
\end{align*}%
where $v=\lambda \tau $.

\begin{proof}
See Appendix \ref{MPsSA}.
\end{proof}
\end{proposition}

We can obtain the result of Proposition \ref{prop: merton short} directly
from the Merton's formula. To see this, let $\mathsf{C}\left( \tau ,\bar{k}%
\tau ;\sigma \right) $ be the normalized BS\ price of a call option
considered as a function of time, time-proportional log-strike, and
volatility,%
\begin{equation*}
\mathsf{C}\left( \tau ,\bar{k}\tau ;\sigma \right) =\Phi \left( \left( -%
\frac{\bar{k}}{\sigma }+\frac{\sigma }{2}\right) \sqrt{\tau }\right) -e^{%
\bar{k}\tau }\Phi \left( -\left( \frac{\bar{k}}{\sigma }+\frac{\sigma }{2}%
\right) \sqrt{\tau }\right) .
\end{equation*}%
To the leading order we can represent the normalized price of an ATM call
option\ on a MP as follows 
\begin{equation*}
\mathsf{C}^{M}\left( \tau ,0\right) \sim \mathsf{C}\left( \tau ,-\lambda
\tau \left( 1-e^{q}\right) ;0\right) +\lambda \tau e^{q}\mathsf{C}\left(
\tau ,-\left( \lambda \tau \left( 1-e^{q}\right) +q\right) ;\frac{\eta }{%
\sqrt{\tau }}\right) .
\end{equation*}%
We have 
\begin{align*}
\mathsf{C}\left( \tau ,-\lambda \tau \left( 1-e^{q}\right) ;0\right) &
=\left( 1-e^{-\lambda \tau \left( 1-e^{q}\right) }\right) ^{+}=\lambda \tau
\left( 1-e^{q}\right) ^{+}, \\
\lambda \tau e^{q}\mathsf{C}\left( \tau ,-\left( \lambda \tau \left(
1-e^{q}\right) +q\right) ;\frac{\eta }{\sqrt{\tau }}\right) & \sim \lambda
\tau e^{q}\left( \Phi \left( \frac{q}{\eta }+\frac{\eta }{2}\right)
-e^{-q}\Phi \left( \frac{q}{\eta }-\frac{\eta }{2}\right) \right) .
\end{align*}%
Accordingly, 
\begin{equation*}
\mathsf{C}^{M}\left( \tau ,0\right) \sim \lambda \tau \left( \left(
1-e^{q}\right) ^{+}+e^{q}\mathsf{C}^{BS}\left( \eta ,-q\right) \right) .
\end{equation*}%
in agreement with our previous result (recall that, from (\ref{eq: cl}) $%
\mathsf{C}=\mathfrak{l}$). Expressions for $\mathfrak{m}$ and $\mathfrak{n}$
can be obtained in the same manner.

\subsection{Local Volatility Processes\label{LVPsS}}

Consider the general local volatility case and assume that an underlying is
governed by the SDE of the form 
\begin{equation}
dF\left( t\right) =\sigma _{loc}^{N}\left( F\left( t\right) \right) dW\left(
t\right) ,\ \ \ \ \ F\left( 0\right) =F_{0},  \label{eq:norm_SDE}
\end{equation}%
where $\sigma _{loc}^{N}\left( F\right) $ is the so-called \textit{Normal
local volatility}. The corresponding \textit{log-normal local volatility} is
given by $\sigma _{loc}^{N}\left( F\right) /F$, for brevity we denote it as $%
\sigma _{loc}\left( F\right) $. We consider options on the underlying driven
by Brownian motion with local volatility described by (\ref{eq:norm_SDE}).
When $F_{t}$ is driven by a Brownian process with local volatility,\ we call
the corresponding process a local volatility process (LVP). It is well-known
(and had been established via asymptotic methods by \cite{gavalas} a long
time ago) that the corresponding short term implied volatility $\sigma
_{imp}\left( \tau ,k\right) $, is independent of $\tau $ to the leading
order while RRs and BFs are proportional to $\tau ^{1/2}$and $\tau $,
respectively. Thus, market observed behavior of FX\ volatilities (see Table %
\ref{tab:inputs}) cannot be explained in the local volatility framework. Let
us assume that 
\begin{equation*}
\sigma _{loc}\left( k\right) =\sum\limits_{n\geq 0}\xi _{n}\frac{k^{n}}{n!}%
,\ \ \ \ \ \sigma _{imp}\left( \tau ,k\right) =\sum\limits_{n\geq 0}\chi _{n}%
\frac{k^{n}}{n!},
\end{equation*}%
and express $\chi _{n}$, RRs and BFs in terms of $\xi _{n}$ and $\tau $.

\begin{proposition}
In the limit of $\tau \rightarrow 0$ the implied volatility $\sigma
_{imp,0}\left( k\right) $ can be written as 
\begin{equation}
\sigma _{imp,0}\left( k\right) =\left\{ 
\begin{array}{cc}
\frac{k}{\int_{0}^{k}\frac{dk^{\prime }}{\sigma _{loc}\left( k^{\prime
}\right) }}=\frac{k}{\int_{F_{0}}^{K}\frac{dK^{\prime }}{\sigma
_{loc}^{N}\left( K^{\prime }\right) }}, & k\neq 0, \\ 
\sigma _{loc}\left( 0\right) , & k=0,%
\end{array}%
\right. \equiv b_{0}\left( k\right) .  \label{eq:sigma0_loc}
\end{equation}%
Accordingly, the coefficients $\chi _{n}^{\left( 0\right) }$ can be
expressed in terms of $\xi _{n}$ as follows:%
\begin{equation}
\chi _{0}^{\left( 0\right) }=\xi _{0},\ \ \ \ \ \chi _{1}^{\left( 0\right) }=%
\frac{\xi _{1}}{2},\ \ \ \ \ \chi _{2}^{\left( 0\right) }=\frac{2\xi _{0}\xi
_{2}-\xi _{1}^{2}}{6\xi _{0}},  \label{eq:xi_of_chi}
\end{equation}%
while RRs and BFs we have the following expressions:%
\begin{equation}
RR\left( \tau \right) \sim -\sqrt{\frac{\pi }{8}}\xi _{0}\xi _{1}\tau ^{1/2},
\label{eq: RR1}
\end{equation}%
\begin{equation}
BF\left( \tau \right) \sim \frac{\pi }{32}\xi _{0}\left( -\frac{1}{2}\xi
_{0}\xi _{1}+\frac{4}{3}\xi _{0}\xi _{2}+\frac{1}{3}\xi _{1}^{2}\right) \tau
.  \label{eq: BF1}
\end{equation}

\begin{proof}
Formula (\ref{eq:sigma0_loc}) is given in \cite{and-bro}, \cite{berestycki}, 
\cite{berestycki1}, \cite{durr1}. Rewriting it in the form%
\begin{equation}
\frac{k}{\sigma _{imp}\left( k\right) }=\int_{0}^{k}\frac{d\xi }{\sigma
_{loc}\left( \xi \right) }.  \label{eq:sigma_loc_imp}
\end{equation}%
and differentiating (\ref{eq:sigma_loc_imp}) with respect to $k$ three
times, we obtain (\ref{eq:xi_of_chi}). Substituting (\ref{eq:xi_of_chi}) in (%
\ref{eq: RR}), (\ref{eq: BF}), we obtain (\ref{eq: RR1}), (\ref{eq: BF1}).
\end{proof}
\end{proposition}

\begin{remark}
More accurate formulas incorporating linear dependence on $\tau $ are also
well-known (see,e.g., \cite{and-bro}, \cite{hagan et al}, \cite{lipton-book}%
, \cite{henry}): 
\begin{equation}
\sigma _{imp,1}\left( \tau ,k\right) =b_{0}\left( k\right) +\tau b_{1}\left(
k\right) ,  \label{eq:sigma1_loc}
\end{equation}%
where 
\begin{equation*}
b_{1}\left( k\right) =\left\{ 
\begin{array}{ll}
\frac{b_{0}^{3}\left( k\right) }{2k^{2}}\ln \left( \frac{\sigma _{loc}\left(
0\right) \sigma _{loc}\left( k\right) }{b_{0}^{2}\left( k\right) }\right) ,
& k\neq 0, \\ 
\frac{\sigma _{loc}^{2}\left( 0\right) b_{0,kk}\left( 0\right) }{4}, & k=0.%
\end{array}%
\right.
\end{equation*}
\end{remark}

QVPs can be used as a convenient test bed for checking the asymptotic
formulas (\ref{eq:sigma0_loc}) and (\ref{eq:sigma1_loc}). For the quadratic
volatility model with negative real roots, say, the terms in (\ref%
{eq:sigma1_loc}) become especially simple: 
\begin{equation*}
b_{0}\left( k\right) =\frac{\mathfrak{aq}k}{\mathfrak{S}\left( k\right) },
\end{equation*}%
\begin{equation*}
b_{1}\left( k\right) =\left\{ 
\begin{array}{cc}
\frac{\mathfrak{a}^{3}\mathfrak{q}^{3}k}{2\mathfrak{S}^{3}\left( k\right) }%
\ln \left( \frac{\left( \left( 1-\mathfrak{p}\right) ^{2}+\mathfrak{q}%
^{2}\right) \left( \left( e^{k}-\mathfrak{p}\right) ^{2}+\mathfrak{q}%
^{2}\right) \mathfrak{S}^{2}\left( k\right) }{e^{k}\mathfrak{q}^{2}k^{2}}%
\right) , & k\neq 0 \\ 
\frac{1}{24}\mathfrak{a}^{3}\left( \left( 1-\mathfrak{p}\right) ^{2}+%
\mathfrak{q}^{2}\right) \left( \left( \left( 1-\mathfrak{p}\right) ^{2}+%
\mathfrak{q}^{2}\right) ^{2}+4\mathfrak{q}^{2}\right) , & k=0.%
\end{array}%
\right.
\end{equation*}%
where%
\begin{equation*}
\mathfrak{S}\left( k\right) =\arctan \left( \frac{\mathfrak{p}-1}{\mathfrak{q%
}}\right) -\arctan \left( \frac{\mathfrak{p}-e^{k}}{\mathfrak{q}}\right) .
\end{equation*}%
The quality of these approximations versus the exact expression obtained
Proposition \ref{Prop:QuadVol} is very good, as is shown in Figure \ref%
{fig:QVPsShort}. However, it is clear that having substantial risk-reversal
and straddles for short maturities is not feasible in this setting when
parameters are of order unity.

\begin{figure}[h]
\subfigure[T=1y] {\includegraphics[width=1.0\textwidth, angle=0]
{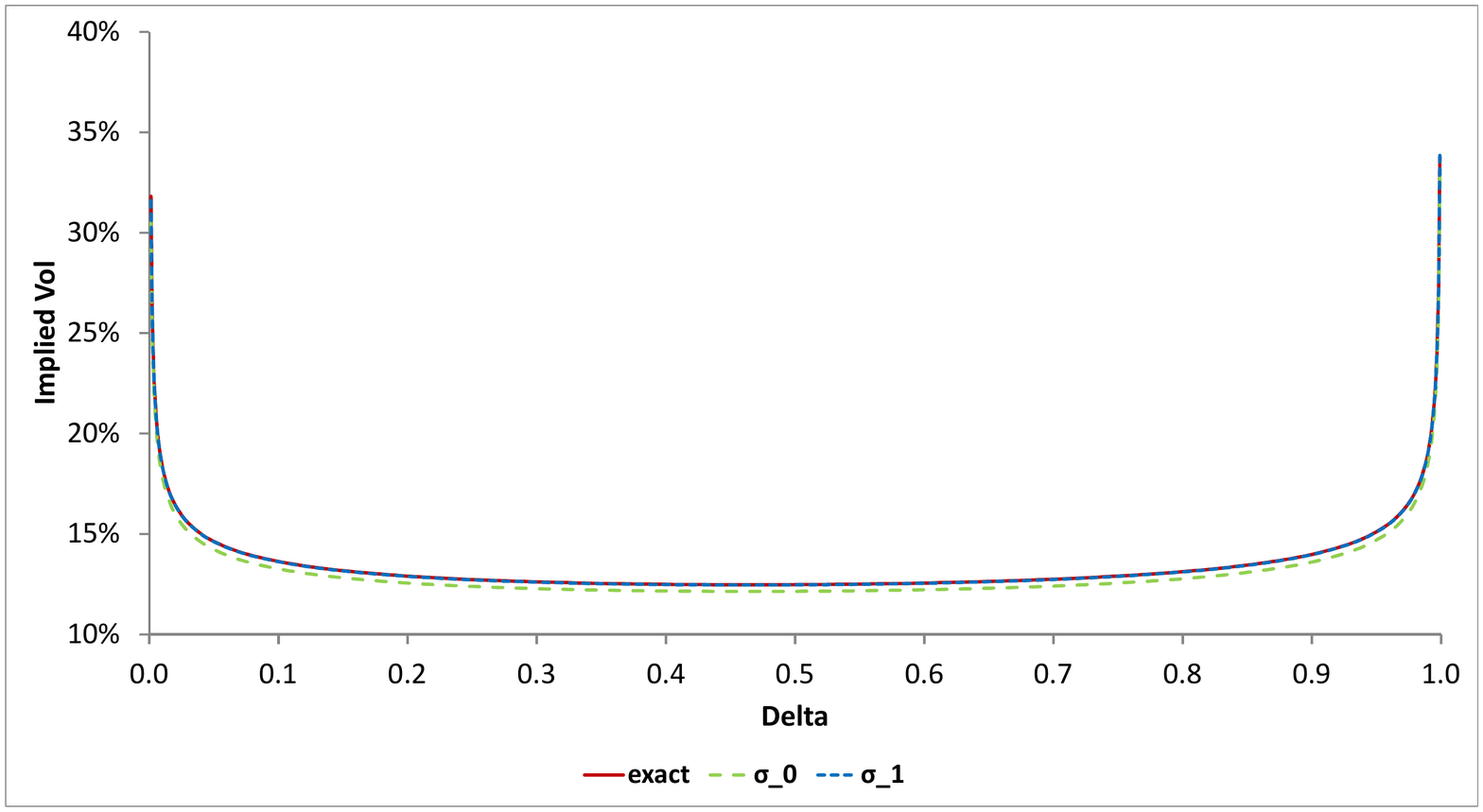}} 
\subfigure[T=5y] {\includegraphics[width=1.0\textwidth, angle=0]
{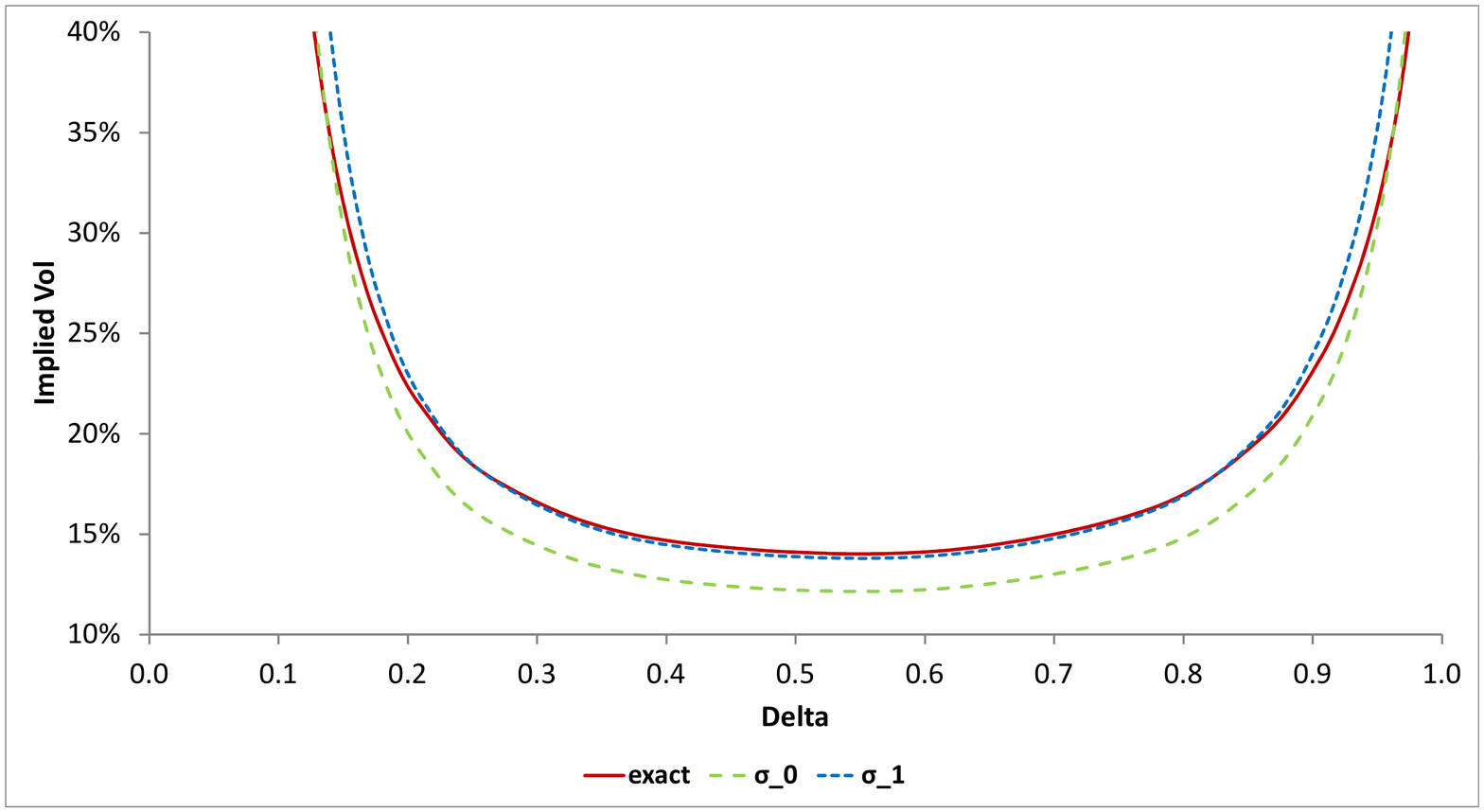}}
\caption{{}Comparison of the exact and asymptotic expressions for $\protect%
\sigma _{imp}$ for the calibrated QVP for different maturities. Here "exact"
denotes the exact volatility calculated by virtue of the formula (\protect
\ref{eq: LLQV}), while $\protect\sigma \_0$, $\protect\sigma \_1$ are given
by (\protect\ref{eq:sigma0_loc}), (\protect\ref{eq:sigma1_loc}).}
\label{fig:QVPsShort}
\end{figure}

\clearpage

\subsection{Heston Stochastic Volatility Processes\label{HSVPsS}}

We can easily analyze the asymptotic behavior of the implied volatility for
the Heston model either by using the LL formula or by direct computation,
see e.g., \cite{lewis-1}, \cite{lewis-3}, \cite{lipton-1}, \cite{lipton-book}%
, \cite{lipton-3}, \cite{medvedev}, \cite{forde1}, \cite{zeliade}, \cite%
{forde2}. Comparison of the asymptotic and exact formulas can be performed
by computing the relevant integrals numerically, which, by now is a
well-understood procedure, see, e.g., \cite{janek}, \cite{schmelzle}. In
view of Proposition \ref{price to vol general}, in order to describe the
asymptotic behavior of the implied volatility in the Heston framework, we
need to evaluate asymptotically the corresponding integrals $\mathfrak{l},%
\mathfrak{m},\mathfrak{n}$.

\begin{proposition}
The asymptotic behavior of the integrals $\mathfrak{\hat{l}},\mathfrak{\hat{m%
}},\mathfrak{\hat{n}}$ defined by (\ref{eq:lmn_hat})is described as follows%
\begin{align}
\mathfrak{\hat{l}}\left( \tau \right) & \sim \sqrt{\varpi _{0}\tau }\left(
1+\lambda _{1}\tau +\lambda _{2}\tau ^{2}+...\right) ,  \label{eq: HSVPexp}
\\
\mathfrak{\hat{m}}\left( \tau \right) & \sim \sqrt{\varpi _{0}\tau }\frac{%
\rho \varepsilon }{4\varpi _{0}}\left( 1+\mu _{1}\tau +...\right) ,  \notag
\\
\mathfrak{\hat{n}}\left( \tau \right) & \sim \frac{1}{\sqrt{\varpi _{0}\tau }%
}\left( 1+\nu _{1}\tau +\nu _{2}\tau ^{2}+...\right) ,  \notag
\end{align}%
where the corresponding $\lambda _{i},\mu _{i},\nu _{i}$ are given by (\ref%
{eq: HSVP lmn}).

\begin{proof}
See Appendix \ref{HSVPsSA}.
\end{proof}
\end{proposition}

\begin{proposition}
\label{price to vol copy SV} For HSVPs, we have the following expressions
for $\chi _{i}$ and $\sigma _{imp}\left( \tau ,k\right) $:%
\begin{align}
\chi _{0}\left( \tau \right) & \sim \sqrt{\varpi _{0}}\left( 1+\chi
_{0,1}\tau +...\right) ,  \label{eq: HSVPxi} \\
\chi _{1}\left( \tau \right) & \sim \frac{\rho \varepsilon }{4\varpi _{0}}%
\sqrt{\varpi _{0}}\left( 1+\chi _{1,1}\tau +...\right) ,  \notag \\
\chi _{2}\left( \tau \right) & \sim \frac{\varepsilon ^{2}}{12\varpi _{0}^{2}%
}\sqrt{\varpi _{0}}\left( \left( 1-\frac{5}{2}\rho ^{2}\right) +\chi
_{2,1}\tau +...\right) .  \notag
\end{align}%
where the corresponding $\chi _{i,1}$ are given by (\ref{eq: HSVP xi_01}).
Accordingly, $\sigma _{imp}\left( \tau ,k\right) $ can be written in the form%
\begin{equation}
\sigma _{imp}\left( \tau ,k\right) \sim \hat{b}_{0}\left( k\right) +\tau 
\hat{b}_{1}\left( k\right) +...,  \label{eq: Hsigmaimp}
\end{equation}%
where%
\begin{align}
\hat{b}_{0}\left( k\right) & =\sqrt{\varpi _{0}}\left( 1+\frac{\rho
\varepsilon k}{4\varpi _{0}}+\frac{\left( 1-\frac{5}{2}\rho ^{2}\right)
\varepsilon ^{2}k^{2}}{24\varpi _{0}^{2}}\right) =\sqrt{\varpi _{0}}\left( 1+%
\frac{\rho l}{4}+\frac{\left( 1-\frac{5}{2}\rho ^{2}\right) l^{2}}{24}%
\right) ,  \label{eq: Hsigmaimp0} \\
\hat{b}_{1}\left( k\right) & =\sqrt{\varpi _{0}}\left( \chi _{0,1}+\frac{%
\chi _{1,1}\rho \varepsilon k}{4\varpi _{0}}+\frac{\chi _{2,1}\varepsilon
^{2}k^{2}}{24\varpi _{0}^{2}}\right) =\sqrt{\varpi _{0}}\left( \chi _{0,1}+%
\frac{\chi _{1,1}\rho l}{4}+\frac{\chi _{2,1}l^{2}}{24}\right) ,  \notag
\end{align}%
and $l=\varepsilon k/\varpi _{0}$.

\begin{proof}
See Appendix \ref{HSVPsSA}.
\end{proof}
\end{proposition}

By using the well-known duality properties of the Brownian motion for $\tau
\rightarrow 0$ and $\tau \rightarrow \infty $, it is possible to analyze the
asymptotic behavior of the call price and the corresponding implied
volatility for fixed $k\neq 0$, i.e., for options which are not ATM. It is
clear that for $\tau \rightarrow 0$ any strike $k\neq 0$ is located "far
away", so that we can analyze the corresponding price via asymptotic
methods, more specifically via the saddlepoint approximation.

\begin{proposition}
Consider $y\in \left( -\mathsf{Y}_{+},\mathsf{Y}_{-}\right) $, where%
\begin{equation}
Y_{\pm }=\frac{\pi \mp \pi \pm 2\phi }{\bar{\rho}\varepsilon },
\label{eq:Y_pm_1}
\end{equation}%
and $\phi =\arctan \left( \bar{\rho}/\rho \right) $, $0<\phi <\pi $, and
define $\Xi _{0}\left( y\right) ,$ $\Xi _{1}\left( y\right) =\Xi
_{0}^{\prime }\left( y\right) ,$ $\Xi _{01}\left( y\right) =\Xi _{0}\left(
y\right) -\Xi _{1}\left( y\right) $ as follows%
\begin{align*}
\Xi _{0}\left( y\right) & =\frac{\varpi _{0}y\sin \left( \mathcal{X}%
_{-}\right) }{\varepsilon \sin \left( \mathcal{X}_{+}\right) }, \\
\Xi _{1}\left( y\right) & =\frac{\varpi _{0}}{\varepsilon }\frac{\left( \sin
\left( \mathcal{X}_{-}\right) \sin \left( \mathcal{X}_{+}\right) +\frac{1}{2}%
\bar{\rho}^{2}\varepsilon y\right) }{\sin ^{2}\left( \mathcal{X}_{+}\right) }%
, \\
\Xi _{01}\left( y\right) & =\frac{\varpi _{0}}{\varepsilon }\frac{\left(
\left( y-1\right) \sin \left( \mathcal{X}_{-}\right) \sin \left( \mathcal{X}%
_{+}\right) -\frac{1}{2}\bar{\rho}^{2}\varepsilon y\right) }{\sin ^{2}\left( 
\mathcal{X}_{+}\right) },
\end{align*}%
where $\mathcal{X}_{\pm }=\left( \bar{\rho}\varepsilon y+\phi \pm \phi
\right) /2$. Then for $k=-\Xi _{1}\left( y\right) $, $k\neq 0$, the
corresponding $\sigma _{imp}\left( \tau ,k\right) $ can be written in the
form%
\begin{equation*}
\sigma _{imp}\left( \tau ,k\left( y\right) \right) =\left( a_{0}\left(
y\right) +a_{1}\left( y\right) \tau +...\right) ^{1/2}=b_{0}\left( y\right)
+b_{1}\left( y\right) \tau +....
\end{equation*}%
Here the leading order terms $a_{0}\left( y\right) ,b_{0}\left( y\right) $
are given by%
\begin{align}
a_{0}\left( y\right) & =-\frac{k^{2}\left( y\right) }{2y\left( \frac{\varpi
_{0}\sin \left( \mathcal{X}_{-}\right) }{\varepsilon \sin \left( \mathcal{X}%
_{+}\right) }+k\left( y\right) \right) }=-\frac{\varpi _{0}l^{2}\left(
y\right) }{2\varepsilon y\left( \frac{\sin \left( \mathcal{X}_{-}\right) }{%
\sin \left( \mathcal{X}_{+}\right) }+l\left( y\right) \right) },
\label{eq: HSVP short a_0 b_0} \\
b_{0}\left( y\right) & =a_{0}^{1/2}\left( y\right) ,  \notag
\end{align}%
where $l=\varepsilon k/\varpi _{0}$, while the higher order terms $%
a_{1}\left( y\right) ,b_{1}\left( y\right) $ are given by (\ref{eq: HSVP
short a_1 b_1}).

\begin{proof}
See Appendix \ref{HSVPsSA}.
\end{proof}
\end{proposition}

In practice, it is convenient to solve the characteristic equation%
\begin{equation*}
\Xi _{1}\left( y\right) +k=0,
\end{equation*}%
for $y$, and express $\sigma _{imp}\left( \tau ,k\right) $ in terms of $k$,
or, equivalently, $l=\varepsilon k/\varpi _{0}$. Define the function $%
\mathfrak{Z}\left( l;\rho \right) $ such that $z^{\ast }=\mathfrak{Z}\left(
l;\rho \right) $ solves the equation%
\begin{equation*}
\frac{\left( \sin \left( \mathcal{X}_{-}^{\ast }\right) \sin \left( \mathcal{%
X}_{+}^{\ast }\right) +\frac{\bar{\rho}^{2}z^{\ast }}{2}\right) }{\sin
^{2}\left( \mathcal{X}_{+}^{\ast }\right) }=-l,
\end{equation*}%
where $\mathcal{X}_{\pm }^{\ast }=\left( \bar{\rho}z^{\ast }+\phi \pm \phi
\right) /2$. Then $y^{\ast }=\mathfrak{Z}\left( l;\rho \right) /\varepsilon $
solves the characteristic equation. While $\mathfrak{Z}$ does not have a
simple analytical form, it is very easy to calculate it numerically. The
corresponding calculation is particularly efficient since parametrically $%
\mathfrak{Z}$ depends only on $\rho $. We can define the following function%
\begin{equation*}
\mathfrak{Y}\left( l;\rho \right) =\mathfrak{Z}\left( l;\rho \right) \left( 
\frac{\sin \left( \mathcal{X}_{-}^{\ast }\right) }{\sin \left( \mathcal{X}%
_{+}^{\ast }\right) }+l\right) .
\end{equation*}%
Then, for $l\neq 0$ and $\tau \rightarrow 0$ the corresponding implied
volatility behaves as follows%
\begin{equation*}
\sigma _{imp}\left( \tau ,l\right) =\frac{\sqrt{\varpi _{0}}\left\vert
l\right\vert }{\sqrt{-2\mathfrak{Y}\left( l;\rho \right) }}\left( 1+O\left(
\tau \right) \right) .
\end{equation*}%
While the above expression is not defined for $l=0$, it is easy to see that%
\begin{equation*}
\frac{\sqrt{\varpi _{0}}\left\vert l\right\vert }{\sqrt{-2\mathfrak{Y}\left(
l;\rho \right) }}\underset{l\rightarrow 0}{\rightarrow }\sqrt{\varpi _{0}},
\end{equation*}%
so that%
\begin{equation*}
\sigma _{imp}\left( \tau ,0\right) \underset{\tau \rightarrow 0}{\rightarrow 
}\sqrt{\varpi _{0}},
\end{equation*}%
as expected. Moreover, it is not difficult to show that%
\begin{align*}
\mathfrak{Z}\left( l;\rho \right) & =-l+\frac{3\rho l^{2}}{4}+O\left(
l^{3}\right) , \\
\mathfrak{Y}\left( l;\rho \right) & =-\frac{l^{2}}{2}\left( 1-\frac{\rho l}{2%
}-\frac{\left( 1-\frac{19}{4}\rho ^{2}\right) l^{2}}{12}\right) +O\left(
l^{5}\right) , \\
\frac{\left\vert l\right\vert }{\sqrt{-2\mathfrak{Y}\left( l;\rho \right) }}%
& =1+\frac{\rho l}{4}+\frac{\left( 1-\frac{5}{2}\rho ^{2}\right) l^{2}}{24}%
+O\left( l^{3}\right) ,
\end{align*}%
in complete agreement with formulas (\ref{eq: Hsigmaimp0}).

The quality of the above approximations is illustrated in Figure \ref%
{fig:HSVPsShort}. As we can see, accuracy is acceptable but not particularly
good. Specifically, for HSVPs short-time asymptotics work well for very
short maturities ($\tau \sim 1m$), but loose their accuracy for moderate
maturities ($\tau \sim 6m$).

\begin{figure}[h]
\subfigure[T=1m] {\includegraphics[width=1.0\textwidth, angle=0]
{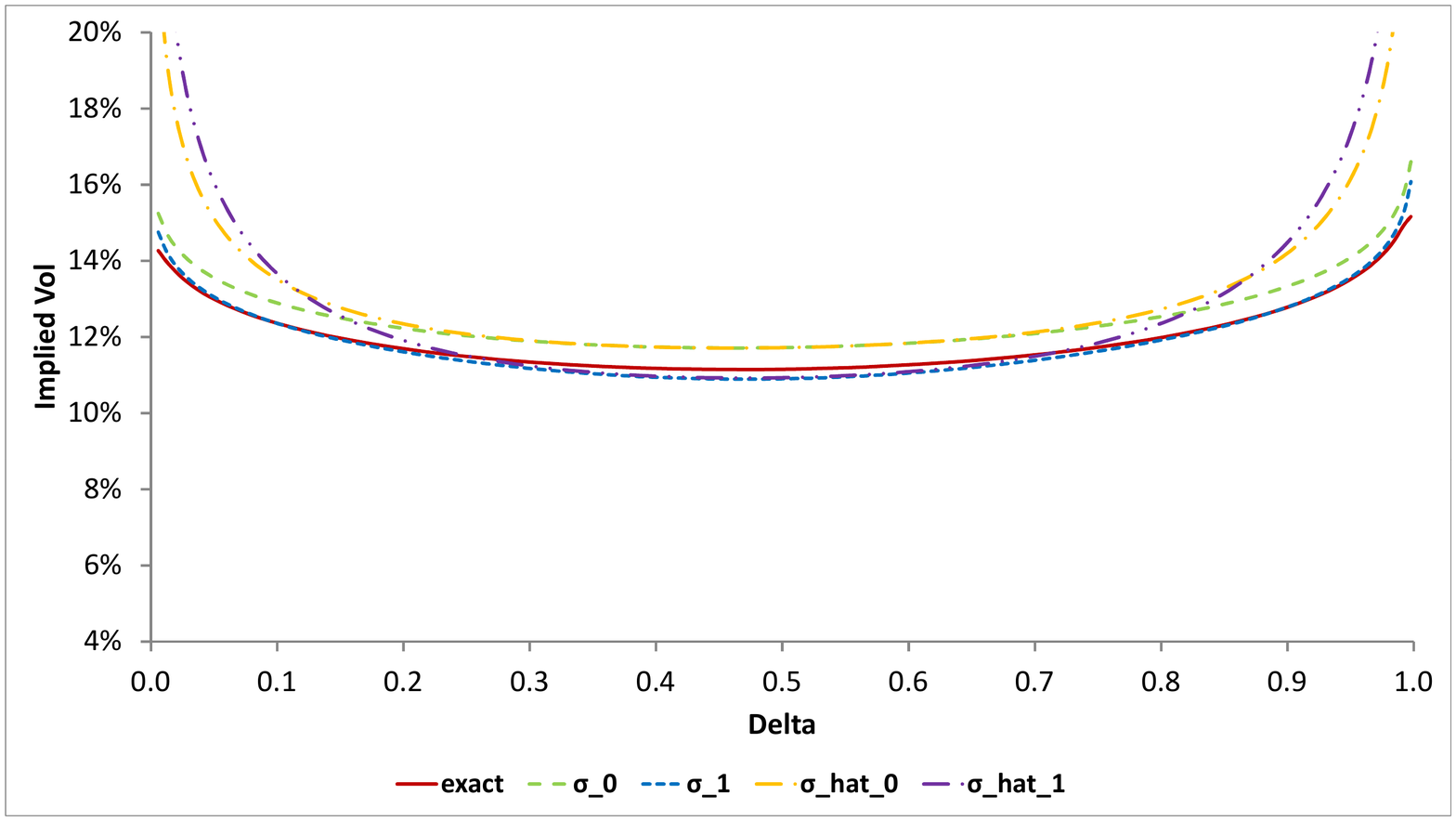}} 
\subfigure[T=6m] {\includegraphics[width=1.0\textwidth, angle=0]
{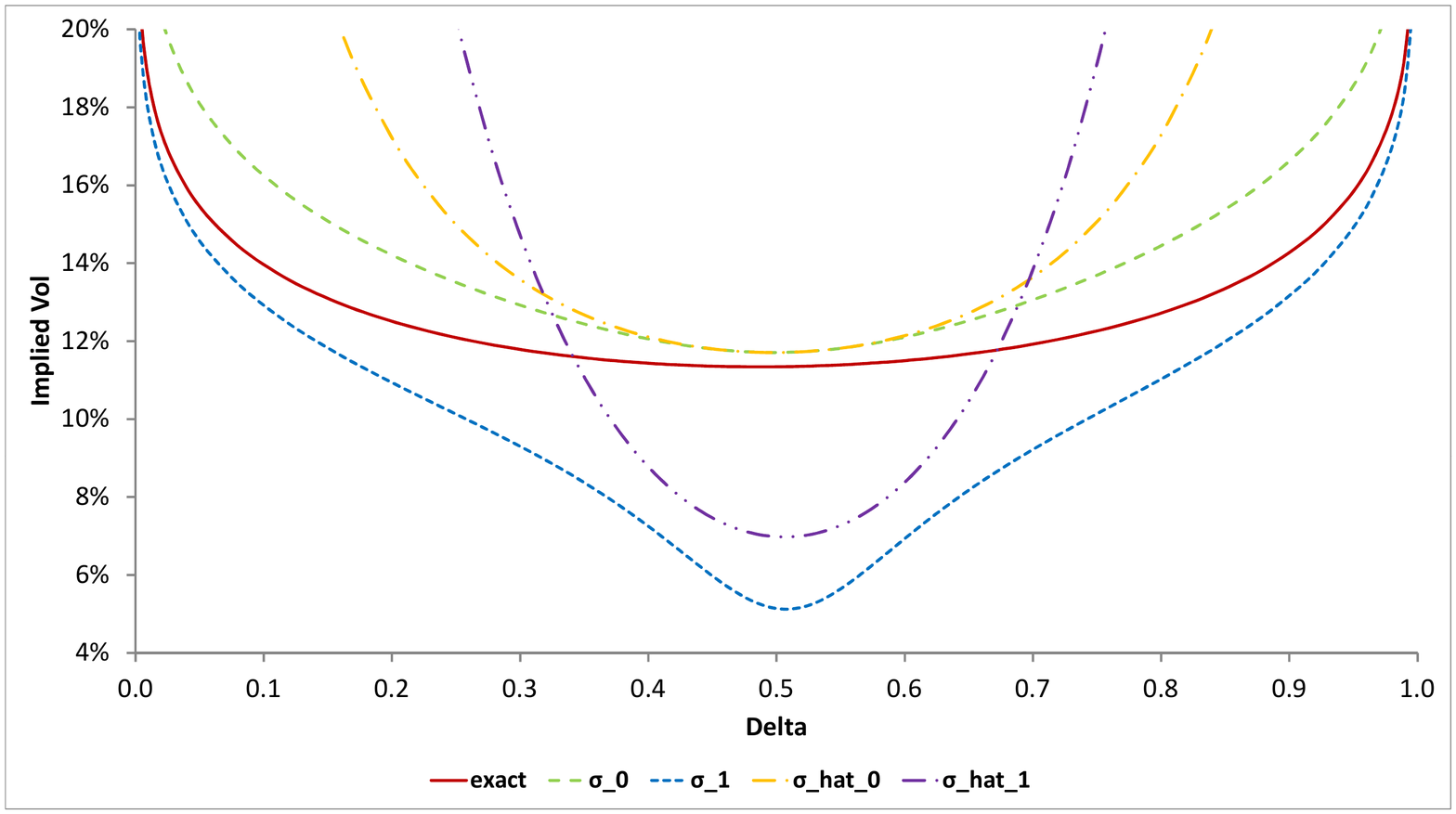}}
\caption{{}Comparison of the exact and asymptotic expressions for $\protect%
\sigma _{imp}$ for the calibrated HSVP for different maturities. Here
"exact" denotes the exact volatility calculated by virtue of the LL formula,
while $\hat{\protect\sigma}_{\_}0$, $\hat{\protect\sigma}_{\_}1$ and $%
\protect\sigma _{\_}0$, $\protect\sigma _{\_}1$ are given by $\hat{\protect%
\sigma}_{\_}0=\hat{b}_{0}$,$\ \hat{\protect\sigma}_{\_}1=\hat{b}_{0}+\protect%
\tau \hat{b}_{1}$, and $\protect\sigma \_0=b_{0}$,$\ \protect\sigma %
_{\_}1=b_{0}+\protect\tau b_{1}$.}
\label{fig:HSVPsShort}
\end{figure}

\clearpage

\section{Wing Asymptotics\label{AsymptoticsWing}}

\subsection{General Remarks\label{GeneralRemarksWings}}

It is well known that the high frequency asymptotics for Fourier integrals
cannot be obtained in closed form unless some assumptions about smoothness
or analyticity of the integrand are made. For instance, if $f\left( u\right) 
$ is $n$ times differentiable and decays at infinity sufficiently rapidly,
one can use integration by parts and show that%
\begin{equation*}
I\left( k\right) =\int_{-\infty }^{\infty }f\left( u\right)
e^{-iku}du=O\left( \left\vert k\right\vert ^{-n}\right) .
\end{equation*}%
Moreover, if $f\left( u\right) $ is meromorphic only in a strip $-\mathsf{Y}%
_{+}<\func{Im}u<\mathsf{Y}_{-}$, $\mathsf{Y}_{\pm }>\frac{1}{2}$, then it
can be shown that%
\begin{equation}
I\left( k\right) \underset{k\rightarrow \pm \infty }{\sim }\Pi _{\pm }\left(
k\right) +\mathsf{c}_{\pm }e^{-\mathsf{Y}_{\pm }\left\vert k\right\vert },
\label{eq: I(k)}
\end{equation}%
where $\Pi _{\pm }\left( k\right) $ represent contributions of the poles in
the lower (upper) half-stip, respectively, and $\mathsf{c}_{\pm }$ are
constants, see, e.g., \cite{maslov}. Below we show how to use this result in
order to calculate the wing asymptotics for the implied volatility.

\subsection{Generic Exponential L\'{e}vy Processes\label{ELPsWings}}

It is easy to apply these general results in the case of ELPs in order to
calculate the wing asymptotics of the implied volatility.

\begin{proposition}
\label{GeneralWingsProp}Assume that the function $E\left( \tau ,u\right) $
in the LL\ formula is analytical in a strip $-\mathsf{Y}_{+}<\func{Im}\left(
u\right) <\mathsf{Y}_{-}$, $\mathsf{Y}_{\pm }>\frac{1}{2}$.\footnote{%
According to the Lukacs's theorem (see, e.g., \cite{lukacs} for a
discussion), $S\left( u\right) $ is singular at $u=\mp iY_{\pm }$.} Then the
asymptotics for the time value of a call option $\mathsf{\delta C}\left(
\tau ,k\right) $ has the form%
\begin{equation}
\mathsf{\delta C}\left( \tau ,k\right) \underset{k\rightarrow \pm \infty }{%
\sim }\mathsf{c}_{\pm }e^{-\left( \mathsf{Y}_{\pm }\mp \frac{1}{2}\right)
\left\vert k\right\vert },  \label{eq:cpm}
\end{equation}%
where $\mathsf{c}_{\pm }$ are (positive) constants in (\ref{eq: I(k)}).
Accordingly,%
\begin{equation}
v_{imp}\left( \tau ,k\right) \underset{k\rightarrow \pm \infty }{\sim }\beta
_{\pm }\left\vert k\right\vert ,\ \ \ \ \ \sigma _{imp}\left( \tau ,k\right) 
\underset{k\rightarrow \pm \infty }{\sim }\sqrt{\frac{\beta _{\pm
}\left\vert k\right\vert }{\tau }},  \label{eq:sigma_imp_wing}
\end{equation}%
where%
\begin{equation*}
\beta _{\pm }=4\left( \mathsf{Y}_{\pm }-\sqrt{R\left( \mathsf{Y}_{\pm
}\right) }\right) ,\ \ \ \ \ 0<\beta _{\pm }<2.
\end{equation*}

\begin{proof}
See Appendix \ref{ELPsWA}.
\end{proof}
\end{proposition}

This result is well-known. For instance, Lee derived it by studying moment
explosions, see \cite{lee}. However, its simple mathematical nature is
seldom emphasized.

We can use these formulas in Proposition \ref{GeneralWingsProp} to obtain
expressions for $\sigma _{imp}\left( \tau ,\Delta \right) $ for deep OTM ($%
\Delta \rightarrow 0$) and deep ITM ($\Delta \rightarrow 1$) options.

\begin{proposition}
\label{DeltaWingsProp} Assume that the assumptions of Proposition \ref%
{GeneralWingsProp} hold. Then, in the OTM case we have%
\begin{align*}
& \sigma _{imp}\left( \tau ,k\right) \underset{k\rightarrow \infty }{\sim }%
\sqrt{\frac{\beta _{+}k}{\tau }}, \\
& \Delta \left( \tau ,k\right) \underset{k\rightarrow \infty }{\sim }\Phi
\left( \left( -\frac{1-\frac{1}{2}\beta _{+}}{\sqrt{\beta _{+}}}\right) 
\sqrt{k}\right) \underset{k\rightarrow \infty }{\sim }\frac{\sqrt{\beta _{+}}%
e^{-\frac{\left( 1-\beta _{+}/2\right) ^{2}k}{2\beta _{+}}}}{\left( 1-\beta
_{+}/2\right) \sqrt{k}}, \\
& \sigma _{imp}\left( \tau ,\Delta \right) \underset{\Delta \rightarrow 0}{%
\sim }\frac{\beta _{+}}{\left( 1-\frac{1}{2}\beta _{+}\right) }\sqrt{-\frac{%
2\ln \left( \Delta \right) }{\tau }}.
\end{align*}%
Similarly, for ITM, we have 
\begin{equation*}
\sigma _{imp}\left( \tau ,\Delta \right) \underset{\Delta \rightarrow 1}{%
\sim }\frac{\beta _{-}}{\left( 1-\frac{1}{2}\beta _{-}\right) }\sqrt{-\frac{%
2\ln \left( 1-\Delta \right) }{\tau }}.
\end{equation*}%
Combination of these formulas yields%
\begin{equation*}
\sigma _{imp}\left( \tau ,\Delta \right) \underset{\Delta \left( 1-\Delta
\right) \rightarrow 0}{\sim }\frac{\beta _{\mathrm{sign}\left( 1/2-\Delta
\right) }}{\left( 1-\frac{1}{2}\beta _{\mathrm{sign}\left( 1/2-\Delta
\right) }\right) }\sqrt{-\frac{2\ln \left( \frac{1}{2}-\mathrm{sign}\left(
1/2-\Delta \right) \left( \frac{1}{2}-\Delta \right) \right) }{\tau }}.
\end{equation*}
\end{proposition}

\subsubsection{Specific Exponential L\'{e}vy Processes\label{TSPWings}}

For specific ELPs, such as TSPs, NIGPs, MPs, etc., the\ corresponding
formulas can be made explicit.

\begin{proposition}
For TSPs and NIGPs Proposition \ref{GeneralWingsProp} holds provided that
the corresponding $\mathsf{Y}_{\pm }$ are defined as follows:%
\begin{align*}
\mathsf{Y}_{\pm }& =\kappa _{\pm }\mp \frac{1}{2},\ \ \ \ \ \beta _{\pm
}=\mp 2+4\left( \kappa _{\pm }-\sqrt{\kappa _{\pm }^{2}\mp \kappa _{\pm }}%
\right) ,\ \ \ TSP, \\
\mathsf{Y}_{\pm }& =\bar{\omega},\ \ \ \ \ \beta _{\pm }=4\left( \bar{\omega}%
-\mathcal{\bar{\varkappa}}\right) ,\ \ \ \ \ NIG.
\end{align*}%
Thus, for NIGPs the wing volatility is symmetric. Moreover, for NIGPs the
corresponding $\mathsf{c}_{\pm }$ in (\ref{eq:cpm}) have the form%
\begin{equation*}
\mathsf{c}_{\pm }=\sqrt{\frac{\bar{\omega}}{2\pi \left\vert k\right\vert ^{3}%
}}\frac{v}{\mathcal{\bar{\varkappa}}}e^{\mathcal{\bar{\varkappa}}^{2}v}.
\end{equation*}

\begin{proof}
See Appendices \ref{TSPsWA}, \ref{NIGPsWA} for details.
\end{proof}
\end{proposition}

The situation with MPs is somewhat different.

\begin{proposition}
For MPs $\mathsf{Y}_{\pm }=\infty $, so that the corresponding strip of
analyticity coincides with the whole axis. Accordingly, tail prices decay
faster than exponential. The asymptotic behavior of the price and implied
volatility for extreme strikes is given by%
\begin{equation*}
\mathsf{\delta C}^{M}\left( \tau ,k\right) \underset{k\rightarrow \pm \infty 
}{\sim }\mathsf{c}_{\pm }\exp \left( -\frac{\sqrt{2\ln \left\vert
k\right\vert }\left\vert k\right\vert }{\eta }\right) ,\ \ \ \ \ \sigma
_{imp}\left( \tau ,k\right) \underset{k\rightarrow \pm \infty }{\sim }\sqrt{%
\frac{\eta \left\vert k\right\vert }{2\sqrt{2\ln \left\vert k\right\vert }%
\tau }},
\end{equation*}%
so that for MPs the wing volatility grows slower than linearly in absolute
strike.
\end{proposition}

\begin{proof}
See Appendix \ref{MPsWA} for details.
\end{proof}

Additional discussion is given in \cite{piterbarg}, \cite{benaim}, \cite%
{benaim1} and \cite{gulisashvili1}, \cite{gulisashvili2} among others. Fast
decay of the call price in the wings for MPs is in agreement with general
results presented in \cite{albin}.

\begin{figure}[h]
\subfigure[TSP, $\alpha$=2/3] {\includegraphics[width=0.75\textwidth, angle=0]
{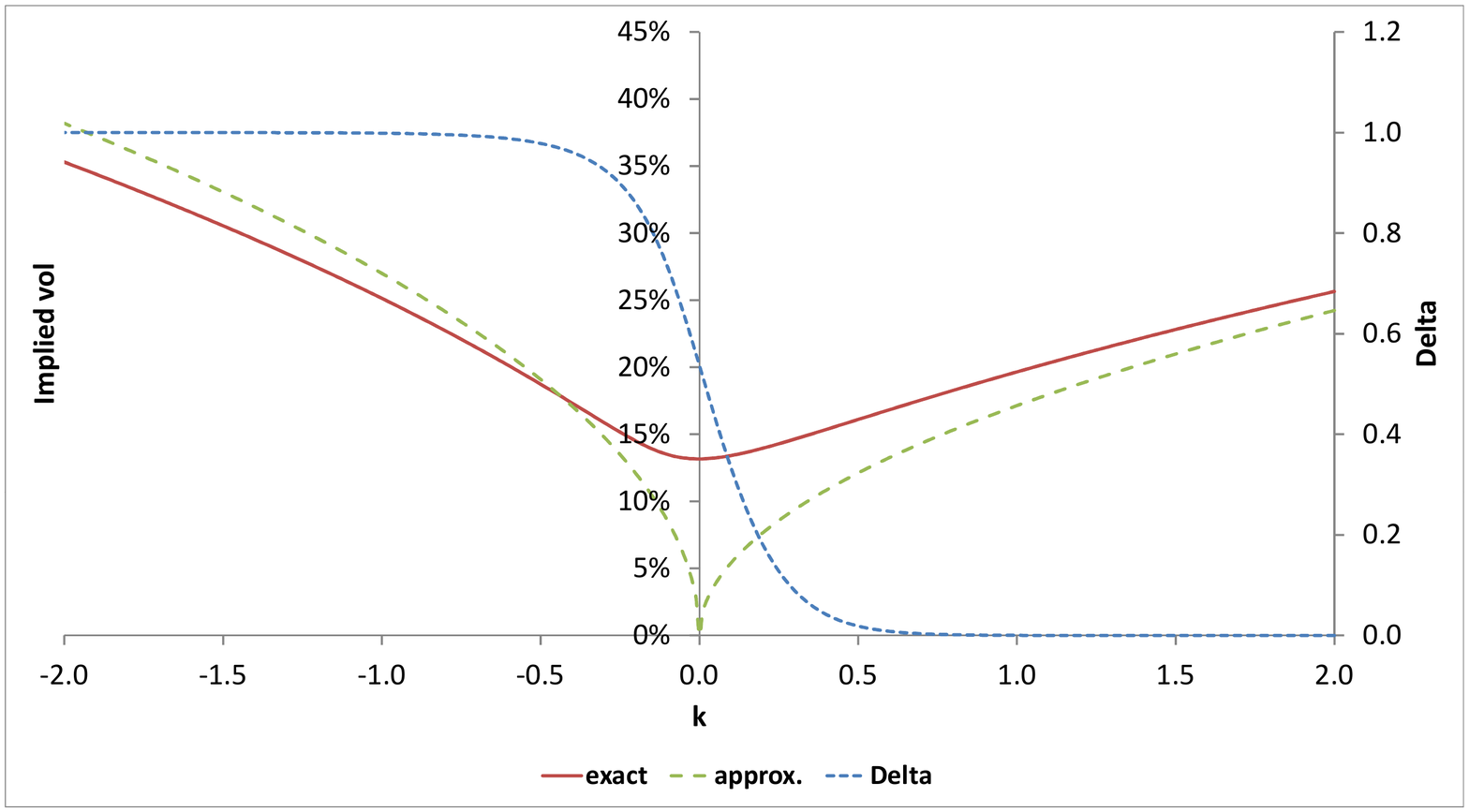}} 
\subfigure[TSP, $\alpha$=3/2] {\includegraphics[width=0.75\textwidth, angle=0]
{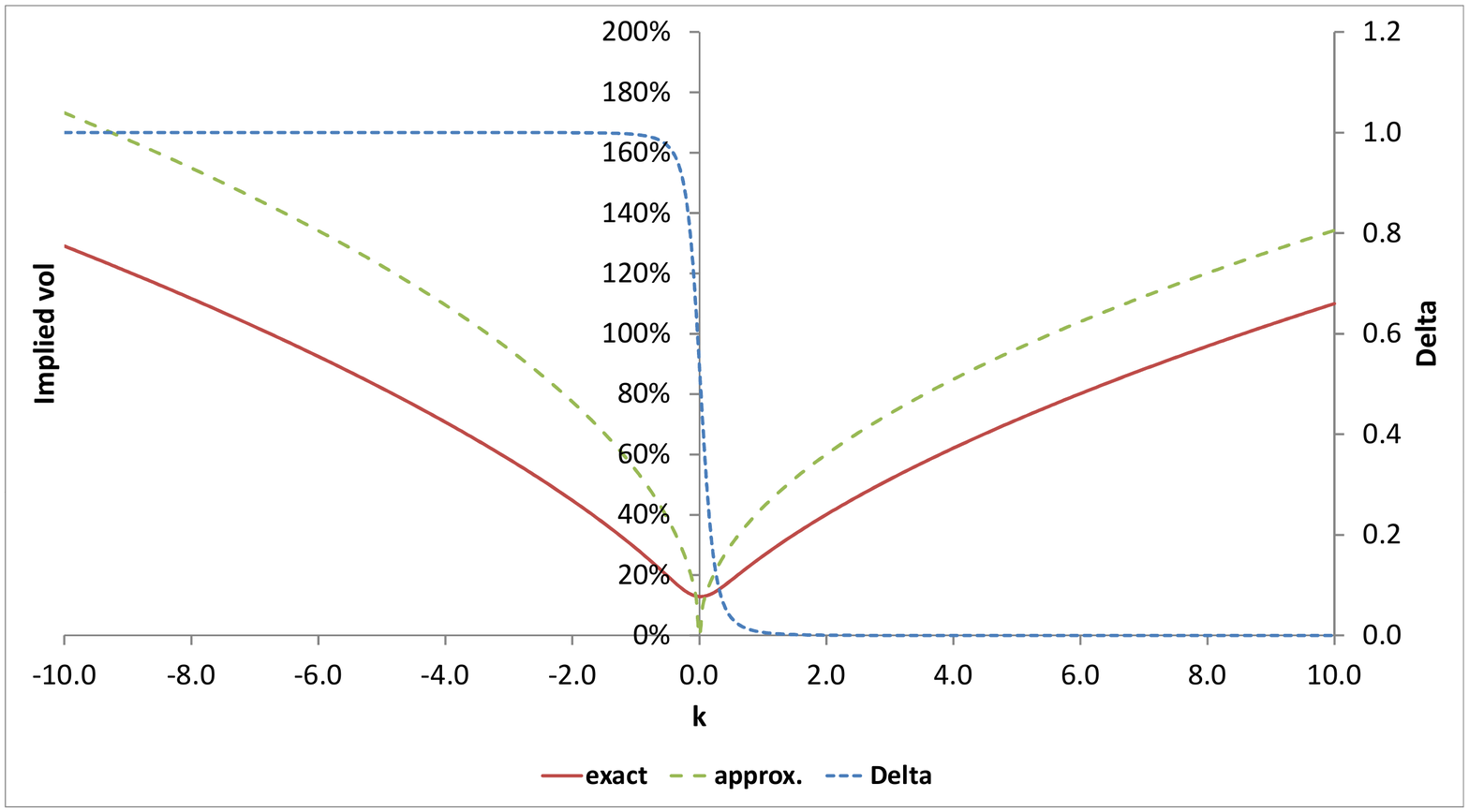}}
\caption{{}Comparison of the exact and asymptotic expressions for $\protect%
\sigma _{imp}\left( k\right) $ of the calibrated TSPs for $\protect\tau =2.0$
and $\left\vert k\right\vert \rightarrow \infty $. Here and below "exact"
denotes the exact volatility calculated by virtue of the LL formula,
"approx" denotes the approximate volatility given by (\protect\ref%
{eq:sigma_imp_wing}), and $\Delta $ denotes the BS\ delta.}
\label{fig:ELPsWing}
\end{figure}

\begin{figure}[h]
\subfigure[NIGP] {\includegraphics[width=1.0\textwidth, angle=0]
{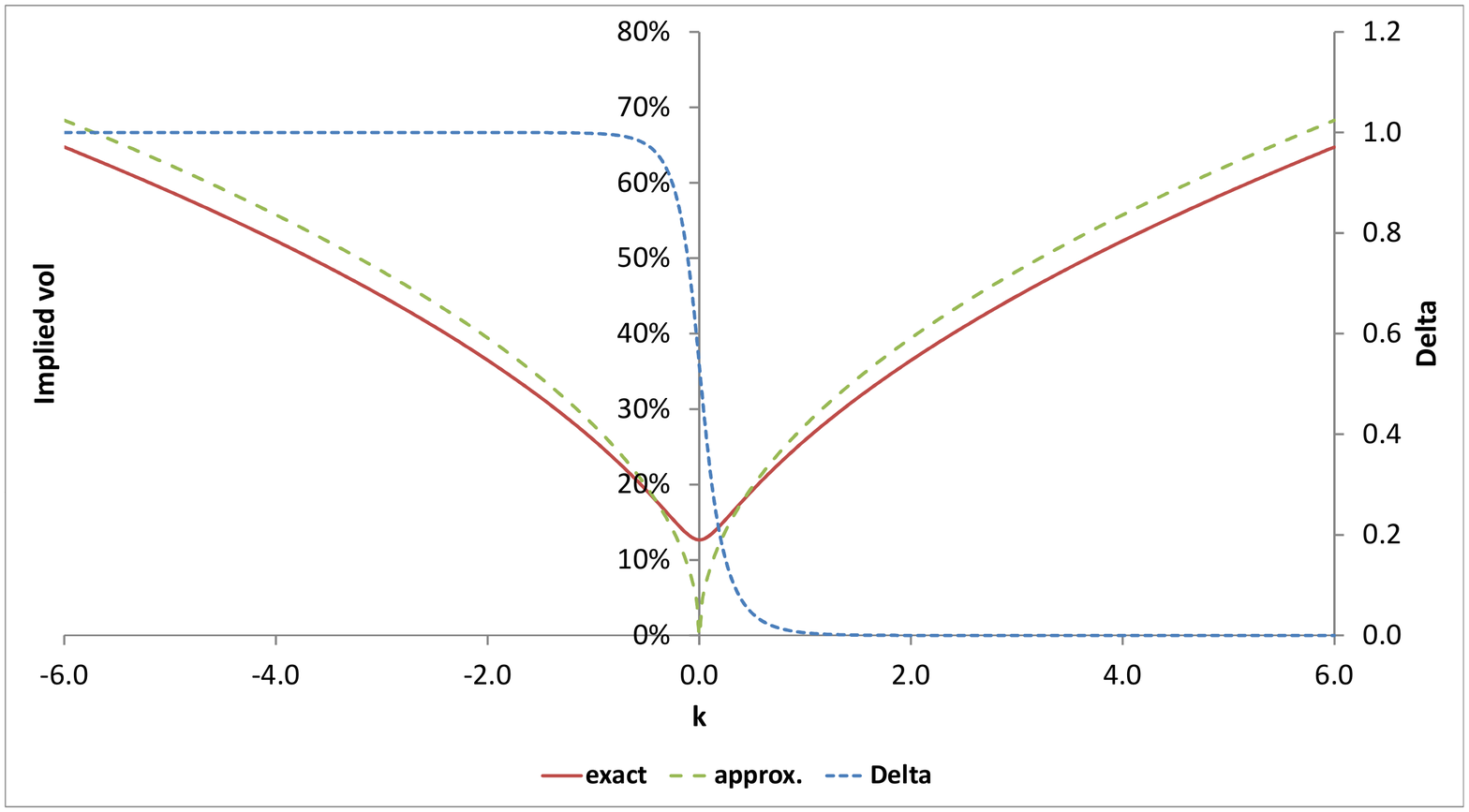}} 
\subfigure[MP] {\includegraphics[width=1.0\textwidth, angle=0]
{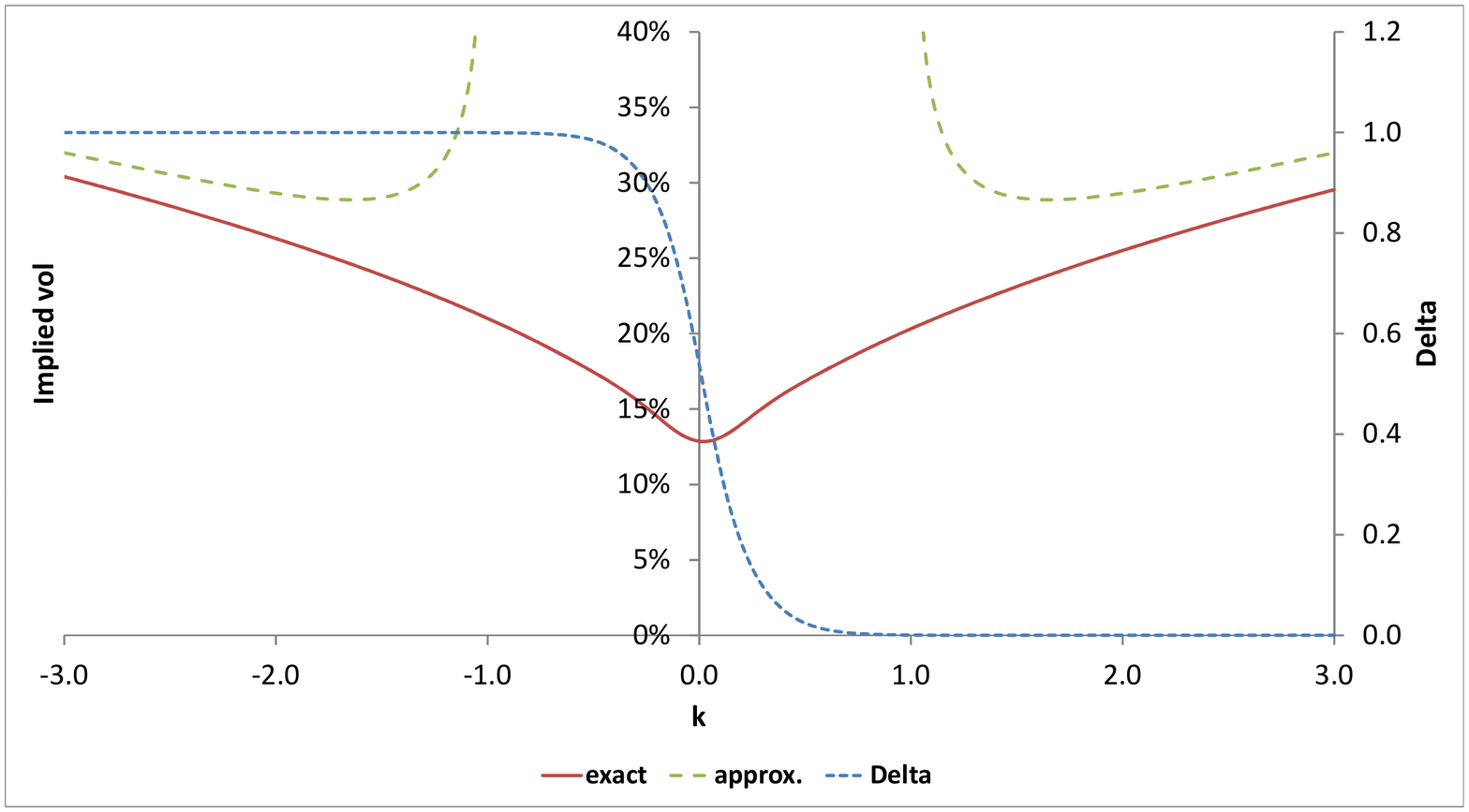}}
\caption{{}Comparison of the exact and asymptotic expressions for $\protect%
\sigma _{imp}\left( k\right) $ of the calibrated NIGP and MP for $\protect%
\tau =2.0$ and $\left\vert k\right\vert \rightarrow \infty $.}
\label{fig:ELPsWing1}
\end{figure}

Typical cross-sections of the corresponding volatility surfaces for $\tau
=2y $ are shown in Figures \ref{fig:ELPsWing}, \ref{fig:ELPsWing1},
respectively. It is clear the quality of the asymptotic approximation is far
from perfect.

\clearpage

\clearpage

\subsection{Quadratic Volatility Processes\label{QVPsWings}}

\begin{proposition}
For QVPs the asymptotic behavior of prices and implied volatility for
extreme strikes is given by%
\begin{equation*}
\mathsf{\delta C}^{QV}\left( \tau ,k\right) \underset{k\rightarrow \pm
\infty }{\sim }\mathsf{c}_{\pm }e^{-\left( \frac{1}{2}\mp \frac{1}{2}\right)
\left\vert k\right\vert },\ \ \ \ \ \sigma _{imp}\left( \tau ,k\right) 
\underset{k\rightarrow \pm \infty }{\sim }\sqrt{\frac{2\left\vert
k\right\vert }{\tau }}.
\end{equation*}

\begin{proof}
See Appendix \ref{QVPsWA}.
\end{proof}
\end{proposition}

A typical cross-section of the volatility surface for $\tau =2y$ is shown in
Figure \ref{fig:ELPsWing2}(a). Once again, the quality of the asymptotic
approximation is poor.

\subsection{Heston Stochastic Volatility Processes\label{SVP Wings}}

It is clear that in order to apply the general formula to HSVPs, we have to
determine the analyticity strip for the function $E\left( \tau ,u\right) $
given by (\ref{eq: Hest Form}). Analyzing the corresponding expression term
by term, one can show that the corresponding strip is defined by the
inequalities $-\mathsf{Y}_{+}<\func{Im}u<\mathsf{Y}_{-}$, where $\mp \mathsf{%
Y}_{\pm }$ are (time-dependent) roots of the equation%
\begin{equation*}
\mathcal{F}_{-}\left( \mp i\mathsf{Y}_{\pm }\right) +\mathcal{F}_{+}\left(
\mp i\mathsf{Y}_{\pm }\right) \exp \left( -\mathcal{Z}\left( \mp i\mathsf{Y}%
_{\pm }\right) \tau \right) =0,
\end{equation*}%
closest to $\mp 1/2$, respectively

\begin{proposition}
The function $E\left( \tau ,u\right) $ is analytical in the strip $\func{Im}%
u\in (-\mathsf{\tilde{Y}}\left( \tau \right) _{+},\mathsf{\tilde{Y}}\left(
\tau \right) _{-})$, where%
\begin{equation*}
\mathsf{\tilde{Y}}_{\pm }\left( \tau \right) \geq \mathsf{Y}_{\pm }>\frac{1}{%
2},
\end{equation*}%
$\mathsf{Y}_{\pm }$ are given by (\ref{eq:Y_pm}), and $\mp \mathsf{\tilde{Y}}%
_{\pm }\left( \tau \right) $ are the largest negative and the smallest
positive real roots of the equation%
\begin{equation*}
\left( \rho y+\check{\kappa}\right) +\mathfrak{\varsigma }\left( y\right)
+\left( -\left( \rho y+\check{\kappa}\right) +\mathfrak{\varsigma }\left(
y\right) \right) e^{-\varepsilon \tau \mathfrak{\varsigma }\left( y\right)
}=0,
\end{equation*}%
where $\mathfrak{\varsigma }\left( y\right) $ is given by (\ref{eq:zeta(y)}%
). The wing asymptotics of call prices and implied volatilities for HSVPs
can be written as follows%
\begin{equation*}
\mathsf{\delta C}^{HSV}\left( \tau ,k\right) \underset{k\rightarrow \pm
\infty }{\sim }\mathsf{c}_{\pm }e^{-\left( \mathsf{\tilde{Y}}_{\pm }\left(
\tau \right) \mp \frac{1}{2}\right) \left\vert k\right\vert },\ \ \ \ \
\sigma _{imp}\left( \tau ,k\right) \underset{k\rightarrow \pm \infty }{\sim }%
\sqrt{\frac{\beta _{\pm }\left\vert k\right\vert }{\tau }},
\end{equation*}%
where $\beta _{\pm }=4\left( \mathsf{\tilde{Y}}_{\pm }\left( \tau \right) -%
\sqrt{R\left( \mathsf{\tilde{Y}}_{\pm }\left( \tau \right) \right) }\right) $%
, $0<\beta _{\pm }<2$.

\begin{proof}
Straightforward calculation.
\end{proof}
\end{proposition}

\begin{figure}[h]
\subfigure[QVP] {\includegraphics[width=1.0\textwidth, angle=0]
{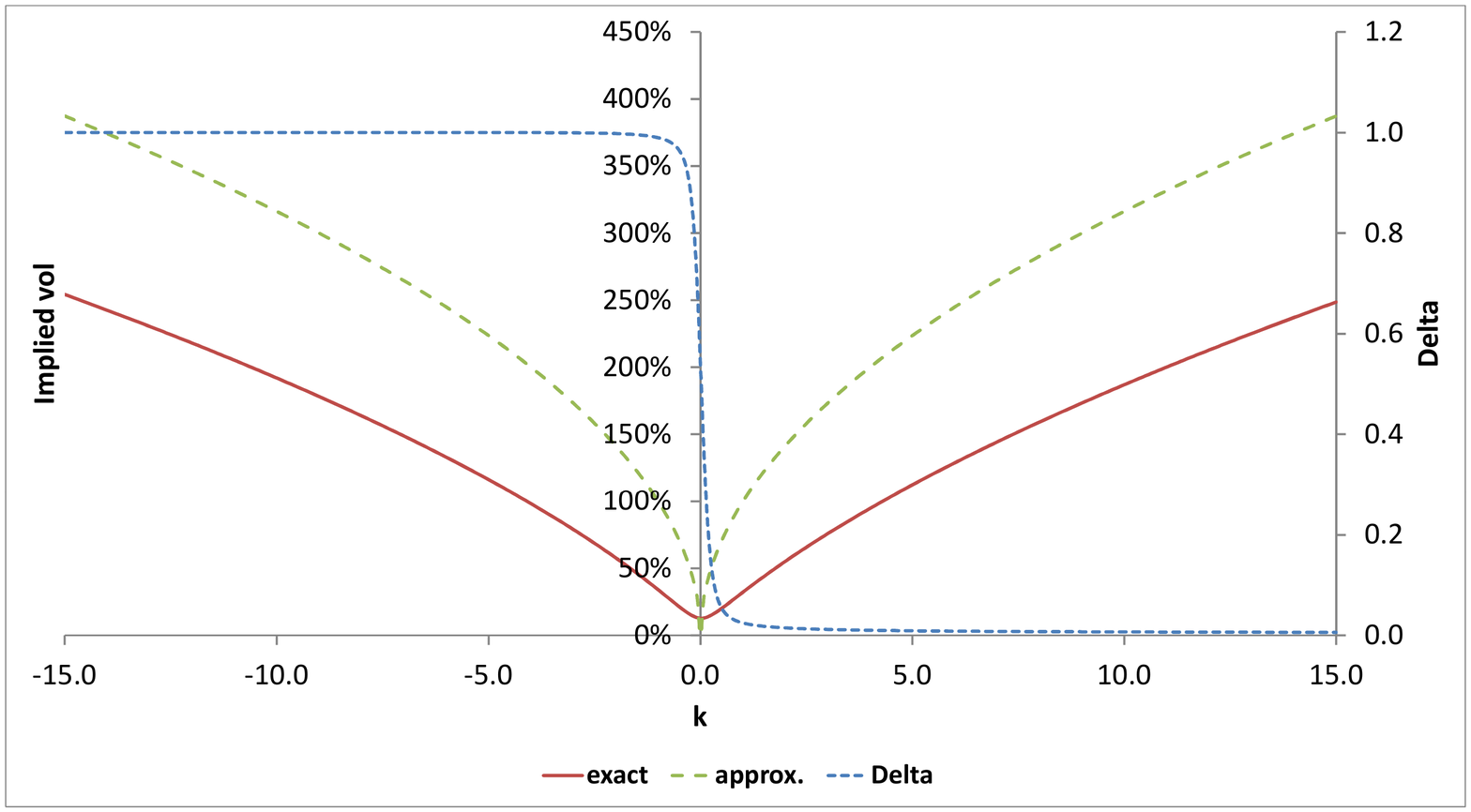}} 
\subfigure[HSVP] {\includegraphics[width=1.0\textwidth, angle=0]
{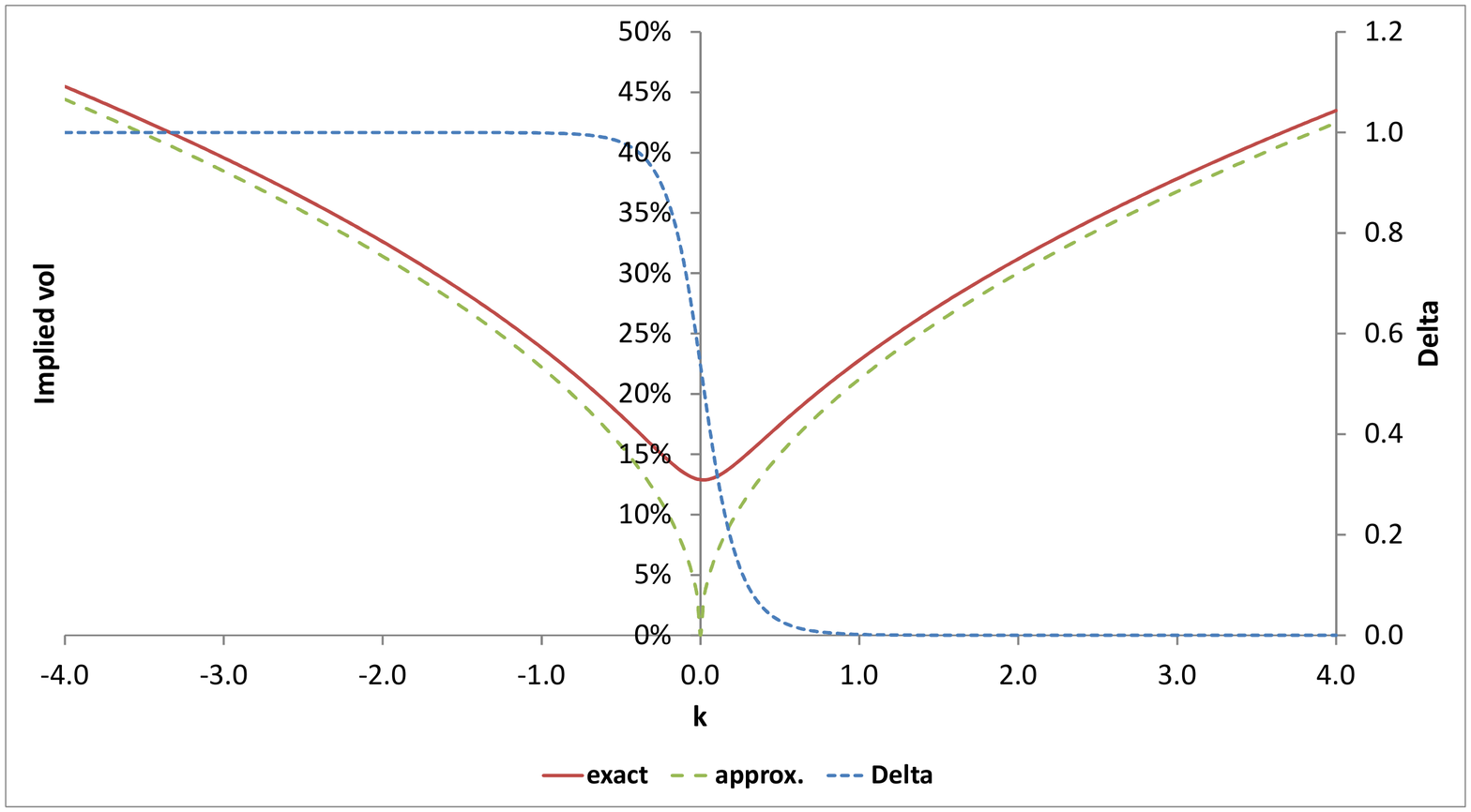}}
\caption{{}Comparison of the exact and asymptotic expressions for $\protect%
\sigma _{imp}\left( k\right) $ of the calibrated QVP and HSVP for $\protect%
\tau =2.0$ and $\left\vert k\right\vert \rightarrow \infty $.}
\label{fig:ELPsWing2}
\end{figure}

A typical cross-section of the volatility surface for $\tau =2y$ is shown in
Figure \ref{fig:ELPsWing2}(b). This Figure shows that the quality of wing
asymptotics is satisfactory but not perfect for HSVPs.

\clearpage

\section{Conclusions\label{Conclusions}}

This paper has been dedicated to the study of implied volatility asymptotics
for a range of processes that allows for the use the Lewis-Lipton (LL)
Fourier-integral representation of call option prices. Of key importance to
us was the class of exponential Levy processes (ELPs), especially the
tempered $\alpha $-stable processes, but we also discussed pure diffusion
processes of the local and stochastic volatility types.

As we have demonstrated, the LL representation is highly conducive for
asymptotic work, as well-established classical methods can be used to carry
out examinations of large-time, large-strike, and small-time regimes --
albeit occasionally with some delicacies and often involving quite laborious
computations. Our work is quite complete in establishing the formulas that
characterize limit behavior for volatility of the processes in question.

While some of the technical results in this paper are known, many existing
results in the literature have been derived using a variety of (often
complex) methods, and our paper provides a clean unification of these
results along with a variety of new formulas. Generally speaking these
results are theoretically appealing and provide definitive answers to a
number of questions, e.g. whether non-zero limits for FX risk reversals and
butterflies exist for the processes in question.

In order to establish the practical relevance for the various asymptotic
results, we have taken substantial time to undertake numerical comparisons
of asymptotics against the exact LL solution. The difficulty of this task
should not be underestimated as the integrals in the exact LL representation
become challenging to handle numerically in the various limits.
Nevertheless, once the analysis is carried out, it becomes clear that the
performance of the various asymptotics are definitely a mixed bag, and
overall rather disappointing. For instance, it is clear that most wing (i.e.
large-strike) asymptotics have domains of validity that are exceedingly
small, rendering the results mostly useless in practice. The same holds for
small-time asymptotics of implied volatility for ELPs, where maturities
generally need to be much less than a day for the asymptotics to be
accurate. On the other hand, small-time asymptotics work well for diffusion
processes with local and stochastic volatility, where it is not uncommon to
find that options with maturities of several years can be successfully
studied in the short-time limit. This discrepancy of performance is perhaps
understandable given that ELPs (with jumps) are fundamentally different from
diffusions, especially when observed at small time scales.

For ELPs, what does seem to work reasonable well in many case are the
long-term asymptotics where the domain of validity is often respectably
large (say, covering maturities larger than one or two years).
Unfortunately, this behavior is neither universal nor particularly robust,
and it is not difficult to find examples of ELP configurations which, while
matching the market well, result in poor long-term asymptotics. For
instance, we have observed that for tempered $\alpha $-stable processes the
region of validity of the long-term asymptotics is sometimes dramatically
reduced when the parameter $\alpha $ is greater than one.

In summary, while ELPs are viable candidates for describing FX market
dynamic, it is relatively difficult to analyze their behavior in pertinent
asymptotic regimes. In order to accomplish this task successfully, one has
to combine analytical and numerical methods, and even then success is not
guaranteed.

\appendix

\section{Fractional PIDEs\label{FracPIDE}}

For $\alpha \in (0,1)$ we can rewrite the regularized integrals in \ the
pricing equation as follows%
\begin{align*}
I_{\alpha ,s}& =\frac{\left( -s\right) ^{\alpha }}{\Gamma \left( -\alpha
\right) }\int_{0}^{\infty }\left( V\left( x+sy\right) -V\left( x\right)
\right) \frac{e^{-\kappa _{s}y}dy}{y^{1+\alpha }} \\
& =\frac{\left( -s\right) ^{\alpha }}{\Gamma \left( -\alpha \right) }%
\int_{0}^{\infty }\left( e^{-\kappa _{s}y}V\left( x+sy\right) -V\left(
x\right) +\left( 1-e^{-\kappa _{s}y}\right) V\left( x\right) \right) \frac{dy%
}{y^{1+\alpha }} \\
& =\frac{\left( -s\right) ^{\alpha }}{\Gamma \left( -\alpha \right) }%
e^{s\kappa _{s}x}\int_{0}^{\infty }\left( \tilde{V}\left( x+sy\right) -%
\tilde{V}\left( x\right) \right) \frac{dy}{y^{1+\alpha }} \\
& +\frac{\left( -s\right) ^{\alpha }}{\Gamma \left( -\alpha \right) }%
\int_{0}^{\infty }\frac{\left( 1-e^{-\kappa _{s}y}\right) dy}{y^{1+\alpha }}%
V\left( x\right) \\
& =e^{s\kappa _{s}x}\mathfrak{D}_{s}^{\alpha }\left( e^{-s\kappa
_{s}x}V\right) -\left( -s\right) ^{\alpha }\kappa _{s}^{\alpha }V,
\end{align*}%
where $\tilde{V}\left( x\right) =e^{-s\kappa _{s}x}V\left( x\right) $.
Similarly, for $\alpha \in (1,2)$ we have%
\begin{align*}
I_{\alpha ,s}& =\frac{\left( -s\right) ^{\alpha }}{\Gamma \left( -\alpha
\right) }\int_{0}^{\infty }\left( V\left( x+sy\right) -V\left( x\right)
-syV_{x}\left( x\right) \right) \frac{e^{-\kappa _{s}y}dy}{y^{1+\alpha }} \\
& =\frac{\left( -s\right) ^{\alpha }}{\Gamma \left( 1-\alpha \right) }%
\int_{0}^{\infty }\left( V\left( x+sy\right) -V\left( x\right)
-syV_{x}\left( x\right) \right) e^{-\kappa _{s}y}d\left( \frac{1}{y^{\alpha }%
}\right) \\
& =\frac{\left( -s\right) ^{\alpha }}{\Gamma \left( 1-\alpha \right) }%
\int_{0}^{\infty }\left( -s\left( V_{x}\left( x+sy\right) -V_{x}\left(
x\right) \right) +\kappa _{s}\left( V\left( x+sy\right) -V\left( x\right)
-syV_{x}\left( x\right) \right) \right) \frac{e^{-\kappa _{s}y}dy}{y^{\alpha
}} \\
& =\frac{\left( -s\right) ^{\alpha -1}}{\Gamma \left( 1-\alpha \right) }%
e^{s\kappa _{s}x}\int_{0}^{\infty }\left( \tilde{V}_{x}\left( x+sy\right) -%
\tilde{V}_{x}\left( x\right) \right) \frac{dy}{y^{1+\left( \alpha -1\right) }%
} \\
& +\frac{\left( -s\right) ^{\alpha -1}}{\Gamma \left( 1-\alpha \right) }%
\int_{0}^{\infty }\frac{\left( 1-e^{-\kappa _{s}y}\right) dy}{y^{1+\left(
\alpha -1\right) }}e^{s\kappa _{s}x}\tilde{V}_{x}\left( x\right) +\frac{%
\left( -s\right) ^{\alpha -1}}{\Gamma \left( 1-\alpha \right) }\kappa
_{s}\int_{0}^{\infty }\frac{e^{-\kappa _{s}y}dy}{y^{\alpha -1}}V_{x}\left(
x\right) \\
& =e^{s\kappa _{s}x}\mathfrak{D}_{s}^{\alpha }\left( e^{-s\kappa
_{s}x}V\right) -\left( -s\right) ^{\alpha -1}\alpha \kappa _{s}^{\alpha
-1}V_{x}\left( x\right) -\left( -s\right) ^{\alpha }\kappa _{s}^{\alpha
}V\left( x\right) ,
\end{align*}%
where $\tilde{V}\left( x\right) =e^{-s\kappa _{s}x}V\left( x\right) $.
Accordingly, PIDEs (\ref{eq:PIDE_alpha_3}), (\ref{eq:PIDE_alpha_4}) can be
written as%
\begin{equation*}
V_{t}+\gamma V_{x}+\frac{1}{2}\sigma ^{2}\left( V_{xx}-V_{x}\right)
+\sum\limits_{s=\pm }\left( -s\right) ^{\alpha }a_{s}e^{s\kappa _{s}x}%
\mathfrak{D}_{s}^{\alpha }\left( e^{-s\kappa _{s}x}V\right)
-\sum\limits_{s=\pm }a_{s}\kappa _{s}^{\alpha }V=0,
\end{equation*}%
and%
\begin{equation*}
V_{t}+\left( \gamma +\sum\limits_{s=\pm }sa_{s}\alpha \kappa _{s}^{\alpha
-1}\right) V_{x}+\frac{1}{2}\sigma ^{2}\left( V_{xx}-V_{x}\right)
+\sum\limits_{s=\pm }\left( -s\right) ^{\alpha }a_{s}e^{s\kappa _{s}x}%
\mathfrak{D}_{s}^{\alpha }\left( e^{-s\kappa _{s}x}V\right)
-\sum\limits_{s=\pm }a_{s}\kappa _{s}^{\alpha }V=0.
\end{equation*}

\section{The Green's Function and Call Prices for Gauss-L\'{e}vy and
Tempered Gauss-L\'{e}vy Processes\label{GreenFun}}

We start with the following assumptions: $\alpha \in (0,1)$, volatility is
zero, $\sigma =0$, the process is maximally skewed to the left and there is
no dumping, $c_{+}=0$, $c_{-}>0$, $\kappa _{\pm }=0$. For brevity we omit
subscripts where possible. The corresponding backward and forward equations
have the form%
\begin{equation*}
V_{t}+\gamma V_{x}+c\int_{0}^{\infty }\left( V\left( x-y\right) -V\left(
x\right) \right) \frac{dy}{y^{1+\alpha }}=0,\ \ \ \ \ V\left( T,x\right)
=V_{T}\left( x\right) ,
\end{equation*}%
\begin{equation*}
G_{t}^{\left( \alpha \right) }+\gamma G_{x}^{\left( \alpha \right)
}-c\int_{0}^{\infty }\left( G^{\left( \alpha \right) }\left( x+y\right)
-G^{\left( \alpha \right) }\left( x\right) \right) \frac{dy}{y^{1+\alpha }}%
=0,\ \ \ \ \ G^{\left( \alpha \right) }\left( 0,x\right) =\delta \left(
x\right) .
\end{equation*}%
where $\gamma =-a=-\Gamma \left( -\alpha \right) c=\sec \left( \alpha \pi
/2\right) \vartheta ^{\alpha }>0$, or, in terms of fractional derivatives,%
\begin{equation*}
V_{t}+\gamma V_{x}+a\mathfrak{D}_{-}^{\alpha }\left( V\right) =0,\ \ \ \ \
V\left( T,x\right) =V_{T}\left( x\right) ,
\end{equation*}%
\begin{equation*}
G_{t}^{\left( \alpha \right) }+\gamma G_{x}^{\left( \alpha \right) }-\left(
-1\right) ^{\alpha }a\mathfrak{D}_{+}^{\alpha }\left( G^{\left( \alpha
\right) }\right) =0,\ \ \ \ \ G^{\left( \alpha \right) }\left( 0,x\right)
=\delta \left( x\right) .
\end{equation*}

\begin{lemma}
Let $G^{\left( \alpha \right) }\left( t,x\right) $ be the Green's function
for a maximally negatively skewed stable process, such that 
\begin{equation*}
\mathrm{P}\left( X(t)\in \lbrack x,x+dx]\right) =G^{\left( \alpha \right)
}\left( t,x\right) .
\end{equation*}%
Then, due to the scaling invariance of the corresponding process, we have%
\begin{equation*}
G^{\left( \alpha \right) }\left( t,x\right) =\frac{1}{\iota ^{1/\alpha }}%
g^{\left( \alpha \right) }\left( \xi \right) ,
\end{equation*}%
where 
\begin{equation*}
\iota =\sec \left( \frac{\alpha \pi }{2}\right) \vartheta ^{\alpha }t,\ \ \
\xi =\frac{\left( x-\iota \right) }{\iota ^{1/\alpha }},
\end{equation*}%
and $g^{\left( \alpha \right) }\left( .\right) $ is an appropriate positive
function of a single variable, which satisfies the following ordinary
integro-differential equation%
\begin{equation}
\xi g^{\left( \alpha \right) }\left( \xi \right) +\frac{1}{\Gamma \left(
1-\alpha \right) }\int_{0}^{\infty }g^{\left( \alpha \right) }\left( \xi
+\xi ^{\prime }\right) \frac{d\xi ^{\prime }}{\xi ^{\prime \alpha }}=0.
\label{eq:green_g}
\end{equation}%
In particular, the Green's function associated with the LGP has the form%
\begin{equation}
G^{\left( 1/2\right) }\left( t,x\right) =\frac{1}{\iota ^{2}}g^{\left(
1/2\right) }\left( \xi \right) ,  \label{eq:LGP_Green}
\end{equation}%
where%
\begin{equation*}
\iota =\sqrt{2\vartheta }t,\ \ \ \xi =\frac{\left( x-\iota \right) }{\iota
^{2}},\ \ \ g^{\left( 1/2\right) }\left( \xi \right) =\frac{\exp \left( 
\frac{1}{4\xi }\right) }{2\sqrt{\pi }\left( -\xi \right) ^{3/2}}\mathbf{1}%
_{\xi <0}.
\end{equation*}%
This density is known as the \emph{L\'{e}vy distribution}.
\end{lemma}

\begin{proof}
In order to calculate $G^{\left( \alpha \right) }$ we use the following
ansatz%
\begin{equation*}
G^{\left( \alpha \right) }\left( t,x\right) =\frac{1}{\iota ^{\alpha
^{\prime }}}g^{\left( \alpha \right) }\left( \frac{x-\iota }{\iota ^{\alpha
^{\prime }}}\right) ,
\end{equation*}%
where $\iota =\gamma t$, and $\alpha ^{\prime }=1/\alpha $. Its validity is
verified below. Since 
\begin{equation*}
\int_{-\infty }^{\infty }G^{\left( \alpha \right) }\left( t,x\right)
dx=\int_{-\infty }^{\infty }g^{\left( \alpha \right) }\left( \frac{x-\iota }{%
\iota ^{\alpha ^{\prime }}}\right) d\left( \frac{x-\iota }{\iota ^{\alpha
^{\prime }}}\right) =\int_{-\infty }^{\infty }g^{\left( \alpha \right)
}\left( \xi \right) d\xi ,
\end{equation*}%
we have to impose the following constraints on $g$: 
\begin{equation}
g^{\left( \alpha \right) }\left( \xi \right) \geq 0,\ \ \ \ \ g^{\left(
\alpha \right) }\left( \xi \right) \underset{\xi \rightarrow \pm \infty }{%
\rightarrow }0,\ \ \ \ \ \int_{-\infty }^{\infty }g^{\left( \alpha \right)
}\left( \xi \right) d\xi =1,  \label{eq:green_cond}
\end{equation}%
which, broadly speaking, play the role of initial and boundary conditions.
We have%
\begin{align*}
0& =G_{t}^{\left( \alpha \right) }\left( t,x\right) +\gamma G_{x}^{\left(
\alpha \right) }\left( t,x\right) -\left( -1\right) ^{\alpha }a\mathfrak{D}%
_{+}^{\alpha }G^{\left( \alpha \right) }\left( t,x\right) \\
& =-\frac{\alpha ^{\prime }\gamma }{\iota ^{\alpha ^{\prime }+1}}\left( \xi
g_{\xi }^{\left( \alpha \right) }\left( \xi \right) +g^{\left( \alpha
\right) }\left( \xi \right) \right) -\frac{\left( -1\right) ^{\alpha }}{%
\iota ^{\alpha ^{\prime }+\alpha \alpha ^{\prime }}}a\mathfrak{D}%
_{+}^{\alpha }g^{\left( \alpha \right) }\left( \xi \right) .
\end{align*}%
It is clear that $t$-dependence disappears and the integro-differential
equation for $g\left( \xi \right) $ becomes 
\begin{equation}
\xi g_{\xi }^{\left( \alpha \right) }\left( \xi \right) +g^{\left( \alpha
\right) }\left( \xi \right) +\left( -1\right) ^{\alpha }\alpha \mathfrak{D}%
_{+}^{\alpha }g^{\left( \alpha \right) }\left( \xi \right) =0.
\label{eq:green_ode}
\end{equation}%
Here $-\infty <\xi <0$. If so desired, we can manipulate this equation
further. Specifically, we can integrate by parts and write 
\begin{equation*}
\left( \xi g^{\left( \alpha \right) }\left( \xi \right) \right) _{\xi }+%
\frac{1}{\Gamma \left( 1-\alpha \right) }\int_{0}^{\infty }g_{\xi }^{\left(
\alpha \right) }\left( \xi +\xi ^{\prime }\right) \frac{d\xi ^{\prime }}{\xi
^{\prime \alpha }}=0,
\end{equation*}%
or, after integration over the interval $\left( -\infty ,\xi \right] $,%
\begin{equation*}
\xi g^{\left( \alpha \right) }\left( \xi \right) +\frac{1}{\Gamma \left(
1-\alpha \right) }\int_{0}^{\infty }g^{\left( \alpha \right) }\left( \xi
+\xi ^{\prime }\right) \frac{d\xi ^{\prime }}{\xi ^{\prime \alpha }}=0,
\end{equation*}%
which is (\ref{eq:green_g}). In particular, it is easy to check directly
that for the LGPs with $\alpha =1/2$, $c_{+}=0$, $c_{-}=\sqrt{2\vartheta }$,
the corresponding $g^{\left( 1/2\right) }$ is given by (\ref{eq:LGP_Green}).
The case of $\alpha \in \left( 1,2\right) $ can be studied in a similar
fashion.
\end{proof}

It is clear that our calculations make sense by virtue of the fact that the
corresponding maximally skewed stable processes are scale-invariant. Since
two-sided stable processes possess this property as well, the above
calculations can be repeated \textit{verbatim} for such processes. It can be
shown that the PDFs of standard stable laws have the following support: (A)
the positive semi-axis if $\alpha \in \left( 0,1\right) ,\ c_{-}=0$; (B) the
negative semi-axis if $\alpha \in \left( 0,1\right) ,\ c_{+}=0$; (C) the
whole axis in all other cases. If $\alpha \in \left( 0,2\right) $, $c_{+}>0$%
,\ $c_{-}\geq 0$, then for $\xi \rightarrow \infty $ we have 
\begin{equation*}
g^{\left( \alpha \right) }\left( \left. \xi \right\vert \alpha ,\beta
\right) \sim \xi ^{-\left( \alpha +1\right) }.
\end{equation*}%
If $\alpha \in \left( 0,2\right) $, $c_{+}\geq 0$,\ $c_{-}>0$, then for $\xi
\rightarrow -\infty $ we have 
\begin{equation*}
g^{\left( \alpha \right) }\left( \left. \xi \right\vert \alpha ,\beta
\right) \sim \left\vert \xi \right\vert ^{-\left( \alpha +1\right) }.
\end{equation*}%
Thus, for all $\alpha \in \left( 0,2\right) $, $c_{+}>0$,\ $c_{-}>0$, both
tails are described by power laws. If $c_{-}=0$ $\left( c_{+}=0\right) $,
then the left (right) tail is described by an exponential law.

It is clear that appropriately transformed tempered stable density is not
scale-invariant unless it is maximally skewed. Because of that, we only deal
with maximally skewed TSPs.\footnote{%
It is clear that two-sided TSPs cannot be treated along similar lines
because of the violation of scaling properties.} Thus, we assume that $%
\alpha \in (0,1)$, $\sigma =0$, $c_{-}>0$, $\kappa _{-}>0$, and $c_{+}=0$.
For brevity, we omit subscripts. In this case, backward equation (\ref%
{eq:PIDE_alpha_3}) simplifies to the form 
\begin{equation*}
V_{t}+\gamma V_{x}+c\int_{0}^{\infty }\left( V\left( x-y\right) -V\left(
x\right) \right) \frac{e^{-\kappa y}dy}{y^{1+\alpha }}=0,
\end{equation*}%
where $\gamma =-$\textsf{$\zeta $}$a=-\Gamma \left( -\alpha \right) \zeta
c>0 $, $\mathsf{\zeta }=\left( \kappa +1\right) ^{\alpha }-\kappa ^{\alpha }$%
. The corresponding adjoint equation for the Green's function $G^{\left(
\alpha ,\kappa \right) }\left( t,x\right) $ is 
\begin{equation*}
G_{t}^{\left( \alpha ,\kappa \right) }+\gamma G_{x}^{\left( \alpha ,\kappa
\right) }-c\int_{0}^{\infty }\left( G^{\left( \alpha ,\kappa \right) }\left(
x+y\right) -G^{\left( \alpha ,\kappa \right) }\left( x\right) \right) \frac{%
e^{-\kappa y}dy}{y^{1+\alpha }}=0.
\end{equation*}%
Equivalently, the backward and forward equations can be written in terms of
fractional derivatives as follows 
\begin{equation*}
V_{t}+\gamma V_{x}+ae^{-\kappa x}\mathfrak{D}_{-}^{\alpha }\left( e^{\kappa
x}V\right) +\delta V=0,\ \ \ \ \ V\left( T,x\right) =v\left( x\right) ,
\end{equation*}%
\begin{equation}
G_{t}^{\left( \alpha ,\kappa \right) }+\gamma G_{x}^{\left( \alpha ,\kappa
\right) }-\left( -1\right) ^{\alpha }ae^{\kappa x}\mathfrak{D}_{+}^{\alpha
}\left( e^{-\kappa x}G^{\left( \alpha ,\kappa \right) }\right) -\delta
G^{\left( \alpha ,\kappa \right) }=0,\ \ \ \ \ G^{\left( \alpha \right)
}\left( 0,x\right) =\delta \left( x\right) ,  \label{eq: Green ak}
\end{equation}%
where $\delta \mathcal{=}-\kappa ^{\alpha }a=-\Gamma \left( -\alpha \right)
\kappa ^{\alpha }c>0$. It turns out that for a maximally skewed TSP can be
converted into a maximally skewed SP via a simple scaling transformation.

\begin{lemma}
Let $G^{\left( \alpha ,\kappa \right) }\left( t,x\right) $ be the Green's
function for a maximally negatively skewed tempered stable process. Then%
\begin{equation}
G^{\left( \alpha ,\kappa \right) }\left( t,x\right) =\frac{\exp \left( \iota
_{1}+\kappa \left( x-\left( \iota _{2}-\iota _{1}\right) \right) \right) }{%
\iota ^{1/\alpha }}g^{\left( \alpha \right) }\left( \xi \right) ,
\label{eq:GreenFunction}
\end{equation}%
where%
\begin{equation*}
\iota =\sec \left( \frac{\alpha \pi }{2}\right) \vartheta ^{\alpha }t,\ \ \
\iota _{1}=\kappa ^{\alpha }\iota ,\ \ \ \iota _{2}=\left( \kappa +1\right)
^{\alpha }\iota ,\ \ \ \xi =\frac{\left( x-\left( \iota _{2}-\iota
_{1}\right) \right) }{\iota ^{1/\alpha }}.
\end{equation*}%
In particular, for $\alpha =1/2$ we have%
\begin{equation}
G^{\left( 1/2,\kappa \right) }\left( t,x\right) =\frac{\exp \left( \iota
_{1}+\kappa \left( x-\left( \iota _{2}-\iota _{1}\right) \right) \right) }{%
\iota ^{2}}g^{\left( 1/2\right) }\left( \xi \right) ,  \label{eq:TLGD}
\end{equation}%
where%
\begin{equation*}
\iota =\sqrt{2\vartheta }t,\ \ \ \iota _{1}=\sqrt{\kappa }\iota ,,\ \ \
\iota _{2}=\sqrt{\kappa +1}\iota ,\ \ \ \xi =\frac{\left( x-\left( \iota
_{2}-\iota _{1}\right) \right) }{\iota ^{2}}.
\end{equation*}
\end{lemma}

\begin{proof}
After some tedious algebra it can be show that solution of (\ref{eq: Green
ak}) can be written in the form%
\begin{equation*}
G^{\left( \alpha ,\kappa \right) }\left( t,x\right) =\mathsf{\zeta }^{-\frac{%
1}{\left( 1-\alpha \right) }}e^{\kappa ^{\alpha }\iota +\kappa \left(
x-\zeta \iota \right) }G^{\left( \alpha \right) }\left( \mathsf{\zeta }^{-%
\frac{\alpha }{\left( 1-\alpha \right) }}\iota ,\mathsf{\zeta }^{-\frac{1}{%
\left( 1-\alpha \right) }}x\right) .
\end{equation*}%
Accordingly, we can represent $G^{\left( \alpha ,\kappa \right) }\left(
t,x\right) $ as follows 
\begin{equation*}
G^{\left( \alpha ,\kappa \right) }\left( t,x\right) =\frac{e^{\kappa
^{\alpha }\iota +\kappa \left( x-\zeta \iota \right) }}{\iota ^{1/\alpha }}%
g^{\left( \alpha \right) }\left( \frac{x-\zeta \iota }{\iota ^{1/\alpha }}%
\right) ,
\end{equation*}%
where $g^{\left( \alpha \right) }$ satisfies (\ref{eq:green_g}) subject to (%
\ref{eq:green_cond}). In particular, for the TLGPs we have (\ref{eq:TLGD}).
\end{proof}

Next, we consider prices of options for maximally skewed stable and tempered
stable process via the Esscher transform technique (see, e.g., \cite%
{gerber-shiu}). Due to the strong localization of the corresponding Green's
function, we can define its Esscher transform as follows 
\begin{equation*}
g^{\left( \alpha \right) }\left( \xi \right) \Rightarrow g^{\left( \alpha
,p\right) }\left( \xi \right) =\frac{e^{p\xi }g^{\left( \alpha \right)
}\left( \xi \right) }{\mathbb{E}\left\{ e^{p\xi }g^{\left( \alpha \right)
}\left( \xi \right) \right\} }=e^{p\xi -\psi \left( -ip\right) }g^{\left(
\alpha \right) }\left( \xi \right) ,
\end{equation*}%
As usual, we denote the corresponding complementary cumulative function as 
\begin{equation*}
\Gamma ^{\left( \alpha ,p\right) }\left( \xi \right) =\int_{\xi }^{\infty
}g^{\left( \alpha ,p\right) }\left( \xi ^{\prime }\right) d\xi ^{\prime }.
\end{equation*}%
With this notation in mind, we can write the price of a call option on a
maximally skewed SP as follows%
\begin{align*}
\mathsf{C}\left( \iota ,k\right) & =\int_{k}^{\infty }\left(
e^{x}-e^{k}\right) \frac{1}{\iota ^{1/\alpha }}g^{\left( \alpha \right)
}\left( \frac{x-\iota }{\iota ^{1/\alpha }}\right) dx \\
& =\int_{\frac{k-\iota }{\iota ^{1/\alpha }}}^{\infty }\left( e^{\iota
+\iota ^{1/\alpha }\xi }-e^{\kappa }\right) g^{\left( \alpha \right) }\left(
\xi \right) d\xi \\
& =e^{\iota +\psi \left( -i\iota ^{1/\alpha }\right) }\Gamma ^{\left( \alpha
,\iota ^{1/\alpha }\right) }\left( \frac{k-\iota }{\iota ^{1/\alpha }}%
\right) -e^{k}\Gamma ^{\left( \alpha ,0\right) }\left( \frac{k-\iota }{\iota
^{1/\alpha }}\right) .
\end{align*}%
In particular, for the LGP, assuming that $k<\iota $, we can represent the
normalized price of a call option as follows%
\begin{align*}
\mathsf{C}\left( \iota ,k\right) & =\frac{\iota }{2\sqrt{\pi }}%
\int_{k}^{\iota }\left( e^{x}-e^{k}\right) \frac{e^{-\frac{\iota ^{2}}{%
4(\iota -x)}}}{\left( \iota -x\right) ^{3/2}}dx \\
& =\frac{1}{2\sqrt{\pi }}\int_{\frac{k-\iota }{\iota ^{2}}}^{0}\left(
e^{\iota +\iota ^{2}\xi }-e^{k}\right) \frac{e^{\frac{1}{4\xi }}}{\left(
-\xi \right) ^{3/2}}d\xi \\
& =e^{\iota }\mathsf{D}\left( v,2\iota \right) -e^{k}\mathsf{D}\left(
v,0\right) ,
\end{align*}%
where $v=2\iota ^{2}/\left( \iota -k\right) $. In order to compute the
corresponding integrals explicitly, we use the change of variables $\xi
=-1/2z^{2}$. For maximally skewed TLGP we can represent the price of a call
option as follows%
\begin{align*}
\mathsf{C}\left( \iota ,k\right) & =\frac{\iota }{2\sqrt{\pi }}%
\int_{k}^{\iota _{2}-\iota _{1}}\left( e^{x}-e^{k}\right) \frac{e^{\iota
_{1}+\kappa \left( x-\left( \iota _{2}-\iota _{1}\right) \right) -\frac{%
\iota ^{2}}{4(\left( \iota _{2}-\iota _{1}\right) -x)}}}{\left( \iota
_{2}-\iota _{1}-x\right) ^{3/2}}dx \\
& =\frac{e^{\iota _{1}}}{2\sqrt{\pi }}\int_{\frac{k-\left( \iota _{2}-\iota
_{1}\right) }{\iota ^{2}}}^{0}\left( e^{\iota _{2}-\iota _{1}-\iota ^{2}\xi
}-e^{k}\right) \frac{e^{\kappa \iota ^{2}\xi +\frac{1}{4\xi }}}{\left( -\xi
\right) ^{3/2}}d\xi \\
& =e^{\iota _{2}}\mathsf{D}\left( v,2\iota _{2}\right) -e^{k+\iota _{1}}%
\mathsf{D}\left( v,2\iota _{1}\right) ,
\end{align*}%
where $v=2\iota ^{2}/\left( \iota _{2}-\iota _{1}-k\right) $.

\section{The Implied Volatility Expansion\label{ImpVolExp}}

By definition,%
\begin{align*}
\mathsf{C}\left( \tau ,k\right) & =\Phi \left( -\frac{k}{\bar{\sigma}_{imp}}+%
\frac{\bar{\sigma}_{imp}}{2}\right) -e^{k}\Phi \left( -\frac{k}{\bar{\sigma}%
_{imp}}-\frac{\bar{\sigma}_{imp}}{2}\right) , \\
\mathsf{C}_{k}\left( \tau ,k\right) & =\phi \left( -\frac{k}{\bar{\sigma}%
_{imp}}+\frac{\bar{\sigma}_{imp}}{2}\right) \partial _{k}\bar{\sigma}%
_{imp}-e^{k}\Phi \left( -\frac{k}{\bar{\sigma}_{imp}}-\frac{\bar{\sigma}%
_{imp}}{2}\right) , \\
\mathsf{C}_{kk}\left( \tau ,k\right) & =\phi \left( -\frac{k}{\bar{\sigma}%
_{imp}}+\frac{\bar{\sigma}_{imp}}{2}\right) \left( \partial _{k}^{2}\bar{%
\sigma}_{imp}+\partial _{k}\bar{\sigma}_{imp}-\frac{\bar{\sigma}_{imp}\left(
\partial _{k}\bar{\sigma}_{imp}\right) ^{2}}{4}\right. \\
& \left. +\frac{1}{\bar{\sigma}_{imp}}+\frac{k^{2}\left( \partial _{k}\bar{%
\sigma}_{imp}\right) ^{2}}{\bar{\sigma}_{imp}^{3}}-\frac{2k\partial _{k}\bar{%
\sigma}_{imp}}{\bar{\sigma}_{imp}^{2}}\right) -e^{k}\Phi \left( -\frac{k}{%
\bar{\sigma}_{imp}}-\frac{\bar{\sigma}_{imp}}{2}\right) .
\end{align*}%
For $k=0$ we have%
\begin{equation}
\mathsf{C}\left( \tau ,0\right) =\Phi \left( \frac{\bar{\chi}_{0}\left( \tau
\right) }{2}\right) -\Phi \left( -\frac{\bar{\chi}_{0}\left( \tau \right) }{2%
}\right) \sim \frac{1}{\sqrt{2\pi }}\left( 1-\frac{\bar{\chi}_{0}^{2}\left(
\tau \right) }{24}\right) \bar{\chi}_{0}\left( \tau \right) ,
\label{eq:c_xi_0}
\end{equation}%
\begin{align}
\mathsf{C}_{k}\left( \tau ,0\right) & =\phi \left( \frac{\bar{\chi}%
_{0}\left( \tau \right) }{2}\right) \bar{\chi}_{1}\left( \tau \right) -\Phi
\left( -\frac{\bar{\chi}_{0}\left( \tau \right) }{2}\right)
\label{eq:c_xi_1} \\
& \sim \frac{1}{\sqrt{2\pi }}\left( 1-\frac{\bar{\chi}_{0}^{2}\left( \tau
\right) }{8}\right) \bar{\chi}_{1}\left( \tau \right) +\frac{1}{2}\left( 
\mathsf{C}\left( \tau ,0\right) -1\right) ,  \notag
\end{align}%
\begin{align}
\mathsf{C}_{kk}\left( \tau ,0\right) & =\phi \left( \frac{\bar{\chi}%
_{0}\left( \tau \right) }{2}\right) \left( \bar{\chi}_{2}\left( \tau \right)
+\bar{\chi}_{1}\left( \tau \right) -\frac{1}{4}\bar{\chi}_{0}\left( \tau
\right) \bar{\chi}_{1}^{2}\left( \tau \right) +\frac{1}{\bar{\chi}_{0}\left(
\tau \right) }\right) -\Phi \left( -\frac{\bar{\chi}_{0}\left( \tau \right) 
}{2}\right)  \label{eq:c_xi_2} \\
& \sim \frac{1}{\sqrt{2\pi }}\left( 1-\frac{\bar{\chi}_{0}^{2}\left( \tau
\right) }{8}\right) \left( \bar{\chi}_{2}\left( \tau \right) -\frac{1}{4}%
\bar{\chi}_{0}\left( \tau \right) \bar{\chi}_{1}^{2}\left( \tau \right) +%
\frac{1}{\bar{\chi}_{0}\left( \tau \right) }\right) +\mathsf{C}_{k}\left(
\tau ,0\right) .  \notag
\end{align}%
Comparison of (\ref{eq: cl}), (\ref{eq:c_xi_0}) - (\ref{eq:c_xi_2}) yields%
\begin{equation*}
\left( 1-\frac{\bar{\chi}_{0}^{2}\left( \tau \right) }{24}\right) \bar{\chi}%
_{0}\left( \tau \right) =\mathfrak{\hat{l}}\left( \tau \right) ,
\end{equation*}%
\begin{equation*}
\left( 1-\frac{\bar{\chi}_{0}^{2}\left( \tau \right) }{8}\right) \bar{\chi}%
_{1}\left( \tau \right) =\mathfrak{\hat{m}}\left( \tau \right) ,
\end{equation*}%
\begin{equation*}
\left( 1-\frac{\bar{\chi}_{0}^{2}\left( \tau \right) }{8}\right) \left( \bar{%
\chi}_{2}\left( \tau \right) -\frac{1}{4}\bar{\chi}_{0}\left( \tau \right) 
\bar{\chi}_{1}^{2}\left( \tau \right) +\frac{1}{\bar{\chi}_{0}\left( \tau
\right) }\right) =\mathfrak{\hat{n}}\left( \tau \right) ,
\end{equation*}%
\begin{equation*}
\bar{\chi}_{0}\left( \tau \right) \sim \mathfrak{\hat{l}}\left( \tau \right)
\left( 1+\frac{\mathfrak{\hat{l}}^{2}\left( \tau \right) }{24}\right) ,
\end{equation*}%
\begin{equation*}
\bar{\chi}_{1}\left( \tau \right) \sim \mathfrak{\hat{m}}\left( \tau \right)
\left( 1+\frac{\mathfrak{\hat{l}}^{2}\left( \tau \right) }{8}\right) ,
\end{equation*}%
\begin{equation*}
\bar{\chi}_{2}\left( \tau \right) \sim -\frac{1}{\mathfrak{\hat{l}}\left(
\tau \right) }+\frac{\mathfrak{\hat{l}}\left( \tau \right) }{24}+\frac{1}{4}%
\mathfrak{\hat{l}}\left( \tau \right) \mathfrak{\hat{m}}^{2}\left( \tau
\right) +\mathfrak{\hat{n}}\left( \tau \right) \left( 1+\frac{\mathfrak{\hat{%
l}}^{2}\left( \tau \right) }{8}\right) .
\end{equation*}%
Noticing that $\bar{\chi}_{i}\left( \tau \right) =\chi _{i}\left( \tau
\right) \tau ^{1/2}$, we obtain the result.

\section{Long-time Asymptotics\label{LongA}}

\subsection{Exponential L\'{e}vy Processes\label{ELPsLA}}

Due to the fact that we have chosen a parametric representation of $\bar{k}$
in terms of $y$, so that we know that the saddlepoint is located on the
imaginary axis at $u=iy$, it is easy to perform the corresponding
saddlepoint calculation. When $-\frac{1}{2}<y<\frac{1}{2}$, we can perform
parallel move of the contour of integration to a new contour given by the
equation%
\begin{equation*}
\func{Im}u=y,
\end{equation*}%
without hitting singularities located at $u=\pm i/2$. Accordingly, we can
approximate the price of the call option as follows%
\begin{equation}
\mathsf{C}\left( \tau ,\bar{k}\left( y\right) \tau \right) =1+\frac{1}{\sqrt{%
2\pi \tau }}e^{\tau \Xi _{01}^{\left( +\right) }\left( y\right) }c_{0}\left(
y\right) \left( 1-\frac{1}{\tau }c_{1}\left( y\right) \right) .
\label{eq: ATM}
\end{equation}%
When $-\kappa _{+}+\frac{1}{2}<y<-\frac{1}{2}$ or $\frac{1}{2}<y<\kappa _{-}+%
\frac{1}{2}$, in the process of moving the contour of integration, we shall
hit the points $u=-i/2$ and $i/2$, respectively. We note that%
\begin{equation*}
S_{+}=S\left( \frac{i}{2}\right) =\bar{k},\ \ \ S_{-}=S\left( -\frac{i}{2}%
\right) =a_{+}\left( \kappa _{+}-1\right) ^{\alpha }+a_{-}\left( \kappa
_{-}+1\right) ^{\alpha }+\gamma +\delta =0,
\end{equation*}%
so that, by virtue of the Cauchy theorem, the corresponding approximate
prices can be written as%
\begin{equation}
\mathsf{C}\left( \tau ,\bar{k}\left( y\right) \tau \right) =\left\{ 
\begin{array}{ll}
1-e^{\bar{k}\left( y\right) \tau }+\frac{1}{\sqrt{2\pi \tau }}e^{\tau \Xi
_{01}^{\left( +\right) }\left( y\right) }c_{0}\left( y\right) \left( 1-\frac{%
1}{\tau }c_{1}\left( y\right) \right) , & \frac{1}{2}<y<\kappa _{-}+\frac{1}{%
2}, \\ 
\frac{1}{\sqrt{2\pi \tau }}e^{\tau \Xi _{01}^{\left( +\right) }\left(
y\right) }c_{0}\left( y\right) \left( 1-\frac{1}{\tau }c_{1}\left( y\right)
\right) , & -\kappa _{+}+\frac{1}{2}<y<-\frac{1}{2}.%
\end{array}%
\right.  \label{eq: OTM-ITM 1}
\end{equation}%
Comparison on expressions (\ref{eq: ATM}), (\ref{eq: OTM-ITM 1}) and (\ref%
{eq: BS ass 1}) immediately yields (\ref{eq: sigma imp}).

Consider $y\in \left( -\mathsf{Y}_{+},\mathsf{Y}_{-}\right) $ and define 
\begin{equation}
\Xi _{2}\left( y\right) =\Xi _{0}^{\prime \prime }\left( y\right) ,\ \ \ \Xi
_{3}\left( y\right) =\Xi _{0}^{\prime \prime \prime }\left( y\right) ,\ \ \
\Xi _{4}\left( y\right) =\Xi _{0}^{\prime \prime \prime \prime }\left(
y\right) ,  \label{eq: Xi_i}
\end{equation}%
\begin{equation}
f_{0}\left( y\right) =\frac{1}{R\left( y\right) },\ \ \ f_{1}\left( y\right)
=f_{0}^{\prime }\left( y\right) =-\frac{2y}{\left( R\left( y\right) \right)
^{2}},\ \ \ f_{2}\left( y\right) =f_{0}^{\prime \prime }\left( y\right) =%
\frac{\left( 6y^{2}+\frac{1}{2}\right) }{\left( R\left( y\right) \right) ^{3}%
},  \label{eq: f_i}
\end{equation}%
\begin{align}
c_{0}\left( y\right) & =\frac{f_{0}\left( y\right) }{\sqrt{\Xi _{2}\left(
y\right) }}=\frac{1}{\sqrt{\Xi _{2}\left( y\right) }R\left( y\right) },
\label{eq: c_i} \\
c_{1}\left( y\right) & =-\frac{1}{2\Xi _{2}\left( y\right) }\left( \frac{%
f_{2}\left( y\right) }{f_{0}\left( y\right) }-\frac{\Xi _{3}\left( y\right)
f_{1}\left( y\right) }{\Xi _{2}\left( y\right) f_{0}\left( y\right) }+\frac{%
5\Xi _{3}^{2}\left( y\right) }{12\Xi _{2}^{2}\left( y\right) }-\frac{\Xi
_{4}\left( y\right) }{4\Xi _{2}\left( y\right) }\right)  \notag \\
& -\frac{1}{2\Xi _{2}\left( y\right) }\left( \frac{\left( 6y^{2}+\frac{1}{2}%
\right) }{\left( R\left( y\right) \right) ^{2}}+\frac{2y\Xi _{3}\left(
y\right) }{R\left( y\right) \Xi _{2}\left( y\right) }+\frac{5\Xi
_{3}^{2}\left( y\right) }{12\Xi _{2}^{2}\left( y\right) }-\frac{\Xi
_{4}\left( y\right) }{4\Xi _{2}\left( y\right) }\right) ,  \notag
\end{align}%
\begin{align}
a_{0}\left( y\right) & =2\left( s_{+}\left( y\right) -s_{-}\left( y\right)
\right) ^{2},  \label{eq: a_i} \\
a_{1}\left( y\right) & =\frac{2\ln \left( a_{0}^{1/2}\left( y\right) r\left(
y\right) c_{0}\left( y\right) \right) }{r\left( y\right) },  \notag \\
a_{2}\left( y\right) & =\frac{2c_{1}\left( y\right) }{r\left( y\right) } 
\notag \\
& +\frac{\left( r\left( y\right) a_{1}\left( y\right) \left( \left( r\left(
y\right) a_{1}\left( y\right) -3\right) \left( r\left( y\right) +\frac{1}{4}%
\right) -\frac{1}{4}\right) +6r\left( y\right) +2\right) }{r^{3}\left(
y\right) a_{0}\left( y\right) },  \notag
\end{align}%
\begin{equation}
b_{0}\left( y\right) =a_{0}^{1/2}\left( y\right) ,\ \ \ b_{1}\left( y\right)
=\frac{a_{1}\left( y\right) }{2a_{0}^{1/2}\left( y\right) },\ \ \
b_{2}\left( y\right) =\frac{4a_{0}\left( y\right) a_{2}\left( y\right)
-a_{1}^{2}\left( y\right) }{8a_{0}^{3/2}\left( y\right) },  \label{eq: b_i}
\end{equation}%
where%
\begin{equation}
s_{\pm }\left( y\right) =\mathrm{sign}\left( y\pm \frac{1}{2}\right) \sqrt{%
-\Xi _{01,\pm }\left( y\right) },\ \ \ r\left( y\right) =R\left( \frac{\Xi
_{1}\left( y\right) }{a_{0}\left( y\right) }\right) .  \label{eq: spm}
\end{equation}

\subsection{Tempered Stable Processes\label{TSPsLA}}

For completeness we present explicit formulas for $\Xi _{i}$:%
\begin{align*}
\Xi _{0}\left( y\right) & =\frac{1}{2}\sigma ^{2}\left( y^{2}-\frac{1}{4}%
\right) +\sum_{s=\pm }a_{s}\left( \kappa _{s}+s\left( y-\frac{1}{2}\right)
\right) ^{\alpha }-\gamma \left( y-\frac{1}{2}\right) +\delta , \\
\Xi _{1}\left( y\right) & =\sigma ^{2}y+\alpha \sum_{s=\pm }sa_{s}\left(
\kappa _{s}+s\left( y-\frac{1}{2}\right) \right) ^{\alpha -1}-\gamma , \\
\Xi _{2}\left( y\right) & =\sigma ^{2}+\alpha \left( \alpha -1\right)
\sum_{s=\pm }a_{s}\left( \kappa _{s}+s\left( y-\frac{1}{2}\right) \right)
^{\alpha -2}, \\
\Xi _{3}\left( y\right) & =\alpha \left( \alpha -1\right) \left( \alpha
-2\right) \sum_{s=\pm }sa_{s}\left( \kappa _{s}+s\left( y-\frac{1}{2}\right)
\right) ^{\alpha -3}, \\
\Xi _{4}\left( y\right) & =\alpha \left( \alpha -1\right) \left( \alpha
-2\right) \left( \alpha -3\right) \sum_{s=\pm }a_{s}\left( \kappa
_{s}+s\left( y-\frac{1}{2}\right) \right) ^{\alpha -4}.
\end{align*}

\subsection{Normal Inverse Gaussian Processes\label{NIGPsLA}}

For completeness, we present explicit expressions for $\Xi _{i}$:%
\begin{align*}
\Xi _{0}\left( y\right) & =\sigma ^{2}\mathcal{\bar{\varkappa}}\left( 
\mathcal{\bar{\varkappa}}-\Lambda \left( y\right) \right) , \\
\Xi _{1}\left( y\right) & =\sigma ^{2}\mathcal{\bar{\varkappa}}y\Lambda
^{-1}\left( y\right) , \\
\Xi _{2}\left( y\right) & =\sigma ^{2}\mathcal{\bar{\varkappa}\omega }%
^{2}\Lambda ^{-3}\left( y\right) , \\
\Xi _{3}\left( y\right) & =3\sigma ^{2}\mathcal{\bar{\varkappa}\omega }%
^{2}y\Lambda ^{-5}\left( y\right) , \\
\Xi _{4}\left( y\right) & =3\sigma ^{2}\mathcal{\bar{\varkappa}\omega }%
^{2}\left( \mathcal{\omega }^{2}+4y^{2}\right) \Lambda ^{-7}\left( y\right) ,
\end{align*}%
where $\Lambda \left( y\right) =\sqrt{\mathcal{\bar{\omega}}^{2}-y^{2}}$.

By using (\ref{eq: CNIG}), we can calculate long-time asymptotics for call
prices directly. For \thinspace $k=\hat{k}v$ and $v\rightarrow \infty $,
where $v=\sigma ^{2}\tau $, we have%
\begin{align*}
\mathsf{C}^{NIG}\left( v,\hat{k}v\right) & =\frac{\bar{\omega}\mathcal{\bar{%
\varkappa}}ve^{\mathcal{\bar{\varkappa}}^{2}v}}{\pi }\int_{\hat{k}v}^{\infty
}\frac{\left( e^{\frac{x}{2}}-e^{\hat{k}v-\frac{x}{2}}\right) K_{1}\left( 
\bar{\omega}\sqrt{x^{2}+\mathcal{\bar{\varkappa}}^{2}v^{2}}\right) }{\sqrt{%
x^{2}+\mathcal{\bar{\varkappa}}^{2}v^{2}}}dx \\
& =1-\frac{\bar{\omega}\mathcal{\bar{\varkappa}}ve^{\mathcal{\bar{\varkappa}}%
^{2}v}}{\pi }\left( \int_{-\frac{\hat{k}}{2\mathcal{\bar{\varkappa}}}%
}^{\infty }\frac{e^{-\mathcal{\bar{\varkappa}}vy}K_{1}\left( 2\bar{\omega}%
\mathcal{\bar{\varkappa}}vQ^{1/2}\left( y\right) \right) }{Q^{1/2}\left(
y\right) }dy\right. \\
& \left. +\int_{\frac{\hat{k}}{2\mathcal{\bar{\varkappa}}}}^{\infty }\frac{%
e^{\hat{k}v-\mathcal{\bar{\varkappa}}vy}K_{1}\left( 2\bar{\omega}\mathcal{%
\bar{\varkappa}}vQ^{1/2}\left( y\right) \right) }{Q^{1/2}\left( y\right) }%
dy\right) \\
& \sim 1-\sqrt{\frac{\bar{\omega}\mathcal{\bar{\varkappa}}v}{4\pi }}e^{%
\mathcal{\bar{\varkappa}}^{2}v}\left( \int_{-\frac{\hat{k}}{2\mathcal{\bar{%
\varkappa}}}}^{\infty }\frac{e^{-\mathcal{\bar{\varkappa}}v\left( y+2\bar{%
\omega}Q^{1/2}\left( y\right) \right) }}{Q^{3/4}\left( y\right) }\left( 1+%
\frac{3}{16\bar{\omega}\mathcal{\bar{\varkappa}}vQ^{1/2}\left( y\right) }%
\right) dy\right. \\
& \left. +e^{\hat{k}v}\int_{\frac{\hat{k}}{2\mathcal{\bar{\varkappa}}}%
}^{\infty }\frac{e^{-\mathcal{\bar{\varkappa}}v\left( y+2\bar{\omega}%
Q^{1/2}\left( y\right) \right) }}{Q^{3/4}\left( y\right) }\left( 1+\frac{3}{%
16\bar{\omega}\mathcal{\bar{\varkappa}}\bar{\tau}Q^{1/2}\left( y\right) }%
\right) dy\right) ,
\end{align*}%
where $x=\pm 2\mathcal{\bar{\varkappa}}vy$. Consider the following function%
\begin{equation*}
\Sigma \left( y\right) =\mathcal{\bar{\varkappa}}\left( y+2\bar{\omega}%
Q^{1/2}\left( y\right) \right) .
\end{equation*}%
It is easy to show that on the real axis this function has a unique global
minimum $y_{\ast }$ such that 
\begin{equation*}
y_{\ast }=-\frac{1}{4\mathcal{\bar{\varkappa}}},\ \ \ \ \ \Sigma _{\ast
}=\Sigma \left( y_{\ast }\right) =\mathcal{\bar{\varkappa}}^{2}.
\end{equation*}%
Assuming for brevity that $\left\vert \hat{k}\right\vert <1/2$, so that this
point lies outside the domain of integration for both integrals in the above
expression, we can use the original Laplace method to evaluate these
integrals asymptotically and get%
\begin{align*}
\mathsf{C}\left( v,\hat{k}v\right) & \sim 1+\sqrt{\frac{\bar{\omega}}{\pi 
\mathcal{\bar{\varkappa}}v}}e^{\mathcal{\bar{\varkappa}}v\left( \mathcal{%
\bar{\varkappa}}+\frac{\hat{k}}{2\mathcal{\bar{\varkappa}}}-2\bar{\omega}%
\hat{Q}^{1/2}\right) }\frac{\hat{Q}^{1/4}}{R\left( \hat{k}\right) } \\
& \times \left( 1-\frac{1}{\mathcal{\bar{\varkappa}}v}\left( \frac{3\bar{%
\omega}\hat{k}^{2}}{2\mathcal{\bar{\varkappa}}R\left( \hat{k}\right) \hat{Q}%
^{1/2}}-\frac{\sqrt{2}\bar{\omega}\hat{Q}^{1/4}\left( R\left( \hat{k}\right)
\left( 1+3\mathcal{\bar{\varkappa}}^{2}\right) +\bar{\omega}^{2}\right) }{%
\mathcal{\bar{\varkappa}}^{2}R^{2}\left( \hat{k}\right) }-\frac{3}{16\bar{%
\omega}\hat{Q}^{1/2}}\right) \right) ,
\end{align*}%
where $\hat{Q}=Q\left( \hat{k}/2\mathcal{\bar{\varkappa}}\right) $. This
expression agrees with the one obtained via the saddlepoint method.

\subsection{Merton Processes\label{MPsLApp}}

For completeness we present explicit formulas for $\Xi _{i}$:%
\begin{align*}
\Xi _{0}\left( y\right) & =\frac{1}{2}\sigma ^{2}R\left( y\right) +\lambda
\left( \bar{\Lambda}\left( y\right) -1-\left( 1-e^{q}\right) \left( y-\frac{1%
}{2}\right) \right) , \\
\Xi _{1}\left( y\right) & =\sigma ^{2}y+\lambda \left( \left( -q+\eta
^{2}y\right) \bar{\Lambda}\left( y\right) -\left( 1-e^{q}\right) \right) , \\
\Xi _{2}\left( y\right) & =\sigma ^{2}+\lambda \left( \eta ^{2}+\left(
-q+\eta ^{2}y\right) ^{2}\right) \bar{\Lambda}\left( y\right) , \\
\Xi _{3}\left( y\right) & =\lambda \left( -q+\eta ^{2}y\right) \left( 3\eta
^{2}+\left( -q+\eta ^{2}y\right) ^{2}\right) \bar{\Lambda}\left( y\right) ,
\\
\Xi _{4}\left( y\right) & =\lambda \left( 3\eta ^{4}+6\eta ^{2}\left(
-q+\eta ^{2}y\right) ^{2}+\left( -q+\eta ^{2}y\right) ^{4}\right) \bar{%
\Lambda}\left( y\right) ,
\end{align*}%
where $\bar{\Lambda}\left( y\right) =e^{-q\left( y-\frac{1}{2}\right) +\frac{%
\eta ^{2}}{2}R\left( y\right) }$.

\subsection{Heston Stochastic Volatility Processes\label{HSVPsLA}}

Substitution of $u=iy$ in (\ref{eq:S_HSVP}) yields

\begin{align*}
\Xi _{0}\left( y\right) & =\frac{\kappa \theta }{\varepsilon }\left( \rho y+%
\check{\kappa}-\varsigma \left( y\right) \right) , \\
f_{0}\left( y\right) & =\frac{e^{\varpi _{0}\left( \rho y+\check{\kappa}%
-\varsigma \left( y\right) \right) /\varepsilon }}{R\left( y\right) \left( 
\frac{\rho y+\check{\kappa}+\varsigma \left( y\right) }{2\varsigma \left(
y\right) }\right) ^{2\kappa \theta /\varepsilon ^{2}}}, \\
\varsigma \left( y\right) & =\sqrt{-\bar{\rho}^{2}y^{2}+2\rho \check{\kappa}%
y+\check{\kappa}^{2}+\frac{1}{4}}.
\end{align*}%
Differentiation of $\Xi _{0}\left( y\right) $ gives explicit expressions for 
$\Xi _{1}\left( y\right) ,\Xi _{2}\left( y\right) $:%
\begin{equation*}
\Xi _{1}\left( y\right) =\frac{\kappa \theta }{\varepsilon }\left( \rho -%
\frac{-\bar{\rho}^{2}y+\rho \check{\kappa}}{\varsigma \left( y\right) }%
\right) ,\ \ \ \Xi _{2}\left( y\right) =\frac{\kappa \theta }{\varsigma
\left( y\right) }\left( \bar{\rho}^{2}+\frac{\left( -\bar{\rho}^{2}y+\rho 
\check{\kappa}\right) ^{2}}{\varsigma ^{2}\left( y\right) }\right) .
\end{equation*}%
Let $y^{\ast }\left( \bar{k}\right) $ is a saddlepoint such that 
\begin{equation*}
\Xi _{1}\left( y^{\ast }\right) +\bar{k}=0.
\end{equation*}%
A simple calculation yields%
\begin{equation*}
\varsigma \left( y^{\ast }\right) =\frac{-\bar{\rho}^{2}y^{\ast }+\rho 
\check{\kappa}}{l},
\end{equation*}%
where $l=\rho +\varepsilon \bar{k}/\kappa \theta $. After squaring this
relation, we arrive at the following quadratic equation 
\begin{equation*}
\bar{\rho}^{2}y^{\ast 2}-2\rho \check{\kappa}y^{\ast }-\check{\kappa}^{2}+%
\frac{\check{\kappa}^{2}-\frac{1}{4}l^{2}}{\bar{\rho}^{2}+l^{2}}=0,
\end{equation*}%
which has two roots%
\begin{equation*}
y_{\pm }^{\ast }=\frac{\rho \check{\kappa}\pm \sqrt{\check{\kappa}^{2}-\bar{%
\rho}^{2}\left( \check{\kappa}^{2}-\frac{1}{4}l^{2}\right) /\left( \bar{\rho}%
^{2}+l^{2}\right) }}{\bar{\rho}^{2}}.
\end{equation*}%
It is easy to show that the relevant root has the form 
\begin{equation*}
y^{\ast }\left( \bar{k}\right) =\frac{\rho \check{\kappa}-\mathrm{sign}%
\left( l\right) \sqrt{\check{\kappa}^{2}-\bar{\rho}^{2}\left( \check{\kappa}%
^{2}-\frac{1}{4}l^{2}\right) /\left( \bar{\rho}^{2}+l^{2}\right) }}{\bar{\rho%
}^{2}},
\end{equation*}%
and%
\begin{equation*}
\Xi _{01,\pm }\left( \bar{k}\right) =\frac{\kappa \theta }{\varepsilon l}%
\left( \left( \bar{\rho}^{2}+\rho l\right) y^{\ast }\left( \bar{k}\right)
+\left( l-\rho \right) \check{\kappa}\right) \pm \frac{1}{2}\bar{k}.
\end{equation*}%
Thus, to the leading order, $\sigma _{imp,0}\left( \bar{k}\right) $ is given
by (\ref{eq: sigma_0 HSVP}).

When $k\sim \tau $, the corresponding option becomes deep OTM or ITM, with
one notable exception, which can be described as follows. Consider the
implied volatility as a function of $\Delta $, and in particular, calculate $%
\Sigma \left( \tau ,0.5\right) $. For brevity, denote $\Sigma \left( \tau
,0.5\right) $ by $\bar{\Sigma}$. It is clear that the corresponding $\bar{k}$
has the form $\bar{k}=\bar{\Sigma}^{2}/2$. We see that $\bar{\Sigma}$ can be
found from the following nonlinear equation 
\begin{equation*}
\bar{\Sigma}=\sigma _{imp,0}\left( \frac{\bar{\Sigma}^{2}}{2}\right) ,
\end{equation*}%
which can be solved numerically via a simple Newton-Raphson method.

\section{Short-time Asymptotics\label{ShortA}}

\subsection{Tempered Stable Processes\label{TSPsSA}}

If $\alpha \in \left( 0,1\right) $, then%
\begin{equation*}
\frac{\partial }{\partial \tau }\mathsf{C}\left( 0,k\right) =-\frac{e^{\frac{%
k}{2}}}{2\pi }\int_{-\infty }^{\infty }\frac{\upsilon \left( u\right)
e^{-iku}}{Q\left( u\right) }du.
\end{equation*}%
Since the last integral converges and no further treatment is needed.

If $\alpha \in \left( 1,2\right) $, then we need to do integration by parts
first (which is possible because $k\neq 0$):%
\begin{align*}
\frac{\partial }{\partial \tau }\mathsf{C}\left( 0,k\right) & =\left. -\frac{%
e^{\frac{k}{2}}}{2\pi }\frac{\partial }{\partial \tau }\left( \int_{-\infty
}^{\infty }\frac{e^{\tau \upsilon \left( u\right) -iku}}{Q\left( u\right) }%
du\right) \right\vert _{\tau =0} \\
& =\left. -\frac{e^{\frac{k}{2}}}{2\pi }\frac{\partial }{\partial \tau }%
\left( \int_{-\infty }^{\infty }\frac{e^{\tau \upsilon \left( u\right) }}{%
Q\left( u\right) }d\left( \frac{e^{-iku}}{-ik}\right) \right) \right\vert
_{\tau =0} \\
& =\left. -\frac{e^{\frac{k}{2}}}{2\pi ik}\frac{\partial }{\partial \tau }%
\left( \int_{-\infty }^{\infty }e^{-iku}d\left( \frac{e^{\tau \upsilon
\left( u\right) }}{Q\left( u\right) }\right) \right) \right\vert _{\tau =0}
\\
& =\left. -\frac{e^{\frac{k}{2}}}{2\pi ik}\frac{\partial }{\partial \tau }%
\left( \int_{-\infty }^{\infty }\left( \frac{\tau \upsilon ^{\prime }\left(
u\right) }{Q\left( u\right) }-\frac{2u}{Q^{2}\left( u\right) }\right)
e^{\tau \upsilon \left( u\right) -iku}du\right) \right\vert _{\tau =0} \\
& =\left. -\frac{e^{\frac{k}{2}}}{2\pi ik}\int_{-\infty }^{\infty }\left( 
\frac{\upsilon ^{\prime }\left( u\right) +\tau \upsilon ^{\prime }\left(
u\right) \upsilon \left( u\right) }{Q\left( u\right) }-\frac{2u\upsilon
\left( u\right) }{Q^{2}\left( u\right) }\right) e^{\tau \upsilon \left(
u\right) -iku}du\right\vert _{\tau =0} \\
& =-\frac{e^{\frac{k}{2}}}{2\pi ik}\int_{-\infty }^{\infty }\left( \frac{%
\upsilon \left( u\right) }{Q\left( u\right) }\right) ^{\prime }e^{-iku}du,
\end{align*}%
where the last integral clearly converges.

We are now ready to consider the ATM case. For convenience, we summarize our
existing notation and introduce additional notation which is used below. We
have%
\begin{align*}
\upsilon \left( u\right) & =\sum_{s=\pm }a_{s}\left( \kappa _{s}-s\left( iu+%
\frac{1}{2}\right) \right) ^{\alpha }+\gamma \left( iu+\frac{1}{2}\right)
+\delta , \\
\tilde{\upsilon}\left( u\right) & =\sum_{s=\pm }a_{s}\left( \kappa
_{s}-s\left( iu+\frac{1}{2}\right) \right) ^{\alpha }, \\
\hat{\upsilon}\left( u\right) & =\sum_{s=\pm }a_{s}\left( -siu\right)
^{\alpha }.
\end{align*}%
It is clear that for real $u$ we have 
\begin{equation*}
\hat{\upsilon}\left( u\right) =\left( p+iq\mathrm{sign}\left( u\right)
\right) \left\vert u\right\vert ^{\alpha }.
\end{equation*}

In case (A) we use change of variables $v=\tau u$ and represent the
integrals $\mathfrak{l},\mathfrak{m},\mathfrak{n}$ as follows%
\begin{equation*}
\mathfrak{l}\left( \tau \right) =\tau \mathfrak{L}\left( \tau \right) ,\ \ \
\ \ \mathfrak{m}\left( \tau \right) =\mathfrak{M}\left( \tau \right) ,\ \ \
\ \ \mathfrak{n}\left( \tau \right) =\tau ^{-1}\mathfrak{N}\left( \tau
\right) ,
\end{equation*}%
where $\mathfrak{L},\mathfrak{M},\mathfrak{N}$ have the form (\ref{eq: LMNE}%
) with%
\begin{align*}
E\left( \tau ,v\right) & =\exp \left( \tau ^{1-\alpha }\sum_{s=\pm
}a_{s}\left( \tau \left( \kappa _{s}-\frac{s}{2}\right) -siv\right) ^{\alpha
}+iv\gamma +\tau \left( \frac{\gamma }{2}+\delta \right) \right) \\
& \equiv \exp \left( \theta _{1}\left( \tau ,v\right) \right) .
\end{align*}%
As before, we represent $\mathfrak{L}$ as follows: 
\begin{equation*}
\mathfrak{L}\left( \tau \right) =\mathfrak{L}_{0}^{\varepsilon }\left( \tau
\right) +\mathfrak{L}_{\varepsilon }^{\infty }\left( \tau \right) ,
\end{equation*}%
where 
\begin{equation*}
\mathfrak{L}_{\alpha }^{\beta }\left( \tau \right) =\frac{1}{\pi }\mathrm{Re}%
\left\{ \int_{\alpha }^{\beta }\frac{1-e^{\theta _{1}\left( \tau ,v\right) }%
}{v^{2}+\frac{1}{4}\tau ^{2}}dv\right\} .
\end{equation*}%
Expansion near the origin for $\varepsilon \rightarrow 0$ yields%
\begin{align*}
\mathfrak{L}_{0}^{\varepsilon }\left( \tau \right) & \sim -\frac{1}{\pi }%
\mathrm{Re}\left\{ \int_{0}^{\varepsilon }\frac{\theta _{1}\left( \tau
,v\right) }{v^{2}+\frac{1}{4}\tau ^{2}}dv\right\} \\
& =-\frac{\tau }{\pi \tau }\mathrm{Re}\left\{ \int_{0}^{\varepsilon /\tau }%
\frac{\tilde{\upsilon}\left( u\right) +\frac{\gamma }{2}+\delta }{Q\left(
u\right) }du\right\} \\
& \sim -\frac{1}{\pi }\mathrm{Re}\left\{ \int_{0}^{\infty }\frac{\tilde{%
\upsilon}\left( u\right) }{u^{2}+\frac{1}{4}}du\right\} -\frac{\gamma }{2}%
-\delta \\
& =-\left( \frac{\gamma }{2}+\delta +\varrho \right) ,
\end{align*}%
while the DCT yields%
\begin{equation*}
\mathfrak{L}_{\varepsilon }^{\infty }\left( \tau \right) \sim \frac{%
\left\vert \gamma \right\vert }{2},
\end{equation*}%
so that 
\begin{equation*}
\mathfrak{L}\left( \tau \right) =-\left( \gamma _{-}+\delta +\varrho \right)
\equiv C_{\mathfrak{L}}^{\left( a\right) }.
\end{equation*}%
Here we use the fact that%
\begin{equation*}
\frac{1}{\pi }\func{Re}\left\{ \int_{0}^{\infty }\frac{\tilde{\upsilon}%
\left( u\right) }{Q\left( u\right) }du\right\} =\varrho ,
\end{equation*}%
which can be verified directly. For $\mathfrak{M}\left( \tau \right) $ we
obtain%
\begin{align*}
\mathfrak{M}\left( \tau \right) & =-\frac{1}{\pi }\mathrm{Im}\left\{
\int_{0}^{\infty }\frac{e^{\theta _{1}\left( \tau ,v\right) }}{v^{2}+\frac{1%
}{4}\tau ^{2}}vdv\right\} \\
& \sim -\frac{1}{\pi }\mathrm{Im}\left\{ \int_{0}^{\infty }e^{\gamma iv}%
\frac{dv}{v}\right\} \\
& =-\frac{1}{\pi }\int_{0}^{\infty }\sin \left( \gamma v\right) \frac{dv}{v}
\\
& =-\frac{1}{2}\text{$\mathrm{sign}$}\left( \gamma \right) \equiv C_{%
\mathfrak{M}}^{\left( a\right) }.
\end{align*}%
For $\mathfrak{N}\left( \tau \right) $ we obtain%
\begin{equation*}
\mathfrak{N}\left( \tau \right) =\frac{1}{\pi }\mathrm{\func{Re}}\left\{
\int_{0}^{\infty }e^{\theta _{1}\left( \tau ,v\right) }dv\right\} \sim \frac{%
1}{\pi }\mathrm{Re}\left\{ \int_{0}^{\infty }e^{\gamma iv}dv\right\} =\delta
\left( \gamma \right) \equiv C_{\mathfrak{N}}^{\left( a\right) }.
\end{equation*}%
In case (B) we use change of variables $v=\tau ^{\alpha ^{\prime }}u$, where 
$\alpha ^{\prime }=1/\alpha $, and rewrite the corresponding integrals as
follows%
\begin{equation*}
\mathfrak{l}\left( \tau \right) =\tau ^{\alpha ^{\prime }}\mathfrak{L}\left(
\tau \right) ,\ \ \ \ \ \mathfrak{m}\left( \tau \right) =\mathfrak{M}\left(
\tau \right) ,\ \ \ \ \ \mathfrak{n}\left( \tau \right) =\tau ^{-\alpha
^{\prime }}\mathfrak{N}\left( \tau \right) ,
\end{equation*}%
where%
\begin{align*}
\mathfrak{L}\left( \tau \right) & =\frac{1}{\pi }\mathrm{Re}\left\{
\int_{0}^{\infty }\frac{1-E\left( \tau ,v\right) }{v^{2}+\frac{1}{4}\tau
^{2\alpha ^{\prime }}}dv\right\} , \\
\mathfrak{M}\left( \tau \right) & =-\frac{1}{\pi }\mathrm{Im}\left\{
\int_{0}^{\infty }\frac{E\left( \tau ,v\right) }{v^{2}+\frac{1}{4}\tau
^{2\alpha ^{\prime }}}vdv\right\} , \\
\mathfrak{N}\left( \tau \right) & =\frac{1}{\pi }\mathrm{Re}\left\{
\int_{0}^{\infty }E\left( \tau ,v\right) dv\right\} ,
\end{align*}%
\begin{align*}
E\left( \tau ,v\right) & =\exp \left( \sum_{s=\pm }a_{s}\left( \tau ^{\alpha
^{\prime }}\left( \kappa _{s}-\frac{s}{2}\right) -siv\right) ^{\alpha
}+i\tau ^{1-\alpha ^{\prime }}v\gamma +\tau \left( \frac{\gamma }{2}+\delta
\right) \right) \\
& \equiv \exp \left( \theta _{\alpha ^{\prime }}\left( \tau ,v\right)
\right) .
\end{align*}%
As before 
\begin{equation*}
\mathfrak{L}\left( \tau \right) =\mathfrak{L}_{0}^{\varepsilon }\left( \tau
\right) +\mathfrak{L}_{\varepsilon }^{\infty }\left( \tau \right) ,
\end{equation*}%
where%
\begin{equation*}
\mathfrak{L}_{0}\left( \tau \right) =-\frac{1}{\pi }\mathrm{Re}\left\{
\int_{0}^{\varepsilon }\frac{\theta _{\alpha ^{\prime }}\left( \tau
,v\right) }{v^{2}+\frac{1}{4}\tau ^{2\alpha ^{\prime }}}du\right\} \underset{%
\tau \rightarrow 0}{\rightarrow }O\left( \varepsilon \right) ,
\end{equation*}%
\begin{align*}
\mathfrak{L}_{\varepsilon }^{\infty }\left( \tau \right) & \sim \frac{1}{\pi 
}\mathrm{Re}\left\{ \int_{0}^{\infty }\frac{1-e^{\hat{\upsilon}\left(
v\right) }}{v^{2}}dv\right\} \\
& =\frac{1}{\pi }\mathrm{Re}\left\{ \int_{0}^{\infty }\frac{1-e^{\left(
p+iq\right) \left\vert v\right\vert ^{\alpha }}}{v^{2}}dv\right\} \\
& =-\frac{\alpha ^{\prime }}{\pi }\Gamma \left( -\alpha ^{\prime }\right) 
\mathrm{Re}\left\{ \left( -p-iq\right) ^{\alpha ^{\prime }}\right\} \\
& =\frac{\Gamma \left( 1-\alpha ^{\prime }\right) }{\pi }r^{\alpha ^{\prime
}}\cos \left( \alpha ^{\prime }\chi \right) \equiv C_{\mathfrak{L}}^{\left(
B\right) }>0.
\end{align*}%
Thus, 
\begin{equation*}
\mathfrak{L}\left( \tau \right) =\frac{1}{\pi }\Gamma \left( 1-\alpha
^{\prime }\right) r^{\alpha ^{\prime }}\cos \left( \alpha ^{\prime }\chi
\right) =C_{\mathfrak{L}}^{\left( B\right) }.
\end{equation*}%
For $\mathfrak{M}\left( \tau \right) $ we have%
\begin{align*}
\mathfrak{M}\left( \tau \right) & =-\frac{1}{\pi }\mathrm{Im}\left\{
\int_{0}^{\infty }\frac{e^{\theta _{\alpha ^{\prime }}\left( \tau ,v\right) }%
}{\left( v^{2}+\frac{1}{4}\tau ^{2\alpha ^{\prime }}\right) }vdv\right\} \\
& \sim -\frac{1}{\pi }\mathrm{Im}\left\{ \int_{0}^{\infty }\frac{e^{\hat{%
\upsilon}\left( v\right) }}{v}dv\right\} \\
& =-\frac{1}{\pi }\mathrm{Im}\left\{ \int_{0}^{\infty }\exp \left( \left(
p+iq\right) v^{\alpha }\right) \frac{dv}{v}\right\} \\
& =-\frac{\alpha ^{\prime }}{\pi }\arg \left( \left( p+iq\right) \right) \\
& =-\frac{\alpha ^{\prime }\chi }{\pi }\equiv C_{\mathfrak{M}}^{\left(
B\right) }.
\end{align*}%
\ Finally, for $\mathfrak{N}\left( \tau \right) $ we have%
\begin{align*}
\mathfrak{N}\left( \tau \right) & =\frac{1}{\pi }\mathrm{Re}\left\{
\int_{0}^{\infty }e^{\theta _{\alpha ^{\prime }}\left( \tau ,v\right)
}dv\right\} \\
& \sim \frac{1}{\pi }\mathrm{Re}\left\{ \int_{0}^{\infty }e^{\hat{\upsilon}%
\left( v\right) }dv\right\} \\
& =\frac{1}{\pi }\mathrm{Re}\left\{ \int_{0}^{\infty }\exp \left( \left(
p+iq\right) v^{\alpha }\right) dv\right\} \\
& =\frac{\Gamma \left( 1+\alpha ^{\prime }\right) }{\pi }\mathrm{Re}\left\{
\left( -\left( p+iq\right) \right) ^{-\alpha ^{\prime }}\right\} \\
& =\frac{\Gamma \left( 1+\alpha ^{\prime }\right) }{\pi }r^{-\alpha ^{\prime
}}\cos \left( \alpha ^{\prime }\chi \right) \equiv C_{\mathfrak{N}}^{\left(
B\right) }<0.
\end{align*}%
In case (C) we use change of variables $v=\tau ^{1/2}u$ and rewrite the
corresponding integrals as follows%
\begin{equation*}
\mathfrak{l}\left( \tau \right) =\tau ^{1/2}\mathfrak{L}\left( \tau \right)
,\ \ \ \ \ \mathfrak{m}\left( \tau \right) =\mathfrak{M}\left( \tau \right)
,\ \ \ \ \ \mathfrak{n}\left( \tau \right) =\tau ^{-1/2}\mathfrak{N}\left(
\tau \right) ,
\end{equation*}%
where%
\begin{align*}
\mathfrak{L}\left( \tau \right) & =\frac{1}{\pi }\mathrm{Re}\left\{
\int_{0}^{\infty }\frac{1-E\left( \tau ,v\right) }{v^{2}+\frac{\tau }{4}}%
dv\right\} , \\
\mathfrak{M}\left( \tau \right) & =-\frac{1}{\pi }\mathrm{Im}\left\{
\int_{0}^{\infty }\frac{E\left( \tau ,v\right) }{v^{2}+\frac{\tau }{4}}%
vdv\right\} , \\
\mathfrak{N}\left( \tau \right) & =\frac{1}{\pi }\mathrm{Re}\left\{
\int_{0}^{\infty }E\left( \tau ,v\right) dv\right\} ,
\end{align*}%
\begin{align*}
E\left( \tau ,v\right) & =\exp \left( \tau ^{1-\alpha /2}\sum_{s=\pm
}a_{s}\left( \tau ^{1/2}\left( \kappa _{s}-\frac{s}{2}\right) -siv\right)
^{\alpha }+i\tau ^{1/2}v\gamma +\tau \left( \frac{\gamma }{2}+\delta \right)
-\frac{\sigma ^{2}}{2}\left( v^{2}+\frac{\tau }{4}\right) \right) \\
& \equiv \exp \left( \theta _{1/2}\left( \tau ,v\right) \right) .
\end{align*}%
As before 
\begin{equation*}
\mathfrak{L}\left( \tau \right) =\mathfrak{L}_{0}^{\varepsilon }\left( \tau
\right) +\mathfrak{L}_{\varepsilon }^{\infty }\left( \tau \right) ,
\end{equation*}%
where%
\begin{align*}
\mathfrak{L}_{0}^{\varepsilon }\left( \tau \right) & \sim -\left( \frac{1}{%
\pi }\mathrm{Re}\left\{ \int_{0}^{\infty }\frac{\hat{\upsilon}\left(
u\right) }{Q\left( u\right) }du\right\} +\frac{\gamma }{2}+\delta \right)
\tau ^{1/2}, \\
\mathfrak{L}_{\varepsilon }^{\infty }\left( \tau \right) & \sim \frac{1}{\pi 
}\int_{0}^{\infty }\frac{1-\exp \left( -\frac{\sigma ^{2}}{2}v^{2}\right) }{%
v^{2}}dv=\frac{\sigma }{\sqrt{2\pi }},
\end{align*}%
so that 
\begin{equation*}
\mathfrak{L}\left( \tau \right) \sim \frac{\sigma }{\sqrt{2\pi }}-\left( 
\frac{\gamma }{2}+\delta +\varrho \right) \tau ^{1/2}=\frac{\sigma }{\sqrt{%
2\pi }}+C_{\mathfrak{L}}^{\left( C\right) }\tau ^{1/2}.
\end{equation*}%
Likewise, for $\mathfrak{M}\left( \tau \right) $ we have%
\begin{align*}
\mathfrak{M}\left( \tau \right) & \sim -\frac{1}{\pi }\mathrm{Im}\left\{
\int_{0}^{\infty }\frac{e^{-\frac{\sigma ^{2}}{2}v^{2}}\left( \gamma \tau
^{1/2}iv+\tau ^{1-\alpha /2}\hat{\upsilon}\left( v\right) \right) }{v}%
dv\right\} \\
& =-\frac{\gamma \tau ^{1/2}}{\pi }\int_{0}^{\infty }e^{-\frac{{\sigma
^{2}v^{2}}}{2}}dv-\frac{q\tau ^{1-\alpha /2}}{\pi }\int_{0}^{\infty }e^{-%
\frac{\sigma ^{2}v^{2}}{2}}v^{\alpha -1}dv \\
& =-\frac{\gamma }{\sqrt{\pi }}\sigma ^{-1/2}\tau ^{1/2}-\frac{2^{\alpha
/2-1}\Gamma \left( \alpha /2\right) q}{\pi }\sigma ^{-\alpha }\tau
^{1-\alpha /2} \\
& \equiv C_{\mathfrak{M}}^{\left( C\right) }\tau ^{1/2}+D_{\mathfrak{M}%
}^{\left( C\right) }\tau ^{1-\alpha /2}.
\end{align*}%
Finally, for $\mathfrak{N}\left( \tau \right) $ we get%
\begin{align*}
\mathfrak{N}\left( \tau \right) & \sim \frac{1}{\pi }\mathrm{Re}\left\{
\int_{0}^{\infty }e^{-\frac{\sigma ^{2}v^{2}}{2}}\left( 1+\gamma \tau
^{1/2}iv+\tau ^{1-\alpha /2}\hat{\upsilon}\left( v\right) \right) dv\right\}
\\
& =\frac{1}{\sqrt{2\pi }\sigma }+\frac{p\tau ^{1-\alpha /2}}{\pi }%
\int_{0}^{\infty }e^{-\frac{\sigma ^{2}v^{2}}{2}}v^{\alpha }dv \\
& =\frac{1}{\sqrt{2\pi }\sigma }+\frac{2^{\alpha /2-1/2}\Gamma \left( \alpha
/2+\frac{1}{2}\right) p}{\pi }\sigma ^{-\left( 1+\alpha \right) }\tau
^{1-\alpha /2} \\
& =\frac{1}{\sqrt{2\pi }\sigma }+C_{\mathfrak{N}}^{\left( C\right) }\tau
^{1-\alpha /2}.
\end{align*}%
In case (D) we use the same change of variables as in case (C) and obtain
the following expressions for the relevant integrals%
\begin{align*}
\mathfrak{L}\left( \tau \right) & \sim \frac{1}{\pi }\mathrm{Re}\left\{
\int_{0}^{\infty }\frac{1-e^{\tau ^{1-\alpha /2}\hat{\upsilon}\left(
v\right) -\frac{\sigma ^{2}v^{2}}{2}}}{v^{2}}dv\right\} \\
& =\frac{1}{\pi }\mathrm{Re}\left\{ \int_{0}^{\infty }\frac{1-e^{-\frac{%
\sigma ^{2}v^{2}}{2}}}{v^{2}}dv\right\} -\frac{\sigma \tau ^{3/2-\alpha /2}}{%
\pi }\mathrm{Re}\left\{ \int_{0}^{\infty }e^{-\frac{\sigma ^{2}v^{2}}{2}}%
\hat{\upsilon}\left( v\right) \frac{dv}{v^{2}}\right\} \\
& =\frac{1}{\pi }\int_{0}^{\infty }\frac{1-e^{-\frac{\sigma ^{2}v^{2}}{2}}}{%
v^{2}}dv-\frac{\sigma p\tau ^{3/2-\alpha /2}}{\pi }\int_{0}^{\infty }\frac{%
e^{-\frac{{\sigma ^{2}v^{2}}}{2}}}{v^{2-\alpha }}dv \\
& =\frac{\sigma }{\sqrt{2\pi }}-\frac{2^{\alpha /2-3/2}\Gamma \left( \alpha
/2-\frac{1}{2}\right) p}{\pi }\sigma ^{-\left( \alpha -1\right) }\tau
^{1-\alpha /2} \\
& \equiv \frac{\sigma }{\sqrt{2\pi }}+C_{\mathfrak{L}}^{\left( D\right)
}\tau ^{1-\alpha /2}.
\end{align*}%
For $\mathfrak{M}\left( \tau \right) $ we get%
\begin{align*}
\mathfrak{M}\left( \tau \right) & \sim -\frac{\tau ^{1-\alpha /2}}{\pi }%
\mathrm{Im}\left\{ \int_{0}^{\infty }e^{-\frac{\sigma ^{2}v^{2}}{2}}\hat{%
\upsilon}\left( v\right) \frac{dv}{v}\right\} \\
& =-\frac{q\tau ^{1-\alpha /2}}{\pi }\int_{0}^{\infty }e^{-\frac{\sigma
^{2}v^{2}}{2}}v^{\alpha -1}dv \\
& =-\frac{2^{\omega -1}\Gamma \left( \alpha /2\right) q}{\pi }\sigma
^{-\alpha }\tau ^{1-\alpha /2} \\
& \equiv C_{\mathfrak{M}}^{\left( D\right) }\tau ^{1-\alpha /2}.
\end{align*}%
Finally, for $\mathfrak{N}\left( \tau \right) $ we get%
\begin{align*}
\mathfrak{N}\left( \tau \right) & \sim \frac{1}{\pi }\mathrm{Re}\left\{
\int_{0}^{\infty }e^{-\frac{\sigma ^{2}v^{2}}{2}}\left( 1+\hat{\upsilon}%
\left( v\right) \right) dv\right\} \\
& =\frac{1}{\sqrt{2\pi }\sigma }+\frac{p\tau ^{1-\alpha /2}}{\pi }%
\int_{0}^{\infty }e^{-\frac{\sigma ^{2}v^{2}}{2}}v^{\alpha }dv \\
& =\frac{1}{\sqrt{2\pi }\sigma }+\frac{2^{\alpha /2-1/2}\Gamma \left( \alpha
/2+\frac{1}{2}\right) p}{\pi }\sigma ^{-\left( \alpha +1\right) }\tau
^{1-\alpha /2} \\
& \equiv \frac{1}{\sqrt{2\pi }\sigma }+C_{\mathfrak{N}}^{\left( D\right)
}\tau ^{1-\alpha /2}.
\end{align*}

\subsection{Normal Inverse Gaussian Processes\label{NIGPsSA}}

First we consider OTM case ($k>0$). We have%
\begin{align*}
\mathsf{C}^{NIG}\left( v,k\right) & =\frac{\bar{\omega}\mathcal{\bar{%
\varkappa}}ve^{\mathcal{\bar{\varkappa}}^{2}v}}{\pi }\int_{k}^{\infty }\frac{%
\left( e^{\frac{x}{2}}-e^{k-\frac{x}{2}}\right) }{\sqrt{x^{2}+\mathcal{\bar{%
\varkappa}}^{2}v^{2}}}K_{1}\left( \bar{\omega}\sqrt{x^{2}+\mathcal{\bar{%
\varkappa}}^{2}v^{2}}\right) dx \\
& \sim \frac{\bar{\omega}\mathcal{\bar{\varkappa}}v}{\pi }\int_{k}^{\infty }%
\frac{\left( e^{\frac{x}{2}}-e^{k-\frac{x}{2}}\right) }{x}K_{1}\left( \bar{%
\omega}x\right) dx\equiv \mathsf{c}_{0}^{NIG}\left( k\right) v.
\end{align*}%
since the corresponding integral converges. Likewise,%
\begin{align*}
\mathsf{C}_{k}^{NIG}\left( v,k\right) & =-\frac{\bar{\omega}\mathcal{\bar{%
\varkappa}}ve^{\mathcal{\bar{\varkappa}}^{2}v}}{\pi }\int_{k}^{\infty }\frac{%
e^{k-\frac{x}{2}}}{\sqrt{x^{2}+\mathcal{\bar{\varkappa}}^{2}v^{2}}}%
K_{1}\left( \bar{\omega}\sqrt{x^{2}+\mathcal{\bar{\varkappa}}^{2}v^{2}}%
\right) dx \\
& \sim -\frac{\bar{\omega}\mathcal{\bar{\varkappa}}v}{\pi }\int_{k}^{\infty }%
\frac{e^{k-\frac{x}{2}}}{x}K_{1}\left( \bar{\omega}x\right) dx\equiv \mathsf{%
c}_{1}^{NIG}\left( k\right) v,
\end{align*}%
\begin{align*}
\mathsf{C}_{kk}^{NIG}\left( v,k\right) & =\frac{\bar{\omega}\mathcal{\bar{%
\varkappa}}ve^{\mathcal{\bar{\varkappa}}^{2}v}e^{\frac{k}{2}}}{\pi \sqrt{%
k^{2}+\mathcal{\bar{\varkappa}}^{2}v^{2}}}K_{1}\left( \bar{\omega}\sqrt{%
k^{2}+\mathcal{\bar{\varkappa}}^{2}v^{2}}\right) +C_{k}\left( v,k\right) \\
& \sim \frac{\bar{\omega}\mathcal{\bar{\varkappa}}ve^{\frac{k}{2}}}{\pi k}%
K_{1}\left( \bar{\omega}k\right) +\mathsf{c}_{1}^{NIG}\left( k\right)
v\equiv \mathsf{c}_{2}^{NIG}\left( k\right) v.
\end{align*}%
It is clear that higher-order expansions can be calculated without too much
effort if required.

For the ITM case ($k<0$) we can use call-put parity and study puts rather
than call. The corresponding results are similar to the OTM case.

For $k=0$ the corresponding price is given by 
\begin{equation*}
\mathsf{C}^{NIG}\left( v,0\right) =\frac{\bar{\omega}\mathcal{\bar{\varkappa}%
}ve^{\mathcal{\bar{\varkappa}}^{2}v}}{\pi }\int_{0}^{\infty }\frac{\left( e^{%
\frac{x}{2}}-e^{-\frac{x}{2}}\right) }{\sqrt{x^{2}+\mathcal{\bar{\varkappa}}%
^{2}v^{2}}}K_{1}\left( \bar{\omega}\sqrt{x^{2}+\mathcal{\bar{\varkappa}}%
^{2}v^{2}}\right) dx.
\end{equation*}%
Since the integrand is potentially singular at $x=0,v=0$, we have to be more
careful than before. We split the price in two, 
\begin{equation*}
\mathsf{C}^{NIG}\left( v,0\right) =\mathsf{C}_{0}^{\varepsilon }\left(
v\right) +\mathsf{C}_{\varepsilon }^{\infty }\left( v\right) ,
\end{equation*}%
where 
\begin{equation*}
\mathsf{C}_{\alpha }^{\beta }\left( v\right) =\frac{\bar{\omega}\mathcal{%
\bar{\varkappa}}ve^{\mathcal{\bar{\varkappa}}^{2}v}}{\pi }\int_{\alpha
}^{\beta }\frac{\left( e^{\frac{x}{2}}-e^{-\frac{x}{2}}\right) }{\sqrt{x^{2}+%
\mathcal{\bar{\varkappa}}^{2}v^{2}}}K_{1}\left( \bar{\omega}\sqrt{x^{2}+%
\mathcal{\bar{\varkappa}}^{2}v^{2}}\right) dx.
\end{equation*}%
Expansion of the integrand for small $x,v$ yields%
\begin{equation*}
\mathsf{C}_{0}^{\varepsilon }\left( v\right) \sim \frac{\mathcal{\bar{%
\varkappa}}v}{\pi }\int_{0}^{\varepsilon }\frac{x}{x^{2}+\mathcal{\bar{%
\varkappa}}^{2}v^{2}}dx=\frac{\mathcal{\bar{\varkappa}}v}{\pi }\int_{0}^{%
\frac{\varepsilon }{\mathcal{\bar{\varkappa}}v}}\frac{y}{y^{2}+1}dy=\frac{%
\mathcal{\bar{\varkappa}}v}{2\pi }\ln \left( 1+\left( \frac{\varepsilon }{%
\mathcal{\bar{\varkappa}}v}\right) ^{2}\right) =\frac{\mathcal{\bar{\varkappa%
}}}{\pi }v\ln \left( \frac{1}{v}\right) .
\end{equation*}%
while the DCT yields%
\begin{equation*}
\mathsf{C}_{\varepsilon }^{\infty }\left( v\right) =O\left( v\right) ,
\end{equation*}%
so that 
\begin{equation*}
\mathsf{C}^{NIG}\left( v,0\right) \sim \frac{\mathcal{\bar{\varkappa}}}{\pi }%
v\ln \left( \frac{1}{v}\right) .
\end{equation*}%
Next, we consider%
\begin{equation*}
\mathsf{C}_{k}^{NIG}\left( v,0\right) =-\frac{\bar{\omega}\mathcal{\bar{%
\varkappa}}ve^{\mathcal{\bar{\varkappa}}^{2}v}}{\pi }\int_{0}^{\infty }\frac{%
e^{-\frac{x}{2}}}{\sqrt{x^{2}+\mathcal{\bar{\varkappa}}^{2}v^{2}}}%
K_{1}\left( \bar{\omega}\sqrt{x^{2}+\mathcal{\bar{\varkappa}}^{2}v^{2}}%
\right) dx=\mathsf{D}_{0}^{\varepsilon }\left( v\right) +\mathsf{D}%
_{\varepsilon }^{\infty }\left( v\right) ,
\end{equation*}%
where%
\begin{equation*}
\mathsf{D}_{\alpha }^{\beta }\left( v\right) =-\frac{\bar{\omega}\mathcal{%
\bar{\varkappa}}ve^{\mathcal{\bar{\varkappa}}^{2}v}}{\pi }\int_{\alpha
}^{\beta }\frac{e^{-\frac{x}{2}}}{\sqrt{x^{2}+\mathcal{\bar{\varkappa}}%
^{2}v^{2}}}K_{1}\left( \bar{\omega}\sqrt{x^{2}+\mathcal{\bar{\varkappa}}%
^{2}v^{2}}\right) dx.
\end{equation*}%
As before, we can show that%
\begin{equation*}
\mathsf{D}_{0}^{\varepsilon }\left( v\right) \sim -\frac{\mathcal{\bar{%
\varkappa}}v}{\pi }\int_{0}^{\varepsilon }\frac{1}{x^{2}+\mathcal{\bar{%
\varkappa}}^{2}v^{2}}dx=-\frac{1}{\pi }\int_{0}^{\frac{\varepsilon }{%
\mathcal{\bar{\varkappa}}v}}\frac{1}{y^{2}+1}dy=-\frac{1}{2},\ \ \ \ \ 
\mathsf{D}_{\varepsilon }^{\infty }\left( v\right) \sim O\left( v\right) ,
\end{equation*}%
\begin{equation*}
\mathsf{C}_{k}^{NIG}\left( v,0\right) =\mathsf{D}_{0}^{\varepsilon }\left(
v\right) +\mathsf{D}_{\varepsilon }^{\infty }\left( v\right) \sim -\frac{1}{2%
}.
\end{equation*}%
Finally,%
\begin{equation*}
\mathsf{C}_{kk}^{NIG}\left( v,0\right) =\frac{\bar{\omega}e^{\mathcal{\bar{%
\varkappa}}^{2}v}}{\pi }K_{1}\left( \bar{\omega}\mathcal{\bar{\varkappa}}%
v\right) +\mathsf{C}_{k}^{NIG}\left( v,0\right) \sim \frac{1}{\pi \mathcal{%
\bar{\varkappa}}v}.
\end{equation*}

\subsection{Merton Processes\label{MPsSA}}

We start with the following expansion of $E\left( \tau ,u\right) $:%
\begin{align*}
E\left( \tau ,u\right) & =\exp \left( \lambda \tau \left( \left( e^{q\left(
iu+1/2\right) -\eta ^{2}\left( u^{2}+1/4\right) /2}-1\right) +\left(
1-e^{q}\right) \left( iu+\frac{1}{2}\right) \right) \right) \\
& =\exp \left( \lambda \tau \left( 1-e^{q}\right) \left( iu+\frac{1}{2}%
\right) \right) \exp \left( \lambda \tau \left( e^{q\left( iu+1/2\right)
-\eta ^{2}\left( u^{2}+1/4\right) /2}-1\right) \right) \\
& =\exp \left( \lambda \tau \left( 1-e^{q}\right) \left( iu+\frac{1}{2}%
\right) \right) \left( 1+\lambda \tau \left( e^{q\left( iu+1/2\right) -\eta
^{2}\left( u^{2}+1/4\right) /2}-1\right) +...\right) .
\end{align*}%
Accordingly, we can represent $\mathfrak{l}\left( \tau \right) $ as follows%
\begin{equation*}
\mathfrak{l}\left( \tau \right) =\frac{1}{2\pi }\dint\limits_{-\infty
}^{\infty }\frac{1-e^{\lambda \tau \left( 1-e^{q}\right) \left(
iu+1/2\right) }}{u^{2}+\frac{1}{4}}du-\frac{\lambda \tau }{2\pi }%
\dint\limits_{-\infty }^{\infty }\frac{\left( e^{q\left( iu+1/2\right) -\eta
^{2}\left( u^{2}+1/4\right) /2}-1\right) }{u^{2}+\frac{1}{4}}du+....
\end{equation*}%
As we know, the above integrals can be computed explicitly. The
corresponding calculation yields%
\begin{align*}
\mathfrak{l}\left( \tau \right) & =1-e^{\lambda \tau \left( 1-e^{q}\right)
^{-}}-\lambda \tau \left( \Phi \left( \frac{q}{\eta }-\frac{\eta }{2}\right)
+e^{q}\Phi \left( -\frac{q}{\eta }-\frac{\eta }{2}\right) -1\right) +... \\
& =-\lambda \tau \left( \left( 1-e^{q}\right) ^{-}+\Phi \left( \frac{q}{\eta 
}-\frac{\eta }{2}\right) +e^{q}\Phi \left( -\frac{q}{\eta }-\frac{\eta }{2}%
\right) -1\right) +... \\
& =-\lambda \tau \left( \left( 1-e^{q}\right) ^{-}+\Phi \left( \frac{q}{\eta 
}-\frac{\eta }{2}\right) +e^{q}\left( \Phi \left( -\frac{q}{\eta }-\frac{%
\eta }{2}\right) -1\right) -\left( 1-e^{q}\right) \right) +... \\
& =-\lambda \tau \left( \left( 1-e^{q}\right) ^{-}+\Phi \left( \frac{q}{\eta 
}-\frac{\eta }{2}\right) -e^{q}\Phi \left( \frac{q}{\eta }+\frac{\eta }{2}%
\right) -\left( 1-e^{q}\right) \right) +... \\
& =\lambda \tau \left( \left( 1-e^{q}\right) ^{+}+e^{q}C^{BS}\left( \eta
,-q\right) \right) +...,
\end{align*}%
as claimed.

Similarly,%
\begin{align*}
\mathfrak{m}\left( \tau \right) & =\frac{1}{2\pi }\dint\limits_{-\infty
}^{\infty }\frac{e^{\lambda \tau \left( 1-e^{q}\right) \left( iu+1/2\right) }%
}{u^{2}+\frac{1}{4}}iudu+\frac{\lambda \tau }{2\pi }\dint\limits_{-\infty
}^{\infty }\frac{\left( e^{q\left( iu+1/2\right) -\eta ^{2}\left(
u^{2}+1/4\right) /2}-1\right) }{u^{2}+\frac{1}{4}}iudu+... \\
& =-\frac{1}{2}\mathrm{sign}\left( \lambda \tau \left( 1-e^{q}\right)
\right) +....
\end{align*}

Finally,%
\begin{align*}
\mathfrak{n}\left( \tau \right) & =\frac{1}{2\pi }\dint\limits_{-\infty
}^{\infty }e^{\lambda \tau \left( 1-e^{q}\right) \left( iu+1/2\right) }du+%
\frac{\lambda \tau }{2\pi }\dint\limits_{-\infty }^{\infty }\left(
e^{q\left( iu+1/2\right) -\eta ^{2}\left( u^{2}+1/4\right) /2}-1\right)
du+... \\
& =\delta \left( \lambda \tau \left( 1-e^{q}\right) \right) +....
\end{align*}

\subsection{Heston Stochastic Volatility Processes\label{HSVPsSA}}

We need to compute the usual integrals $\mathfrak{l},\mathfrak{m},\mathfrak{n%
}$ in the short-time limit $\tau \rightarrow 0$. In this limit we have%
\begin{align*}
\mathcal{A}\left( \tau ,u\right) & =-\frac{\kappa \theta }{\varepsilon ^{2}}%
\left( \mathcal{F}_{+}\left( u\right) \tau +2\ln \left( \frac{\mathcal{F}%
_{-}\left( u\right) +\mathcal{F}_{+}\left( u\right) \exp \left( -\mathcal{Z}%
\left( u\right) \tau \right) }{2\mathcal{Z}\left( u\right) }\right) \right)
\\
& =-\frac{1}{4}\kappa \theta \tau ^{2}Q\left( u\right) \left( 1+\frac{1}{3}%
\left( \rho \varepsilon iu-\hat{\kappa}\right) \tau -\frac{1}{24}\left(
2\left( \rho \varepsilon iu-\hat{\kappa}\right) ^{2}-\varepsilon ^{2}Q\left(
u\right) \right) \tau ^{2}+...\right) , \\
\mathcal{B}\left( \tau ,u\right) & =\frac{1-\exp \left( -\mathcal{Z}\left(
u\right) \tau \right) }{\mathcal{F}_{-}\left( u\right) +\mathcal{F}%
_{+}\left( u\right) \exp \left( -\mathcal{Z}\left( u\right) \tau \right) } \\
& =\frac{1}{2}\tau \left( 1+\frac{1}{2}\left( \rho \varepsilon iu-\hat{\kappa%
}\right) \tau +\frac{1}{12}\left( 2\left( \rho \varepsilon iu-\hat{\kappa}%
\right) ^{2}-\varepsilon ^{2}Q\left( u\right) \right) \tau ^{2}\right. \\
& \left. +\frac{1}{24}\left( \rho \varepsilon iu-\hat{\kappa}\right) \left(
\left( \rho \varepsilon iu-\hat{\kappa}\right) ^{2}-2\varepsilon ^{2}Q\left(
u\right) \right) \tau ^{3}+...\right) , \\
\mathcal{C}\left( \tau ,u\right) & =-\frac{1}{2}\varpi _{0}\tau \left( 1+%
\frac{\kappa \theta }{2\varpi _{0}}\tau \right) Q\left( u\right) +\mathcal{%
\tilde{C}}\left( \tau ,u\right) , \\
\mathcal{\tilde{C}}\left( \tau ,u\right) & =-\frac{1}{4}\tau ^{2}Q\left(
u\right) \left( \left( \varpi _{0}+\frac{1}{3}\kappa \theta \tau \right)
\left( \rho \varepsilon iu-\hat{\kappa}\right) \right. \\
& +\frac{1}{6}\left( \varpi _{0}-\frac{1}{4}\kappa \theta \tau \right)
\left( 2\left( \rho \varepsilon iu-\hat{\kappa}\right) ^{2}-\varepsilon
^{2}Q\left( u\right) \right) \tau \\
& \left. +\frac{1}{12}\varpi _{0}\left( \rho \varepsilon iu-\hat{\kappa}%
\right) \left( \left( \rho \varepsilon iu-\hat{\kappa}\right)
^{2}-2\varepsilon ^{2}Q\left( u\right) \right) \tau ^{2}+...\right) , \\
E\left( \tau ,u\right) & =e^{-\frac{1}{2}\varpi _{0}\tau \left( 1+\frac{%
\kappa \theta }{2\varpi _{0}}\tau \right) Q\left( u\right) }e^{\mathcal{%
\tilde{C}}\left( \tau ,u\right) } \\
& =e^{-\frac{1}{2}\varpi _{0}\tau \left( 1+\frac{\kappa \theta }{2\varpi _{0}%
}\tau \right) Q\left( u\right) }+\frac{1}{4}\tau ^{2}Q\left( u\right) e^{-%
\frac{1}{2}\varpi _{0}\tau \left( 1+\frac{\kappa \theta }{2\varpi _{0}}\tau
\right) Q\left( u\right) } \\
& \times \left( -\varpi _{0}\left( \rho \varepsilon iu-\hat{\kappa}\right) -%
\frac{1}{3}\left( \kappa \theta \left( \rho \varepsilon iu-\hat{\kappa}%
\right) +\varpi _{0}\left( -2\left( \rho \varepsilon iu-\hat{\kappa}\right)
^{2}+\varepsilon ^{2}Q\left( u\right) \right) \right) \tau \right. \\
& -\frac{1}{24}\left( 2\kappa \theta \left( \left( \rho \varepsilon iu-\hat{%
\kappa}\right) ^{2}-2\varepsilon ^{2}Q\left( u\right) \right) -\varpi
_{0}\left( \rho \varepsilon iu-\hat{\kappa}\right) \right. \\
& \times \left( \left( \rho \varepsilon iu-\hat{\kappa}\right) \left(
3\varpi _{0}Q\left( u\right) +2\hat{\kappa}-2\rho \varepsilon iu\right)
+4\varepsilon ^{2}Q\left( u\right) \right) .
\end{align*}%
It is easy to see that the relevant integrals can be calculated explicitly
because the nontrivial part of the integrand is divisible by $Q\left(
u\right) $ and that the corresponding result is given by (\ref{eq: HSVPexp}%
). Moreover,

\begin{align}
\lambda _{1}& =\frac{1}{24\varpi _{0}}\left( 6\left( \kappa \theta -\hat{%
\kappa}\varpi _{0}\right) -\varpi _{0}^{2}-\left( 1-\frac{1}{4}\rho
^{2}\right) \varepsilon ^{2}\right)  \label{eq: HSVP lmn} \\
& +\frac{1}{1920\varpi _{0}^{2}}(-60\left( \kappa ^{2}\theta ^{2}-\hat{\kappa%
}\varpi _{0}^{3}\right) +3\varpi _{0}^{4}+20(1-\frac{3}{4}\rho ^{2})\hat{%
\kappa}\varepsilon ^{2}\varpi _{0},  \notag \\
\lambda _{2}& =-(1+\frac{88}{16}\rho ^{2}-\frac{59}{16}\rho ^{4})\varepsilon
^{4}-20\kappa \theta (3\varpi _{0}^{2}+2\hat{\kappa}\varpi _{0}-2(1-\frac{5}{%
8}\rho ^{2})\varepsilon ^{2})  \notag \\
& +10\varpi _{0}^{2}(10\hat{\kappa}^{2}-(1-\frac{5}{4}\rho ^{2})\varepsilon
^{2})),  \notag \\
\mu _{1}& =\frac{1}{24\varpi _{0}}\left( -10\kappa \theta +2\hat{\kappa}%
\varpi _{0}-3\varpi _{0}^{2}+3\left( 1-\frac{3}{4}\rho ^{2}\right)
\varepsilon ^{2}\right) ,  \notag \\
\nu _{1}& =\frac{1}{8\varpi _{0}}\left( -2\left( \kappa \theta -\hat{\kappa}%
\varpi _{0}\right) -\varpi _{0}^{2}+\left( 1-\frac{7}{4}\rho ^{2}\right)
\varepsilon ^{2}\right) ,  \notag \\
\nu _{2}& =\frac{1}{6144\varpi _{0}^{2}}(16(36\kappa ^{2}\theta ^{2}+3\varpi
_{0}^{4}+2\varepsilon ^{2}\varpi _{0}^{2}+11\varepsilon ^{4}+12\hat{\kappa}%
\varpi _{0}(\varpi _{0}^{2}+\varepsilon ^{2})+4\hat{\kappa}^{2}\varpi
_{0}^{2}  \notag \\
& -4\kappa \theta (3\varpi _{0}^{2}+12\varepsilon ^{2}+10\hat{\kappa}\varpi
_{0}))+8\rho ^{2}\varepsilon ^{2}(198\kappa \theta +\varpi
_{0}^{2}-89\varepsilon ^{2}-54\hat{\kappa}\varpi _{0})+491\rho
^{4}\varepsilon ^{4}).  \notag
\end{align}

Expressions for $\chi _{i}$ can be derived from the the general case,%
\begin{eqnarray}
\chi _{0,1} &=&\frac{1}{24\varpi _{0}}\left( 6\left( \kappa \theta -\hat{%
\kappa}\varpi _{0}\right) -\left( 1-\frac{1}{4}\rho ^{2}\right) \varepsilon
^{2}\right) ,  \label{eq: HSVP xi_01} \\
\chi _{1,1} &=&\frac{1}{24\varpi _{0}}\left( -10\kappa \theta +2\hat{\kappa}%
\varpi _{0}+3\left( 1-\frac{3}{4}\rho ^{2}\right) \varepsilon ^{2}\right) , 
\notag \\
\chi _{2,1} &=&\frac{1}{120\varpi _{0}}\left( -29\varpi _{0}^{4}/\varepsilon
^{2}+30\left( 1-\frac{7}{2}\rho ^{2}\right) \hat{\kappa}\varpi _{0}\right. 
\notag \\
&&\left. +2\left( -60\left( 1-\frac{23}{8}\rho ^{2}\right) \kappa \theta
+19\left( 1-\frac{679}{152}\rho ^{2}+\frac{1883}{608}\rho ^{4}\right)
\varepsilon ^{2}\right) \right) .  \notag
\end{eqnarray}%
Their calculation is straightforward and is omitted for brevity.

As many times before, we start with the LL formula (\ref{eq: LLH}) and
introduce a new independent variable $u=\tau ^{-1}v$. A simple calculation
yields%
\begin{equation*}
\mathsf{C}\left( \tau ,k\right) =1-\frac{\tau }{2\pi }\int_{-\infty
}^{\infty }\frac{e^{\frac{\left( \Psi \left( v\right) -kiv\right) }{\tau }+%
\tilde{\Psi}\left( \tau ,v\right) }}{\left( v^{2}+\frac{1}{4}\tau
^{2}\right) }dv,
\end{equation*}%
where%
\begin{equation*}
\Psi \left( v\right) =-\frac{\varpi _{0}}{\varepsilon }\frac{iv}{\rho +i\bar{%
\rho}\coth \left( \frac{\bar{\rho}\varepsilon v}{2}\right) },
\end{equation*}%
and $\tilde{\Psi}\left( \tau ,v\right) $ is non-singular when $\tau
\rightarrow 0$. It is clear that this integral can be analyzed via the
saddlepoint approximation. The phase $S\left( v,k\right) $ has the form%
\begin{equation*}
S\left( v,k\right) =-\left( \frac{\varpi _{0}}{\rho \varepsilon +i\bar{\rho}%
\varepsilon \coth \left( \frac{\bar{\rho}\varepsilon v}{2}\right) }+k\right)
iv,
\end{equation*}%
so that the corresponding saddlepoint is located on the imaginary axis.
Accordingly, we can introduce $u=iy$ and define%
\begin{equation*}
\Xi _{01}\left( y,k\right) =\Xi _{0}\left( y\right) +ky,
\end{equation*}%
where%
\begin{equation*}
\Xi _{0}\left( y\right) =\frac{\varpi _{0}}{\varepsilon }\frac{y\sin \left( 
\mathcal{X}_{-}\right) }{\sin \left( \mathcal{X}_{+}\right) },
\end{equation*}%
and $\mathcal{X}_{\pm }=\left( \bar{\rho}\varepsilon y+\phi \pm \phi \right)
/2$. Expressions for $f_{0},\Xi _{1}\left( y\right) =\Xi _{0}^{\prime
}\left( y\right) $, $\Xi _{2}\left( y\right) =\Xi _{0}^{\prime \prime
}\left( y\right) $, $c_{0}\left( y\right) $ can be obtained after some
tedious but straightforward algebra%
\begin{align*}
f_{0}\left( y\right) & =\frac{e^{\frac{\kappa \theta \rho \varepsilon y}{%
\varepsilon ^{2}}+\frac{\hat{\kappa}\varpi _{0}\left( \sin \left( \mathcal{X}%
_{-}\right) \cos \left( \mathcal{X}_{+}\right) -\frac{\rho \bar{\rho}%
\varepsilon y}{2}\right) }{\varepsilon ^{2}\bar{\rho}\sin ^{2}\left( 
\mathcal{X}_{+}\right) }}}{y^{2}\left( \frac{\sin \left( \mathcal{X}%
_{+}\right) }{\bar{\rho}}\right) ^{2\kappa \theta /\varepsilon ^{2}}}, \\
\Xi _{1}\left( y\right) & =\frac{\varpi _{0}}{\varepsilon }\frac{\left( \sin
\left( \mathcal{X}_{-}\right) \sin \left( \mathcal{X}_{+}\right) +\frac{1}{2}%
\bar{\rho}^{2}\varepsilon y\right) }{\sin ^{2}\left( \mathcal{X}_{+}\right) }%
, \\
\Xi _{2}\left( y\right) & =\varpi _{0}\bar{\rho}^{2}\frac{\left( \sin \left( 
\mathcal{X}_{+}\right) -\frac{1}{2}\cos \left( \mathcal{X}_{+}\right) \bar{%
\rho}\varepsilon y\right) }{\sin ^{3}\left( \mathcal{X}_{+}\right) }, \\
c_{0}\left( y\right) & =\frac{f_{0}\left( y\right) }{\sqrt{\Xi _{2}\left(
y\right) }}.
\end{align*}

For $k=-\Xi _{1}\left( y\right) $, and $\Xi _{01}\left( y\right) =\Xi
_{0}\left( y\right) -y\Xi _{1}\left( y\right) $, to the leading order we have%
\begin{equation}
\delta \mathsf{C}\left( \tau ,k\right) \sim c_{0}\left( y\right) \tau
^{3/2}\exp \left( \frac{\Xi _{01}\left( y\right) }{\tau }+\frac{k}{2}\right)
.  \label{eq: dc1}
\end{equation}%
On the other hand, as we know, for $\tau \rightarrow 0$, 
\begin{equation}
\delta \mathsf{C}\left( \tau ,k\right) \sim \frac{\tau ^{3/2}}{k^{2}}\exp
\left( -\frac{k^{2}}{2\sigma _{imp}^{2}\left( \tau ,k\right) \tau }+\frac{k}{%
2}\right) .  \label{eq: dc2}
\end{equation}%
Comparison of (\ref{eq: dc1}), (\ref{eq: dc2}) shows that the corresponding $%
\sigma _{imp}\left( \tau ,k\right) $ can be written in the form%
\begin{equation*}
\sigma _{imp}\left( \tau ,k\left( y\right) \right) =\left( a_{0}\left(
y\right) +a_{1}\left( y\right) \tau +...\right) ^{1/2}=b_{0}\left( y\right)
+b_{1}\left( y\right) \tau +...,
\end{equation*}%
with $a_{0},b_{0}$ given by (\ref{eq: HSVP short a_0 b_0}), and $a_{1},b_{1}$
of the form%
\begin{equation}
a_{1}\left( y\right) =\frac{2a_{0}^{2}\left( y\right) \ln \left( \frac{%
k^{2}\left( y\right) c_{0}\left( y\right) }{a_{0}^{3/2}\left( y\right) }%
\right) }{k^{2}\left( y\right) }=\frac{l^{2}\left( y\right) \ln \left( \frac{%
\varpi _{0}^{2}l^{2}\left( y\right) c_{0}\left( y\right) }{\varepsilon
^{2}a_{0}^{3/2}\left( y\right) }\right) }{2y^{2}\left( \frac{\sin \left( 
\mathcal{X}_{-}\right) }{\sin \left( \mathcal{X}_{+}\right) }+l\left(
y\right) \right) ^{2}},\ \ \ b_{1}\left( y\right) =\frac{a_{1}\left(
y\right) }{2a_{0}\left( y\right) }.  \label{eq: HSVP short a_1 b_1}
\end{equation}

\section{Wing Asymptotics\label{WingA}}

\subsection{Exponential L\'{e}vy Processes\label{ELPsWA}}

First, we consider OTM asymptotics, $k\rightarrow \infty $. In this case, in
order to find the relevant asymptotics, the corresponding Fourier integral
initially defined on the real axis has to be pushed in the lower half-plane.
It is clear that there is a simple pole at $u=-i/2$ whose contribution is
equal to $1$. Assuming that the corresponding shifted L\'{e}vy exponent $%
\upsilon \left( u\right) $ is regular in the strip $-\mathsf{Y}_{+}\leq 
\func{Im}u\leq 0$, we obtain 
\begin{equation}
\mathsf{C}\left( \tau ,k\right) \underset{k\rightarrow \infty }{\sim }%
\mathsf{c}_{+}e^{-\left( \mathsf{Y}_{+}-\frac{1}{2}\right) k}.
\label{eq: C_wing}
\end{equation}%
Comparison of (\ref{eq:BS_asym_C}) and (\ref{eq: C_wing}) yields 
\begin{equation*}
\frac{k}{2\hat{v}}-\frac{k}{2}+\frac{\hat{v}k}{8}=\left( \mathsf{Y}_{+}-%
\frac{1}{2}\right) k.
\end{equation*}%
It is clear that lhs and rhs terms can be balanced iff 
\begin{equation*}
\hat{v}=\beta _{+},\ \ \ \ \ v=\beta _{+}k,\ \ \ \ \ \sigma _{imp}\underset{%
k\rightarrow \infty }{\sim }\sqrt{\frac{\beta _{+}k}{\tau }},\ \ \ \ \
0<\beta _{+}<2,
\end{equation*}%
where 
\begin{equation*}
\frac{1}{2\beta _{+}}-\frac{1}{2}+\frac{\beta _{+}}{8}=\left( \mathsf{Y}_{+}-%
\frac{1}{2}\right) ,
\end{equation*}%
so that%
\begin{equation*}
\beta _{+}=4\left( \mathsf{Y}_{+}-\sqrt{R\left( \mathsf{Y}_{+}\right) }%
\right) .
\end{equation*}%
A similar expression is valid for ITM case with $\mathsf{Y}_{+}$ replaced by 
$\mathsf{Y}_{-}$. In this case the upper limit of regularity of the L\'{e}vy
exponent is given by $\func{Im}u\leq \mathsf{Y}_{-}$, with the contribution
of the pole $u=i/2$ being $e^{k}$, so that%
\begin{equation*}
\mathsf{C}\left( \tau ,k\right) \underset{k\rightarrow -\infty }{\sim }%
1-e^{k}+\mathsf{c}_{-}e^{\left( \mathsf{Y}_{-}+\frac{1}{2}\right) k}.
\end{equation*}%
Thus%
\begin{equation*}
\frac{\left\vert k\right\vert }{2\hat{v}}+\frac{\left\vert k\right\vert }{2}+%
\frac{\hat{v}\left\vert k\right\vert }{8}=\left( \mathsf{Y}_{-}+\frac{1}{2}%
\right) \left\vert k\right\vert ,
\end{equation*}%
\begin{equation*}
\hat{v}=\beta _{-},\ \ \ \ \ v=\beta _{-}\left\vert k\right\vert ,\ \ \ \ \
\sigma _{imp}\underset{k\rightarrow -\infty }{\sim }\sqrt{\frac{\beta
_{-}\left\vert k\right\vert }{\tau }},\ \ \ \ \ 0<\beta _{-}<2,
\end{equation*}%
\begin{equation*}
\frac{1}{2\beta _{-}}+\frac{1}{2}+\frac{\beta _{-}}{8}=\left( \mathsf{Y}_{-}+%
\frac{1}{2}\right) ,
\end{equation*}%
\begin{equation*}
\beta _{-}=4\left( \mathsf{Y}_{-}-\sqrt{R\left( \mathsf{Y}_{-}\right) }%
\right) .
\end{equation*}

\subsection{Tempered Stable Processes\label{TSPsWA}}

All we need to do in order to be able to apply general formulas in the case
under consideration is to establish the strip in which the corresponding
integrand is meromorphic. For TSPs it is particularly simple. The
corresponding strip is defined by the following conditions%
\begin{equation*}
-\mathsf{Y}_{+}<\func{Im}u<\mathsf{Y}_{-},\ \ \ \ \ \mathsf{Y}_{\pm }=\left(
\kappa _{\pm }\mp \frac{1}{2}\right) \mathsf{Y}_{-}.
\end{equation*}%
Accordingly,%
\begin{equation*}
\sigma _{imp}\sim \left\{ 
\begin{array}{ccc}
\sqrt{\frac{\beta _{+}k}{\tau }}, & \beta _{+}=-2+4\left( \kappa _{+}-\sqrt{%
\kappa _{+}^{2}-\kappa _{+}}\right) & k\rightarrow +\infty , \\ 
\sqrt{\frac{\beta _{-}\left\vert k\right\vert }{\tau }}, & \beta
_{-}=2+4\left( \kappa _{-}-\sqrt{\kappa _{-}^{2}+\kappa _{-}}\right) , & 
k\rightarrow -\infty .%
\end{array}%
\right.
\end{equation*}

\subsection{Normal Inverse Gaussian Processes\label{NIGPsWA}}

For deep OTM options ($k\rightarrow \infty $) we have%
\begin{align*}
\mathsf{C}\left( v,k\right) & =\frac{\bar{\omega}\mathcal{\bar{\varkappa}}%
ve^{\mathcal{\bar{\varkappa}}^{2}v}}{\pi }\int_{k}^{\infty }\frac{\left( e^{%
\frac{x}{2}}-e^{k-\frac{x}{2}}\right) K_{1}\left( \bar{\omega}\sqrt{x^{2}+%
\mathcal{\bar{\varkappa}}^{2}v^{2}}\right) }{\sqrt{x^{2}+\mathcal{\bar{%
\varkappa}}^{2}v^{2}}}dx \\
& =\frac{\bar{\omega}\mathcal{\bar{\varkappa}}ve^{\mathcal{\bar{\varkappa}}%
^{2}v}}{\pi }\int_{\frac{k}{2\mathcal{\bar{\varkappa}}v}}^{\infty }\frac{%
\left( e^{\mathcal{\bar{\varkappa}}vy}-e^{k-\mathcal{\bar{\varkappa}}%
vy}\right) K_{1}\left( 2\bar{\omega}\mathcal{\bar{\varkappa}}vQ^{1/2}\left(
y\right) \right) }{Q^{1/2}\left( y\right) }dy \\
& \sim \sqrt{\frac{\bar{\omega}\mathcal{\bar{\varkappa}}v}{4\pi }}e^{%
\mathcal{\bar{\varkappa}}^{2}v}\int_{\frac{k}{2\mathcal{\bar{\varkappa}}v}%
}^{\infty }y^{-\frac{3}{2}}\left( e^{\mathcal{\bar{\varkappa}}vy}-e^{k-%
\mathcal{\bar{\varkappa}}vy}\right) e^{-2\bar{\omega}\mathcal{\bar{\varkappa}%
}vy}dy \\
& =\sqrt{\frac{\bar{\omega}}{2\pi }}\mathcal{\bar{\varkappa}}ve^{\mathcal{%
\bar{\varkappa}}^{2}v}\left( \sqrt{\bar{\omega}-\frac{1}{2}}\int_{\left( 
\bar{\omega}-\frac{1}{2}\right) k}^{\infty }z^{-\frac{3}{2}}e^{-z}dz-\sqrt{%
\bar{\omega}+\frac{1}{2}}e^{k}\int_{\left( \bar{\omega}+\frac{1}{2}\right)
k}^{\infty }z^{-\frac{3}{2}}e^{-z}dz\right) \\
& \sim \sqrt{\frac{\bar{\omega}}{2\pi }}\mathcal{\bar{\varkappa}}ve^{%
\mathcal{\bar{\varkappa}}^{2}v-\left( \bar{\omega}-\frac{1}{2}\right)
k}\left( \frac{1}{\left( \bar{\omega}-\frac{1}{2}\right) k^{\frac{3}{2}}}-%
\frac{1}{\left( \bar{\omega}+\frac{1}{2}\right) k^{\frac{3}{2}}}\right) \\
& =\sqrt{\frac{\bar{\omega}}{2\pi k^{3}}}\frac{v}{\mathcal{\bar{\varkappa}}}%
e^{\mathcal{\bar{\varkappa}}^{2}v-\left( \bar{\omega}-\frac{1}{2}\right) k}.
\end{align*}%
Thus,%
\begin{equation*}
\mathsf{C}\left( v,k\right) \underset{k\rightarrow \infty }{\sim }\sqrt{%
\frac{\omega }{2\pi \left\vert k\right\vert ^{3}}}\frac{v}{\mathcal{\bar{%
\varkappa}}}e^{\mathcal{\bar{\varkappa}}^{2}v-\left( \bar{\omega}-\frac{1}{2}%
\right) k}.
\end{equation*}%
Similarly,%
\begin{equation*}
\mathsf{C}\left( v,k\right) \underset{k\rightarrow -\infty }{\sim }1-e^{k}+%
\sqrt{\frac{\omega }{2\pi \left\vert k\right\vert ^{3}}}\frac{v}{\mathcal{%
\bar{\varkappa}}}e^{\mathcal{\bar{\varkappa}}^{2}v+\left( \bar{\omega}+\frac{%
1}{2}\right) k}.
\end{equation*}%
Accordingly, 
\begin{equation*}
\kappa _{\pm }=\bar{\omega}\pm \frac{1}{2},\ \ \ \ \ \beta _{\pm }=4\left( 
\bar{\omega}-\mathcal{\bar{\varkappa}}\right) ,
\end{equation*}%
so that for NIG the wing volatility is symmetric

\subsection{Merton Processes\label{MPsWA}}

As we know, the price of a call option on a MP is given by%
\begin{equation*}
\mathsf{C}\left( \tau ,k\right) =1-\frac{1}{2\pi }\int_{-\infty }^{\infty }%
\frac{E\left( \tau ,u\right) }{Q\left( u\right) }du,
\end{equation*}%
where%
\begin{equation*}
E\left( \tau ,u\right) =\exp \left( \tau \left( -\frac{1}{2}\sigma
^{2}Q\left( u\right) +\lambda \left( e^{q\left( iu+\frac{1}{2}\right) -\frac{%
\eta ^{2}}{2}Q\left( u\right) }-1+\left( 1-e^{q}\right) \left( iu+\frac{1}{2}%
\right) \right) \right) -k\left( iu-\frac{1}{2}\right) \right) .
\end{equation*}%
We wish to use the saddlepoint approximation in order to evaluate the
corresponding integral when $\left\vert k\right\vert \rightarrow \infty $.
It is clear that this approximation cannot be applied directly since the
exponent does not have a suitable form. In order to rectify the situation,
we introduce a new independent variable $v$, such that%
\begin{equation*}
u=\frac{v}{\mathsf{h}\left( k\right) }-i\mathsf{h}\left( k\right) ,\ \ \ \ \ 
\mathsf{h}\left( k\right) \underset{k\rightarrow \pm \infty }{\rightarrow }%
\pm \infty ,
\end{equation*}%
and notice that%
\begin{equation*}
Q\left( u\right) \sim -2iv-\mathsf{h}^{2},\ \ \ \ \ -k\left( iu-\frac{1}{2}%
\right) \sim -\frac{kiv}{\mathsf{h}}-k\mathsf{h},
\end{equation*}%
\begin{equation*}
\tau \left( -\frac{1}{2}\sigma ^{2}Q\left( u\right) +\lambda \left(
e^{q\left( iu+\frac{1}{2}\right) -\frac{\eta ^{2}}{2}Q\left( u\right)
}-1+\left( 1-e^{q}\right) \left( iu+\frac{1}{2}\right) \right) \right) \sim
\lambda \tau e^{\frac{\eta ^{2}\mathsf{h}^{2}}{2}}e^{\eta ^{2}iv}.
\end{equation*}%
Thus, if we choose $\mathsf{h}$ in such a way that%
\begin{equation}
e^{\frac{\eta ^{2}\mathsf{h}^{2}\left( k\right) }{2}}=\frac{k}{\mathsf{h}%
\left( k\right) }\gg 1,  \label{eq:balance}
\end{equation}%
we shall be able to balance terms, and represent the integrand (to the
leading order) in the form%
\begin{equation*}
\exp \left( \frac{k}{\mathsf{h}\left( k\right) }S\left( v\right) -k\mathsf{h}%
\right) ,\ \ \ \ \ S\left( v\right) =\lambda \tau e^{\eta ^{2}iv}-iv.
\end{equation*}%
We can introduce introduce $y$, such that $v=iy$, and obtain the following
expression for $S$ on the imaginary axis:%
\begin{equation*}
S\left( y\right) =\lambda \tau e^{-\eta ^{2}y}+y,\ \ \ \ \ S^{\prime }\left(
y\right) =-\lambda \tau \eta ^{2}e^{-\eta ^{2}y}+1.
\end{equation*}%
Thus, the saddlepoint is located at%
\begin{equation*}
y^{\ast }=\frac{\ln \left( \lambda \tau \eta ^{2}\right) }{\eta ^{2}},
\end{equation*}%
and 
\begin{equation*}
\mathsf{\delta C}\left( \tau ,k\right) \underset{\left\vert k\right\vert
\rightarrow \infty }{\sim }\exp \left( -k\mathsf{h}\right) .
\end{equation*}%
In order to compute $\mathsf{h}$, we need to solve (\ref{eq:balance}). It
can be checked that%
\begin{equation*}
\mathsf{h}\underset{\left\vert k\right\vert \rightarrow \infty }{\sim }\frac{%
\mathrm{sign}\left( k\right) }{\eta }\left( \sqrt{2\ln \left\vert \eta
k\right\vert }-\frac{\ln \left( 2\ln \left\vert \eta k\right\vert \right) }{2%
\sqrt{2\ln \left\vert \eta k\right\vert }}\right) \underset{\left\vert
k\right\vert \rightarrow \infty }{\sim }\frac{\mathrm{sign}\left( k\right) 
\sqrt{2\ln \left\vert k\right\vert }}{\eta },
\end{equation*}%
is the required solution, so that%
\begin{equation*}
\mathsf{\delta C}\left( \tau ,k\right) \underset{\left\vert k\right\vert
\rightarrow \infty }{\sim }\exp \left( -\frac{\sqrt{2\ln \left\vert
k\right\vert }\left\vert k\right\vert }{\eta }\right) ,
\end{equation*}%
as stated.

\subsection{Quadratic Volatility Processes\label{QVPsWA}}

In order to describe the wing asymptotics for the QV\ model we consider
formula (\ref{eq: LLQV}) in some detail. We need to analyze the asymptotics
of%
\begin{equation*}
\mathfrak{c}\left( l,K\right) =\mathfrak{\zeta }^{s}\sin \left(
2k_{l}X_{K}\right) -2k_{l}\mathfrak{\zeta }^{c}\cos \left(
2k_{l}X_{K}\right) ,
\end{equation*}%
in the large and small strike limits. We start with deep OTM options and
assume that $K\rightarrow \infty $. For $K\rightarrow \infty $ and $l\sim 1$
a simple but very tedious calculation yields%
\begin{equation*}
\mathfrak{c}\left( l,K\right) \underset{K\rightarrow \infty ,l\sim 1}{\sim }%
\left( -1\right) ^{l+1}k_{l}\sqrt{\mathfrak{p}^{2}+\mathfrak{q}^{2}}%
\mathfrak{q}.
\end{equation*}%
As usual, large $l$ are not relevant for our purposes since the
corresponding exponential factors decay very rapidly. Substitution of this
expression in formula (\ref{eq: LLQV}) yields%
\begin{align*}
\frac{C\left( t;\tau ,K\right) }{F_{t}}\underset{K\rightarrow \infty }{\sim }%
& 1+\frac{\sqrt{\mathfrak{p}^{2}+\mathfrak{q}^{2}}\mathfrak{q}}{F_{t}\left( 
\mathfrak{p}\sin \left( X_{F}\right) +\mathfrak{q}\cos \left( X_{F}\right)
\right) }\sum\limits_{l=1}^{\infty }\left( -1\right) ^{l}\frac{k_{l}e^{-%
\frac{1}{2}vR\left( k_{l}\right) }}{R\left( k_{l}\right) }\sin \left(
2k_{l}X_{F}\right) \\
=& \mathsf{c}_{+},
\end{align*}%
This formula shows that, in general, in the large strike limit the value of
the call option converges to a constant. For the calibrated parameters $%
\mathfrak{a}=1.322$, $\mathfrak{p}=0.967$, $\mathfrak{q}=0.301$, we have $%
\mathsf{c}_{+}=0.003283$. Next, we consider deep ITM options and assume that 
$K\rightarrow 0$. For $K\rightarrow 0$ and $l\sim 1$, we have%
\begin{equation*}
\mathfrak{c}\left( l,K\right) \underset{K\rightarrow 0,l\sim 1}{\sim }k_{l}K%
\mathfrak{q},
\end{equation*}%
so that 
\begin{align*}
\frac{C\left( t;\tau ,K\right) }{F_{t}}\underset{K\rightarrow 0}{\sim }& 1-%
\frac{K\mathfrak{q}}{F_{t}\left( \mathfrak{p}\sin \left( X_{F}\right) +%
\mathfrak{q}\cos \left( X_{F}\right) \right) }\sum\limits_{l=1}^{\infty }%
\frac{k_{l}e^{-\frac{1}{2}vR\left( k_{l}\right) }}{R\left( k_{l}\right) }%
\sin \left( 2k_{l}X_{F}\right) \\
=& 1-\frac{K}{F_{t}}+\mathsf{c}_{-}\frac{K}{F_{t}}.
\end{align*}%
For the calibrated set of parameters we have $\mathsf{c}_{-}=0.004896$.
These are the asymptotics we need.

\end{document}